\documentclass[ip,twocolumn]{jpsj3}

\usepackage{graphicx}
\usepackage{comment}

\usepackage{multirow}
\usepackage{xspace}

\newcommand{\sub}[1]{$_{\mathrm {#1}}$}
\newcommand{\subm}[1]{_{\mathrm {#1}}}
\newcommand{\sps}[1]{$^{\mathrm {#1}}$}
\newcommand{\spsm}[1]{^{\mathrm {#1}}}
\newcommand{\etal}{$et~al.$\xspace}

\newcommand{\Tc}{T\subm{c}}
\newcommand{\TTRSB}{T\subm{TRSB}}
\newcommand{\Hc}{H\subm{c}}
\newcommand{\Hcc}{H\subm{c2}}
\newcommand{\Hccab}{H_{\mathrm{c2}\parallel ab}}
\newcommand{\Hccc}{H_{\mathrm{c2}\parallel c}}

\newcommand{\sro}{Sr$_2$RuO$_4$\xspace}
\newcommand{\Hp}{H\subm{P}}

\newcommand{\Horb}{H\subm{c2}^{\mathrm{orb}}}
\newcommand{\Horbc}{H\subm{c2\parallel c}^{\mathrm{orb}}}
\newcommand{\Horbab}{H\subm{c2\parallel ab}^{\mathrm{orb}}}

\usepackage{xcolor}

\newcommand{\SRO}{Sr$_2$RuO$_4$\xspace}

\renewcommand{\arraystretch}{1.5}

\def\bk{{\mathbf{k}}}

\def\m1{{^{-1}}}

\title{Still mystery after all these years \\
--- Unconventional superconductivity of Sr$_2$RuO$_4$ --- }

\author{Yoshiteru Maeno$^1$, Shingo Yonezawa$^{2,3}$, Aline Ramires$^4$}

\inst{$^1$Toyota Riken – Kyoto University Research Center (TRiKUC), Kyoto 606-8501, Japan\\
$^2$Department of Electronic Science and Engineering, Graduate School of Engineering, Kyoto University, Kyoto 615-8510, Japan \\
$^3$Department of Physics, Graduate School of Science, Kyoto University, Kyoto 606-8502, Japan\\
$^4$Condensed Matter Theory Group, Paul Scherrer Institute, CH-5232 Villigen PSI, Switzerland\\}

\abst{
\begin{center}
(Dated: \today)\\
\end{center}

This review describes recent significant research developments made on the layered perovskite \SRO and discusses current issues from both experimental and theoretical perspectives. 
Since the discovery of superconductivity in \SRO in 1994, studies using high-quality single crystals quickly revealed it to be an archetypal unconventional superconductor among strongly correlated electron systems. 
In particular, it was thought that the spin-triplet chiral $p$-wave superconducting state, which breaks time-reversal symmetry, was a prominent possibility.
In 2019, however, a new development overturned the past experimental results, and spin-singlet-like behavior became conclusive. 
Furthermore, innovation in uniaxial strain devices has stimulated researchers to explore changes in the superconducting state by controlling the symmetry and dimensionality of the Fermi surfaces and enhancing the superconducting transition temperature $T\subm{c}$ from 1.5 K to 3.5 K. 
A spin-singlet chiral $d$-wave superconducting state is consistent with most of these recent experimental results. 
Nevertheless, there are still unnatural aspects that remain to be explained. The focus of this review is on unraveling this mystery.
Unlike many other unconventional superconductors, the normal state of \SRO exhibits typical Fermi liquid behavior. 
Nevertheless, to elucidate its superconducting state, it may be essential to go beyond the traditional framework of unconventional superconductivity and recast the theory explicitly considering the multi-orbital aspects of its electronic states. In this review, we describe the frontiers of superconductivity research in \SRO and discuss how the remaining issues may be resolved.}

\begin{document}
\maketitle

\section*{CONTENTS}

\newcommand{\SECTIONINTRODUCTION}{Introduction}
\section{\SECTIONINTRODUCTION \hfill\pageref{Sec:Introduction}}

\newcommand{\SECTIONUNCONVENTIONAL}{Unconventional superconductivity in multi-orbital systems}
\section{\SECTIONUNCONVENTIONAL \hfill\pageref{Sec:Unconventional}}

\newcommand{\SECTIONSYMMETRY}{Symmetry considerations}
\subsection{\SECTIONSYMMETRY \hfill\pageref{Sec:Symmetry}}

\newcommand{\SECTIONNORMAL}{Normal-state Hamiltonian in the orbital basis}
\subsection{\SECTIONNORMAL \hfill\pageref{Sec:Normal}}

\newcommand{\SECTIONORDERPARAMETER}{Superconducting order parameter}
\subsection{\SECTIONORDERPARAMETER \hfill\pageref{Sec:OrderParameter}}

\newcommand{\SECTIONSCORBITAL}{Order parameter in the orbital basis}
\subsection{\SECTIONSCORBITAL \hfill\pageref{Sec:SCOrbital}}

\newcommand{\SECTIONNORMALPROPERTIES}{Normal state properties}
\section{\SECTIONNORMALPROPERTIES \hfill\pageref{Sec:NormalProperties}}

\newcommand{\SECTIONFLPROPERTIES}{Fermi-liquid properties}
\subsection{\SECTIONFLPROPERTIES \hfill\pageref{Sec:FLProperties}}

\newcommand{\SECTIONFS}{Fermi surfaces}
\subsection{\SECTIONFS \hfill\pageref{Sec:FS}}

\newcommand{\SECTIONSTRAIN}{Properties under strain and the Lifshitz transition}
\subsection{\SECTIONSTRAIN\\ \mbox{}\hfill\pageref{Sec:Strain}}

\newcommand{\SECTIONSCPROPERTIES}{Spin susceptibility and the behavior near $H\subm{c2}$}
\section{\SECTIONSCPROPERTIES\\ \mbox{}\hfill\pageref{Sec:SCProperties}}

\newcommand{\SECTIONTABLE}{Summary of superconducting properties}
\subsection{\SECTIONTABLE \hfill\pageref{Sec:Table}}

\newcommand{\SECTIONNMR}{Revised spin susceptibility by NMR}
\subsection{\SECTIONNMR \hfill\pageref{Sec:NMR}}

\newcommand{\SECTIONHAB}{Behavior near the upper critical field}
\subsection{\SECTIONHAB \hfill\pageref{Sec:Hab}}

\newcommand{\SECTIONFFLO}{FFLO state near $\Hcc$}

\newcommand{\SECTIONFOT}{$\Hcc$ suppression and first-order transition}

\newcommand{\SECTIONOUTOFPLANE}{Out-of-plane superconducting anisotropy}

\newcommand{\SECTIONINPLANE}{In-plane superconducting anisotropy}

\newcommand{\SECTIONSCGAP}{Superconducting gap structure}
\section{\SECTIONSCGAP \hfill\pageref{Sec:SCGap}}

\newcommand{\SECTIONQP}{Low-energy quasiparticle excitations}
\subsection{\SECTIONQP \hfill\pageref{Sec:QP}}

\newcommand{\SECTIONTHERMAL}{Thermal conductivity}
\subsection{\SECTIONTHERMAL \hfill\pageref{Sec:Thermal}}

\newcommand{\SECTIONSPECIFICHEAT}{Field-angle-resolved specific heat}
\subsection{\SECTIONSPECIFICHEAT \hfill\pageref{Sec:SpecificHeat}}

\newcommand{\SECTIONSTM}{Quasiparticle interference}
\subsection{\SECTIONSTM \hfill\pageref{sec:STM_SRO}}

\newcommand{\SECTIONMYSTERY}{Mystery of the horizontal line node}
\subsection{\SECTIONMYSTERY \hfill\pageref{sec:mystery_1}}
\mbox{}

\newcommand{\SECTIONPHASESENSITIVE}{Phase sensitive experiments}
\section{\SECTIONPHASESENSITIVE \hfill\pageref{Sec:PhaseSensitive}}

\newcommand{\SECTIONFILMS}{Superconducting thin films}
\subsection{\SECTIONFILMS \hfill\pageref{Sec:Thin films}}

\newcommand{\SECTIONTUNNELING}{Quasiparticle tunneling}
\subsection{\SECTIONTUNNELING \hfill\pageref{qp tunneling}}

\newcommand{\SECTIONDYNAMICAL}{Dynamical behavior of the 1.5-K phase}
\subsection{\SECTIONDYNAMICAL \hfill\pageref{sec:dynamical 1.5 K}}

\newcommand{\SECTIONJUNCTIONS}{Superconducting junctions with Ru inclusions}
\subsection{\SECTIONJUNCTIONS \hfill\pageref{Sec:Junctions}}

\newcommand{\SECTIONPROXIMITY}{Proximity effects into a ferromagnet}
\subsection{\SECTIONPROXIMITY \hfill\pageref{Sec:Proximity}}

\newcommand{\SECTIONSQUID}{SQUID and half quantum fluxoid (HQF)}
\subsection{\SECTIONSQUID \hfill\pageref{Sec:SQUID}}

\newcommand{\SECTIONMULTIOP}{Two-Component Order Parameters}
\section{\SECTIONMULTIOP \hfill\pageref{Sec:MulticomponentOP}}

\newcommand{\SECTIONLANDAU}{Landau theory of phase transitions}
\subsection{\SECTIONLANDAU \hfill\pageref{Sec:Landau}}

\newcommand{\SECTIONULTRASOUND}{Two-component order parameters from ultrasound}
\subsection{\SECTIONULTRASOUND\\ \mbox{}\hfill\pageref{Sec:2CompUS}}

\newcommand{\SECTIONTRSB}{Time-reversal symmetry breaking}
\section{\SECTIONTRSB \hfill\pageref{Sec:TRSB}}

\newcommand{\SECTIONENHANCEMENT}{Enhancement of $\Tc$ by strain}
\subsection{\SECTIONENHANCEMENT \hfill\pageref{Sec:Enhancement}}

\newcommand{\SECTIONMSR}{Split and unsplit transitions by $\mu$SR}
\subsection{\SECTIONMSR \hfill\pageref{Sec:2CompMSR}}

\newcommand{\SECTIONSECONDTRANSITION}{Searching for the second transition}
\subsection{\SECTIONSECONDTRANSITION \hfill\pageref{Sec:SecondTransition}}

\newcommand{\SECTIONOTHER}{Other experiments in search of TRS breaking}
\subsection{\SECTIONOTHER \hfill\pageref{Sec:Other}}

\newcommand{\SECTIONREVISITED}{Mystery of the horizontal line node, revisited}
\subsection{\SECTIONREVISITED \hfill\pageref{sec:mystery_2}}

\newcommand{\SECTIONTHEORETICAL}{Theoretical Review}
\section{\SECTIONTHEORETICAL \hfill\pageref{Sec:Theoretical}}

\newcommand{\SECTIONTHEORIES}{Previous and recent theories}
\subsection{\SECTIONTHEORIES \hfill\pageref{Sec:Theories}}

\newcommand{\SECTIONTHEOACCIDENTAL}{Accidental degeneracy}
\subsection{\SECTIONTHEOACCIDENTAL \hfill\pageref{Sec:TheoAccidental}}

\newcommand{\SECTIONDEGENERACY}{Symmetry protected degeneracy}
\subsection{\SECTIONDEGENERACY \hfill\pageref{Sec:Degeneracy}}

\newcommand{\SECTIONEXTRINSIC}{Extrinsic TRS breaking}
\subsection{\SECTIONEXTRINSIC \hfill\pageref{Sec:Extrinsic}}

\newcommand{\SECTIONREEVALUATION}{Discussion: Re-evaluation of the available results in a new perspective}
\section{\SECTIONREEVALUATION \hfill\pageref{Sec:Reevaluation}}
\mbox{}

\newcommand{\SECTIONPARADIGM}{Paradigm shift from the chiral $p$-wave scenario}
\subsection{\SECTIONPARADIGM \hfill\pageref{Sec:Paradigm}}

\newcommand{\SECTIONUNRESOLVED}{Unresolved mysteries and how they may be solved}
\subsection{\SECTIONUNRESOLVED\\ \mbox{}\hfill\pageref{Sec:Unresolved}}

\newcommand{\SECTIONCONCLUSION}{Conclusion}
\section{\SECTIONCONCLUSION \hfill\pageref{Sec:Conclusion}}

\newcommand{\APPENDIXNORMAL}{Derivation of the normal state Hamiltonian}
\newcommand{\APPENDIXORBITALBAND}{From the orbital-spin basis to the band-pseudospin basis}

{
\appendix
\section{\APPENDIXNORMAL \hfill\pageref{Sec:NormalDetails}}
\section{\APPENDIXORBITALBAND \hfill\pageref{Sec:OrbitalBand}}
}
\mbox{}

\setcounter{section}{0}
\setcounter{subsection}{0}

\section{\SECTIONINTRODUCTION}\label{Sec:Introduction}

\begin{figure}[ht]
\begin{center}
\includegraphics[width=7cm]{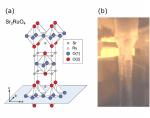}
        \end{center}
\caption{(a) Crystal structure of \SRO, a body-centered tetragonal structure with space group $I4/mmm$ (No. 139). 
Adopted from Ref.~\citen{Kinjo2022.Science.376.397}~($\copyright$~2022 The American Association for the Advancement of Science). 
(b) Single crystal of \SRO grown by a floating-zone method with an infrared image furnace. Shiny flat surface is the basal facet parallel to the RuO$_2$ plane. The feed rod (top) has a diameter of 5.5~mm.}
\label{Fig:structure and crystal}
\end{figure}

The study of unconventional superconductivity plays a central role in modern fundamental research on condensed matter. 
Among archetypal examples is the superconductivity in \SRO, discovered in 1994~\cite{Maeno1994}, that is a layered perovskite isostructural to a high-temperature copper oxide superconductor as shown in Fig. \ref{Fig:structure and crystal} (a). 
Studies using high-quality single crystals grown by a floating-zone method (Fig. \ref{Fig:structure and crystal} (b)) quickly revealed that this is a strongly correlated electron system and clearly exhibits unconventional superconductivity. 
In particular, the spin-triplet chiral $p$-wave superconducting state, which breaks time-reversal symmetry, was considered a prominent candidate state, and extensive experimental and theoretical studies on \SRO contributed to widening the recognition of the concepts of spin-triplet order parameters (OPs) and time-reversal-symmetry breaking (TRS breaking) in the field of superconductivity~\cite{Mackenzie2003RMP, Sigrist2000.JPhysSocJpn.69.1290}. 

Although the normal state of \SRO is well-understood and quantitatively characterized as a highly-correlated multi-band system, the observed superconducting properties have not allowed conclusive determination of its superconducting OP. 
In the last decade, dramatic changes in the superconducting properties under uniaxial strains have provided game-changing clues toward identifying the OP: the superconducting transition temperature $\Tc$ systematically increases from 1.5 K to 3.5 K.~\cite{Steppke2017.Science.355.eaaf9398}
This is a controllable realization of the previously observed phenomenon in the eutectic crystals of \SRO containing ruthenium-metal micro-domains, referred as ``3-K phase" superconductivity.~\cite{Maeno1998.PRL.3-K-phase}
In 2019, it was revealed that previous NMR Knight-shift results, which had been considered as strong evidence for spin-triplet pairing, had a technical problem due to overheating by the radio frequency (RF) pulses used in the measurements~\cite{Pustogow2019.Nature.574.72}.
These new results forced the community to re-examine previous interpretations. 
Nevertheless, the consensus has not been reached yet on the superconducting OP of \SRO.
This paper reviews recent progress, especially since previous reviews [eg. Refs.~\citen{Maeno2011.JPhysSocJpn.81.011009, Mackenzie2017.npjQuantumMater.2.40}] were published, and presents the current status and open questions. 
We also revisit previous key experimental results and discuss their implication from the standpoint of recent progress.

The scope of this review is as follows. 
In Sec. 2, we provide a theoretical background with a particular emphasis on the roles of crystalline symmetry and multiple orbitals in unconventional superconductivity. 
This section is intended as a self-contained tutorial on modern aspects of the theory of superconductivity. 
Depending on their background, readers may first go directly to Sec. 3 and come back to Sec. 2 whenever necessary.
Extending the introductory theory of Sec. 2, Appendices~A and B provide a description of the normal state including the orbital degrees of freedom and discuss the explicit relation between orbital and band bases. 

In Sec. 3, we review the normal state properties of \SRO. 
Its highly correlated three-band Fermi surfaces have been quantitatively well-characterized by state-of-the-art angle-resolved photoemission spectroscopy (ARPES) and band-structure calculations based on dynamical mean-field theory (DMFT). 
Dramatic changes in the electronic states under uniaxial strain that induce a Lifshitz transition are also discussed. 

In Sec. 4, the revised spin-susceptibility and the first-order superconducting transition near the upper critical field $H\subm{c2}$ are discussed. 
In addition, the evidence for the Fulde-Ferrell-Larkin-Ovchinnikov (FFLO) state is introduced. 
These results support a spin-singlet-like pairing state. 
In Sec. 5, results on the superconducting gap structure deduced from a variety of experimental probes are summarized. 
A controversy over the presence of a horizontal line node suggested by the spin-singlet scenario is briefly discussed. Section 6 summarizes the results and interpretation of phase-sensitive experiments in junctions and microstructures, as well as the current status of thin-film growth. 

Section 7 introduces the theoretical basis and key results of ultrasound experiments corroborating the scenario of a two-component OP. 
Section 8 discusses the TRS breaking in the superconducting state. 
Muon spin resonance ($\rm{\mu}$SR) under uniaxial strain has revealed a splitting of the superconducting transition into phase with and without TRS. 
According to complementary $\rm{\mu}$SR studies under hydrostatic pressure or disorder, such splitting does not occur as long as the tetragonal crystalline symmetry is preserved, suggesting that the two components must be symmetry related. 
Nevertheless, the splitting of the superconducting and TRS breaking transitions has not been probed by specific heat nor elastocaloric measurements. 

In Sec. 9, we review recent theoretical models toward identifying the OP of the superconducting state of \SRO. 
Among the models, we describe in some detail a spin-triplet, even-parity, orbital-antisymmetric pairing model, which leads to a spin-singlet-like susceptibility consistent with the experiments. 
Finally, Sec. 10 summarizes the unresolved mysteries with a table of candidate superconducting symmetries (Table~\ref{tab:sc states}) and discusses how the current controversies may be resolved. Sec. 11 concludes this review.


\section{\SECTIONUNCONVENTIONAL}\label{Sec:Unconventional}

Superconductivity is considered ``multi-band" when it emerges in metals with an electronic structure derived from multiple atomic orbitals (or other internal degrees of freedom) contributing to several bands close to the Fermi energy. 
Multi-band superconductivity has been a topic of theoretical research in the context of conventional superconductors such as MgB$_2$ \cite{Mazin2003,Xi2008}, in which bands develop superconducting order parameters of different magnitudes; and in the context of unconventional superconductors such as Fe-pnictides \cite{Fernandes2022},  in which the superconducting order parameters develop with opposite signs in different Fermi surfaces.  
For a long time, it was generally thought that there was no fundamental difference between superconductivity emerging from a single- or multi-band electronic structure, and superconductivity within multi-band systems was understood from the perspective of multiple weakly-coupled single-band systems. 
The picture started to change with the discovery of Fe-based superconductors, for which the multi-orbital description was necessary to properly encode all possible interaction parameters, including interorbital Coulomb interaction and Hund's coupling. 
Similar treatment for \SRO has recently brought us many insights. In this review, we aim at highlighting that the treatment of superconductivity in complex, multi-orbital electronic systems can be better understood from the microscopic point of view by constructing models and superconducting order parameters in the orbital-spin basis, and not directly in the band-pseudospin basis (the definition of these bases is given in the following subsections).

Below, we take the reader over the necessary steps for the construction of a reliable normal-state Hamiltonian of a multi-orbital system using as example Sr$_2$RuO$_4$, and discuss the classification of the superconducting order parameters. 
We start highlighting the symmetries of the crystalline structure of Sr$_2$RuO$_4$, introducing the necessary concepts of group theory  in Sec. \ref{Sec:Symmetry}. 
Such treatment has already been recognized as essential in the past, but it is becoming increasingly important in modern theory of superconductivity and especially to properly interpret various experimental results.
We discuss the most general normal-state Hamiltonian based on the microscopic orbital and spin degrees of freedom in Sec. \ref{Sec:Normal}, giving the details of its derivation in Appendix\ref{Sec:NormalDetails}. 
In Sec. \ref{Sec:OrderParameter}, we discuss the general classification of superconducting order parameters. In Sec. \ref{Sec:SCOrbital} we extend the superconducting order parameter classification to explicitly include the orbital degree of freedom, and discuss in detail the explicit relation between the orbital and band bases in Appendix\ref{Sec:OrbitalBand}.

\subsection{\SECTIONSYMMETRY}\label{Sec:Symmetry}
The crystalline structure of Sr$_2$RuO$_4$, as depicted in Fig. \ref{Fig:structure and crystal}, has multiple symmetries. 
The set of  symmetry operations (translations, rotations, reflections, inversion, and combinations of these) that leave a crystalline structure invariant form a mathematical structure known as a \emph{space group}. 
If we impose that one point must remain invariant under the symmetry operations, we eliminate translations from this set and find what is called a \emph{point group}. Using the standard nomenclature from group theory, 
Sr$_2$RuO$_4$ is associated with the space group $I4/mmm$ ($\#$139) and point group $4/mmm$, most commonly referred to as $D_{4h}$. This point group is composed of sixteen symmetry operations: 

\begin{itemize}
\item $I$, the identity; 

\item $C_{4z}$ and $C_{4z}^{-1}$, rotations by $\pm \pi/2$ around the $z$-axis; 

\item $C_{2x}$, $C_{2y}$, and $C_{2z}$, rotations by $\pi$ around the $x$-, $y$-, and $z$-axis, respectively; 

\item $C_{2d}$ and $C_{2\bar{d}}$, rotations by $\pi$ around the diagonal $d$-axis (x=y for z=0), and anti-diagonal $\bar{d}$-axis (x=-y for z=0), respectively; 

\item $P$, inversion (or parity); 

\item $S_4$ and $S_4^{-1}$, screw rotations consisting of $\pi/2$ rotations along the $z$-axis followed by a $xy$-mirror plane reflection; 

\item $\sigma_h$, $xy$-mirror plane reflection; 

\item $\sigma_{xz}$ and $\sigma_{yz}$, two vertical mirror reflections on the $xz$-, and $yz$-plane, respectively; 

\item $\sigma_{d}$ and $\sigma_{\bar{d}}$, two vertical mirror planes on the $dz$- and $\bar{d}z$-plane, respectively. 
\end{itemize}

The explicit form of the transformation of Cartesian coordinates under each of these operations is given explicitly in Table \ref{Tab:Transformations}. The existence of sixteen symmetry operations can be understood from the independent symmetry under inversion of coordinates $x\rightarrow -x$, $y\rightarrow -y$, and $z\rightarrow -z$, which together account for $2^3=8$ transformations, combined with the equivalence between the $x$ and $y$ axes, which allows for symmetry transformations interchanging $x\leftrightarrow y$, which doubles the number of transformations to $2\times 2^3 = 16$.

\begin{table}[ht]
\begin{center}
  \caption{Symmetry operations and corresponding Cartesian coordinate transformation for the $D_{4h}$ point group. The transformations should be read as $(x,y,z)\rightarrow (a,b,c)$, with $(a,b,c)$ given by the entries in the second column. }
    \label{Tab:Transformations}
    \begin{tabular}{cc}
    \hline\hline
Operation & Coordinate Transformation  \\ \hline
I & (x, y, z) \\ 
$C_{4z}$ & (-y, x, z) \\
$C_{4z}^{-1}$ & (y, -x, z) \\
$C_{2x}$ & (x, -y, -z) \\
$C_{2y}$ & (-x, y, -z) \\
$C_{2z}$ & (-x, -y, z) \\
$C_{2d}$ & (y, x, -z) \\
$C_{2\bar{d}}$ & (-y, -x, -z) \\
$P$ & (-x, -y, -z) \\
$S_4$ & (-y, x, -z) \\
$S_4^{-1}$ & (y, -x, -z) \\
$\sigma_h$ & (x, y, -z) \\
$\sigma_{xz}$ & (x, -y, z) \\
$\sigma_{yz}$ & (-x, y, z) \\
$\sigma_{d}$ & (y, x, z) \\
$\sigma_{\bar{d}}$ & (-y, -x, z) \\
  \hline\hline
    \end{tabular}
        \end{center}
\end{table}

These symmetry operations are divided into ten \emph{conjugacy classes}. Conjugacy classes correspond to sets of symmetry operations of the same ``type", that are related by complementary symmetry operations through conjugation. For example, $C_{2x}$ and $C_{2y}$ belong to the same conjugacy class as they are related by conjugation with $C_{4z}$ as the conjugating element: $C_{4z}.C_{2x}.C_{4z}^{-1} = C_{2y}$. Here the dot $(.)$ corresponds to the composition of operations, which are usually applied from right to left. Examining all possible conjugation relations, the ten conjugacy classes in $D_{4h}$ can be identified as:  $\mathbf{I} = \{I\}$, $\mathbf{2C_{4z}} = \{C_{4z}, C_{4z}^{-1}\}$, $\mathbf{C_{2z}} = \{C_{2z}\}$; $\mathbf{2C_{2x}}= \{C_{2x},C_{2y}\}$; $\mathbf{2C_{2d}}= \{C_{2d},C_{2\bar{d}}\}$; $\mathbf{P} = \{P\}$; $\mathbf{2S_4} = \{S_4,S_4^{-1}\}$; $\boldsymbol{\sigma_h} = \{\sigma_h\}$; $\boldsymbol{2\sigma_v} = \{\sigma_{xz},\sigma_{yz}\}$; $\boldsymbol{2\sigma_d} = \{\sigma_{d},\sigma_{\bar{d}}\}$. Here we label the conjugacy classes by taking one representative element in the class (in bold face), preceded by the number of symmetry operations in the class. Note that the brackets correspond to the set of elements; these do not not indicate anticommutation.

A well known theorem in group theory states that the number of conjugacy classes determines the number of \emph{irreducible representations}, or ``irreps" for short \cite{Hamermesh1989,Bradley2009}, so there are ten irreps in $D_{4h}$.  Irreducible representations can be thought of as families of objects or functions which transform in a well defined manner when acted upon by the symmetry operations. The first column of Table \ref{Tab:D4hCharacter} labels the ten irreducible representations, which carry information about how different objects and symmetries transform according to symmetry operations in each class. We start considering inversion symmetry. An object or a function that does not change under inversion is said to be even parity, while one that picks up a minus sign under inversion is said to be odd parity. The standard notation for the labelling of objects or functions according to their parity follows from the German language, with a subindex $g$ (\emph{gerade}) for even and a subindex $u$ (\emph{ungerade}) for odd parity. Note that there are five even- and five odd-parity irreps in Table \ref{Tab:D4hCharacter}.

\begin{table}[ht]
\begin{center}
  \caption{Simplified character table for the $D_{4h}$ point group. The first column enumerates the ten irreps of $D_{4h}$, and the second column displays the character associated with the conjugacy class of the identity operation, which gives as entries the dimension of the corresponding irreps. The third to fifth columns give the character of three conjugacy classes associated with the three generators of the group, $C_{4z}$, $C_{2x}$, and $P$. The last column displays representative basis functions in Cartesian coordinates that transform according to the corresponding irrep.}
    \label{Tab:D4hCharacter}
    \begin{tabular}{c|ccccc}
    \hline\hline
\textbf{ irrep } & $\boldsymbol{I}$ & $\boldsymbol{2C_{4z}}$  & $\boldsymbol{2C_{2x}}$  & $\boldsymbol{P}$ & \textbf{basis} \\ \hline
 $\boldsymbol{A_{1g}}$ & 1 &  1 & 1 & 1 & 1 \\ 
  $\boldsymbol{A_{2g}}$ & 1 &  1 & -1 &  1 & $xy (x^2-y^2)$\\ 
   $\boldsymbol{B_{1g}}$ & 1 & -1 & 1 &  1 & $x^2-y^2$ \\ 
    $\boldsymbol{B_{2g}}$ & 1 & -1 &  -1 & 1 & $xy$ \\ 
      $\boldsymbol{E_{g}}$ & 2 & 0 &  0 &  2 & $\{xz,yz\}$ \\ 
       $\boldsymbol{A_{1u}}$ & 1 & 1 & 1 &  -1 & $xyz(x^2-y^2)$\\ 
         $\boldsymbol{A_{2u}}$ & 1 & 1  & -1 &  -1 & $z$\\ 
           $\boldsymbol{B_{1u}}$ & 1 & -1  & 1 &  -1 & $xyz$\\ 
             $\boldsymbol{B_{2u}}$ & 1 & -1  & -1 & -1 & $z(x^2-y^2)$\\ 
               $\boldsymbol{E_{u}}$ & 2 & 0  & 0 &  -2 & $\{x,y\}$\\ \hline\hline
    \end{tabular}
        \end{center}
\end{table}

We can refine the irrep labelling considering how objects or functions transform under all point group operations in a similar fashion. Before continuing on this path, it is useful to point out that not all symmetry operations in the group are independent. In particular, $C_{2z} = C_{4z}.C_{4z}$, $C_{2d} = C_{2x}.C_{4z}$, $\sigma_h = P. C_{2z}$, $S_4 = C_{4z}.\sigma_h$, $\sigma_{xz} = P. C_{2y}$, and $\sigma_d = C_{2d}.P$. These suggest that all elements in the group can be found by compositions of three elementary operations (usually referred to as the \emph{generators} of the group): $C_{4z}$, $C_{2x}$ and $P$. These relations allow us to simplify the discussion and to uniquely classify the irreps in terms of the generators of the group.

Continuing now with the labelling of the irreps, let's consider the conjugacy class labelled as $\mathbf{2C_{4z}}$. The function $f_1 \sim z$ goes back into itself under a $C_{4z}$ or a $C_{4z}^{-1}$ transformation, while the function $f_2 \sim x^2-y^2$ picks a minus sign. The standard notation to label objects or functions according to how they transform with respect to rotations along the principal axis of rotation (here the $z$-axis as it is the only axis with 4-fold rotational symmetry) uses $A$ for the objects that go back into themselves, and $B$ for objects that pick a minus sign under these operations. If objects or functions transform differently under other symmetry operations, these are distinguished by an extra numerical subindex. For example, both $f_2 \sim x^2-y^2$ and $f_3 \sim xy$ functions pick a minus sign under a $C_{4z}$ rotation, but transform differently under a $C_{2x}$ rotation: $f_2$ goes back into itself, while $f_3$ gets a minus sign. In this context, the irreps associated with $f_2$ and $f_3$ are labelled as $B_{1g}$ and $B_{2g}$, respectively.  One exception to this scheme concerns higher-dimensional irreps. In these cases we do not have a single object or function that transforms into itself up to a minus sign under the point-group operations, but we have sets of objects or functions that transform into each other under these operations. One example is the pair of functions $f_4 \sim x $ and $f_5 \sim y$. Under $C_{4z}$ these functions transform into each other (up to a minus sign), so this is an example of a two-dimensional irrep. The standard label for two-dimensional irreps is the letter $E$. Table \ref{Tab:D4hCharacter}, known as the \emph{character table}, summarizes this discussion. The entries correspond to the properties of objects or functions in a given symmetry class. Using the example of $f_2 \sim x^2-y^2$ in $B_{1g}$, the $-1$ entry in the fourth line of the third column tells us that this function changes sign under $C_{4z}$ rotations. Note that for the one-dimensional irreps the entries are always $\pm1$, as these correspond to single objects or functions that go back into themselves or pick a minus sign under the symmetry operations. For the two-dimensional irreps, the entries are zero or $\pm 2$. These entries are the traces of the two-dimensional matrices corresponding to the symmetry operation acting on the space formed by the two basis functions. For example, for the $E_u$ irrep, a $C_{2x}$ rotation takes $x\rightarrow x$ and $y \rightarrow -y$, such that the matrix acting on the basis $(x,y)$ in this case would be the Pauli matrix $\sigma_3$, with zero trace. As a second example, inversion takes $x\rightarrow -x$ and $y \rightarrow -y$, such that the corresponding matrix  is minus the two-dimensional identity matrix, with trace equal to $-2$.  Note that the entries in the second column, associated with the conjugacy class of the identity, $\mathbf{I}$, gives us the dimension of the irrep.

\begin{figure}[ht]
\begin{center}
\includegraphics[scale=0.55, keepaspectratio]{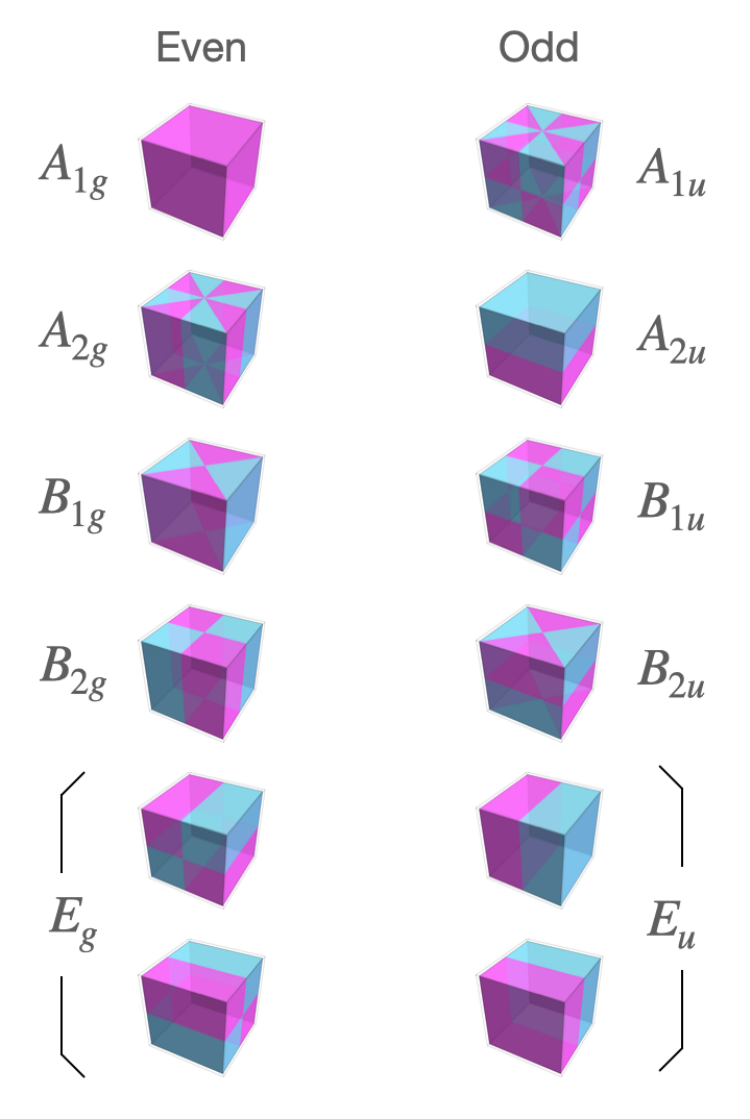}
        \end{center}
\caption{\label{Tab:D4hFigs} Representative objects transforming according to the different irreps of $D_{4h}$. The pink and blue colours correspond to the sign of the lowest order basis functions in Cartesian coordinates  indicated in the last column of Table \ref{Tab:D4hCharacter} ($\copyright$~2019  American Physical Society).}
\label{Fig:D4hFigs}
\end{figure}

Table \ref{Tab:D4hCharacter} also gives examples of functions that transform according to each irrep. 
The displayed functions correspond to the lowest order functions in Cartesian coordinates that capture the symmetries associated with the corresponding irreps. 
These can be thought of as \emph{basis functions}, and, in principle, one should also consider higher order functions with the same symmetry properties. 
Figure \ref{Fig:D4hFigs} gives a visual rendition of the lowest order basis functions.

Group theory also gives us useful rules concerning the product of two functions belonging to specific irreps. 
This is encoded in what is known as the product table, displayed in Table \ref{Tab:D4hProduct}. 
For example, the product of $f_1$ in $A_{2u}$ and $f_2$ in $B_{1g}$ gives a function $f_6 \sim z(x^2-y^2)$ which belongs to $B_{2u}$. 
We note that for the product for $(p,q) = (g,g)$, all the subscripts in this table become $g$.
As we are going to discuss below, this information is going to be key for the construction of Landau theories for multi-component superconductors in presence of external perturbations \cite{Sigrist1991}.

\begin{table}[ht]
\begin{center}
   \caption{Simplified product table for the irreps of the $D_{4h}$ point group. The indices $(p,q,r)=\{g,u\}$. The product rules enforces $r=g$ for $(p,q) = \{(g,g), (u,u)\}$ or $r=u$ for $(p,q) = \{(u,g), (g,u)\}$. The order of the direct product is not relevant. For example: $A_{2u}\otimes B_{1g}  = B_{1g} \otimes A_{2u} = B_{2u}$.}
    \label{Tab:D4hProduct}
    \begin{tabular}{ c| ccccc}
    \hline\hline
&  $\boldsymbol{A_{1p}}$&   $\boldsymbol{A_{2p}}$ &    $\boldsymbol{B_{1p}}$ &     $\boldsymbol{B_{2p}}$&       $\boldsymbol{E_{p}}$  \\ \hline
 $\boldsymbol{A_{1q}}$ & ${A_{1r}}$&   ${A_{2r}}$ &    ${B_{1r}}$ &     ${B_{2r}}$&       ${E_{r}}$ \\ 
  $\boldsymbol{A_{2q}}$ & ${A_{2r}}$&   ${A_{1r}}$ &    ${B_{2r}}$ &     ${B_{1r}}$&       ${E_{r}}$ \\ 
   $\boldsymbol{B_{1q}}$ & ${B_{1r}}$&   ${B_{2r}}$ &    ${A_{1r}}$ &     ${A_{2r}}$&       ${E_{r}}$ \\ 
    $\boldsymbol{B_{2q}}$ & ${B_{2r}}$&   ${B_{1r}}$ &    ${A_{2r}}$ &     ${A_{1r}}$&       ${E_{r}}$ \\ 
      $\boldsymbol{E_{q}}$ & ${E_{r}}$  & ${E_{r}}$  & ${E_{r}}$ & ${E_{r}}$  & ${A_{1r}} \oplus {A_{2r}} \oplus {B_{1r}}  \oplus {B_{2r}}$\\ \hline\hline
    \end{tabular}
        \end{center}
 
\end{table}

\subsection{\SECTIONNORMAL} \label{Sec:Normal}

\begin{figure}[ht]
\begin{center}
\includegraphics[scale=0.4, keepaspectratio]{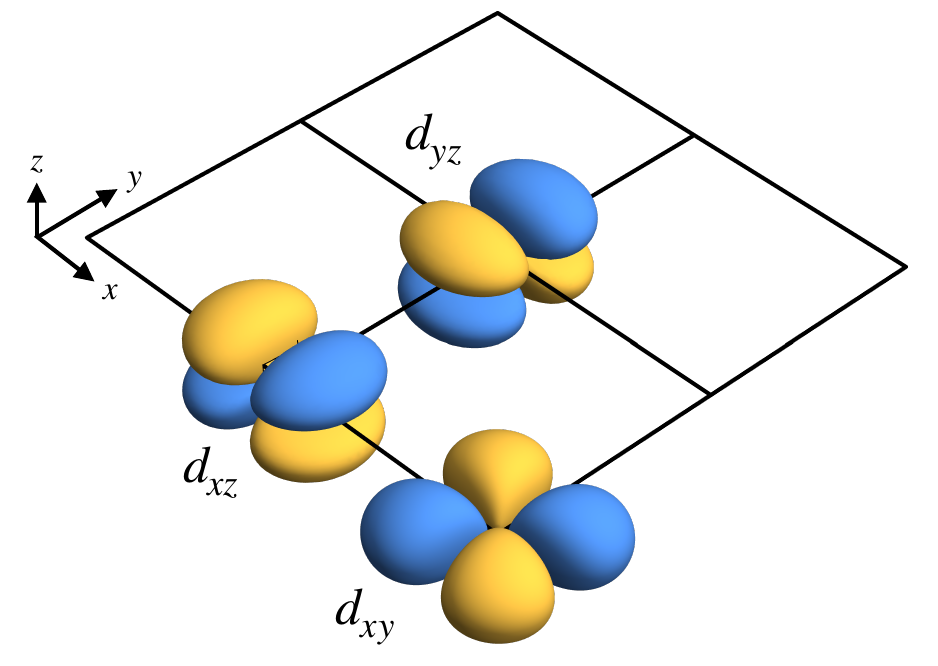}
        \end{center}
\caption{\label{Fig:Orbitals} Representation of $t_{2g}$ orbitals on the square lattice formed by the Ru ions in Sr$_2$RuO$_4$. The blue and yellow colors correspond to opposite signs of the orbital wave functions.}
\end{figure}

In this section, we construct the most general normal-state Hamiltonian to describe Sr$_2$RuO$_4$, in the microscopic orbital-spin basis \cite{Ramires2019}. It is well known that the important electronic degrees of freedom (DOF) for the description of Sr$_2$RuO$_4$, are electrons in Wannier orbitals with $t_{2g}$ symmetry centered at the Ru atoms. These Wannier orbitals are formed by a linear superposition of Ru-d and O-p orbitals, but from here on we refer to them simply as d-electrons and label them as $d_{yz}$, $d_{xz}$ and $d_{xy}$  \cite{Tamai2019}, as depicted in Fig. \ref{Fig:Orbitals}. Choosing the basis $\Psi^\dagger_\bk  = (c_{yz\uparrow}^\dagger, c_{yz\downarrow}^\dagger,  c_{xz\uparrow}^\dagger, c_{xz\downarrow}^\dagger, c_{xy\uparrow}^\dagger, c_{xy\downarrow}^\dagger)_\bk$,  one can construct the most general three-orbital spinfull single-particle Hamiltonian describing the normal state as:
\begin{align} \label{Eq:H}
\mathcal{H}_0(\bk)=\Psi^\dagger_\bk H_0(\bk) \Psi_\bk,
\end{align}
with
\begin{align}\label{Eq:H0}
H_0(\bk)=\sum_{(a,b)}  h_{ab}(\bk) \lambda_a \otimes \sigma_b ,
\end{align}
where $h_{ab}(\bk)$ are thirty-six functions of momenta $\bk$. Here $\lambda_{a=1,...,8}$ are the Gell-Mann (GM) matrices, and $\lambda_0$ the three-dimensional identity matrix standing for the orbital DOF, and $\sigma_{b=1,2,3}$  are Pauli matrices, and $\sigma_0$ the two-dimensional identity matrix standing for the spin DOF.  Note that the Hermitian nature of the Hamiltonian constraints the functions $h_{ab}(\bk)$ to be real, given the choice of Hermitian basis matrices. 

The explicit form of the Pauli and GM matrices is, respectively:
\begin{align}
\sigma_1 = 
\setlength\arraycolsep{2pt}
\begin{pmatrix}
 0 & 1 \\
1 & 0 
\end{pmatrix},  \hspace{0.1cm}
\sigma_2 = \begin{pmatrix}
0 & -i  \\
i & 0 
\end{pmatrix},  \hspace{0.1cm}
\sigma_3 = \begin{pmatrix}
1 & 0  \\
0 & -1  \\
\end{pmatrix},
\nonumber\\
\end{align}
and
\begin{gather} \nonumber
\lambda_1 = 
\setlength\arraycolsep{2pt}
\begin{pmatrix}
0 & 1 & 0 \\
 1 & 0 & 0\\
 0 & 0 & 0
\end{pmatrix}, \hspace{0.2cm}
\lambda_2 = 
\setlength\arraycolsep{2pt}
\begin{pmatrix}
0 & 0 & 1 \\
 0 & 0 & 0\\
 1 & 0 & 0
\end{pmatrix}, \\ \nonumber
\lambda_3 = 
\setlength\arraycolsep{2pt}
\begin{pmatrix}
0 & 0 & 0 \\
 0 & 0 & 1\\
 0 & 1 & 0
\end{pmatrix},
\hspace{0.2cm}
\lambda_4 = 
\setlength\arraycolsep{2pt}
\begin{pmatrix}
0 & -i & 0 \\
 i & 0 & 0\\
 0 & 0 & 0
\end{pmatrix}, \\ \nonumber
\lambda_5 = 
\setlength\arraycolsep{2pt}
\begin{pmatrix}
0 & 0 & -i \\
 0 & 0 & 0\\
 i & 0 & 0
\end{pmatrix}, \hspace{0.2cm}
\lambda_6 =
\setlength\arraycolsep{2pt}
\begin{pmatrix}
0 & 0 & 0 \\
 0 & 0 & -i\\
 0 & i & 0
\end{pmatrix}, \\ 
\lambda_7 = 
\setlength\arraycolsep{2pt}
\begin{pmatrix}
 1 & 0 & 0 \\
 0 & -1 & 0\\
 0 & 0 & 0
    \end{pmatrix}, \hspace{0.2cm}
\lambda_8 = \frac{1}{\sqrt{3}}
\setlength\arraycolsep{2pt}
\begin{pmatrix}
1 & 0 & 0 \\
 0 & 1 & 0\\
 0 & 0 & -2
\end{pmatrix}.
\end{gather}
Note that the indices for the Pauli matrices convey some notion of directionality considering a quantization axis for the spin DOF, with the indices $1$, $2$, and $3$ generally also read as $x$, $y$, and $z$, respectively. Conversely, the indices for the GM matrices do not bring such a strong physical intuition and should be simply taken as labels. Note though, that GM matrices with indices 0, 7, and 8 are diagonal; GM with indices 1, 2, and 3 are off-diagonal but symmetric (can be thought of as three-dimensional versions of the Pauli matrix $\sigma_1$); while GM matrices with indices 4, 5, and 6 are off-diagonal and anti -symmetric (can be thought of as three-dimensional versions of the Pauli matrix $\sigma_2$). 

Requiring the Hamiltonian to be invariant under inversion
 and time-reversal, we find restrictions on the allowed pairs of indices $(a,b)$ which label the functions $h_{ab}(\bk)$ and the basis matrices $\lambda_a \otimes \sigma_b$. Inversion symmetry is implemented as $P=\lambda_0\otimes \sigma_0$ accompanied by $\bk\rightarrow -\bk$, and time-reversal symmetry is implemented by $\Theta = \lambda_0 \otimes (i\sigma_2)\mathcal{K}$, where $\mathcal{K}$ stands for complex conjugation and is accompanied by $\bk\rightarrow -\bk$. Requiring $P H_0(-\bk) P^{-1} = H_0(\bk)$, imposes that all $h_{ab}(\bk)$ are even functions of momentum. Requiring $\Theta H_0(-\bk) \Theta^{-1} = H_0(\bk)$, implies that $h_{ab}(\bk) = \pm h_{ab}^*(-\bk)$, with the plus (minus) sign for combinations of real GM matrices with $\sigma_0$ (three Pauli matrices) and imaginary GM matrices with the three Pauli matrices ($\sigma_0$). Together, the combination of Hermiticity, inversion symmetry, and time-reversal symmetry imposes that only the combination of the six real GM matrices with $\sigma_0$ or the combination of the three imaginary GM matrices with the three Pauli matrices, totalling fifteen basis matrices, are symmetry allowed. These are summarized in the Table \ref{Tab:H0}. A detailed explanation of how these terms are derived can be found in Appendix \ref{Sec:NormalDetails}.
 
\begin{table}[b]
\begin{center}
\caption{List of the fifteen symmetry-allowed $(a,b)$ terms in the
  normal-state Hamiltonian. For each $(a,b)$, the basis function
  $h_{ab}(\bk)$ must belong to the same irrep of $D_{4h}$ as the
  matrix $\lambda_a \otimes \sigma_b$, see details in Appendix \ref{Sec:NormalDetails}. The table gives the irrep associated to each term in the first column, and the
  associated physical process ( or ``Type'') in the last column, the latter can be inferred by the explicit form of the Gell-Mann and Pauli matrices. For the two-dimensional irrep $E_g$, the entries are organized such that the first transforms as $yz$ and the second as $xz$. Here $\bk$-SOC stands for momentum-dependent spin-orbit coupling (SOC).}
\label{Tab:H0}
  \renewcommand\arraystretch{1.25}
    \begin{tabular}{c | ccc}
      \hline\hline
      Irrep in $D_{4h}$ & $(a,b)$ & Type \\
      \hline 
      \multirow{4}{*}{$A_{1g}$} & $(0,0)$ & intraorbital hopping  \\
      & $(8,0)$ & intraorbital hopping  \\
      & $(4,3)$ & atomic SOC  \\
      & $(5,2)-(6,1)$ & atomic SOC \\
      \hline
      $A_{2g}$ &  $(5,1)+(6,2)$ & $\bk$-SOC \\
      \hline
      \multirow{2}{*}{$B_{1g}$}  & $(7,0)$ & intraorbital hopping  \\
      & $(5,2)+(6,1)$ & $\bk$-SOC \\
      \hline
      \multirow{2}{*}{$B_{2g}$} &       $(1,0)$ & interorbital hopping  \\
      & $(5,1)-(6,2)$ & $\bk$-SOC  \\
      \hline
      \multirow{3}{*}{$E_g$} & $\{(3,0),-(2,0)\}$ & interorbital hopping  \\
      & $\{(4,2),-(4,1)\}$ & $\bk$-SOC  \\
      & $\{(5,3),(6,3)\}$ & $\bk$-SOC  \\
      \hline\hline
    \end{tabular}
\end{center}
\end{table}

\subsection{\SECTIONORDERPARAMETER} \label{Sec:OrderParameter}

Superconductivity is usually introduced and classified within the context of a single spin-degenerate electronic band \cite{Sigrist1991}. In its simplest form, the superconducting order parameter can be encoded in two-by-two matrices:
\begin{align}
\Delta_{1Band} (\bk) = \begin{pmatrix}
\Delta_{\uparrow \uparrow}(\bk) & \Delta_{\uparrow \downarrow}(\bk) \\
\Delta_{\downarrow \uparrow}(\bk) & \Delta_{\downarrow \downarrow}(\bk)
\end{pmatrix},
\end{align}
where $\Delta_{\alpha\beta}(\bk)$ indicate all possible types of spin configurations of the superconducting pairs. Given fermionic antisymmetry,  any superconducting order parameter should follow
\begin{align}\label{Eq:FAS}
\Delta(\bk) = - \Delta^T(-\bk).
\end{align}
This identity indicates that if the order parameter is an even function of momenta, it should be accompanied by an antisymmetric matrix. Among the two-by-two matrices, the only antisymmetric matrix is the Pauli matrix $\sigma_2$, which is associated with a spin-singlet configuration of the superconducting pair. Conversely, if the order parameter is an odd function of momenta, it should be accompanied by a symmetric matrix. Among the two-by-two matrices, there are three symmetric matrices, the identity $\sigma_0$, and the Pauli matrices $\sigma_1$ and $\sigma_3$. These three matrices can be more conveniently written in the form of a three-dimensional vector $i\sigma_2\boldsymbol{\sigma} = i\sigma_2(\sigma_1,\sigma_2,\sigma_3)$. These basis matrices for the order parameter encode the three distinct spin configurations of the spin of the superconducting pair in the triplet sector. More concisely, we can write the most general form for the superconducting order parameter in the single band scenario as:
\begin{align}
\Delta_{1Band} (\bk) = i\sigma_2[\phi (\bk)  + \mathbf{d}(\bk) \cdot \boldsymbol{\sigma}],
\end{align}
where
\begin{align}
\phi(\bk) = [ \Delta_{\uparrow \downarrow}(\bk)  -  \Delta_{\downarrow \uparrow}(\bk) ]/2
\end{align}
is an even function of momentum, and 
$ \mathbf{d}(\bk) = (d_x(\bk),d_y(\bk),d_z(\bk))$ is a three-dimensional vector, usually referred to as the \emph{d-vector}, and an odd function of momentum, with
\begin{align}
    d_{x}(\bk) &= -[ \Delta_{\uparrow \uparrow}(\bk)  - \Delta_{\downarrow \downarrow}(\bk) ]/2, \\ \nonumber
    d_{y}(\bk) &= -i [ \Delta_{\uparrow \uparrow}(\bk)  + \Delta_{\downarrow \downarrow}(\bk) ]/2, \\ \nonumber 
    d_z(\bk) &= [ \Delta_{\uparrow \downarrow}(\bk)  +  \Delta_{\downarrow \uparrow}(\bk) ]/2.
\end{align}
Note that, in presence of inversion symmetry, the superconducting order parameters can be classified as even or odd parity. As the spin degree of freedom transforms trivially under inversion, we find that spin-singlet order parameters are even parity and spin triplets are odd parity within the single-band picture. 

\begin{figure*} [ht]
\begin{center}
\includegraphics[width=18cm]{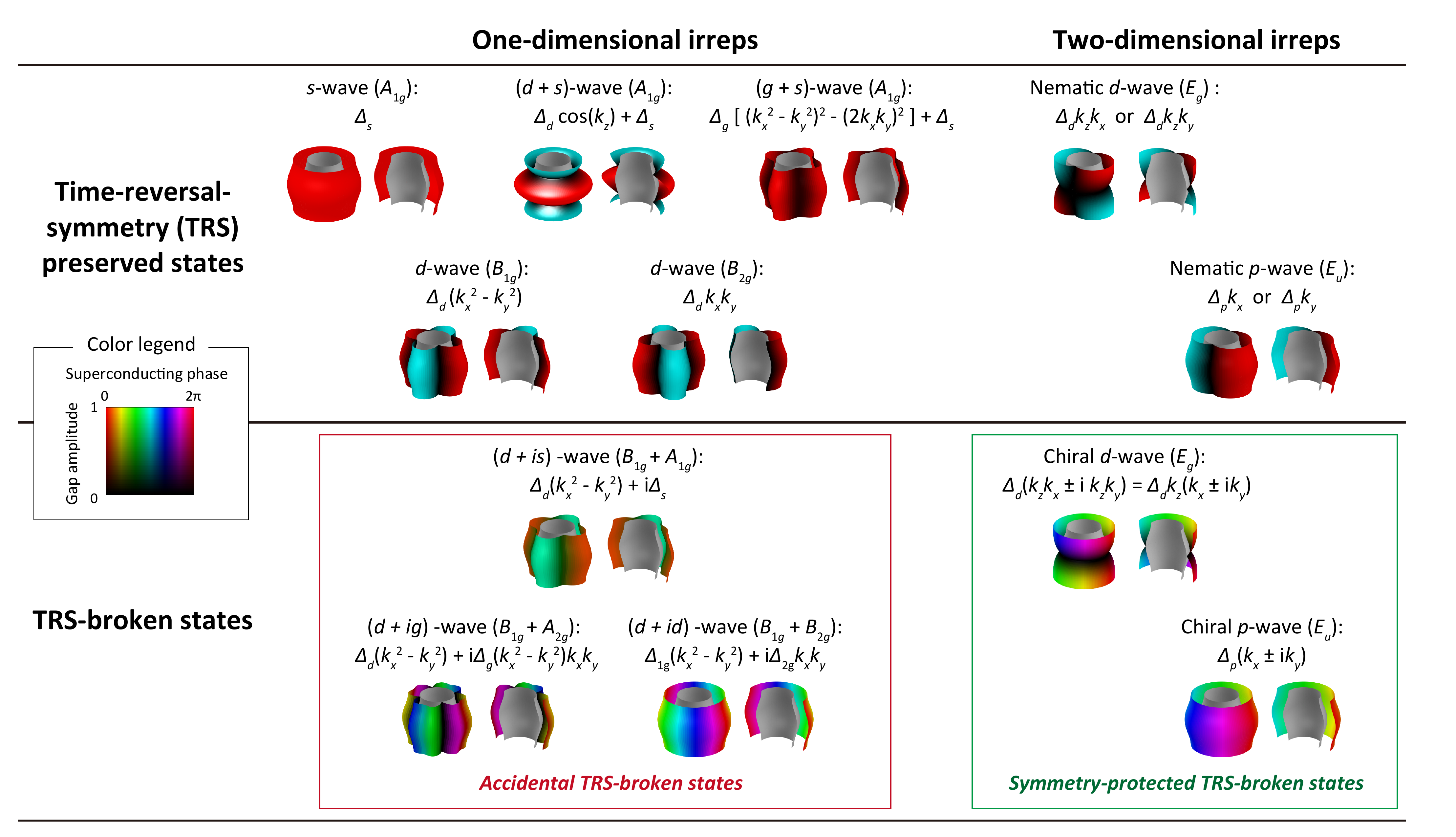}
\end{center}
\caption{Examples of superconducting order parameter candidates for \SRO. 
The order parameters are divided into one dimensional/component (left) and two-dimensional/component (right), and into time-reversal symmetry (TRS) preserving ones (top) and TRS broken ones (bottom). 
The color legend on the left gives information about the magnitude and phase of the gap. 
Note that the superconducting gaps in the top half display only red and cyan colors, indicating that the order parameter does not have a complex phase. 
For these order parameters the darker regions indicate gap minima (for gaps of red color only, such as $s$- or ($g+s$)-wave), or gap nodes (for gaps that change sign and have both red and cyan colors, such as ($d+s$)-, $d$-wave, and $p$-wave). 
In contrast, the superconducting order parameters in the bottom half display multiple colors, indicating the complex phase winding of the order parameter around the Fermi surface, a manifestation of time-reversal symmetry breaking. 
Note that some of these gaps do not have nodes, or dark regions, such as the accidentally degenerate ($d+id$)-wave or the symmetry-protected chiral $p$-wave, while other gaps do display nodes, such as the accidentally degenerate ($d+ig$)-wave or the symmetry protected chiral $d$-wave. 
 We note that the interorbital pairing states described in the next subsection are beyond this traditional picture based on a single-band classification of the conventional ($s$-wave) and unconventional (non $s$-wave) SC states. 
}
\label{fig:SC-gaps_yonezawa}
\end{figure*}

One can refine the classification of the single-band order parameter further based on the point-group symmetry of the crystalline structure in terms of irreps, as introduced in the context of the normal state Hamiltonian above.  The spin component of a spin-singlet superconducting order parameter always transforms trivially under all point-group operations (namely, it goes back into itself after each operation), so all information about the symmetry of the order parameter (or its associated irrep) is encoded in the function $\phi(\bk)$. 
In principle, this is a periodic function of momentum in reciprocal space, but usually the lowest order expansion in momentum is used to describe it. 
For example, an order parameter with $A_{1g}$ symmetry in $D_{4h}$ could have $\phi(\bk) = d_0$, where $d_0$ is a constant. 
Order parameters that are constant in momentum are usually referred to as \emph{s-wave}, in analogy to atomic orbitals with $\ell=0$. 
This type of even-parity, spin-singlet order parameter with a gap that transforms trivially under all point-group operations (associated with the $A_{1g}$ irrep) is usually referred to as \emph{conventional}. Examples of this type of order parameters are presented in the first row under ``one-dimensional irreps" in Fig. \ref{fig:SC-gaps_yonezawa}.

Spin singlets can also be \emph{unconventional} if the associated $\phi(\bk)$ transforms according to a non-trivial irrep (for \SRO, with $D_{4h}$ point-group symmetry, these are $A_{2g}, B_{1g}$, $B_{2g}$, or $E_g$). For example, an even-parity order parameter with $B_{1g}$ symmetry in $D_{4h}$ could have $\phi(\bk)  = d_0 [\cos(k_x a) - \cos(k_y a)]$ (plus higher harmonics), where $a$ is the in-plane lattice constant. 
Generally, this order parameter is also written as $\phi(\bk) \approx d_0 (k_x^2-k_y^2)$, and referred to as \emph{d-wave}, in analogy to the nomenclature of atomic orbitals with angular momentum $\ell = 2$. 
Note that both the momentum-periodic function and its approximation in second order in momenta have nodes along the diagonals $k_x = \pm k_y$, as imposed by the symmetry of the $B_{1g}$ irrep. Superconducting order parameters with $B_{1g}$ and $B_{2g}$ symmetry are displayed in the second row under ``one-dimensional irreps" in Fig. \ref{fig:SC-gaps_yonezawa}.

Order parameters in the triplet sector have a more complex classification which depend on the spin configuration (more details are given below), but all should be accompanied by odd-momentum vectors $\mathbf{d}(\bk)$. 
As one example, we can mention the functions $\{\sin(k_x a), \sin(k_y a)\}$, which form a basis for the two-dimensional irrep $E_u$. 
These periodic momentum-dependent functions are usually approximated by  $\{k_x, k_y\}$, and superconducting order parameters with these momentum dependence are referred to as \emph{p-wave}, again in analogy to the nomenclature of atomic orbitals with angular momentum $\ell = 1$. 
Note that triplet order parameters in the single band context are always odd with respect to inversion symmetry, so these always classify as unconventional.

Both even- and odd-parity order parameter sectors can be associated with one- and two-dimensional irreps. 
Order parameters associated with one-dimensional irreps can have nodes but their amplitude does not reduce the symmetry of the lattice (see order parameters in the left column of Fig. \ref{fig:SC-gaps_yonezawa}). 
In contrast, order parameters associated with two-dimensional irreps have two components that can appear in different configurations.
If the order parameter develops with a real superposition of the two components, the order parameter amplitude breaks some of the lattice symmetries and is called \emph{nematic} (see the top two order parameters in the right column in Fig. \ref{fig:SC-gaps_yonezawa}). 
More interestingly, the two components can appear with a non-trivial relative phase and the order parameter breaks time-reversal symmetry. 
If the relative phase if of $\pi/2$ and the amplitudes of the two components are the same, these are called \emph{chiral} as these carry a finite angular momentum. 
Examples are the chiral $p$-wave and the chiral $d$-wave order parameters, which can be called \emph{symmetry-protected time-reversal symmetry breaking} order parameters. 
Examples of order parameters associated with such states are depicted in the bottom right box in Fig. \ref{fig:SC-gaps_yonezawa}. 
Note that chiral order parameters are not the only type of superconducting order parameters that break time-reversal symmetry. 
If two order parameters associated with different symmetries (or irreps) are accidentally (or almost) degenerate, they can both nucleate with a non-trivial relative phase, leading to \emph{accidental time-reversal symmetry broken states}. 
Examples of order parameters associated with such states are given in the bottom left box in Fig. \ref{fig:SC-gaps_yonezawa}.


The introduction of the superconducting order parameter in the context of single-orbital superconductors above suffices to discuss much of the phenomenology of unconventional superconductors. Recently, an extension of this classification based on multiple-orbitals has been pursued, and brought us many insights toward the apparently contradicting phenomenology of \SRO. We provide an introduction to this extended framework below. For the reader not interested on the technical details, the following subsection can be skipped as these concepts are not necessary for the discussion of the experimental results presented in the next sections.

\subsection{\SECTIONSCORBITAL}\label{Sec:SCOrbital}

Given the introductory discussion to the superconducting order parameter in the context of a single orbital in Sec. \ref{Sec:OrderParameter}, below we extend the classification of the superconducting order parameters for multi-orbital systems from the perspective of the microscopic DOFs. 

Using the same orbital-spin basis employed in the normal state Hamiltonian, we can write the most general superconducting order parameter in terms of thirty-six functions $d_{ab} (\bk)$:
\begin{eqnarray}\label{Eq:OP}
\Delta (\bk) = \sum_{a,b}  d_{ab} (\bk) \lambda_a \otimes \sigma_b (i \sigma_2).
\end{eqnarray}
Here we factor out the matrix $(i \sigma_2)$ such that a spin-singlet state correspond to $b=0$, and spin triplets correspond to $b=\{1,2,3\}.$ In analogy to the \emph{d-vector} parametrizing the triplet order parameter for a single band superconductor, here we introduce a \emph{d-tensor} notation, with $d_{ab} (\bk)$ carrying two indices. We recover the simplest form of the \emph{d-vector} and its properties \cite{Mackenzie2003RMP} if we restrict the discussion to diagonal GM matrices $\lambda_{a=\{0,7,8\}}$. In particular, $a=0$ corresponds to intraorbital pairing with the same amplitude for all orbitals.

\begin{table}[ht]
\caption{Basis matrices $\lambda_a\otimes\sigma_b (i\sigma_2)$, labelled as $[a,b]$, associated with even-parity superconducting order parameters $\Delta(\bk)$, as defined in Eq. \ref{Eq:OP}.  The irrep associated with even-parity superconducting order parameters is given by the product of the irreps for the basis matrices listed below and the irrep of the accompanying $d_{ab}(\bk)$, which must be even parity. Here Inter (S/AS) corresponds to interorbital pairing which is symmetric/anti-symmetric under orbital exchange.}
 \label{Tab:EvenOP}
\begin{center}
    \begin{tabular}{c | ccc}
    \hline\hline
     Irrep  & $[a,b]$ &    Orbital & Spin \\ \hline
    \multirow{4}{*}{$A_{1g}$} & $[0,0]$ &  Intra & Singlet  \\ 
    & $[4,3]$ &   Inter (AS) & Triplet  \\ 
    & $[8,0]$ &    Intra & Singlet  \\ 
    & $[5,2]-[6,1]$ & Inter (AS) & Triplet   \\ \hline
    $A_{2g}$ & $[5,1]+[6,2]$ &   Inter (AS) & Triplet   \\ \hline
    \multirow{2}{*}{$B_{1g}$} & $[7,0]$ &    Intra & Singlet  \\ 
    & $[5,2]+[6,1]$ &  Inter (AS) & Triplet   \\ \hline
    \multirow{2}{*}{$B_{2g}$}& $[1,0]$   &  Inter  (S) & Singlet  \\ 
    & $[5,1]-[6,2]$ & Inter (AS)& Triplet   \\ \hline
        \multirow{3}{*}{$E_g$ } &  $\{[3,0],-[2,0]\}$&  Inter (S) & Singlet  \\ 
    & $\{[4,2],-[4,1]\}$ &   Inter (AS) & Triplet  \\ 
    & $\{[5,3],[6,3]\}$ &    Inter (AS) & Triplet  \\ \hline\hline
    \end{tabular}
\end{center}
 \end{table}

In order to satisfy fermionic antisymmetry, the order parameter should follow Eq. \ref{Eq:FAS}, such that we can separate the functions $d_{ab} (\bk)$ in even or odd in momentum if the accompanying matrix product $\lambda_a \otimes \sigma_b (i\sigma_2)$ is anti-symmetric or symmetric under transposition, respectively. Note that, as inversion acts trivially  on the orbital and spin DOFs, the parity of the order parameter is directly determined by the parity of the $d_{ab} (\bk)$ functions. There are fifteen basis matrices associated with even-parity order parameters [these are the same as the symmetry allowed matrices in the normal state Hamiltonian, multiplied by ($i\sigma_2$)], and twenty one basis matrices associated with odd-parity order parameters. These are summarized in Tables \ref{Tab:EvenOP} and \ref{Tab:OddOP}, respectively. In these tables, we introduce the label $[a,b]$ corresponding to the indexes in the basis matrices $\lambda_a \otimes \sigma_b (i\sigma_2)$. Note the rectangular brackets, $[a,b]$, label the basis matrices associated with the superconducting order parameters, while the round brackets, $(a,b)$, label the basis matrices present in the normal state Hamiltonian. In Table \ref{Tab:EvenOP}, we can identify the order parameter with indices $[a,b] = [0,0]$ as an even-parity spin-singlet order parameter that has the same amplitude for all orbitals. In Table \ref{Tab:OddOP}, we can identify the order parameter with indices $[a,b] = [0,i=1,2,3]$ as odd-parity spin-triplet order parameters with the same amplitude for all orbitals.

 \begin{table}[ht]
 \caption{Basis matrices $\lambda_a\otimes\sigma_b (i\sigma_2)$, labelled as $[a,b]$, associated with odd-parity superconducting order parameters,  $\Delta(\bk)$, as defined in Eq. \ref{Eq:OP}. Note that the irrep associated with the basis matrices is always  even parity, as both the spin and the orbitals transform trivially under inversion. The irrep associated with odd-parity superconducting order parameters is given by the product of the irreps of the basis matrices listed below and the irrep of the accompanying $d_{ab}(\bk)$, which must be odd parity.}
\label{Tab:OddOP}
\begin{center}
    \begin{tabular}{c | ccc}
    \hline\hline
    Irrep  &  $[a,b]$ &    Orbital & Spin \\ \hline
    $A_{1g}$ &  $[2,2]-[3,1]$ &  Inter (S) & Triplet   \\ \hline
    \multirow{3}{*}{ $A_{2g}$} & $[0,3]$ &  Intra & Triplet  \\ 
    & $[4,0]$ &    Inter (AS) & Singlet  \\ 
    & $[8,3]$ &   Intra & Triplet  \\ 
    & $[2,1]+[3,2]$ & Inter (S) & Triplet   \\ \hline
   \multirow{2}{*}{$B_{1g}$ } & $[1,3]$ &  Inter (S) & Triplet  \\ 
    & $[2,2]+[3,1]$  & Inter (S) & Triplet   \\ \hline    
     \multirow{2}{*}{$B_{2g}$ } & $[7,3]$ &  Intra & Triplet  \\ 
    & $[2,1]-[3,2]$ &  Inter (S) & Triplet   \\ \hline
       \multirow{6}{*}{$E_{g}$ } & $\{[0,1],[0.2]\}$ & Intra & Triplet  \\ 
    & $\{[1,2],[1,1]\}$ &   Inter (S) & Triplet  \\ 
    & $\{[2,3],[3,3]\}$ &   Inter (S) & Triplet  \\ 
    & $\{[6,0],-[5,0]\}$ &  Inter (AS) & Singlet \\ 
    & $\{[7,1],-[7,2]\}$ &  Intra & Triplet  \\ 
    & $\{[8,1],[8,2]\}$ &    Intra & Triplet  \\ \hline\hline
    \end{tabular}  
\end{center}
\end{table}

Note that the classification of order parameters from a microscopic perspective brings some surprises \cite{Ramires2018,Kaba2019}. Focusing on even-momentum order parameters, enumerated in Table \ref{Tab:EvenOP}, we find that these can be either intraorbital spin-singlet, such as $[0,0]$, $[7,0]$, and $[8,0]$; symmetric inter orbital spin singlet, such as $[1,0]$, $[2,0]$, and $[3,0]$; and antisymmetric inter orbital spin triplet, such as $[4,3]$, $[5,1]$, $[5,2]$, among others. The last scenario shows that, in the orbital-spin basis classification, the one-to-one correspondence between parity and spin configuration of the electrons in the pairs is lost. In the single-band scenario this correspondence is guaranteed by fermionic anti-symmetry, as the only matrix-part of the superconducting order parameter is associated with the spin DOF. In case the matrix structure encodes other DOFs, the matrix can be antisymmetrized w.r.t. the spin or other DOFs, generating the exotic possibility of an even-parity spin-triplet superconductor if pairing is anti-symmetric in the orbitals.

We can further classify the order parameters considering other point-group transformations. The analysis for the superconducting order parameter takes the same steps as the classification for the terms in the normal state Hamiltonian. Note, though, that the superconducting order parameter does not need to transform trivially under the point-group operations, as it can spontaneously break some of the lattice symmetries, what characterizes it as \emph{unconventional} \cite{Sigrist1991}. The classification of the superconducting order parameters is therefore much richer than the classification of the terms in the normal state Hamiltonian. More precisely, considering the symmetry operation $C_{4z}$, the basis matrices for the order parameter transforms as $C_{4z}\lambda_a\otimes \sigma _b(i\sigma_2)[C_{4z}^{-1}]^*$, with $C_{4z}$ encoding the transformations in both the spin and orbital sectors discussed above. Note the complex conjugation on the last factor \cite{Ramires2016, Ramires2018}. The result of the analysis of the transformations of the basis matrices $\lambda_a\otimes\sigma_b (i\sigma_2)$ with respect to $C_{4z}$ and $C_{2x}$ allows us to associate these with specific irreps of $D_{4h}$, as shown in Tables \ref{Tab:EvenOP} and \ref{Tab:OddOP}. The specific form of $d_{ab}(\bk)$ stems from the underlying pairing interaction. For on-site interactions it is momentum independent and associated with the $A_{1g}$ irrep. If associated with long-range interactions, it can develop a higher angular momentum and be associated with one of the non-trivial irreps. Note that, if the parameters $d_{ab}(\bk)$ are constants or belong to the trivial irrep, $A_{1g}$, the irrep of the complete order parameter given by Eq. \ref{Eq:OP} is the listed irrep in Tables   \ref{Tab:EvenOP} and \ref{Tab:OddOP}. If the function $d_{ab}(\bk)$ transforms nontrivially under the point-group operations, the order parameter is associated with the irrep corresponding to the product of the irrep of the basis matrices and the irrep of the $d_{ab}(\bk)$ function following Table \ref{Tab:D4hProduct}. For example, if the order parameter basis matrix is $[1,0]$, associated with the $B_{2g}$ irrep, and the momentum dependent function $d_{10}(\bk)\propto k_x^2-k_y^2$ belongs to the $B_{1g}$ irrep, the complete order parameter ${\Delta}(\bk) = d_{10}(\bk)\lambda_1\otimes \sigma_0 (i\sigma_2)$ belongs to the $A_{2g}$ irrep.

Order parameters associated with a one-dimensional irreducible representation are said to be \emph{single component}, while order parameters associated with higher dimensional irreps are said to be \emph{multi component}. 
In the context of $D_{4h}$, order parameters associated with $A_{1g/u}$, $A_{2g/u}$, $B_{1g/u}$, and $B_{2g/u}$ are single component, while order parameters associated with the irreps $E_{g/u}$ are two components. 
These two types of order parameters are distinguishable by experiments such as polar Kerr effect~\cite{Xia2006.PhysRevLett.97.167002} and muon spin resonance~\cite{Grinenko2021.NatPhys.17.748}, which can probe time-reversal symmetry breaking in case the two components develop with a complex relative phase, or by ultrasound attenuation experiments, which can probe the two-component nature of the superconducting order parameter by the coupling of gauge-invariant combinations of the order parameter components to shear lattice vibrational modes~\cite{Benhabib2021.NatPhys.17.194, Ghosh2021.NatPhys.17.199}, as discussed in detail in Sec. \ref{Sec:MulticomponentOP}. 

Some of the order parameters, when constructed in the orbital basis, have unexpected features, such as being an even parity spin triplet (necessarily antisymmetric in the orbital DOF),  or being nodal even though the order parameter in the orbital basis is momentum-independent \cite{Ramires2022}. 
Comparing the classification of superconducting order parameters in the single-band scenario with order parameters in the orbital-spin basis we could say that the classification in the latter case includes cases that are ``beyond unconventional", see summary in Table \ref{Tab:Beyond}. 

\begin{table*}[ht]
\caption{Proposed classification of superconducting order parameters in the orbital-spin basis. Here $b=1,2,3$. We label as \emph{conventional} superconducting order parameters that are spin singlets and even parity with a momentum-dependent function in $A_{1g}$ (generally a constant), corresponding to intraorbital pairing, which in the single-orbital scenario correspond to intraband pairing. We label as \emph{unconventional} intraorbital  order parameters with $d_{ab}(\bk)$ functions that do not belong to the $A_{1g}$ irrep. We label as \emph{beyond unconventional} order parameters associated with interorbital pairing, which can be symmetric (S) or anti-symmetric (AS) in the orbital degree of freedom. If the order parameter is symmetric in the orbitals, the association of spin singlet with even parity and spin triplet with odd parity is preserved, as it is known for the single band scenario. If the order parameter is anti-symmetric (AS) in the orbitals, spin singlets are associated with odd parity and spin triplets with even parity. *The case of $[7,0]$ is subtle, as the basis matrix belongs to $B_{1g}$, so generally this order parameter should be classified as unconventional, but if $d_{ab}(\bk)$ also belongs to $B_{1g}$, the overall order parameter belongs to $A_{1g}$ and could be classified as \emph{conventional}.}
\label{Tab:Beyond}
\begin{center}
    \begin{tabular}{c | ccccc}
    \hline\hline
Pairing &  $d_{ab}(\bk)$ & Spin  & Orbital  & Parity & Examples \\ \hline
Conventional & In $A_{1g}$ & Singlet & Intra & Even & $[0,0],  [8,0]$ \\ \hline
\multirow{2}{*}{Unconventional }& Not in $A_{1g}$ & Singlet & Intra & Even & $[0,0], [8,0]$ \\ 
 & Any$^*$ & Singlet & Intra & Even & [7,0]\\ 
& Any & Triplet & Intra & Odd & $[0,b],[7,b], [8,b]$ \\ \hline
\multirow{4}{*}{Beyond Unconventional} &  \multirow{4}{*}{Any}&  Singlet & Inter(S) & Even & $[1,0],[2,0],[3,0]$ \\ 
  &  & Triplet  & Inter(AS) & Even & $[4,b],[5,b],[6,b]$ \\ 
 &  & Triplet  &  Inter(S) & Odd & $[1,b],[2,b],[3,b]$ \\ 
&  & Singlet  &  Inter(AS) & Odd & $[4,0],[5,0],[6,0]$ \\ \hline\hline
    \end{tabular}
        \end{center}
\end{table*}

\section{\SECTIONNORMALPROPERTIES}\label{Sec:NormalProperties}

\subsection{\SECTIONFLPROPERTIES}\label{Sec:FLProperties}

\SRO is an ideal material to examine the effects of strong correlations on the electronic states of a multi-orbital, multi-band system.
It maintains a body-centered tetragonal crystal symmetry (point group $D_{4h}$, space group $I4/mmm$) of a layered perovskite structure down to the lowest temperatures (Fig. \ref{Fig:structure and crystal}). 
Large-size ultra-pure single crystals that retain chemical stability in air over many years are available~\cite{Bobowski2019, kikugawa2021}. 
These characteristics have led to consistent results among researchers using the same probing techniques.

The electronic structure of \SRO is characterized by quasi-two-dimensional (Q2D) Fermi surfaces derived from the ruthenium 4d-t$_{2g}$ orbitals, $d_{xy}$, $d_{yz}$ and $d_{zx}$, hybridized with oxygen 2p-orbitals. 
Table \ref{tab:NormaStateParameters} summarizes the basic normal-state parameters. Details of quantum oscillations revealed substantial enhancements of the thermodynamic cyclotron masses of 3.3 ($\alpha$ band), 7.0 ($\beta$ band), and 16 ($\gamma$ band) over the bare electron mass~\cite{Bergemann2003.AdvPhys.52.639}. 
The in-plane resistivity shows the residual-resistivity ratio (RRR) exceeding 1000, corresponding to the low-temperature mean-free-path of more than one micron.

\begin{table*}[ht]
\begin{center}
\caption{Normal state properties of \sro.}
\label{tab:NormaStateParameters}
\begin{tabular}{rcrc}
\hline\hline 
Quantity & Symbol & Values & Remarks\\ 
\hline
Lattice parameters @300 K & $a=b$  & 0.387073(2) nm  & Ref.~\citen{Chmaissem1998.PhysRevB.57.5067}.  \\ 
    & $c$     & 1.27397(1) nm  &   \\ 
Unit cell volume @300 K     & $V$     & 0.190873(3) nm$^3$  & containing two formula units  \\ 
Density @300 K (15 K)     & $\rho$     & 5.921 (5.956) g/cm$^3$ &  formula-unit atomic mass: 340.31  \\ 
Linear thermal expansion & $\delta_a$ & $-2.28\%$ & [$L$(15 K)-$L$(300 K)]/$L$(300 K) \\ 
    & $\delta_c$     & $-1.35\%$  &   \\ 

Compressibility @300 K   & $\kappa_a$     & $2.24\times10^{-3}$ GPa$^{-1}$ & $-(dL/dP)/L\subm{0}$ \\ 
    & $\kappa_c$     & $2.56\times10^{-3}$ GPa$^{-1}$  &  \\ 
Bulk modulus @300 K &  $B$     & 142 GPa  &   \\ 
\hline

Sommerfeld coefficient   & $\gamma\subm{e}$   & 38.5~mJ/K\sps{2}mol & Ref.~\citen{NishiZaki2000JPhysSocJpn} \\ 
Debye temperature   & $\Theta\subm{D}$   &  410 K  &  \\ 
   &  &  &  Ref.~\citen{Mackenzie2003RMP}, after core-diamagnetism subtraction  \\
Molar magnetic susceptibility  & $\chi\subm{spin}$  &  $1.1\times10^{-8}$~m\sps{3}/mol    & $0.9\times 10^{-3}$~emu/mol\\ 
Volume magnetic susceptibility & $\chi\subm{spin}\spsm{vol}$   & $2.0\times 10^{-4}$  & $1.6\times 10^{-5}$~emu/cm\sps{3} \\  
\hline

Resistivity @300 K   & $\rho_a$     & 120~$\mu\Omega\rm{cm}$  & Ref.~\citen{Barber2019}.  \\ 
          @2 K     &      & < 0.100~$\mu\Omega\rm{cm}$ & RRR > 1200  \\ 
          @300 K  & $\rho_c$     & $\sim 30~\rm{m}\Omega\rm{cm}$  & Ref.~\citen{Yoshida1998}.  \\
          @2 K     &      & $\sim 2.0~\rm{m}\Omega\rm{cm}$ & RRR $\sim15$  \\ 

\hline
\hline
\end{tabular}
\end{center}
\end{table*}


\subsection{\SECTIONFS}\label{Sec:FS}

\begin{figure}[t]
\begin{center}
\includegraphics[width=5cm]{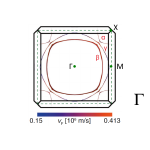}
\end{center}
\caption{Brillouin zone (BZ) and the calculated Fermi surfaces (FSs) of \SRO projected on the $k_xk_y$-plane.~\cite{Jerzembeck2022}. 
The solid lines show the BZ boundaries of the body-centred tetragonal unit cell; the peripheral line is at $k_z =0$ and the central square is at $k_z = \pi$. Under uniaxial strain along the [100], inequivalent M points may be designated as M$_1$ and M$_2$ points as shown in Fig. \ref{Fig:ARPES_under-uni-P} below. 
The dashed green line is the 2D zone boundary of the RuO$_2$ sheet, which is often used in the literature on \SRO. 
The BZ points are added to the figure from Ref.~\citen{Jerzembeck2022} ($\copyright$~2022 The Authors).
The width of the FS lines indicates the warping along $k_z$.}
\label{fig:BZ} 
\end{figure}

 Fermi surface parameters have been determined from quantum oscillations in magnetic susceptibility (the de Haas van Alphen (dHvA) effect) and in resistivity (the Shubnikov-de Haas (SdH) effect) as described in previous reviews~\cite{Mackenzie2003RMP, Bergemann2003.AdvPhys.52.639}. 
 These quantum oscillation measurements were able to determine precise details of the three-dimensional dispersions of the quasiparticles in all three Fermi surfaces (FSs). 
 The Brillouin zone of \SRO with three FSs of $\alpha, \beta$, and $\gamma$ bands is shown in Fig. \ref{fig:BZ}.~\cite{Jerzembeck2022} 

\begin{figure}[ht]
\begin{center}
\includegraphics[width=8.5cm]{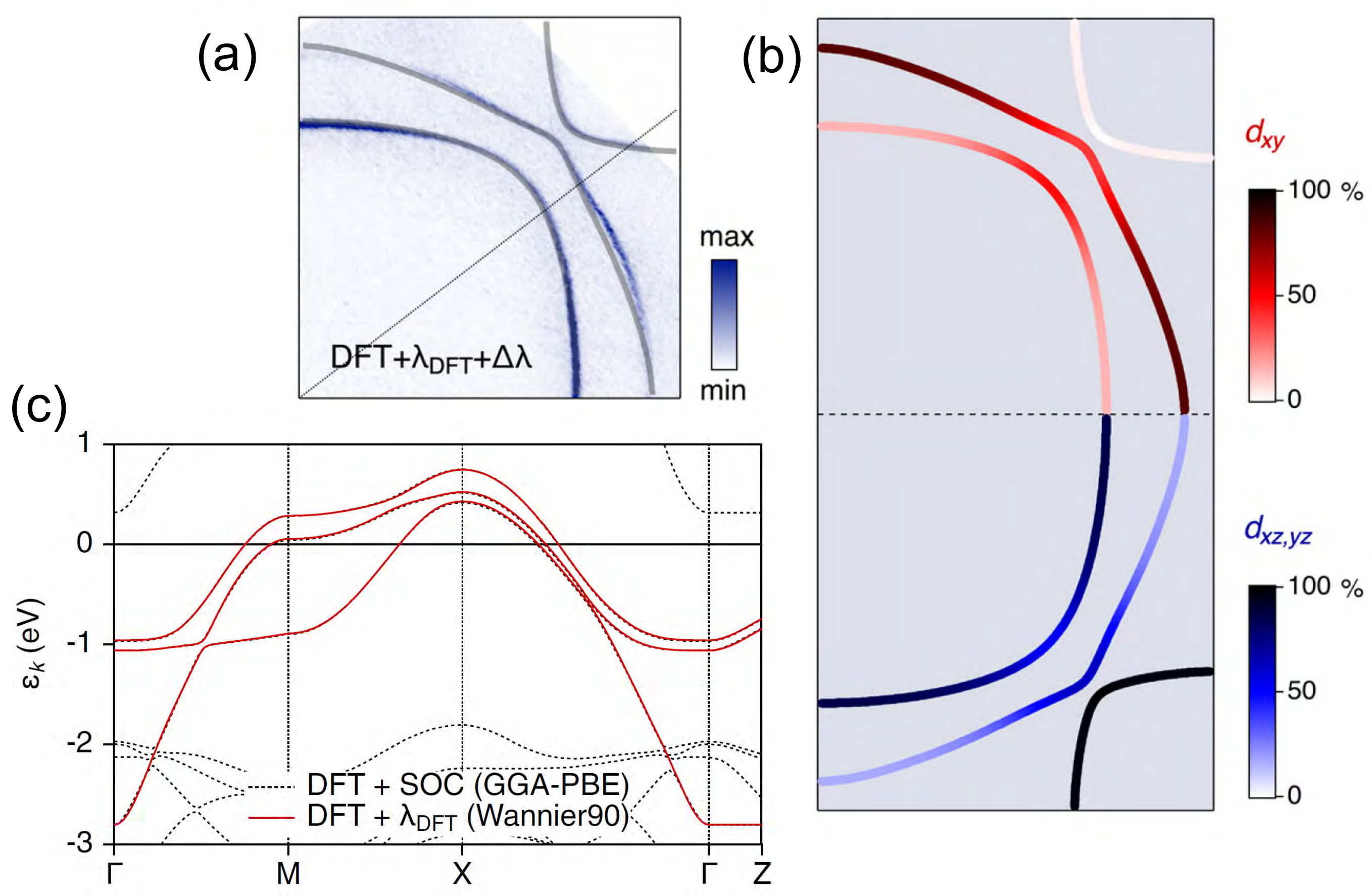}
\end{center}
\caption{(a) Fermi surface of \SRO obtained from high-resolution ARPES (blue). The calculated Fermi surface using DFT
with enhanced SOC (gray) gives an excellent agreement in fine de-
tails of the observation ($\copyright$~2019  American Physical Society).
(b) Orbital character of the eigenstates along the Fermi surface.
On the $\gamma$ FS, the orbital contribution is almost equal between $d_{xy}$ and $\{d_{xz}, d_{yz}\}$  along the diagonal $\Gamma$X direction; the contribution from $\{d_{xz}, d_{yz}\}$ is substantial even along the $\Gamma$M direction. 
(c) DFT band structure for the three bands. 
The $\gamma$ band contains a VHS slightly above $E\subm{F}$. 
Reproduced from Ref. \citen{Tamai2019}.}
\label{Fig:FS}
\end{figure}

Recent high-resolution ARPES measurements and the fitting with dynamical mean-field theory (DMFT) revealed quantitatively that the effective spin-orbit coupling (SOC) is strongly enhanced from the density-functional theory (DFT) values by electron correlations as shown in Fig. \ref{Fig:FS} (a), and the orbital mixing induced by the enhanced SOC is substantial as clearly shown in Fig. \ref{Fig:FS} (b)~\cite{Tamai2019}. 
An estimated minimal mixing between $d_{xy}$ and $\{d_{xz}, d_{yz}\}$ for the $\gamma$ and $\beta$ bands is 20\% : 80\% along the $\Gamma$M direction with a monotonic increase to approximately 50\% along the Brillouin zone (BZ) diagonal $\Gamma$X.  
The multi-orbital character of each band is significant, and the orbitals retain a considerable physical importance at low temperatures. 
An important feature in the band structure shown in Fig. \ref{Fig:FS} (c) is the presence of the Van Hove singularity (VHS) of the $\gamma$ band at the M point, only 35 meV above $E\subm{F}$~\cite{Jerzembeck2022}3

\begin{figure}[ht]
\begin{center}
\includegraphics[width=8cm]{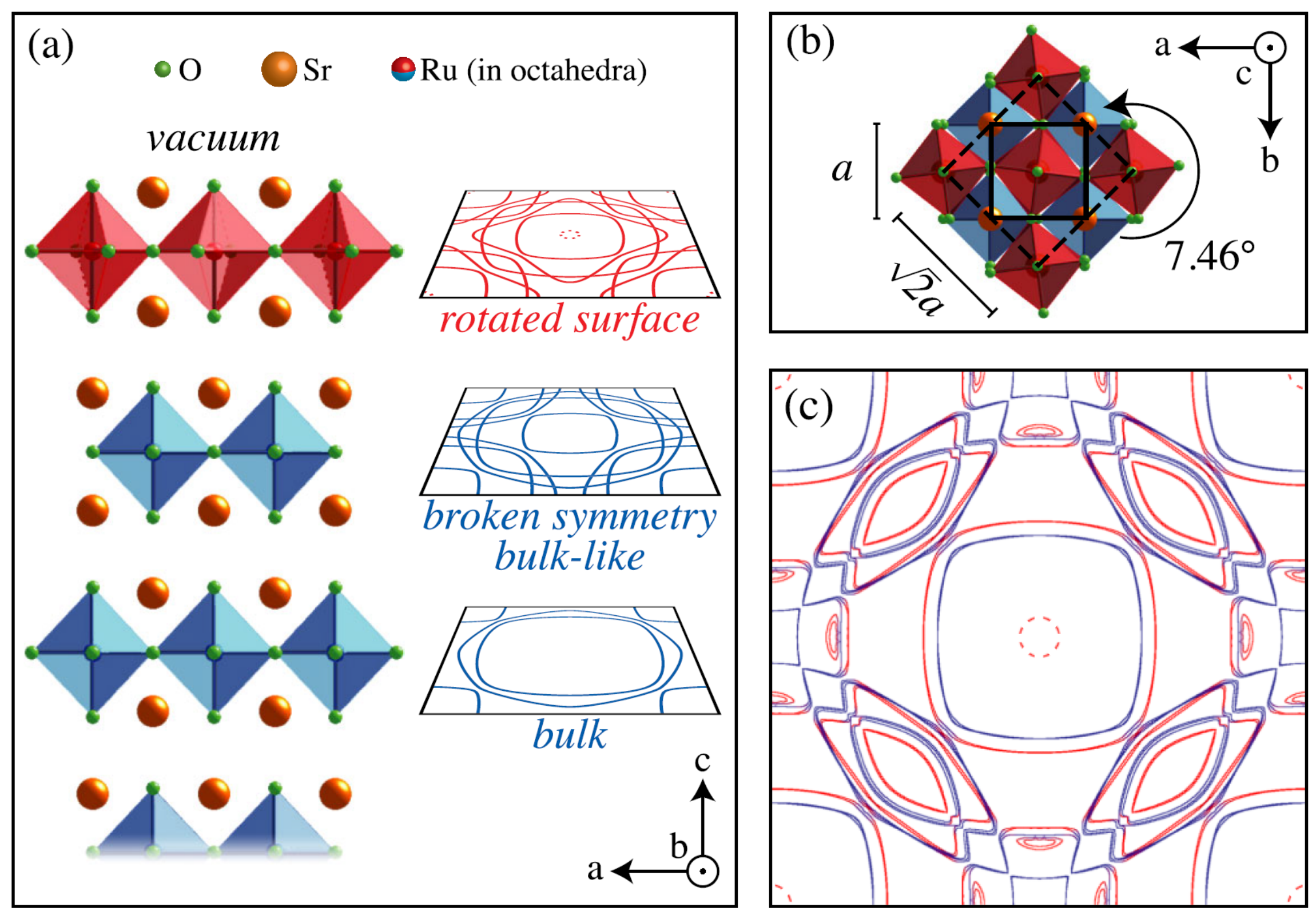}
\end{center}
\caption{Surface reconstruction and the corresponding Fermi surfaces. (a) Partial structure of the \SRO near the surface highlighting the rotated layer (top), broken symmetry layer (middle), and bulk-like layer (bottom) with corresponding FSs. 
(b) Top view of the reconstructed surface of \SRO,
with the original undistorted (rotated) unit cell marked by solid (dashed) lines. 
(c) FS of \SRO with shifted chemical potential to match the slight averaged hole doping of the surface region observed by ARPES. Surface and bulk character are shown in red and blue, respectively.
Reproduce from Ref.~\citen{Veenstra2013} ($\copyright$~2013  American Physical Society).
}
\label{Fig:Surface_reconstruction} 
\end{figure}

Cleaved surface of \SRO is known to exhibit atomic reconstruction. 
Angle-resolved photoemission spectroscopy (ARPES) measurements revealed a $\sqrt{2}\times\sqrt{2}$ surface reconstruction~\cite{Damascelli2000.PhysRevLett.85.5194}, associated with 7.46$^{\circ}$ rotation and about $1^{\circ}$ tilting of the surface RuO$_3$ octahedra. 
The surface-to-bulk progression of the electronic structure was revealed by LDA+SO (local-density approximation plus SOC) calculation shown in Fig. \ref{Fig:Surface_reconstruction} that matches well the ARPES results. 
Near the surface, the electronic structure is expected to consist of a hole-like $\gamma$ FS, in contrast with the electron-like $\gamma$ FS in the bulk. 
In addition, VHSs cross at all four M points within a few layers from the surface. 
Recently, by using low-energy muon spin spectroscopy, exotic surface magnetism has been found in the normal state of \SRO, and interpreted as due to orbital loop currents.~\cite{Fittipaldi2021}

\subsection{\SECTIONSTRAIN}\label{Sec:Strain}

\begin{figure}[ht]
\begin{center}
\includegraphics[scale=0.25, keepaspectratio]{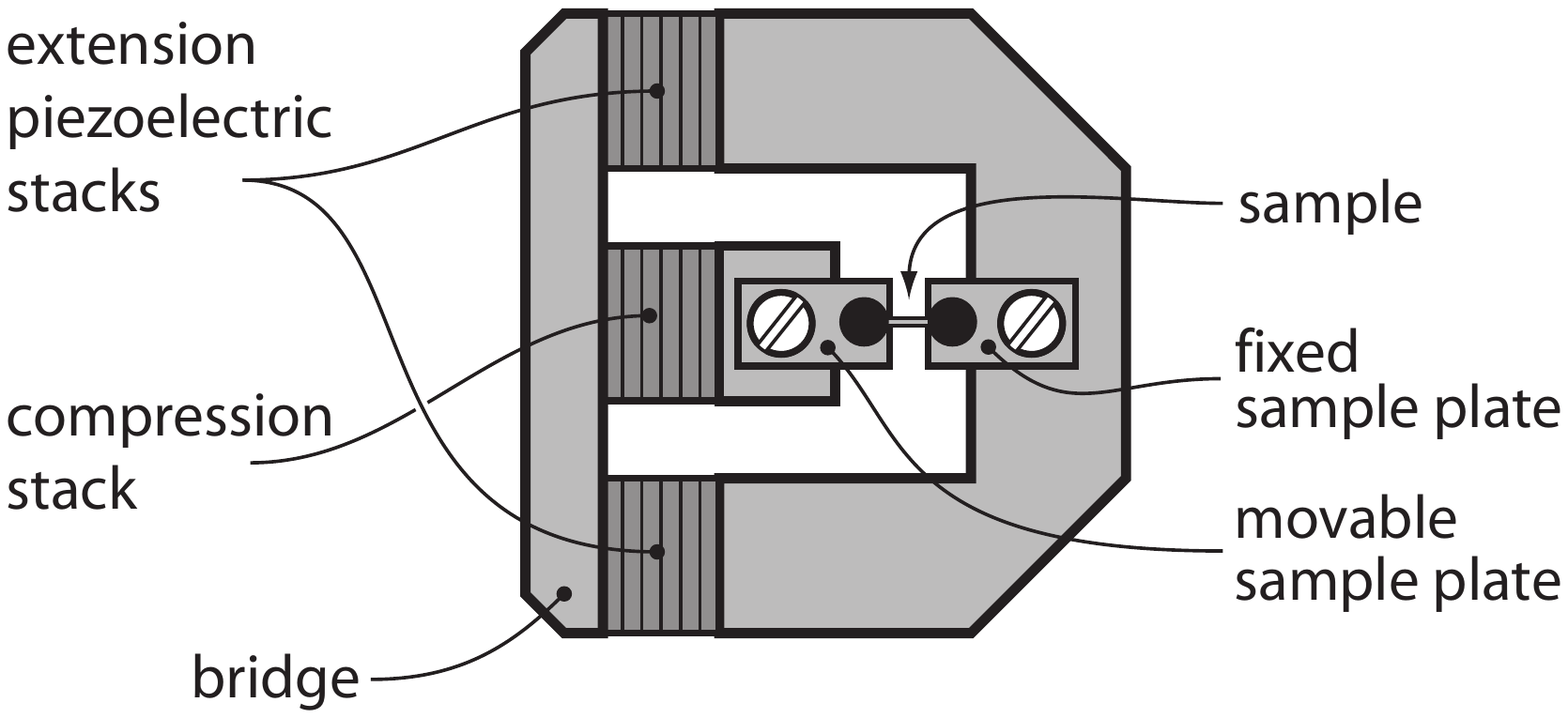}
        \end{center}
\caption{A schematic of the strain apparatus introduced by Hicks \etal~\cite{Hicks2014.Science.344.283, Hicks2014.RSI} Upon voltage application, the inner piezo stacks compress the sample, while the outer pair of piezo stacks stretches the sample. The displacement of a long piezo-stack unit is transferred to a sample with a much shorter length, thereby amplifying the strain in the sample to over 1\% ($\copyright$~2014 The American Association for the Advancement of Science).}
\label{Fig:Hicks_mechanism}
\end{figure}

As a notable recent progress, it has become possible to control the Fermi-surface topology of \SRO in-situ by an innovative uniaxial-strain apparatus~\cite{Hicks2014.Science.344.283, Hicks2014.RSI}. As depicted in Fig. \ref{Fig:Hicks_mechanism}, the apparatus consists of two sets of piezo stacks, one compressing and the other stretching a sample crystal. 
The inner and outer piezo units help compensating the differential thermal contraction of the structural material on cooling. 
This Hicks-type uniaxial strain device is now applied to control a variety of physical properties in quantum materials.~\cite{Kim2018.Science.YBCO, Sun2021.NJP.PdCrO2, Ikhlas2022.NatPhys.Mn3Sn}.

Using this device, it was found that the $\Tc$ of \SRO is remarkably enhanced by both compression (negative strain) and expansion (positive strain) along the [100] direction. 
In order to minimize the effect of strain inhomogeneity especially near the edge of the sample, $\Tc$ is usually measured by alternating-current (AC) susceptibility using a small pickup coil placed near the center of the sample.
Under the compression it was demonstrated that the $\Tc$ is enhanced from 1.5 K to 3.5 K with a maximum at about 0.44\% strain~\cite{Li2022}. 
In contrast, $\Tc$ exhibits a mild linear dependence on strain along the [110] direction, with a decreasing $\Tc$ under compression~\cite{Hicks2014.Science.344.283}.  

\begin{figure}[ht]
\begin{center}
\includegraphics[scale=0.13, keepaspectratio]{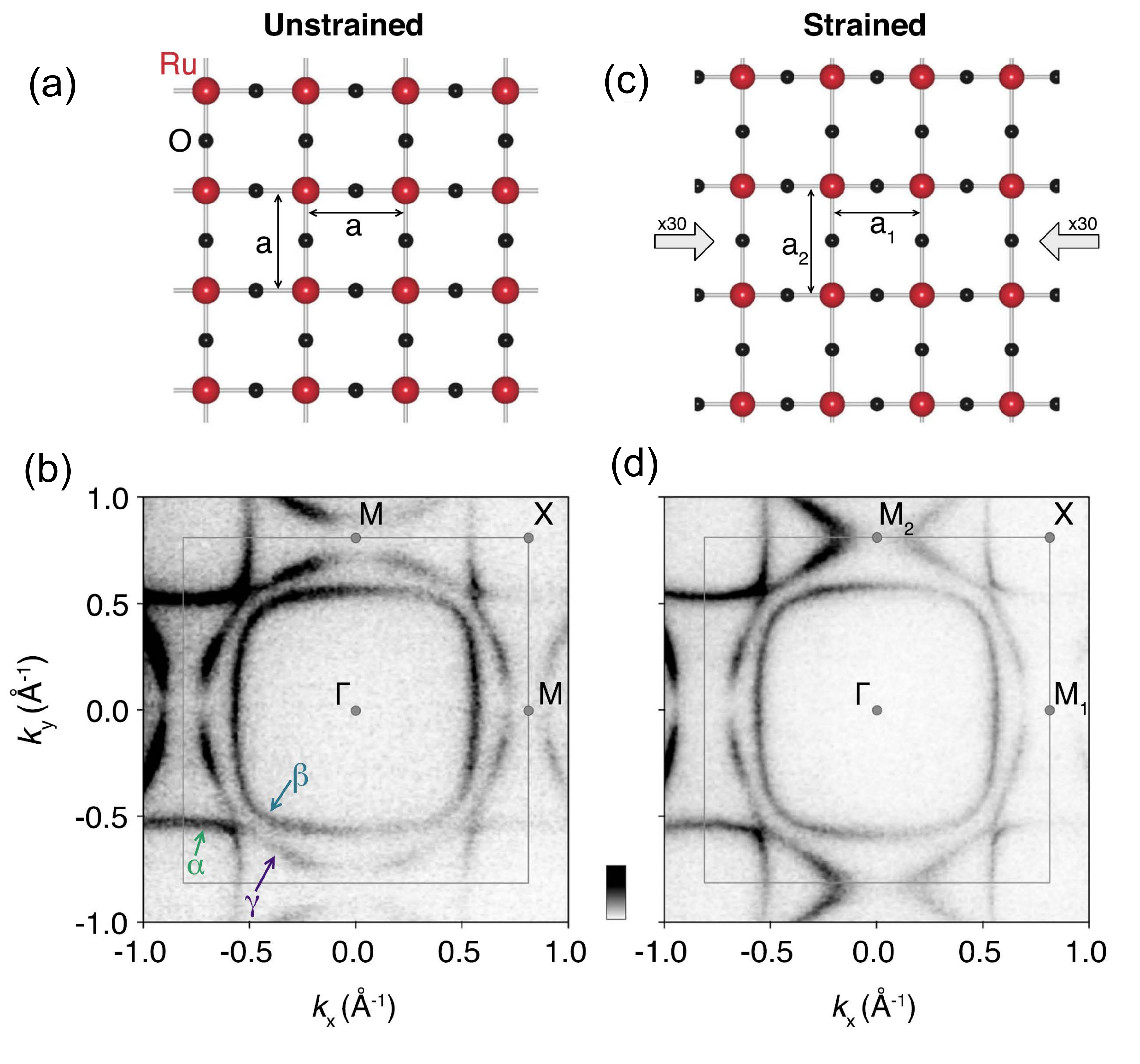}
        \end{center}
\caption{\label{Fig:ARPES_under-uni-P} Unstrained and uniaxially strained \SRO. (a) and (c) are schematic in-plane lattice configurations; (b) and (d) are Fermi surfaces observed by ARPES.~\cite{Sunko2019} ($\copyright$~2019 The Authors).} 
\end{figure}

As a distinct feature of the application of such uniaxial stress over the hydrostatic pressure, the symmetry of the lattice and therefore the electronic states of a material may be readily controlled in-situ by just varying the voltage applied to the piezo stacks. 
The variation of the $\Tc$ of \SRO is understood in terms of the Lifshitz transition associated with the Fermi-level crossing of the VHSs of the $\gamma$ band at two of the four M-points in the BZ. 
Such a transition becomes possible as a consequence of the breaking the tetragonal symmetry.
With a thin sample glued on a substrate, which is strongly compressed by differential thermal contraction, ARPES measurements clearly confirm the strain-induced Lifshitz transition at the critical strain $\varepsilon_\mathrm{VHS}\sim-$0.5\% (Fig. \ref{Fig:ARPES_under-uni-P})~\cite{Sunko2019}. 
According to a simple tight-binding model, only the $\gamma$ band shows a sharp peak in the DOS at the Lifshitz transition; $\alpha$ and $\beta$ bands are affected only weakly (Fig. 2 of Barber \textit{et~al.}~\cite{Barber2019}). 
However, the NMR measurements~\cite{Luo2019.normal-state-NMR} suggest that there are two effects: in addition to the DOS enhancement associated with the $\gamma$ band Fermi energy passing the VHS points, the uniform susceptibility, namely the Stoner factor, is also substantially enhanced toward the Lifshitz transition. 

As an excellent example of demonstrating the importance of the multi-band character of \SRO, a novel bulk acoustic plasmon was recently found in momentum-resolved electron energy-loss spectroscopy (M-EELS).~\cite{Husain2023.Nature.Pines_demon} This zero-gap plasmon is attributable to the oscillation of the electron density between the $\beta$ and $\gamma$ bands and perhaps the first realization of the Pines' demon predicted in 1956.~\cite{Pines1956.Pines_demons}

\section{\SECTIONSCPROPERTIES}\label{Sec:SCProperties}

\subsection{\SECTIONTABLE}\label{Sec:Table}

The superconducting properties of \sro\ are summarized in Table~\ref{tab:SCparameters}.
Note that the $c$-axis coherence length $\xi_{\mathrm{GL}\parallel c}$ is as small as 1.1-1.6~nm,
which is comparable to, but still larger than, the spacing between neighboring RuO\sub{2} planes ($c/2 = 0.64$~nm).
This means that, in spite of substantially large out-of-plane anisotropy of $\Gamma\sim 60$, the interlayer coherence of Cooper pairs is still maintained in \sro. 
Indeed, in \sro, bulk superconductive properties are rather mean-field like, nearly free from phenomena attributable to fluctuations. 
This is in clear contrast with Bi-based cuprates~\cite{Matsubara1992.PhysRevB.45.7414} or ET-based organics~\cite{Lang1994.PhysRevB.49.15227}, where two-dimensional superconductivity with very weak out-of-plane coherence associated with dominant superconducting fluctuations is realized.

The absolute value of the penetration depth has been reported using neutron scattering (Ref.~\citen{Riseman1998.Nature.396.242}; 
$\lambda_{ab} =  190$~nm), $\mu$SR (Ref.~\citen{Luke2000.PhysicaB.289-290.373}; 
$\lambda_{ab} =  190$~nm), and microwave impedance measurements (Ref.~\citen{Ormeno2006.PhysRevB.74.092504}; 
$\lambda_{ab} =  160$~nm).
Here, $\lambda_{ab}$ corresponds to the decay length of in-plane current as a function of in-plane distance, measured under $c$-axis magnetic fields.
Correspondingly, $\lambda_c$, the current decay length along the $c$-axis, is obtained by $\lambda_c = \Gamma\lambda_{ab}$ using the intrinsic superconducting anisotropy $\Gamma$.
Note that different notations for $\lambda_c$ are used among the literature.
The Ginzburg--Landau (GL) parameter $\kappa$ for field parallel to the $c$-axis is then written as $\kappa_{ab} = \lambda_{ab}/\xi_{ab} = 2.7$, which is close to the border between the type-I and type-II superconductivity $\kappa = 1/\sqrt{2}$.
Thus, non-local effects may need to be taken into account for this field direction.
In contrast, under in-plane fields, \sro\ behaves as a strongly type-II superconductor with $\kappa_{c}$ in the range of 81 -- 160.

\begin{table*}[ht]
\begin{center}
\caption{ Superconducting properties of \sro\ evaluated by taking into account the Pauli pair-breaking effect.}
\label{tab:SCparameters}
\begin{tabular}{ccccc}
\hline\hline
Quantity & Symbol & Direction & Values & Remarks     \\
\hline\hline 
Zero-field critical temperature         & $\Tc$                 &  -                         &  1.5~K     & Ref. \citen{Mackenzie1998.PhysRevLett.80.161}  \\ \hline
Thermodynamic critical field            & $\mu_0\Hc$            &  -                         &  0.0194~T   & From $\mu_0 \Hc^2 /2  = \int C\subm{e}(T) dT - \gamma\subm{e}\Tc^2/2$~\cite{Akima1999.JPhysSocJpn.68.694}  \\ \hline
SC condensation energy density          & $E\subm{cond}$        &  -                         &  150~J/m\sps{3}      & From $E\subm{cond}=\mu_0 \Hc^2 /2$ \\ \hline
\multirow{3}{*}{Experimental upper critical field for $T\to 0$} & \multirow{3}{*}{$\mu_0\Hcc$ }    &  $H\parallel c$            &  0.075~T    &   \\ 
                                        &                       &  $H\parallel$ [100]         &  1.45~T     &   \\ 
                                        &                       &  $H\parallel$ [110]         &  1.50~T     &   \\ \hline
\multirow{2}{*}{Estimated orbital limiting field for $T\to 0$}  & \multirow{2}{*}{$\mu_0\Horb$}    &  $H\parallel c$            &  0.075~T    &   \\ 
                                        &                       &  $H\parallel ab$           &  3.0 -- 4.5~T &   \\ \hline
Intrinsic SC anisotropy                 & $\Gamma$              &  -                         &  30 -- 60     &   \\ \hline
Estimated Pauli limiting field          & $\mu_0\Hp$            &  -                         &  1.4~T      & From $E\subm{cond} = \chi\subm{spin}\spsm{vol}\mu_0\Hp^2 /2$~\cite{Yonezawa2013.PhysRevLett.110.077003} \\ \hline
Critical temperature of FOT         & $T\subm{FOT}$         &  $H\parallel ab$           &  0.6 -- 0.8~K   & Ref. \citen{Yonezawa2013.PhysRevLett.110.077003,Yonezawa2014.JPhysSocJpn.83.083706,Kittaka2014.PhysRevB.90.220502}         \\ \hline
\multirow{2}{*}{GL coherence length}    & \multirow{2}{*}{$\xi$}
                                                                &  $\xi_{ab}$          &  66~nm        & From $\Horbc = \Phi_0/(2\pi \xi_{ab}^2)$  \\ 
                                        &                       &  $\xi_{c}$         &  1.1 -- 1.6~nm  & From $\Horbab = \Phi_0/(2\pi \xi_{ab} \xi_c)$ \\ \hline
\multirow{2}{*}{London penetration depth} & \multirow{2}{*}{$\lambda$} 
                                                                &  $\lambda_{ab}$   & 160 -- 190~nm        &  In plane decay of in-plane current~\cite{Ormeno2006.PhysRevB.74.092504} \\ 
                                        &                       &  $\lambda_{c}$  & 11 -- 15~$\mu$m    & 
                                      From $\lambda_c = \Gamma\lambda_{ab}$  \\ \hline
\multirow{2}{*}{GL parameter}           & \multirow{2}{*}{$\kappa$}         
                                                                &  $H\parallel c$            & 2.7      & From $\kappa_{ab} = \lambda_{ab}/\xi_{ab}$ \\ 
                                        &                       &  $H\parallel ab$           & 81 -- 160   & From $\kappa_c = (\lambda_{ab}\lambda_{c} / \xi_{ab}\xi_c)^{1/2} = \Gamma\kappa_{ab}$ 
 \\ \hline
\hline
\end{tabular}
\end{center}
\end{table*}

\subsection{\SECTIONNMR}\label{Sec:NMR}

Up to year 2019, the invariant spin susceptibility across $\Tc$ obtained from NMR Knight shift and polarized neutron scattering was considered as direct evidence for spin-triplet pairing.
Brown \etal at UCLA, while measuring NMR signals of \SRO with elevated $\Tc$ (the 3-K phase) under uniaxial strain, realized that the measured Knight shift depends on the pulsating conditions (S. Brown: private communication).
Using the free-induction decay (FID) method with a reduced RF-pulse energy and working with enhanced $\Tc$ up to 3.5 K, they were able to recognize that exceptionally high conductivity of \SRO induces large self-heating due to eddy current, thus requiring an unexpectedly low RF-pulse energy to keep the sample below $\Tc$.

\begin{figure}[ht]
\begin{center}
\includegraphics[scale=0.16, keepaspectratio]{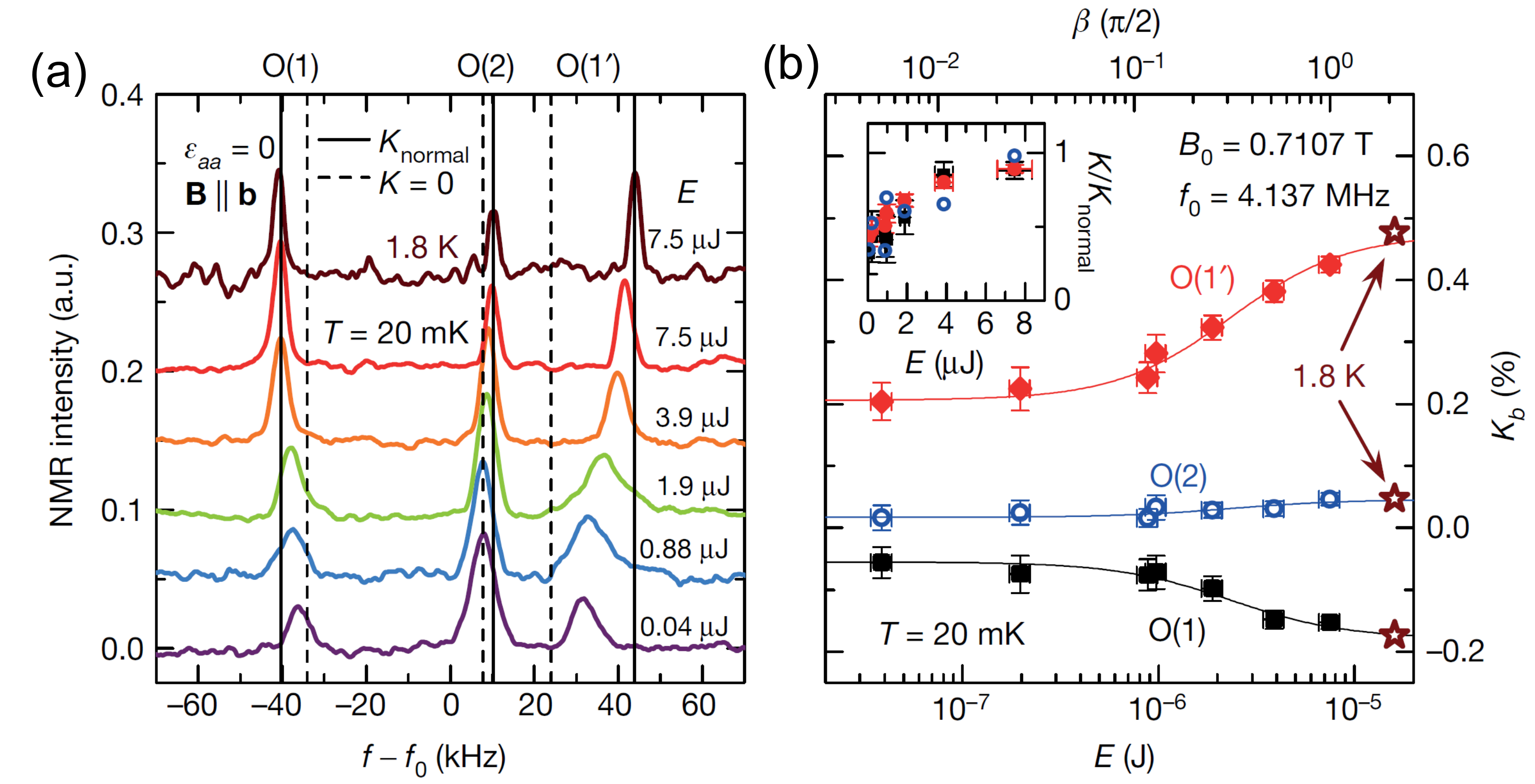}
        \end{center}
\caption{\label{Fig:Ks_Pustogow_Nature} Importance of using small NMR pulse energy to reveal the intrinsic Knight shift. 
(a) NMR spectra for various pulse energy. 
Reduction of the Knight shift at low temperature is observed only for small RF pulse energy $E$. 
(b) Pulse-energy dependence of the Knight shifts of different oxygen nuclear sites. 
Saturation of the Knight shift at low energy indicates the need for reducing the RF pulse energy to that level.
Reproduced from Ref.~\citen{Pustogow2019.Nature.574.72} ($\copyright$~2019 The Authors).}
\end{figure}

\begin{figure}[ht]
\begin{center}
\includegraphics[scale=0.5, keepaspectratio]{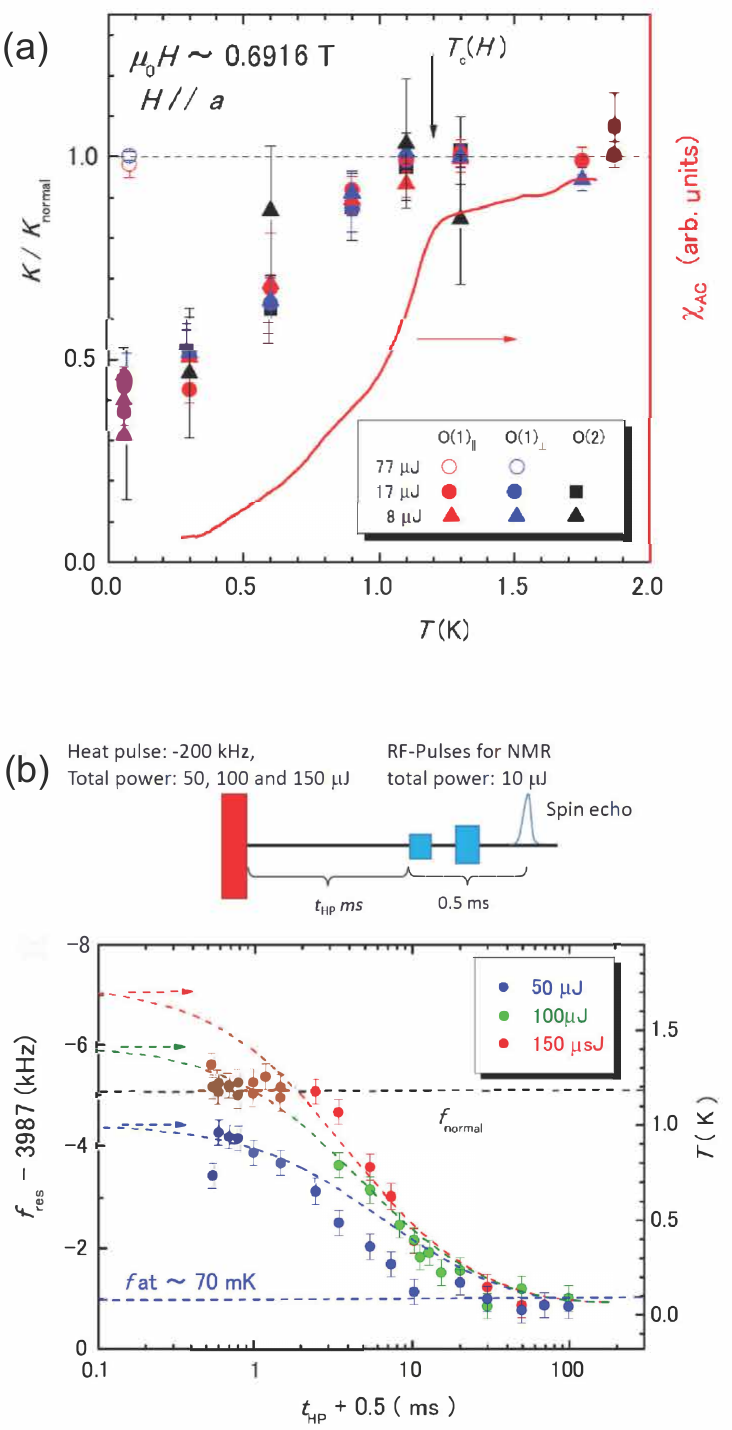}
\caption{(a) Temperature dependence of the Knight shift. 
(b) Characterization of the heating effect by RF pulses used in the NMR. 
The pulse sequence consisting of a strong heat pulse and the subsequent weak spin-echo pulses after varying time $\tau\subm{HP}$ is used to extract the time-evolution of the temperature of the electron-spin system. 
The electron temperature exceeds $T\subm{c}$ immediately after large-energy pulses and recovers the original temperature only after ca. 100 ms. 
Thus, the Knight shift measured immediately after the main pulse gave erroneous information.
The pulse-energy values in Figs. \ref{Fig:Ks_Pustogow_Nature} and this figure cannot be directly compared since their effectiveness depends on individual experimental setup. 
Reproduced from Ref.~\citen{Ishida2020.JPSJ.89.034712} ($\copyright$~2020 The Authors).
}
\label{Fig:Ks-Heating_Ishida_JPSJ} 
\end{center}
\end{figure}

\begin{figure}[ht]
\begin{center}
\includegraphics[scale=0.15, keepaspectratio]{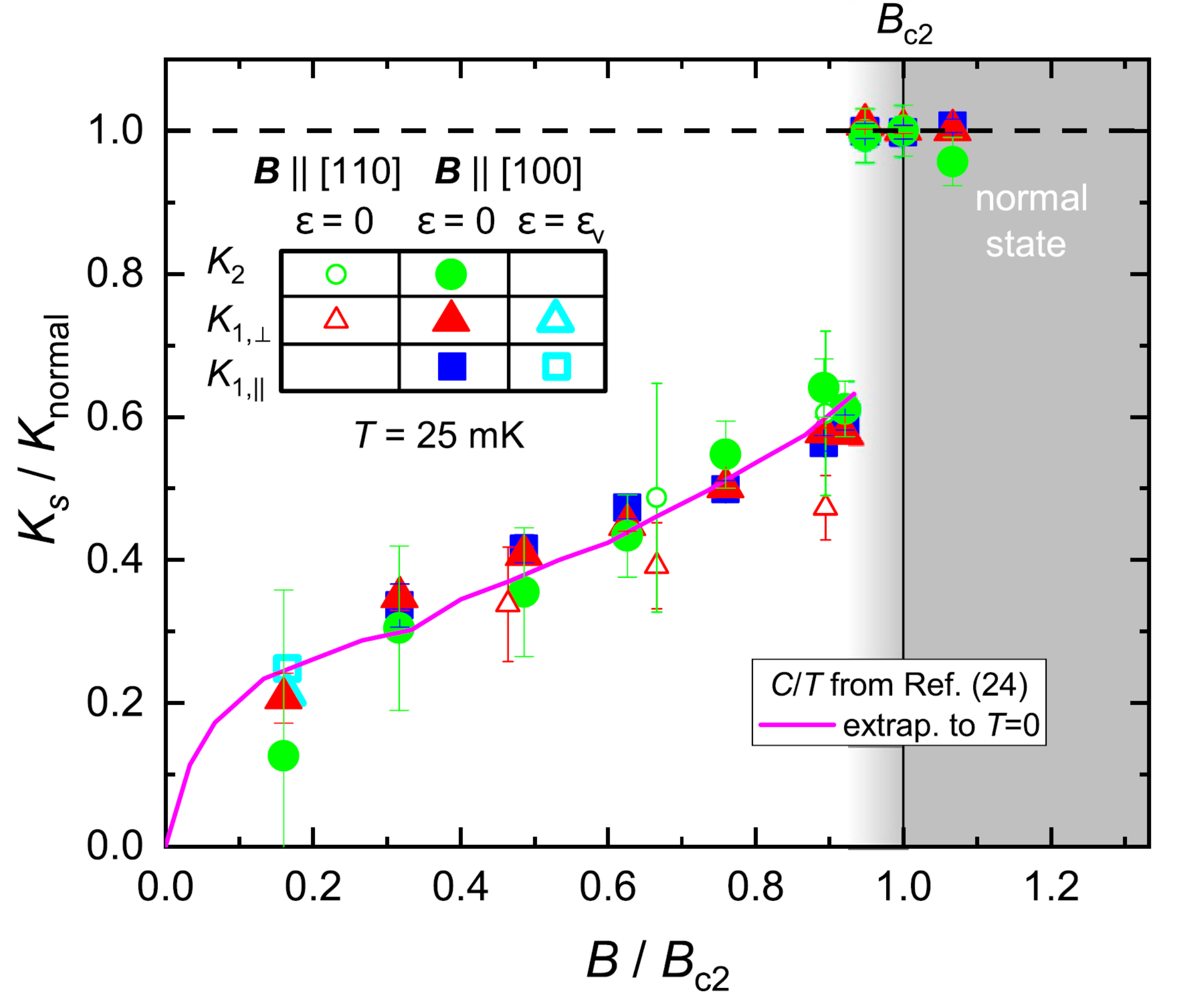}
        \end{center}
\caption{\label{Fig:Ks110_Chronister_PNAS}
Reduction of the Knight shift at low temperature under magnetic field in different in-plane directions. 
The large reduction of the Knight shift in both in-plane directions is incompatible with the $d$-vector pointing in the $c$-direction (proposed spin-triplet chiral p-wave state), the $d$-vector having two components in the $ab$-plane (spin-triplet helical p-wave state), nor a spin-triplet state with the $d$-vector pinned in the $ab$ plane. 
Reproduced from Ref.~\citen{Chronister2021.PNAS.118.25} ($\copyright$~2021  National Academy of Sciences).}
\end{figure}

As shown in Fig. \ref{Fig:Ks_Pustogow_Nature},  they were able to reveal that the 1.5-K phase without strain exhibits reduction in the Knight shift for the field along [100], by reducing the RF energy sufficiently~\cite{Pustogow2019.Nature.574.72}. 
These results were immediately reproduced by the Kyoto group~\cite{Ishida2020.JPSJ.89.034712}, which confirmed a large reduction of the spin-part of the oxygen Knight shift on cooling as shown in Fig.~\ref{Fig:Ks-Heating_Ishida_JPSJ} (a). 
The field dependence of the Knight shift for field along the [110] direction was also measured~\cite{Chronister2021.PNAS.118.25}. 
After taking account of the expected quasi-particle excitations as a function of the applied magnetic fields, these revised results indicate that the reduction in the spin part of the Knight shift (Fig.~\ref{Fig:Ks110_Chronister_PNAS}) are as large as 90\%, which is sufficient to make the spin-triplet scenarios with and without the \emph{d-vector} pinning as well as the spin-triplet helical scenario unlikely. 
The Kyoto group also characterized the process of overheating in detail and deduced that in the previous setup the sample crystal was heated to the normal state for several milliseconds after the RF pulses (Fig.~\ref{Fig:Ks-Heating_Ishida_JPSJ} (b)). 

\begin{figure}[ht]
\begin{center}\includegraphics[scale=0.15, keepaspectratio]{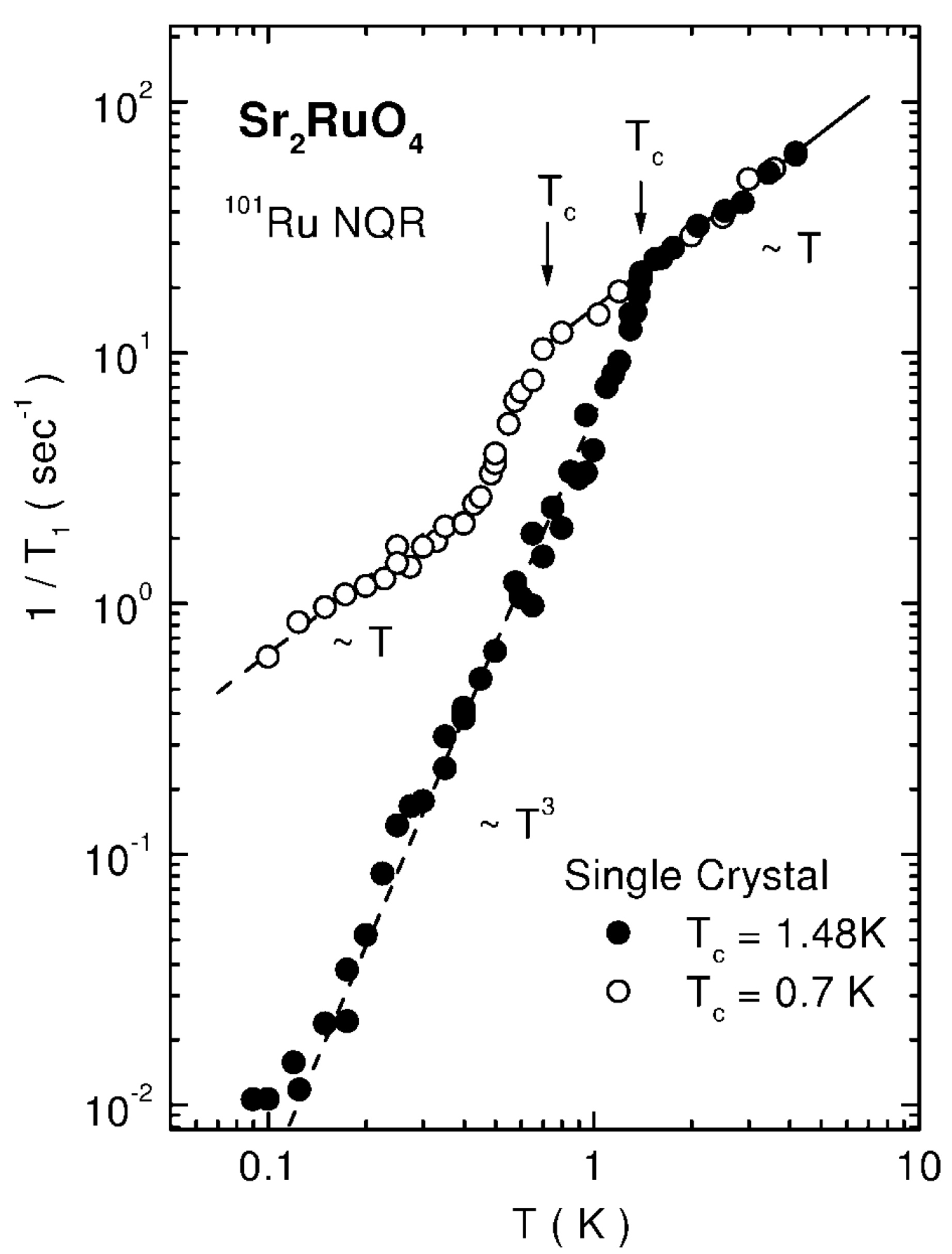}
        \end{center}
\caption{\label{Fig:T1_Ishida} Temperature dependence of the nuclear relaxation rate $1/T_{1}$. 
Since $T_{1}$ is substantially longer than 10 ms even at $\Tc$, this data does not suffer from the effect of overheating.
Reproduced from Ref.~\citen{Ishida2000.PhysRevLett.84.5387} ($\copyright$~2000  American Physical Society).}
\end{figure}

Let us note that the measurements of the spin-lattice relaxation time, $T_1$ (Fig. \ref{Fig:T1_Ishida}) are expected to suffer much less from such initial overheating within 10 ms.
The temperature of the nuclear spin always remains nearly at the initial temperature as the $T_1$ time scale is much longer. It is 30 ms even at $\Tc$.~\cite{Ishida2020.JPSJ.89.034712}. 

\begin{figure*}[ht]
\begin{center}
\includegraphics[scale=0.18, keepaspectratio]{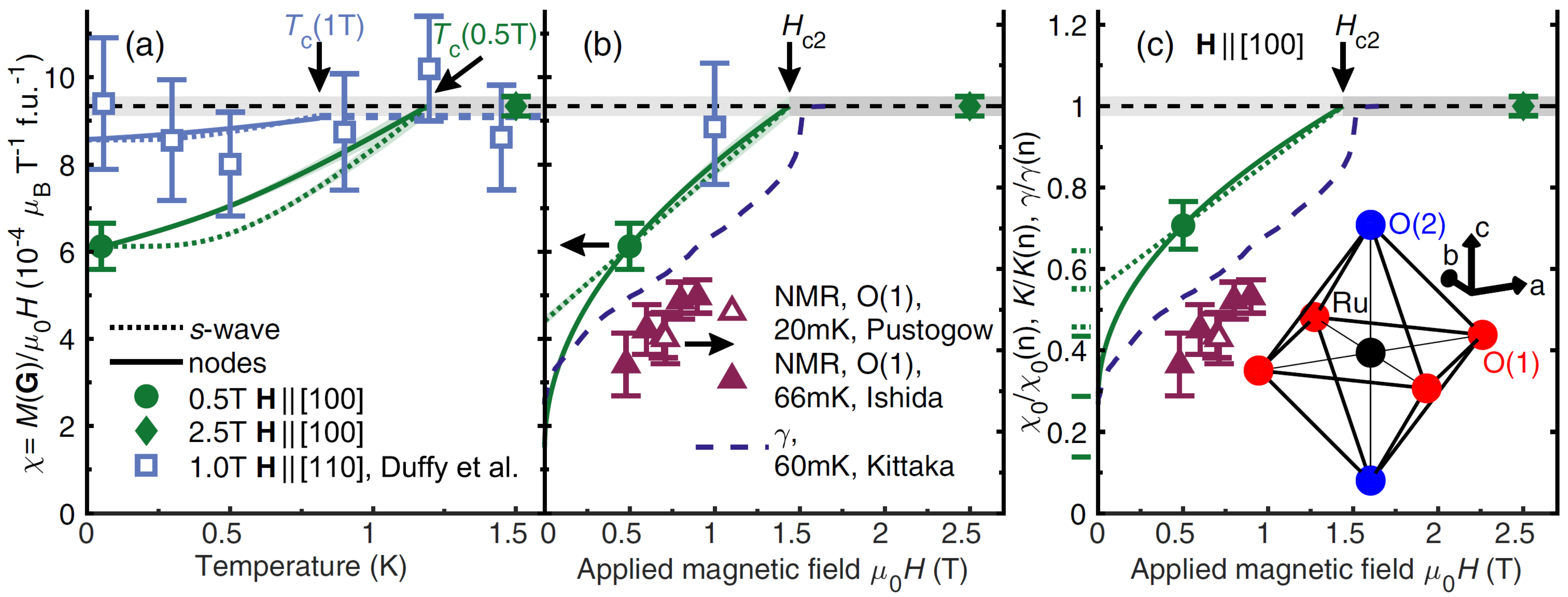}
        \end{center}
\caption{\label{Fig:Neutron_Petsch_PRL} Magnetic susceptibility, $\chi = M(G)/H$, measured by polarized neutron scattering at $G =$~(011) under small magnetic fields. (a) Temperature dependence at different field values and directions, (b) Field dependence at 60 mK. (c) Field dependence of the scaled spin susceptibility. Bars on the vertical scale show expected values by two models with confidence intervals.
Reproduced from. Ref.~\citen{Petsch2020.PhysRevLett.125.217004} ($\copyright$~2020  American Physical Society).}
\end{figure*}

After the revised NMR results, polarized neutron scattering experiments were repeated under magnetic fields parallel to [100] reduced from 1 T to 0.5 T ~\cite{Petsch2020.PhysRevLett.125.217004}. 
Fig. \ref{Fig:Neutron_Petsch_PRL} compares previous and recent results. Although it was difficult to resolve any change in the previous results at 1 T, a clear reduction of 34 $\pm$6\% in the magnetic susceptibility was resolved at 0.5 T. 
These neutron scattering results rule out the odd-parity chiral $p$-wave superconducting state with \emph{d-vector}  $\sim \hat{z} (k_x \pm ik_y)$. 
A spin-triplet helical pairing state may exhibit a substantial drop in the spin susceptibility~\cite{Gupta2020}. 
However, more recent results of a large reduction of the NMR Knight shift shown in Fig.~\ref{Fig:Ks110_Chronister_PNAS} make the helical triplet scenario unlikely.~\cite{Chronister2021.PNAS.118.25} 

\subsection{\SECTIONHAB}\label{Sec:Hab}

\subsubsection{\SECTIONFFLO}\label{Sec:FFLO}

NMR remains an indispensable technique to identify the spin configuration of the pairs in the superconducting state. With sufficiently small RF-pulse energy, a new phase evidencing the Fulde-Ferrell-Larkin-Ovchinnikov (FFLO) state near the upper critical fields was identified~\cite{Kinjo2022.Science.376.397}. 
The FFLO state was theoretically proposed 60 years ago as a superconducting state with finite-momentum Cooper pairs. Under strong spin imbalance due to spin polarization, electrons may pair with a finite total momentum, resulting in a spatial oscillation of the superconducting OP. 
Such an exotic state breaking translational symmetry is possible in an ultraclean system with a large Pauli spin susceptibility. It has been reported only in a handful of superconductors~\cite{Matsuda2007JPhysSocJpnReview, Kinjo2022.Science.376.397}. 

\begin{figure*}[ht]
\begin{center}
\includegraphics[scale=0.17, keepaspectratio]{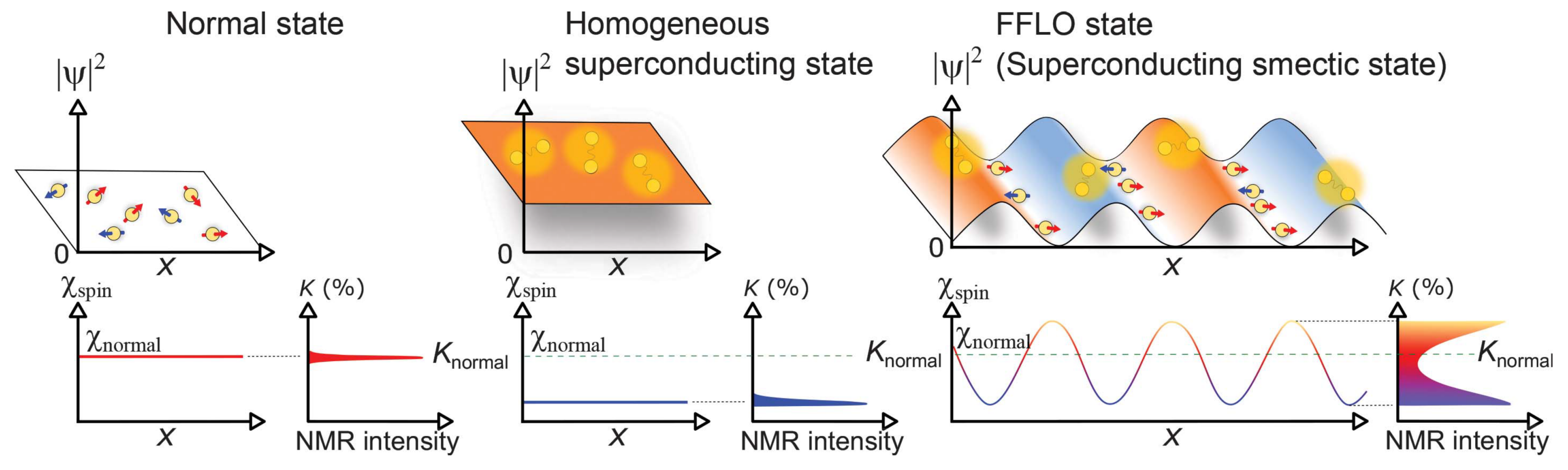}
        \end{center}
\caption{\label{Fig:FFLO_Kinjo_Science} Anticipated features of the NMR spectra in the normal state, homogeneous superconducting state, and superconducting smectic state (FFLO state). Reproduced from Ref.~\citen{Kinjo2022.Science.376.397} 2022 The American Association for the Advancement of Science).}
\end{figure*}

\begin{figure}[ht]
\begin{center}
\includegraphics[scale=0.12, keepaspectratio]{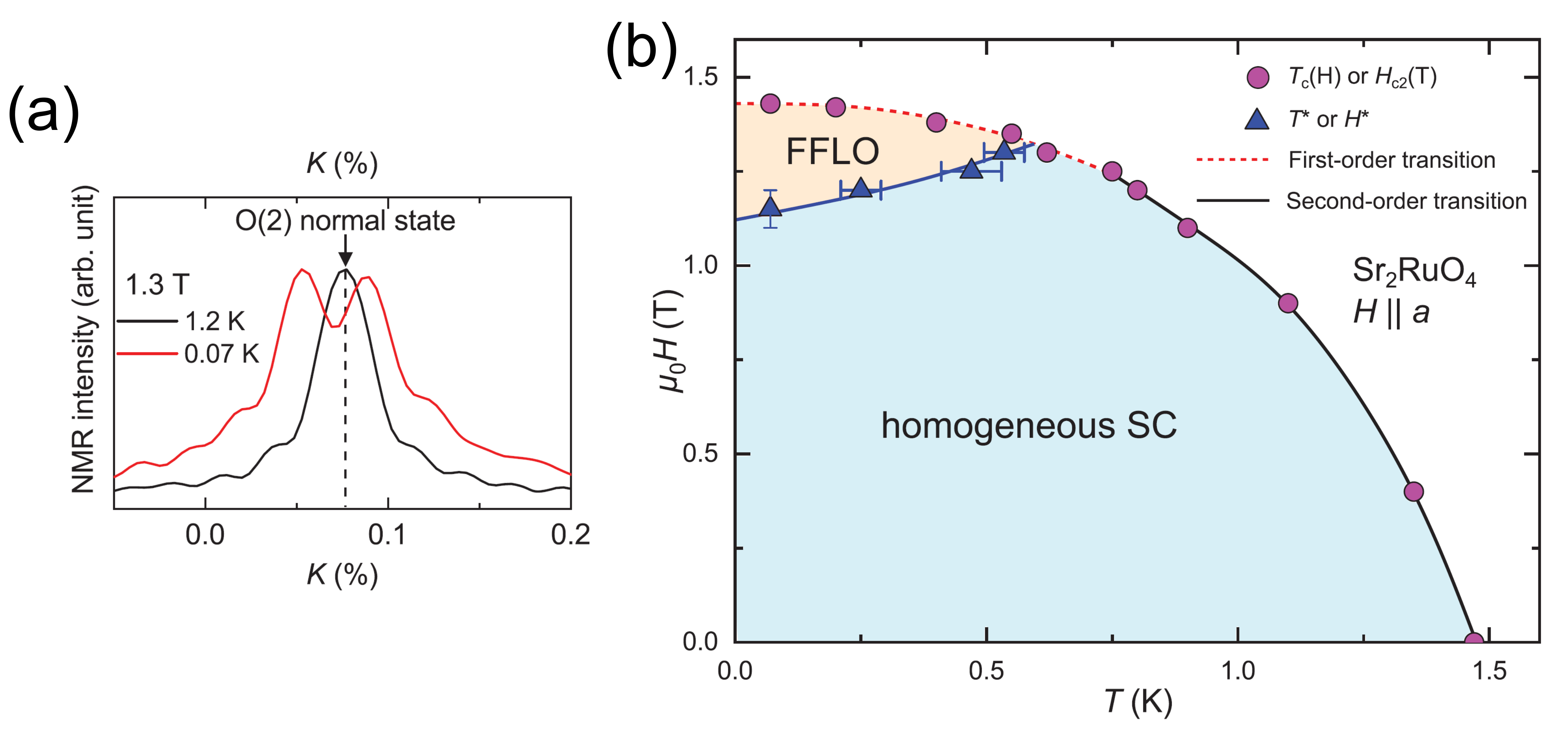}
        \end{center}
\caption{\label{Fig:FFLO-phase-diagram_Kinjo_Science} (a) A Double-horn structure in the NMR spectra. 
(b) Phase diagram under in-plane magnetic field, revealing the superconducting smectic state near the upper critical fields. Reproduced from Ref.~\citen{Kinjo2022.Science.376.397} ($\copyright$~2022 The American Association for the Advancement of Science).}
\end{figure}

As illustrated in Fig. \ref{Fig:FFLO_Kinjo_Science}, the FFLO state with an amplitude modulation in a spin-singlet superconducting state leads to a spatial oscillation of the local spin density. 
Thus, such an electronic fluid state with directional spin modulation is considered as a spin-smectic state, in analogy to a smectic liquid crystal.
In such a state, the NMR Knight shift would be characterized by a double-horn spectrum with one of the peak positions at the enhanced spin susceptibility over the normal state value. 
Indeed, such spectra were observed in \SRO, as shown in Fig. \ref{Fig:FFLO-phase-diagram_Kinjo_Science} (a). The phase diagram in Fig. \ref{Fig:FFLO-phase-diagram_Kinjo_Science} (b) identifies the regions in the $H-T$ phase diagram for $H$//[100] where the spin-smectic phase emerges. 
On cooling at $\mu_0H$ = 1.25 T for example, the Knight shift first decreases on entering the SC state, but on further cooling the peak splits into two, with one of the peaks reflecting the enhanced spin polarization.
From the comparison between the NMR spectrum and the simulation\cite{Ichioka2007.PhysRevB.76.014503}, the modulation wavelength is estimated to be $\lambda\subm{FFLO}\sim 30~\xi_{ab} \sim 2.0~\mu$m.
A crude theoretical estimation of the modulation period gives a reasonable correspondence: $\lambda\subm{FFLO} = 2\pi \hbar v\subm{F}/(g \mu\subm{B} H) = 2\pi^2 \Delta \xi/(g \mu\subm{B} H) = 29.8~\xi = 2.0 ~\mu$m. 

To obtain the definitive evidence of the FFLO state, it is quite important to probe the FFLO phase boundary by other thermodynamic measurements. Imaging of the spatial modulation of the SC order parameter directly with scanning tunneling microscopy is a highly promising method. Since the FFLO state is observed under magnetic fields of one order of magnitude smaller than in other candidate materials, \SRO offers an ideal platform to study such an exotic SC state. 

\subsubsection{\SECTIONFOT}\label{Sec:FOT}

In \sro, the behavior of the upper critical field, $\Hcc$, for fields parallel to the $ab$-plane had been a long mystery.~\cite{Maeno2011.JPhysSocJpn.81.011009, Mackenzie2017.npjQuantumMater.2.40} 
Now, with the correct knowledge on the spin susceptibility~\cite{Pustogow2019.Nature.574.72,Ishida2020.JPSJ.89.034712,Petsch2020.PhysRevLett.125.217004}, the $\Hcc$ behavior is well understood by the Pauli pair breaking effect. 
Below, we give a brief historical overview of these developments.


Before year 2000, it was already found that the temperature dependence of $\Hcc$ for $H \parallel ab$ exhibits strong suppression~\cite{Akima1999.JPhysSocJpn.68.694,NishiZaki2000JPhysSocJpn}: the $\Hcc(T)$ curve at low temperature is substantially suppressed compared to the ordinary Wertharmer-Helfand-Hohenberg (WHH) curve~\cite{Werthamer1966PhysRev} accounting for the orbital pair breaking mechanism. 
In contrast, $\Hcc$ parallel to the $c$-axis matches well the WHH curve. 
This has been reproduced in subsequent studies of specific heat~\cite{Deguchi2002} and AC susceptibility~\cite{Yaguchi2002.PhysRevB.66.214514,Kittaka2009.PhysRevB.80.174514}. 
Related to this $\Hcc$ suppression, thermal conductivity, specific heat, and magnetization have been reported to exhibit rapid recovery toward normal-state values near $\Hcc$~\cite{Izawa2001.PhysRevLett.87.057002,Tanatar2001.PhysRevB.63.064505,Deguchi2002,Tenya2006.JPhysSocJpn.75.023702}.


Similar suppression is also revealed in the polar angle $\theta$ dependence of $\Hcc$~\cite{Deguchi2002,Kittaka2009.PhysRevB.80.174514}:
when the field direction is tilted from the $ab$-plane, the $\Hcc(\theta)$ curve is well fitted with the ordinary anisotropic-mass model, except for the field direction within about 2$^\circ$ from the $ab$-plane.
We should comment here that the noticeable kink in the $\Hcc(\theta)$ curve reported in Ref.~\citen{Deguchi2002} has not been observed in subsequent studies~\cite{Kittaka2009.PhysRevB.80.174514,Yonezawa2013.PhysRevLett.110.077003,Yonezawa2014.JPhysSocJpn.83.083706}; but deviation from the anisotropic-mass model near $H\parallel ab$ has been certainly reproduced.

These facts indicate that ordinary orbital pair breaking effect cannot explain the behavior of $\Hcc$ along the $ab$-plane, and another pair breaking mechanism must play a role. 
Given the substantial decrease of the spin susceptibility~\cite{Pustogow2019.Nature.574.72,Ishida2020.JPSJ.89.034712}, we can now understand that the Pauli effect is responsible for the $\Hcc$ suppression.

\begin{figure} 
\begin{center}
\includegraphics[width=7.5cm]{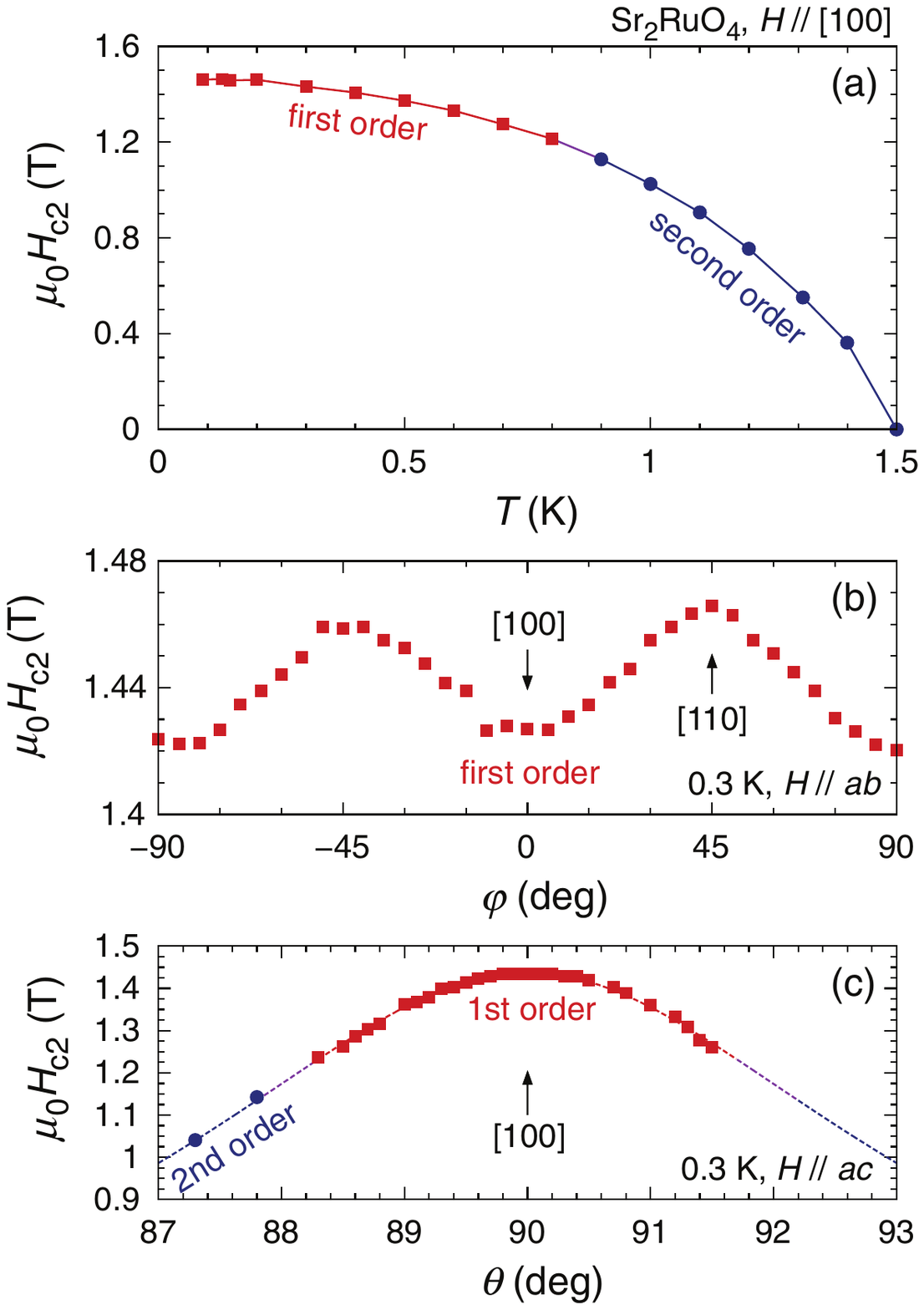}
\end{center}
\caption{Superconducting phase diagrams of \sro revealed by specific-heat measurements using an ultra-clean sample.
First-order phase transition is indicated by red squares and second-order transition by blue circles.
(a) $H$-$T$ phase diagram for $H\parallel ab$.
(b) In-plane $\Hcc$ as a function of the in-plane field angle $\varphi$.
(c) Polar angle $\theta$ dependence of $\Hcc$.
Reproduced from Ref.~\citen{Yonezawa2014.JPhysSocJpn.83.083706} ($\copyright$~2014 The Authors).
} 
\label{fig:H-T_phase-diagram_Yone}
\end{figure}

In 2013, it has been revealed that the superconducting-normal transition at $\Hcc$ is a first-order transition (FOT) when the field is very close to the $ab$-plane and temperature is less than 0.8~K~\cite{Yonezawa2013.PhysRevLett.110.077003}, as depicted in Fig. \ref{fig:H-T_phase-diagram_Yone}.
In the initial report, the magneto-caloric effect, i.e. change of the sample temperature induced by magnetic field sweeps, revealed clear divergent-like peak in the field derivative of the entropy $S$, together with a small but clear hysteresis in $\Hcc$ in ultra-clean samples.
The FOT becomes an ordinary second-order transition by tilting the magnetic field away from the $ab$-plane only by 2.5$^\circ$.
The first-order nature has also been reported in subsequent specific heat~\cite{Yonezawa2014.JPhysSocJpn.83.083706}, magnetization~\cite{Kittaka2014.PhysRevB.90.220502}, and magnetic-torque studies~\cite{Kittaka2016.JMagMagMater.400.81}.
The Pauli effect, in general, has been known to cause FOT at $\Hcc$ at low temperatures~\cite{Clogston1962}.
Indeed, even before 2019, several studies have correctly attributed the FOT to the Pauli effect~\cite{Machida2008.PhysRevB.77.184515}, 
admitting apparent inconsistency with the early reports on invariant spin susceptibility.
We should also comment that several attempts have been made to explain the FOT and $\Hcc$ suppression without the spin paramagnetism~\cite{Gomes-da-Silva2014.PhysLettA.378.1396,Ramires2016,Ramires2017}.



Here, it would be valuable to characterize and discuss the FOT in more detail.
Firstly, $\Hcc$ value for $H\parallel ab$ is comparable to the Pauli-limiting field $\Hp$.
Assuming only the Zeeman effect, $\Hp$ is given by $\Hp = H\subm{c}/(\chi\subm{spin,N}-\chi\subm{spin,SC})^{1/2}$, where $H\subm{c}$ is the thermodynamic critical field ($\mu_0 H\subm{c} = 19.4$~mT for \sro)~\cite{Akima1999.JPhysSocJpn.68.694}, and $\chi\subm{spin,N}$ and $\chi\subm{spin,SC}$ are spin susceptibilities in the normal and superconducting states.
Using the experimental value of the bulk susceptibility $\chi\subm{bulk} = 0.9\times 10^{-3}$~emu/mol~\cite{Mackenzie2003RMP} as $\chi\subm{spin,N}$ and assuming $\chi\subm{spin,SC}= 0$, we obtain $\mu_0\Hp = 1.4$~T, which is in close agreement with the experimental $\Hcc$.
Moreover, the assumption $\chi\subm{spin,SC}=0$ is consistent with the NMR study showing $\chi\subm{spin,SC}$ extrapolating to nearly zero for the limit of $H, T \to 0$.
Secondly, the critical temperature $T\subm{FOT}$, below which the transition becomes FOT, is in the range 0.6--0.8~K (40--50\% of the zero-field $\Tc$), depending on literature~\cite{Yonezawa2013.PhysRevLett.110.077003,Yonezawa2014.JPhysSocJpn.83.083706,Kittaka2014.PhysRevB.90.220502,Kittaka2016.JMagMagMater.400.81}. 
This is comparable to the simple theoretical prediction, where $T\subm{FOT}$ is 56\% of the zero-field $\Tc$, when only the Pauli effect is taken into account.
The entropy jump $\Delta S/T$ at $\Hcc$ revealed by magnetocaloric effect is $-4.7\pm 1.2$~mJ/K\,mol at 0.2~K~\cite{Yonezawa2013.PhysRevLett.110.077003}, amounting to $12.5\pm 3.2\%$ of the normal-state entropy $S\subm{N}/T = \gamma\subm{e} = 37.5$~mJ/K\,mol.
This value is confirmed by the specific heat data~\cite{Yonezawa2014.JPhysSocJpn.83.083706}.
Furthermore, the observed magnetization jump $\mu_0\Delta M$ of $0.074\pm 0.015$~mT at $\Hcc$~\cite{Kittaka2014.PhysRevB.90.220502} satisfies the Clausius-Clapeyron relation for first-order transitions $\Delta S/\Delta M = -d\Hcc(T)/dT$.
More recent NMR study revealed a similar 40\%-jump in the Knight shift~\cite{Chronister2021.PNAS.118.25} at the lowest measurement temperatures. 
Thus, the jump in magnetization is dominated by spin paramagnetism and the contribution of superconducting diamagnetism is rather minor.
Finally, we should comment that features of the FOT, such as the sharpness of the transition and the existence of hysteresis, are highly dependent on sample quality. 
In samples with slightly lower quality, these features are immediately blurred~\cite{Yonezawa2013.PhysRevLett.110.077003,Yonezawa2014.JPhysSocJpn.83.083706}. 
This is analogous to the freezing of water, where supercooling (hysteresis) effect is observed only for pure water stored in a clean container.
This explains why it took around 20 years to reveal the first-order nature of the transition, 
although anomalies reported earlier~\cite{Deguchi2002,Tenya2006.JPhysSocJpn.75.023702} suggested the FOT nature.


\subsubsection{\SECTIONOUTOFPLANE}\label{Sec:OutOfPlane}

As discussed above, $\Hcc$ for the $ab$-plane is dominated by the Pauli pair-breaking mechanism.
Thus, the out-of-plane superconducting anisotropy $\Gamma = \xi_{ab}/\xi_c$ should not be evaluated using the simple formulae
$\Hccc(T = 0) = \Phi_0/(2\pi \xi_{ab}^2)$ and $\Hccab(T = 0) = \Phi_0/(2\pi \xi_{ab}\xi_c)$ for orbital pair-breaking cases, where $\Phi_0~\approx$~2.07~mT$(\mu\mathrm{m})^2$ is the superconducting flux quantum. .
This fact has been recognized before the Pauli pair-breaking is identified as the origin of the $\Hcc$ suppression, and the intrinsic $\Gamma$ value was estimated by evaluating the orbital-limited $\Horbab$ in various ways.
Fitting $\Hccab(T)$ data near $\Tc$ using the WHH formula, $\Horbab$ for $T\to 0$ was estimated to be $\mu_0\Horbab(0) \sim 2.5$~T~\cite{Kittaka2009.PhysRevB.80.174514}.
This value corresponds to the intrinsic $\Gamma= 33$, which is substantially larger than the naive evaluation of 20 from the ratio $\Hccab/\Hccc$.
Extrapolation of the polar angle $\theta$ dependence of $\Hcc$ toward the $ab$-plane yields a temperature-dependent $\Gamma = 22$-$46$~\cite{Deguchi2002,Kittaka2009.PhysRevB.80.174514}.
The ratio $\Hccab/\Hccc$ near $\Tc$ has also been used since $\Hccab$ should be close to $\Horbab$ near $\Tc$.
This evaluation leads to the intrinsic $\Gamma$ value of $\sim 60$ in the limit of $T\to \Tc$~\cite{Kittaka2009.JPhysConfSer.150.052112}.
Analysis of the magnetic torque near $H\parallel ab$ also concluded $\Gamma \sim 60$~\cite{Kittaka2014.PhysRevB.90.220502}.


Intrinsic out-of-plane superconducting anisotropy has been microscopically studied using small-angle neutron scattering (SANS)~\cite{Rastovski2013.PhysRevLett.111.087003}, which characterized the vortex-lattice structure for $H$ along the $ab$-plane.
For this field configuration, observation of the vortex lattice Bragg peak is extremely difficult since the field modulation due to vortices is as small as 3~$\mu$T due to the long $\lambda_{c}$.
This difficulty has been overcome by measuring the field component perpendicular to the magnetic field that is intentionally tilted by a few degrees away from the $ab$-plane.
In addition, among the six expected Bragg peaks, only the ones along the reciprocal $c^\ast$-axis are detectable in principle.
Nevertheless, by assuming the flux quantization within the unit cell of the vortex lattice, the authors of Ref.~\citen{Rastovski2013.PhysRevLett.111.087003} succeeded in evaluating the intrinsic anisotropy $\Gamma$, which amounts to 60 in the limit of $H\parallel{ab}$-plane.
This value is in close agreement with some of the evaluations above.
In addition, the field-angle dependence of the vortex form factor does not follow the ordinary orbital-limited behavior.
These findings are consistent with the Pauli-limiting mechanism.


\subsubsection{\SECTIONINPLANE}\label{Sec:InPlane}

Compared with the large out-of-plane anisotropy, the in-plane superconducting anisotropy is much weaker,
as expected for tetragonal crystal symmetry.
Studies with accurate two-axis field control have reported that in-plane anisotropy is nearly absent in $\Hcc$ above 0.8~K, whereas noticeable anisotropy emerges below this temperature~\cite{Mao2000.PhysRevLett.84.991,Tanatar2001.PhysRevLett.86.2649,Kittaka2009.PhysRevB.80.174514,Yonezawa2014.JPhysSocJpn.83.083706}.
For $T\to 0$,  $\Hcc$ along the [110] direction is larger than $\Hcc$ along [100] by around 3\%.
As a function of in-plane angle $\phi$, $\Hcc$ exhibits a four-fold sinusoidal oscillation~\cite{Mao2000.PhysRevLett.84.991,Tanatar2001.PhysRevLett.86.2649,Yonezawa2014.JPhysSocJpn.83.083706} as exemplified in Fig. \ref{fig:H-T_phase-diagram_Yone}(b).


Interestingly, the temperature below which the in-plane $\Hcc$ anisotropy emerges coincides with the onset temperature of the first-order transition $T\subm{FOT}\simeq 0.8$~K (see Fig. \ref{fig:H-T_phase-diagram_Yone}(a)).
This fact indicates that the four-fold in-plane $\Hcc$ oscillation is intimately related to the Pauli pair-breaking effect.
Indeed, theories predict the emergence of $\Hcc$ anisotropy as a consequence of the formation of the FFLO state~\cite{Croitoru2017.CondMat.2.30}, which is indeed recently discovered by NMR as described earlier in this section~\cite{Kinjo2022.Science.376.397}.

\section{\SECTIONSCGAP}\label{Sec:SCGap}

\subsection{\SECTIONQP}\label{Sec:QP}

The superconducting gap structure in \sro\ has been studied from the temperature dependence of quasiparticle excitations. 
Experiments show power-law $T$ dependence in nuclear-lattice relaxation rate, electronic specific heat, penetration depth and ultrasound attenuation, indicating the presence of low energy excitations due to a nodal superconducting gap structure.

NMR/NQR nuclear-lattice relaxation rate $1/T_1$ is, roughly speaking, proportional to the square of thermally excited quasiparticle density of states at $T\ll \Tc$, whereas just below $\Tc$ it provides information of the order parameter symmetry through existence or absence of the coherence peak.
In \sro, absence of the coherence peak below $\Tc$ shown in Fig.~\ref{Fig:T1_Ishida} clearly indicates non-$s$-wave superconductivity~\cite{Ishida2000.PhysRevLett.84.5387}. 
Moreover, $1/T_1$ exhibits $T^3$ dependence at $T \ll \Tc$ for clean samples whereas $1/T_1 \propto T$ for less clean samples.
This is in good agreement with thermally excited quasiparticle density of states in superconductors with nodes, together with impurity-induced density of states.

The electronic specific heat coefficient, $C\subm{e}/T$, in the superconducting state provides another important measure of the density of thermally-excited quasiparticles.
Various studies revealed line-nodal behavior ($C\subm{e}/T \propto T$) with negligible or small residual values for $T\to 0$~\cite{NishiZaki2000JPhysSocJpn,Deguchi2004.JPhysSocJpn.73.1313,Kittaka2018.JPhysSocJpn.87.093703}.
Nodal superconductivity is also made evident by the size of the specific-heat jump at $\Tc$, which is only 70\% of the normal-state electronic specific heat $\gamma\Tc$ and is much smaller than the BCS-theory prediction of 143\%.
Such a reduction in the specific-heat jump is a common feature in line-nodal weak-coupling superconductors~\cite{HasselbachK1993.PhysRevB.47.509}.

The penetration depth $\lambda(T)$ has been studied using various techniques.
Measurements using tunneling diode oscillators (28~MHz) reveals $\Delta \lambda(T)\equiv \lambda(0) - \lambda(T)\propto T^2$ for samples with $\Tc = 1.3$ -- 1.4~K, whereas $\Delta \lambda(T) \propto T^3$ for an impure sample with $\Tc = 0.8$~K~\cite{Bonalde2000.PhysRevLett.85.4775}.
Since $\lambda(T)^{-2}$ is proportional to the so-called "superfluid density" $n\subm{s}$/$m^\ast$, where $n\subm{s}$ is the superconducting carrier density and $m^\ast$is the effective mass, for $T \ll \Tc$ one can approximate the expected quasiparticle density to be $n=(n_0-n\subm{s}) \propto T^2$.
This is actually different from the canonical line-nodal behavior ($\Delta\lambda(T)\propto T$ and $n \propto T$).
Considering the cleanness of the higher-$\Tc$ samples, the quadratic $T$ dependence was interpreted as a combination of non-local electrodynamic effect and the presence of line-nodal quasiparticles, rather than impurity-induced filling of the gap node.
Non-local effects emerge as the electromagnetic response in a certain position in space is determined by properties distant from that position, resulting in a weaker temperature dependence of $\Delta\lambda(T)$~\cite{Kosztin1997.PhysRevLett.79.135}. 
Indeed, the Ginzburg-Landau parameter $\kappa\subm{GL}$ of \sro\ amounts to only 2 -- 3 for $H\parallel c$, indicating that \sro\ in $c$-axis field is located near the border between type-I and type-II superconductivity~\cite{Ray2014.muSR-vortex-lattice}.
In this case, non-local effects become significant below $T < \kappa\subm{GL} \Delta_0/k\subm{B} \sim 0.6\Tc$.
Results of measurements using dielectric resonators utilizing GHz-range electromagnetic waves using high-quality samples with $\Tc \ge 1.4$~K are somewhat more complicated because of the non-classical skin effect:
due to the cleanness of the sample, the electron scattering rate $1/\tau$ and the  electromagnetic wave frequency $\omega$ become comparable (i.e. $\omega\tau > 1$) and a non-trivial $\omega$ dependence emerges~\cite{Ormeno2006.PhysRevB.74.092504}.
Nevertheless, the conclusion is the same as that of experiments in the MHz-range ($\Delta\lambda(T)\propto T^2$).
In-field muon-spin rotation/relaxation ($\mu$SR) technique has also been used to study the temperature dependences of the penetration depth and superfluid density.
Aegerter~\etal reported that $\Delta\lambda^{-2}(T)$ is well fitted either with functions of $T$-linear or $T^{2.5}$ dependences~\cite{Aegerter1998.JPhysCondensMatter.10.7445}.
Subsequently, Luke~\etal~\cite{Luke2000.PhysicaB.289-290.373} reported $\lambda^{-2}(T) \propto 1-(T/T\subm{c})^\alpha$ with $\alpha=2.78$.
More recently, Khasanov~\etal~\cite{Khasanov2023} reported the canonical $T$-linear line-nodal dependence by $\mu$SR at low temperatures but the fitting requires line-nodal gap with a higher angular harmonics.
We also note that the recent penetration-depth measurements with a scanning SQUID probe~\cite{Mueller2023} and by AC susceptibility~\cite{Landaeta2023} both show $\lambda(T)-\lambda(0)~\propto~T^2$. The latter study explains the observed behavior as due to non-local effect with vertical-line-node gap.

Ultrasound attenuation $\alpha(T)$ is also a sensitive probe to quasiparticle excitations.
In \sro a power-law dependence $\alpha(T)\propto T^{n}$ with $n = 1.4$ -- 1.8 has been observed for all of the measured longitudinal (L) and transverse (T) waves along [100] and [110] directions (L100: $c_{11}$ mode, T100: $c_{66}$ mode, L110: $(c_{11} + c_{12} + 2c_{66})/2$ mode, and T110: $(c_{11}-c_{12})/2$ mode) down to 40~mK using samples with $\Tc = 1.37$~K~\cite{Lupien2001.PhysRevLett.86.5986}. 
Similar results with $\alpha(T)\propto T^{2}$ down to 170~mK have been observed in the $(c_{11}-c_{12})/2$ mode with a sample with $\Tc \sim 1.4$~K~\cite{Matsui2002.PhysRevB.63.060505}. 
In principle, ultrasound can also give us information about nodal directions.
However, the large anisotropy of the power-law exponent expected for vertical-line nodes (i.e. $\alpha(T) \propto T^{1.5}$ for ultrasound that couples well to quasiparticles and $\alpha(T) \propto T^{3.5}$ for ultrasound that does not directly couple to quasiparticles) has not been revealed~\cite{Lupien2001.PhysRevLett.86.5986}.
This might be related to the large attenuation anisotropy existing in the normal state.
It is also worth mentioning that an additional horizontal line node could explain the absence of anisotropy in $\alpha(T)$.

\subsection{\SECTIONTHERMAL}\label{Sec:Thermal}

Thermal conductivity is a powerful and directional probe to investigate quasiparticle excitations in the superconducting state.
Around the year 2000, a number of thermal-conductivity measurements have been performed.
In-plane thermal conductivity $\kappa_{ab}$, which is dominated by electronic carriers, exhibits power-law temperature dependence $\kappa_{ab}/T \propto T$ at zero field~\cite{Tanatar2001.PhysRevB.63.064505,Izawa2001.PhysRevLett.87.057002}.
Moreover, the residual thermal conductivity obtained by the extrapolation $\kappa_{ab}/T$ for $T \to 0$ remains finite even for the clean limit~\cite{Suzuki2002.PhysRevLett.88.227004}.
This is well interpreted as a ``universal'' thermal conductivity, which originates from the perfect cancellation of the increase in the quasiparticle density and in the impurity scattering rate for nodal superconductors due to unitary scatterers~\cite{Graf1996.PhysRevB.53.15147}.
These facts strongly support the existence of line nodes.

\begin{figure} 
\begin{center}
\includegraphics[width=6.5cm]{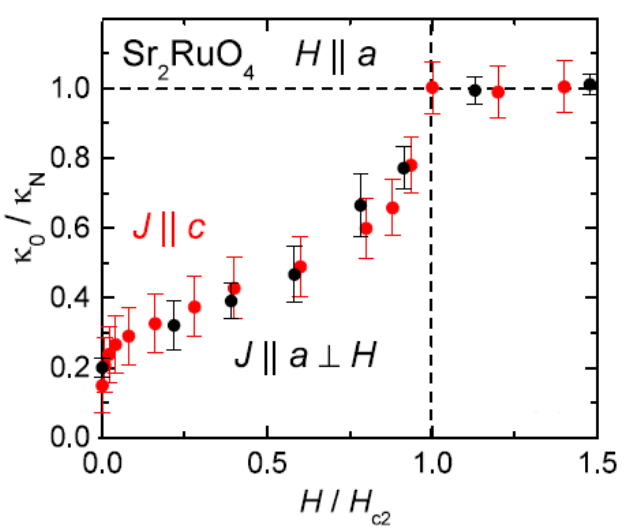}
\end{center}
\caption{Magnetic-field dependence of thermal conductivity $\kappa$ divided by temperature for the $T\to 0$ limit. For the field along the in-plane $a$ axis, $\kappa_c$ exhibits the $\sqrt{H}$-like rapid increase indicative of nodal quasiparticles with the velocity with the $c$-axis component.
Reproduced from Ref.~\citen{Hassinger2017.PhysRevX.7.011032} ($\copyright$~2017  American Physical Society).
} 
\label{fig:thermal-conductivity_Hassinger}
\end{figure}

The field dependence of the thermal conductivity can provide information about another important aspect of superconducting gap structures.
In nodal superconductors, supercurrents flow around magnetic vortices causes a Doppler shift of the quasiparticle energy spectrum.
This shift then excites quasiparticles around gap nodes or minima where the excitation energy is very small.
Such quasiparticles tend to delocalize out of vortex cores, and are detectable via thermal conductivity as discussed below, or specific heat as discussed in the next subsection.
In particular, at low temperatures and fields, it is expected that the quasiparticle density of states exhibits $\sqrt{H}$ behavior. 
This phenomenon is also called the Volovik effect~\cite{Volovik1993.JETPLett.58.469}.
In \sro, for temperatures down to 0.3~K, a non-monotonic field dependence in $\kappa_{ab}$, i.e. a small decrease at low field followed by a $H$-linear increase at higher fields, has been observed~\cite{Tanatar2001.PhysRevB.63.064505,Izawa2001.PhysRevLett.87.057002}.
The observed behavior is interpreted as a result of a delicate balance between the increase in quasiparticle excitations due to the Volovik effect and the increase in scattering by vortices.
The $c$-axis thermal transport $\kappa_c$, which is dominated by phonons at finite temperature, is rather free from possible cancellations of various mechanisms and should be solely sensitive to the quasiparticle excitation density.
Nevertheless, the $\sqrt{H}$ behavior at low fields characteristic of the Volovik effect is not clearly seen even at 0.3~K~\cite{Tanatar2001.PhysRevLett.86.2649}.
More recently, the gap structure has been studied using both $c$-axis and $ab$-plane thermal transport~\cite{Hassinger2017.PhysRevX.7.011032}.
In this study, the quasiparticle excitation contribution is extracted by a careful $T\to 0$ extrapolation.
The extrapolated residual $\kappa/T$ in Fig. \ref{fig:thermal-conductivity_Hassinger} exhibits $\sqrt{H}$-like rapid increase for $\kappa_c$ in low fields along the $a$ axis. 
This rapid increase indicates that the field-induced nodal quasiparticles have noticeable $c$-axis component in their velocity and thus vertical line nodes are likely to exist. This behavior is not compatible with the presence of only symmetry-protected horizontal line nodes at $k_z = 0$ or $\pi/2c$, where $c$-axis velocity is zero. Therefore, the overall results indicate the existence of at least vertical line nodes.

Aiming to determine the position of the nodes, in-plane field-angle dependence of the thermal conductivity has also been tested~\cite{Izawa2001.PhysRevLett.87.057002}.
When quasiparticle excitations due to the Volovik effect are dominant, the excitation is actually field-direction dependent since the Doppler shift is proportional to the inner product of the Fermi velocity and the superfluid current velocity, the latter being perpendicular to the applied magnetic field~\cite{Vekhter1999}.
Such field-orientation dependent shift results in oscillations in the quasiparticle density of states, which are again accessible via thermal conductivity or specific heat as experimentally demonstrated in various unconventional superconductors~\cite{Matsuda2006.JPhysCondensMatter18.R705, Sakakibara2007.JPhysSocJpn.76.051004.review, Sakakibara2016.RepProgPhys.79.094002}.
In \sro, a small four-fold anisotropy in $\kappa_{ab}$ ($\sim 0.5$\% in peak-to-peak, $\kappa_{ab}(H\parallel [110])<\kappa_{ab}(H\parallel [100])$), superimposed to a relatively large two-fold anisotropy originating from the angle between the thermal current and the magnetic field, has been detected in the field range away from in-plane $\Hcc$~\cite{Izawa2001.PhysRevLett.87.057002}. 
The observed four-fold component is very small compared to theoretical predictions and thus it is concluded that it originates from the anisotropy of $\Hcc$.
In $\kappa_{c}$, in-plane anisotropy is not detectable with the accuracy of $0.5$\%~\cite{Tanatar2001.PhysRevLett.86.2649}, except for fields very close to $\Hcc$.
From the smallness of the in-plane anisotropy, both studies favor the presence of horizontal line nodes~\cite{Tanatar2001.PhysRevLett.86.2649,Izawa2001.PhysRevLett.87.057002}.

\subsection{\SECTIONSPECIFICHEAT}\label{Sec:SpecificHeat}

Information about the nodal structure has also been studied based on field-angle-resolved calorimetry.
The principle is similar to the Volovik effect for the thermal transport described in the previous subsection;
When magnetic field is applied perpendicularly to the Fermi velocity at a line node, quasiparticles are immediately excited leading to a $\sqrt{H}$ behavior in $C(H)/T$ at low temperatures.
Oscillation in the specific heat as a function of magnetic-field angle originates from low-energy excitations around nodes.
Compared with thermal transport, specific heat is not a directional probe and thus kinetic information of the quasiparticles is not detectable. Nevertheless, it has several advantages, such as not being influenced by various scattering mechanisms, or by extrinsic anisotropy due to applied thermal current.

For \sro, Deguchi~\etal studied the field-strength and angle dependence of the specific heat~\cite{Deguchi2004.PhysRevLett.92.047002}.
The field-strength dependence indeed shows $\sqrt{H}$ behavior at low fields.
Moreover, the $T$ and $H$ dependence of the specific heat under $H\parallel c$ obeys the scaling law for nodal superconductors.
Most importantly, a four-fold oscillation in the in-plane field-angle dependence of the specific heat deep inside the superconducting state was observed~\cite{Deguchi2004.PhysRevLett.92.047002,Deguchi2004.JPhysSocJpn.73.1313}.
The observed oscillation with specific heat smaller for $H\parallel [100]$ than [110] is attributed to the Volovic effect due to vertical gap minima located in the [100] directions on the $\gamma$ Fermi surface.
Note that the direction of the gap minima would be in agreement with a chiral $p$-wave theory by Nomura \etal~\cite{Nomura2002}, but $45^\circ$ rotated from that of the $d_{x^2-y^2}$ state. 
However, the gap in the $\gamma$ band for even-parity superconductivity (indicated by recent NMR studies~\cite{Pustogow2019.Nature.574.72}) would be large for the [100] direction in order to utilize the large density of states near the van-Hove singularity. 
This is closely related to the large enhancement of $\Tc$ under uniaxial strain along [100]~\cite{Hicks2014.Science.344.283,Steppke2017.Science.355.eaaf9398}.
Indeed, the nodal directions are now discussed to be [110] in a quasiparticle interference experiment~\cite{Sharma2020.ProcNatlAcadSci.117.5222} (discussed below) and in a orbital-singlet spin-triplet pairing theory (discussed in Sec.~\ref{Sec:Degeneracy} and Appendix~\ref{Sec:OrbitalBand}).

Later, it was theoretically revealed that the specific-heat oscillation due to the Volovik effect should exhibit a sign reversal at high temperatures or fields~\cite{VorontsovA2006.PhysRevLett.96.237001}, which is indeed experimentally observed in e.g. CeCoIn\sub{5}~\cite{An2010.PhysRevLett.104.037002}.
However, such $T$-induced sign reversal has not been resolved in \sro~\cite{Deguchi2004.PhysRevLett.92.047002,Deguchi2004.JPhysSocJpn.73.1313}.
This absence of sign reversal has been confirmed in a subsequent study by Kittaka~\etal performed down to 0.06~K~\cite{Kittaka2018.JPhysSocJpn.87.093703}. 
Because of this absence, the oscillation is attributed to a combination of horizontal line nodes and substantial Fermi-velocity anisotropy ($v\subm{F}\parallel [100] \ll v\subm{F}\parallel [110]$ for the $\gamma$ band due to the proximity to the van-Hove singularity), rather than the canonical Volovik effect from vertical line nodes.

To summarize these thermal measurements, the location of the node is not fully understood, although the existence of nodal quasiparticles is consistently revealed in a number of studies.
The existence of vertical line nodes is supported by the quasiparticle contribution obtained from the $\kappa(T\to 0)$ extrapolation.
The in-plane field-angle-dependent specific heat was first attributed to vertical line nodes in the $[100]$ direction but later interpreted as horizontal line nodes with substantial Fermi-velocity anisotropy.
The near absence of in-plane anisotropy in $\kappa$ suggests horizontal line nodes as well.
The interpretation of the $\kappa$ anisotropy, however, might be worth being revisited with updated knowledge.
For example, it is now empirically known that, in many superconductors, the experimentally observed quasiparticle density-of-states modulation due to the Doppler shift tends to be much smaller than the prediction by quasi-classical theories~\cite{Sakakibara2007.JPhysSocJpn.76.051004.review,Sakakibara2016.RepProgPhys.79.094002}.
The multi-band nature of many materials would add more complications toward an accurate theoretical calculation.
In addition, since the in-plane $\Hcc$ in \sro\ is now known to be determined by the Pauli effect, it is not trivial how a Pauli-limited $\Hcc$ anisotropy affects the low-field $\kappa_{ab}$ anisotropy.
Modern theoretical methodologies taking into account the realistic band structures and plausible superconducting gap models would be valuable to resolve these issues.

\begin{figure*}[htb]
\begin{center}
\includegraphics[width=15cm]{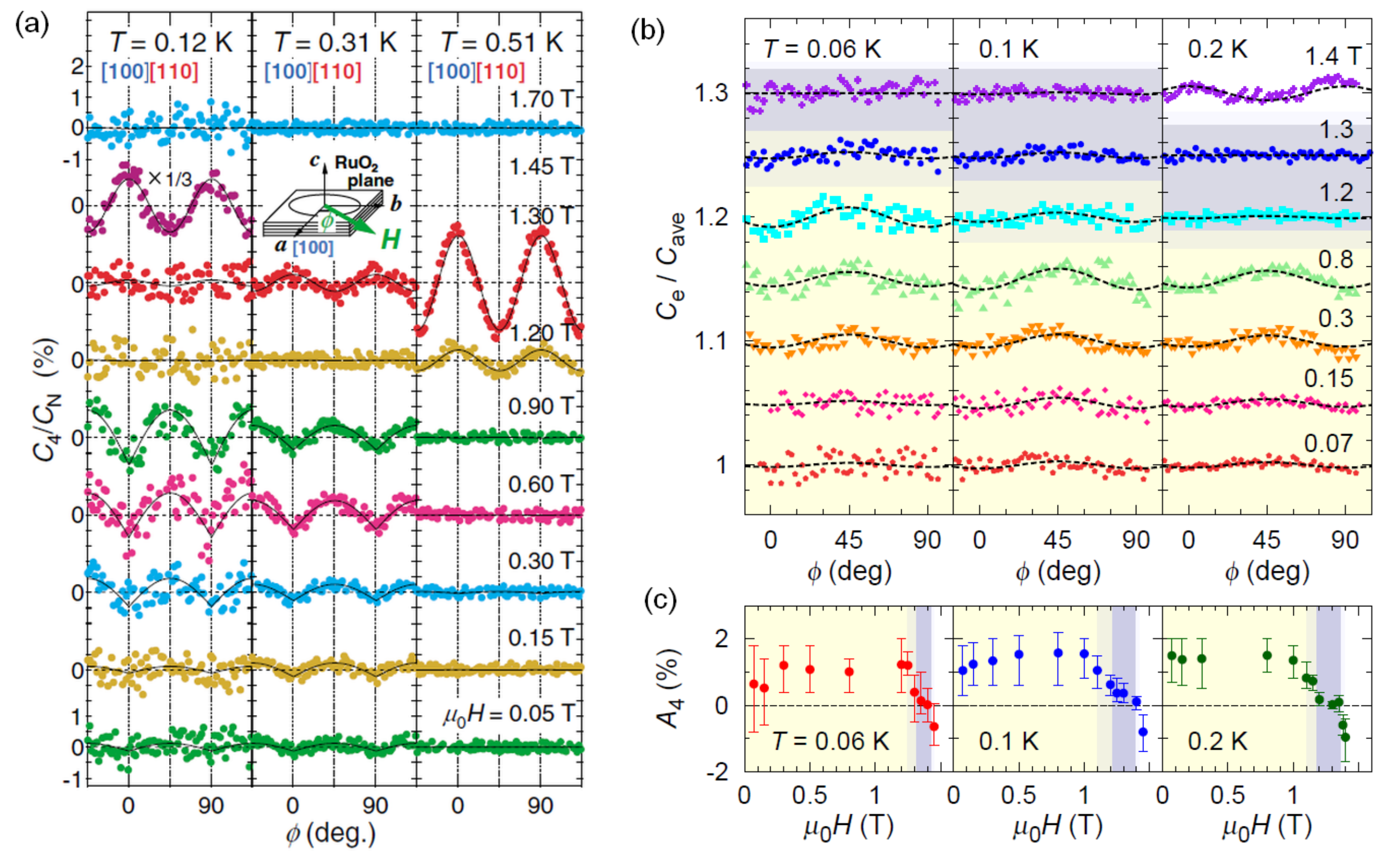}
\end{center}
\caption{In-plane field-angle dependence of the specific heat of \sro. 
(a) Specific-heat oscillation reported in Ref.~\citen{Deguchi2004.PhysRevLett.92.047002}. Here $C_4$ is the oscillatory component in the electronic specific heat and $C_{\mathrm{N}}$ is the normal-state electronic specific heat.
(b) Specific-heat oscillation reported in Ref.~\citen{Kittaka2018.JPhysSocJpn.87.093703}. Here $C_{\mathrm{e}}$ is the $\phi$-dependent electronic specific heat  while $C_{\mathrm{ave}}$ is its avaraged value over $\phi$.
They are essentially characterizing the same quantity, but notice the differences in the definitions of the vertical axes in (a) and (b). The results in (b) is consistent with those in (a), but importantly extended to lower temperatures.
(c) Field dependence of the oscillation amplitude normalized by field-induced part of the electronic specific heat, i.e. $A_4 = \Delta C\subm{e}/ [C\subm{e}(H)-C\subm{e}(H=0)]$, where $\Delta C\subm{e}$ is the oscillation amplitude, deduced from the data in (b).
(a) is reproduced from Refs.~\citen{Deguchi2004.PhysRevLett.92.047002} ($\copyright$~2004  American Physical Society); (b) and (c) from~\citen{Kittaka2018.JPhysSocJpn.87.093703} 
($\copyright$~2018 The Authors).
} 
\label{fig:thermal-conductivity_Deguchi}
\end{figure*}

\subsection{\SECTIONSTM}
\label{sec:STM_SRO}

Microscopic information on the nodal structure can also be obtained via  quasiparticle interference (QPI). 
The density of states oscillations around a lattice defect observed by scanning tunneling microscopy (STM) reflect  quasiparticle scattering processes. 
The Fourier transform of the oscillation pattern indicates the dominant scattering vectors at different energies.
For \sro, STM studies have been quite challenging because the crystalline surface tends to exhibit non-superconductive surface states originating from RuO\sub{6} octahedral rotations.
Nevertheless, this issue has been overcome by a low-temperature high-vacuum cleaving technique~\cite{Sharma2020.ProcNatlAcadSci.117.5222, Mueller2023}.
In these recent studies, superconducting properties are investigated on a cleaved SrO plane.
Note that the result is sensitive to the $\alpha$ and $\beta$ Fermi surfaces, which consist primarily of $d_{xz}$ and $d_{yz}$ orbitals with wavefunctions extending out of the RuO\sub{2} plane; whereas information on the $\gamma$ Fermi surface is not directly accessible via $c$-axis tunneling.
In the superconducting state, five distinct interference wavenumbers of Bogoliubov quasiparticles have been observed.
The results are well interpreted as line nodes along the $k_x \pm k_y = 0$ planes, corresponding to a gap of the $d_{x^2-y^2}$ type in the $\alpha$ and $\beta$ bands.

\begin{figure} 
\begin{center}
\includegraphics[width=8.5cm]{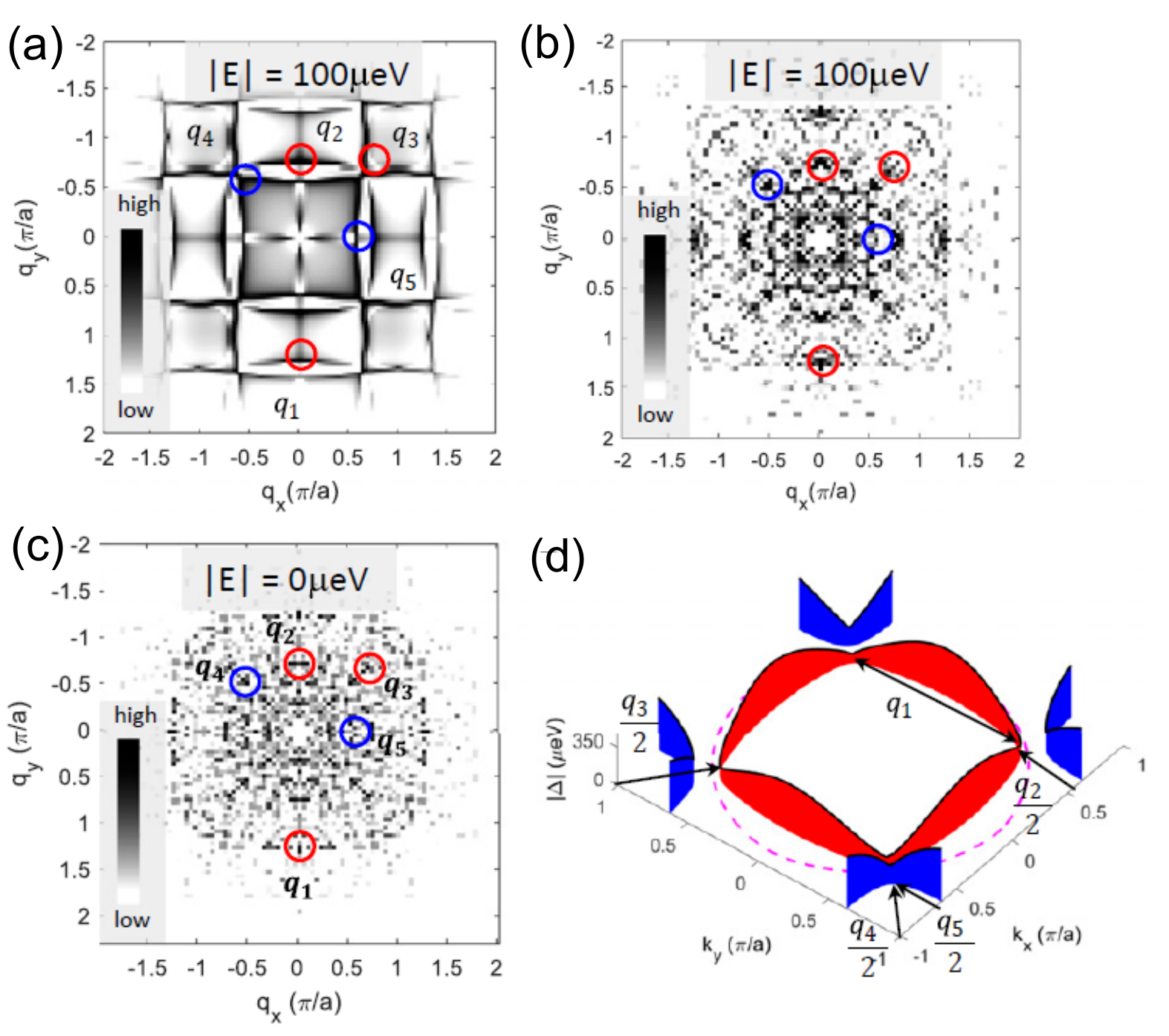}
\end{center}
\caption{QPI of \sro. 
(a) Predicted QPI pattern assuming dominant superconductivity in the $\alpha$ and $\beta$ bands as illustrated in the panel (d).
(b,c) Experimental QPI patterns measured at (b) $E = 100~\mu$V and (c) $0~\mu$V. 
The red and blue circles in (a-c) are dominant interference features predicted in the calculation (a) and indeed detected experimentally (b,c). 
(d) SC gap structure in the $\alpha$ (blue) and $\beta$ (red) bands suggested in this study.
Reproduced from Ref.~\citen{Sharma2020.ProcNatlAcadSci.117.5222} ($\copyright$~2020  National Academy of Sciences).}
\label{fig:H-T_phase-diagram_Sharma}
\end{figure}

\subsection{\SECTIONMYSTERY}
\label{sec:mystery_1}

Here we summarize the experimental studies on gap anisotropy in \sro. 
Almost all experiments are convincingly in agreement with non $s$-wave unconventional superconductivity and a spin-singlet-like character.
In contrast, the detailed momentum dependence of the gap structure and the presence and location of nodes have remained controversial.
Some experiments are in favor of a gap structure with vertical line nodes, other experiments are in favor of horizontal line nodes.
However, symmetry-protected horizontal line node at $k_z = 0$ naively implies that intralayer electrons cannot pair and is thus unnatural for a quasi-2D system like \SRO. We will come back to this mystery in Sec. \ref{sec:mystery_2}.
Nevertheless, we comment here that the vertical and horizontal nodal structures are not mutually exclusive. 
Thus, models with both vertical and horizontal line nodes, as described in next chapters, could be leading candidates to describe the apparent paradox.

\section{\SECTIONPHASESENSITIVE}\label{Sec:PhaseSensitive}

Quasiparticle and Josephson tunneling experiments may provide crucial information on the phase and magnitude of the superconducting energy gap.
It was recognized at an early stage that superconducting junctions of \SRO with conventional superconductors exhibit unusual behavior.
Unlike $d$-wave superconductors with $d_{x^2-y^2}$ symmetry such as the high-temperature superconductor YBa$_2$Cu$_3$O$_{7-\delta}$ (YBCO), the critical current $I\subm{c}$ of a junction of \SRO with a conventional superconductor shows irreproducibility or dynamical behavior as long as the junction size is more than a few micrometers~\cite{Kidwingira2006.Science.314.1267}.
It was also recognized that the $I\subm{c}$ of a SQUID device based on \SRO exhibits behavior suggesting odd parity~\cite{Nelson2004.Science.306.1151}. 

Thin films are used to investigate the superconducting junction properties of many superconductors, but it is a challenge to obtain \SRO thin films exhibiting the intrinsic $T\subm{c}$ of 1.5 K as for bulk crystals.
Nevertheless, there has been substantial recent progress in producing \SRO superconducting thin films.
Up to now, single crystals were mainly used for junction studies and the focused ion beam (FIB) technique was used for micro-fabrication of junctions and other micro-structured devices.

In this section, we review some key results from a variety of tunneling experiments using \SRO and, when necessary, attempt to reinterpret the results from the perspeective of an even-parity superconducting state.
Additional details on tunneling experiments with \SRO can be found in the review papers by
Liu and Mao ($\copyright$~2015)~\cite{Liu2015.PhysicaC.514.339},
Leggett and Liu ($\copyright$~2019)~\cite{Leggett2020},
and Anwar and Robinson ($\copyright$~2021)~\cite{Anwar2021}.

\subsection{\SECTIONFILMS}
\label{Sec:Thin films}

Perhaps the first report of successful thin film growth of \SRO was in 2010 by using pulse laser deposition (PLD)~\cite{Krockenberger2010}. 
Several groups reported improved superconducting films by molecular beam epitaxy (MBE) and PLD as summarized in Table \ref{tab:Thin films}. 
Uchida \etal~\cite{Uchida2017, Uchida2019} concluded that suppression of the ruthenium deficiency in the film is crucially important. 
Nair \etal~\cite{Nair2018} clarified detailed conditions for a successful MBE growth. 
It is interesting that the $T\subm{c}$ becomes higher than 1.5 K probably because the (110) substrate provides anisotropic in-plane strain to the \SRO film: the misfit strain at room temperature is $-$0.39\% along the [001] direction of the substrate and $-$0.16\% along the [1\={1}0] direction of the substrate.
Goodge \etal~\cite{Goodge2022} also suggest that ruthenium vacancies play a more important role than inclusion of local higher-$n$ Sr$_{n+1}$Ru$_n$O$_{3n+1}$ syntactic intergrowths in suppressing superconductivity. 

Thin film growth by PLD has been improved recently. Garcia \etal~\cite{Garcia2020} identified mosaic twist as a key in-plane defect that suppresses superconductivity. 
Thin films exhibiting a sharp resistive transition with onset at 1.15 K and a transition width of 0.1 K were grown by PLD by Kim \etal~\cite{Kim2021}
By using single-crystalline Sr$_3$Ru$_2$O$_7$ ($n=2$ member of the Ruddlesden-Popper (R-P) series) as a target, they were able to induce a $n=2$ intergrowth near the substrate interface that blocks the formation of out-of-phase boundaries between the LSAT substrate ($n=\infty$) and \SRO ($n=1$) film. 
Superconducting thin films by hybrid MBE using a metal--organic precursor of ruthenium were also reported~\cite{Choudhary2023}.

Various phase-sensitive experiments using these superconducting films are awaited at present.

\begin{table*}[ht]
\begin{center} 
\caption{Superconductive thin films of \sro. Those reporting the zero-resistivity films are listed.
LSAT: La$_{1-x}$Sr$_x$Al$_{1-y}$Ti$_y$O$_3$. RRR = $R$(300 K)/$R$(4.2 K) is the residual resistivity ratio.}
\label{tab:Thin films}
\begin{tabular}{lccccl}
\hline\hline 
Authors (Year) & Method  & Substrate & RRR  & $\Tc$(onset)\slash $\Tc$(${R=0}$)  & Remarks  \\
\hline
Krockenberger \etal (2010)~\cite{Krockenberger2010}  & PLD  & LSAT(001)  & 82  & 0.9 K \slash 0.6 K  & 96 nm thick  \\ 

Uchida \etal (2017)~\cite{Uchida2017}  & MBE  & LSAT(001)  & 30   & 1.1 K \slash 0.8 K  &   \\ 

Uchida \etal (2019)~\cite{Uchida2019}  & MBE  & LSAT(001)  & -   & 1.3 K \slash 0.8 K  &   \\ 

Nair \etal (2018)~\cite{Nair2018}  & MBE  & NdGaO$_3$(110)  & 69   & 2.0 K \slash 1.6 K  & anisotropic in-plane misfit  \\ 

Goodge \etal (2022)~\cite{Goodge2022}   & MBE  & NdGaO$_3$(110)  & 69   & 2.0 K \slash 1.6 K   & 55 nm thick  \\ 

Garcia \etal (2021)~\cite{Garcia2020}  & PLD  & LSAT(001)  & 100   & 1.1 K \slash 0.9 K  & single-crystalline Sr$_3$Ru$_2$O$_7$ target  \\ 

Kim JK \etal (2021)~\cite{Kim2021}  & PLD  & LSAT(001)  & 34   & 1.15 K \slash 1.05 K  & single-crystalline Sr$_3$Ru$_2$O$_7$ target  \\ 

Choudhary \etal (2023)~\cite{Choudhary2023}  & hybrid MBE  & LSAT(001)  & 22   & 0.95 K \slash 0.75 K  & metal-organic precursor  \\ 
\hline
\hline
\end{tabular}
\end{center}
\end{table*}


\subsection{\SECTIONTUNNELING}
\label{qp tunneling}

\begin{figure} [ht]
\begin{center}
\includegraphics[width=8.0cm]{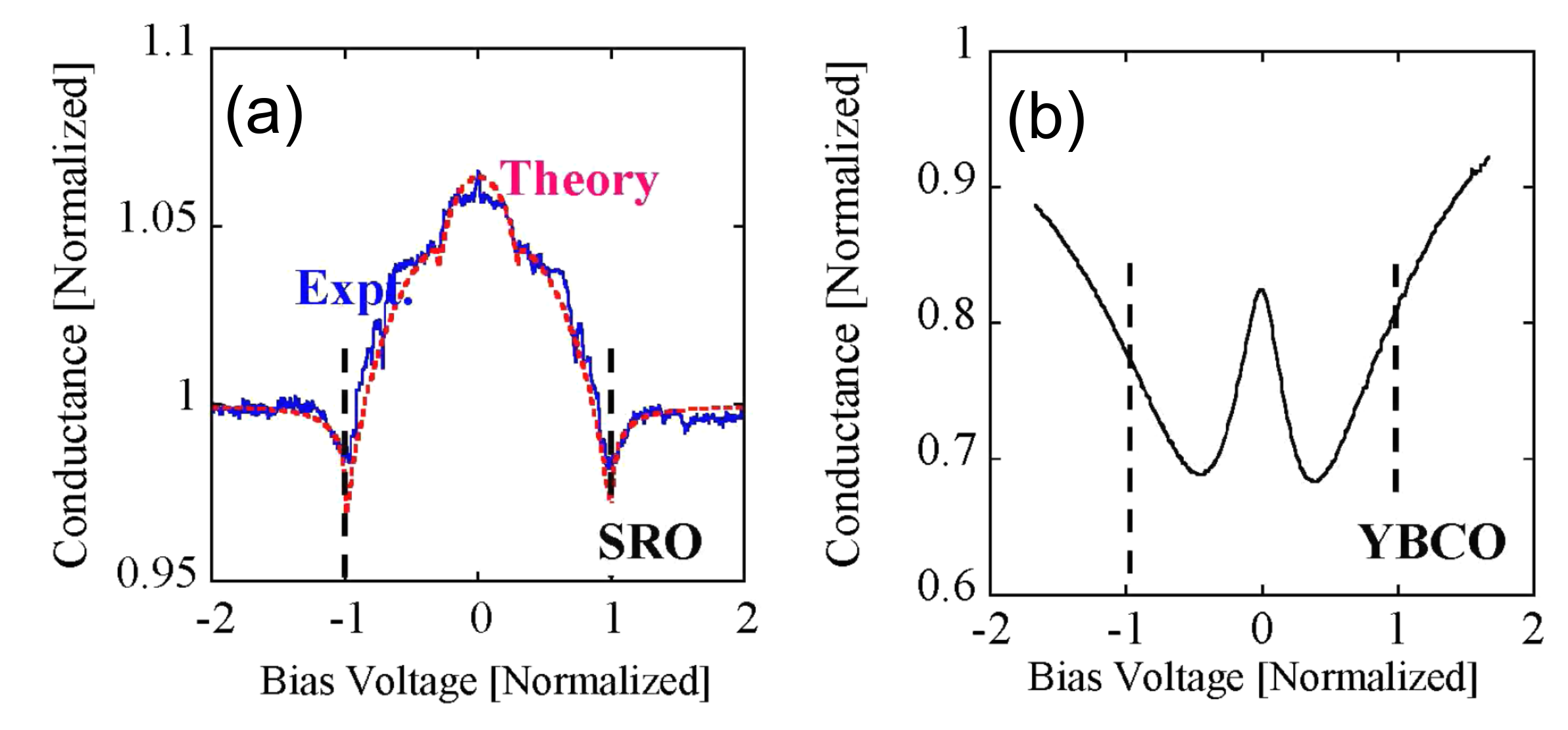}
\end{center}
\caption{Contrasting zero-bias conductance peaks (ZBCPs) in the quasiparticle tunneling experiments reflecting topological edge states. In \SRO, the broad ZBCP is interpreted as due to chiral or helical edge state; the sharp ZBCP along the [110] direction in the high-$\Tc$ cuprate YBCO is interpreted as due to the flat-band edge state.
Reproduced from Ref.~\citen{Kashiwaya2011.PhysRevLett.107.077003} ($\copyright$~2011  American Physical Society).} 
\label{fig:Topo-edge-ZBCP}
\end{figure}

The gap structure determination with STM, suggesting vertical line nodes in the gap, was already discussed in section \ref{sec:STM_SRO}.
Andreev bound states on the surface of \SRO were investigated with S/I/N junctions consisting of Au deposited on the surface of \SRO~\cite{Kashiwaya2011.PhysRevLett.107.077003}.
These junctions exhibit an unusually broad zero-bias conductance peak (ZBCP) with the width comparable to the gap width as shown in Fig. \ref{fig:Topo-edge-ZBCP}.
This wide ZBCP is in sharp contrast with the narrow ZBCP in YBCO, which represents the Andreev bound states formed at the edge of the $d_{x^2-y^2}$ superconductor~\cite{Kashiwaya2000}.
The broad ZBCP in the \SRO junctions was interpreted as due to chiral edge states, but later argued that it could also be explained by helical edge states of a spin-triplet helical superconductor~\cite{Tanaka2009, Tanaka2020}. We add that there are several recent theoretical studies examining the surface Andreev bound states and tunneling spectra of \SRO~\cite{Ando2022, Suzuki2022, Ikegaya2021, Tamura2017}.

\begin{figure} [ht]
\begin{center}
\includegraphics[width=6cm]{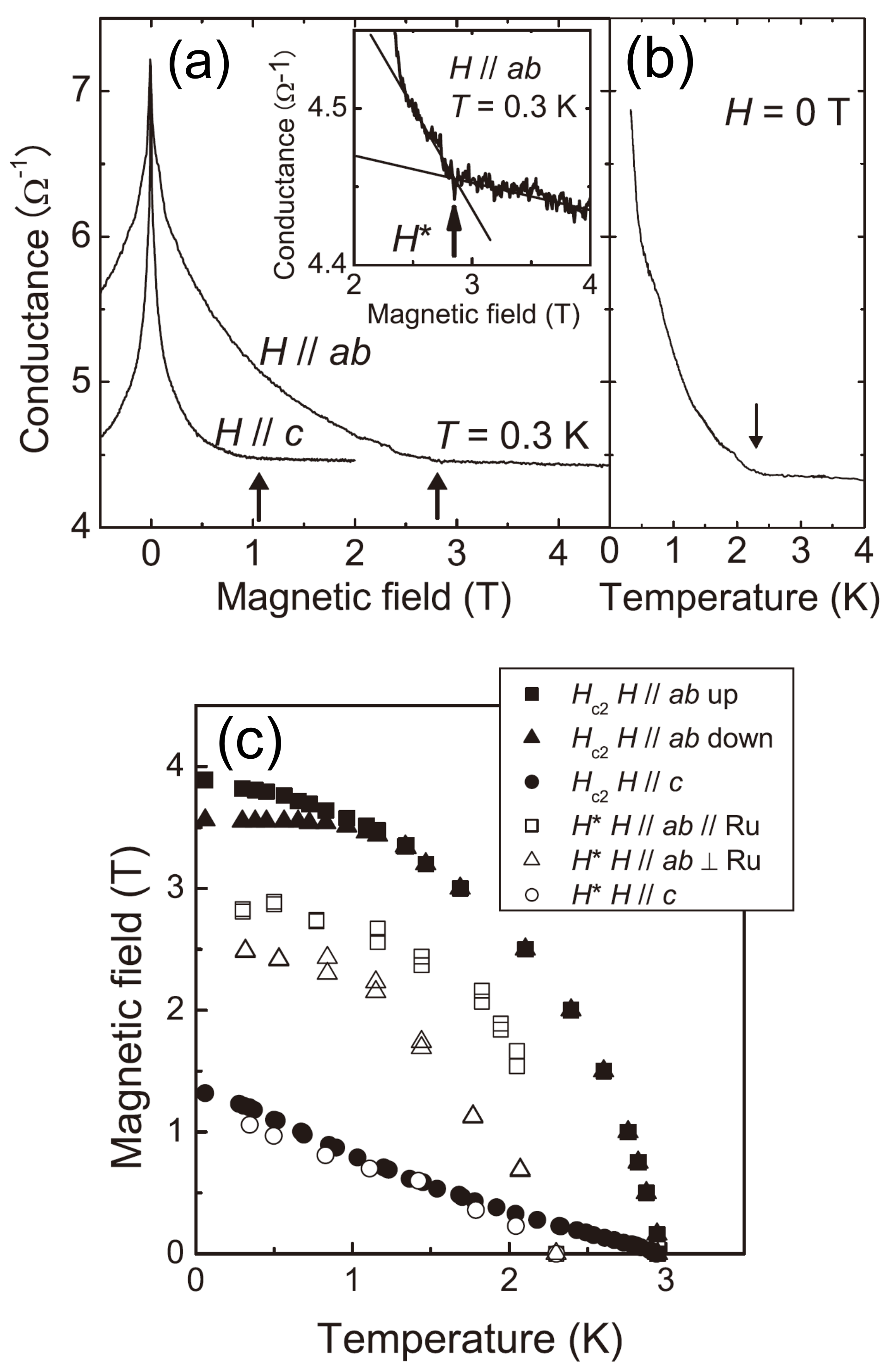}
\caption{Emergence of the zero-bias conductance peak (ZBCP) in a \SRO/Ru junction. 
(a) The onset of ZBCP is at field $H^\ast$ that is somewhat smaller than $H\subm{c2} (T)$. 
(b) The onset temperature of the ZBCP in zero field. 
(c) In the field-temperature phase diagram, the onset of ZBCP shown by the open symbols occurs in the $H-T$ region inside the superconducting phase.
Reproduced from Ref.~\citen{Kawamura2005} ($\copyright$~2005 The Authors).
}
\label{fig:H-T_phase-diagram_Kawamura}
\end{center}
\end{figure}

The ZBCP in the enhanced-$T\subm{c}$ phase of \SRO-Ru eutectic system was investigated using junctions consisting of the bulk \SRO and a micron-size ruthenium metal inclusion~\cite{Kawamura2005}. 
It was found that the onset of the ZBCP is at somewhat smaller values of $H(T)$ compared to the 3-K superconductivity $H\subm{c2}(T)$ (Fig. \ref{fig:H-T_phase-diagram_Kawamura}).
One general criterion of ZBCP is the presence of in-gap edge states. 
For example, for YBCO with flat-band edge states associated with the vertical line nodes along the <110> directions, the ZBCP becomes particularly strong along the <110> directions and absent only in the <100> directions~\cite{Tanaka1995PRL, Iguchi2000}. 
This behavior can be viewed in terms of the flat-band topological edge state~\cite{Sato2011}. For chiral gapless SC states, the edge states appears in all directions. 
So the interpretation of this experiment is that it captured ZBCP spreading in all $k$-directions.
Considering that the ZBCP is a consequence of chiral superconducting OP, the observation of split transition temperatures in Fig.~\ref{fig:H-T_phase-diagram_Kawamura} suggests that the time-reversal symmetry breaking (TRS breaking) emerges in the $H-T$ region somewhat inside the $H\subm{c2} (T)$ of the 3-K superconductivity, as predicted by Kaneyasu \etal~\cite{Kaneyasu2010}
The splitting of $\Tc$ and the onset of TRS breaking is an important current issue that will be discussed further in Sec. 8.

\subsection{\SECTIONDYNAMICAL}
\label{sec:dynamical 1.5 K}

\begin{figure*} [ht]
\begin{center}
\includegraphics[width=14cm]{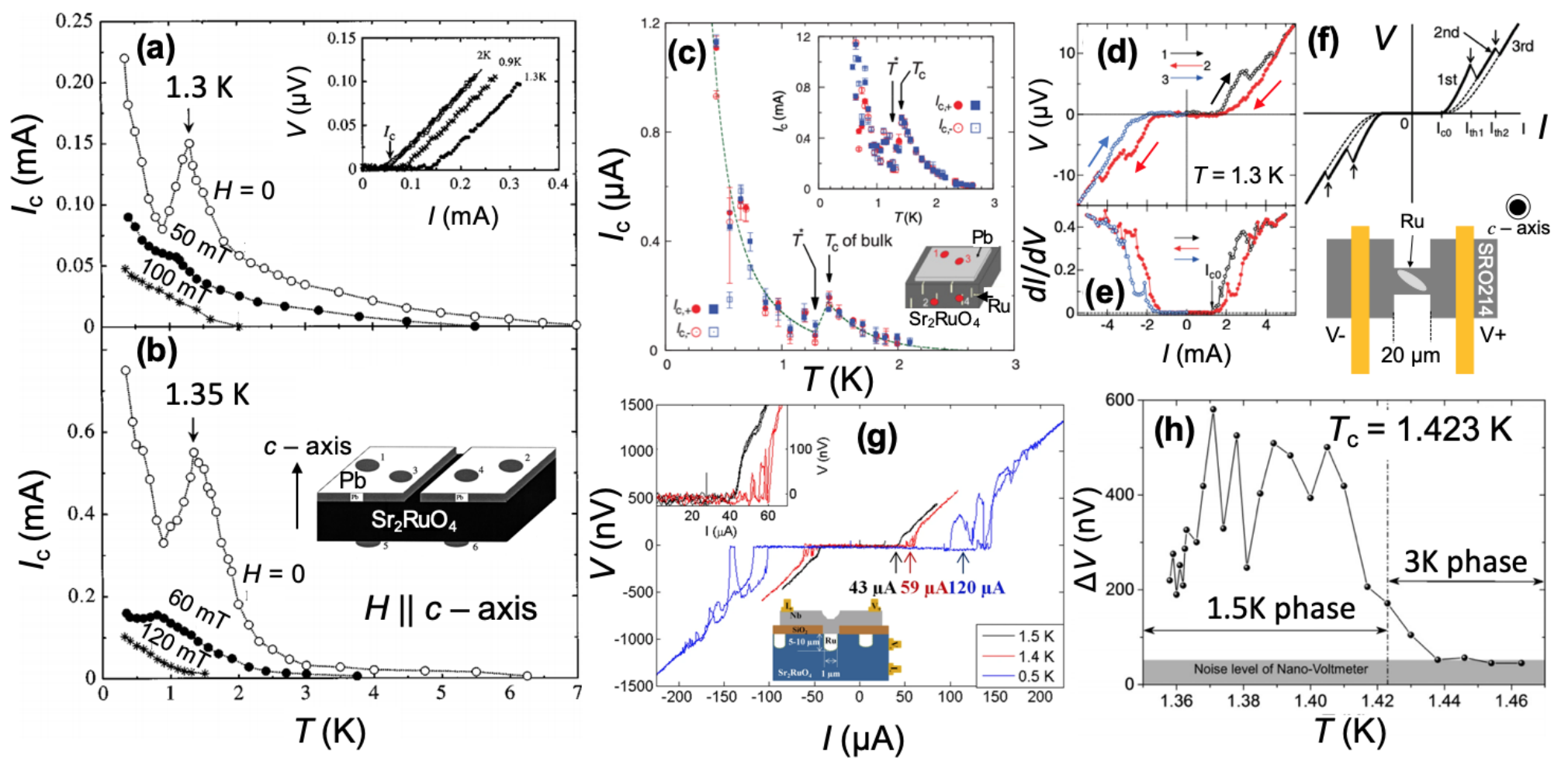}
\end{center}
\caption{Stable $I\subm{c}$ above 1.5 K, and unstable behavior below 1.5 K in various junctions involving eutectic \SRO-Ru crystals. See text for brief descriptions of these figures.
Reproduced from Ref.~\citen{Anwar2021} ($\copyright$~2021 The Authors).
} 
\label{fig:Anwar_tunneling-review}
\end{figure*}

As another feature different from other superconducting junctions including those of high-$\Tc$ cuprates, a dynamical behavior of tunneling junctions consisting of \SRO was first reported by Kidvingra \etal~\cite{Kidwingira2006.Science.314.1267}
In that study, the critical current under external magnetic field, the $I\subm{c}$-vs-$H$ patterns, of in-plane S/N/S’ junctions of SRO/Cu/Pb varies significantly with thermal cycling: changing from Fraunhofer-like to more complicated patters. 
Hysteric patterns in the $H$-sweep, abrupt changes in $I_c$, and time-dependent telegraphic noise behavior are also observed. 
Such behavior was interpreted as due to the formation of superconducting domains with different order-parameters, and ascribed in terms of chiral domain boundaries of a chiral $p$-wave superconductor with the OP $p_x \pm i p_y$. 
From today's  viewpoint, interpretations with $d \pm id$ states may also be applicable. 
Such dynamical junction behavior was observed by other groups as summarized in Fig. \ref{fig:Anwar_tunneling-review}.~\cite{Nakamura2011.PhysRevB.84.060512R, Nakamura2012, Anwar2013.SciRep.3.2480, Anwar2017, Leggett2020}  
In junctions of SRO/Ru/$s$-wave superconductor~\cite{Anwar2017} (Figs. \ref{fig:Anwar_tunneling-review} (g and h)), the critical current $I\subm{c}$ emerging below 3 K is stable down to 1.5 K, but becomes unstable when the same device is cooled below 1.5 K. 
This behavior suggests phase competition due to superconducting domain formation in the 1.5-K phase. We will describe such behavior in more detail in the next subsection. 

\subsection{\SECTIONJUNCTIONS}\label{Sec:Junctions}

An unusual drop in $I\subm{c}(T)$ on cooling across 1.5 K was accidentally discovered in junctions of \SRO with lead as shown in Fig.\ref{fig:Anwar_tunneling-review} (a) and (b)~\cite{Jin1999}.
It was later demonstrated that the essential junction configuration for this behavior is \SRO/Ru/($s$-wave SC) involving eutectic segregation of Ru metal platelets, Fig.\ref{fig:Anwar_tunneling-review} (c)~\cite{Nakamura2012}. 
Due to surface reconstruction of \SRO (Fig. \ref{Fig:Surface_reconstruction}), a proper junction contact cannot be made directly on the $ab$-plane. 
Metallic Ru inclusions exposed on the surface of \SRO, instead, bridge between in-plane conduction of \SRO and the conduction of the $s$-wave superconductor. 
A topological competition between proximity induced $s$-wave in a Ru micro-island and the superconductivity of \SRO \textit{surrounding the Ru micro-island} explains the observed unusual behavior of $I\subm{c}$, provided that the 3-K and 1.5-K phases have different phase-winding numbers~\cite{Nakamura2011.PhysRevB.84.060512R}. 
Similar “topological junctions” of \SRO/Ru/Nb in which only one or a few Ru micro-islands are surrounded by \SRO (Fig.\ref{fig:Anwar_tunneling-review} (d-h)) exhibit poor reproducibility among different cooldowns, hysteresis behavior ascribable to the superconducting domain-wall motion~\cite{Kambara2008.PhysRevLett.101.267003}, asymmetry between positive and negative current directions (the superconducting diode effect)~\cite{Anwar2023}, and time-dependent telegraphic noise only below 1.5~K~\cite{Anwar2017}. 
These experiments highlight the qualitative difference of superconducting OPs between the 1.5-K and 3-K phases of \SRO.

\subsection{\SECTIONPROXIMITY}\label{Sec:Proximity}

Superconducting proximity effects into a ferromagnet have recently been actively studied~\cite{Eschrig2008}.
In particular, the so-called long-range proximity effect by the injection of spin-triplet even-parity odd-frequency pairs into a ferromagnet has been demonstrated experimentally by a number of groups as reviewed in Ref.~\citen{Eschrig2015.Rep.Prog.Phys.}.
To realize such a state using an $s$-wave superconductor and a ferromagnet, an insertion of a non-uniformly magnetized ferromagnetic layer is essentially required.
In contrast, Anwar \etal~\cite{Anwar2016.Nat.Commun.} reported that the long-range proximity effect into the ferromagnet SrRuO$_3$ is realized in epitaxial films of SrRuO$_3$ grown on a \SRO crystal as a substrate, without any inhomogeneous magnetic layer.
The tunneling spectra of the Au/SrRuO$_3$/\SRO junctions suggest the presence of a superconducting gap across SrRuO$_3$ with a proximity coherence length of about 9 nm.
This is much longer than the coherence length of 1 nm expected for a spin-singlet proximity effect into SrRuO$_3$.

\subsection{\SECTIONSQUID}\label{Sec:SQUID}

The superconducting phases of \SRO have also been investigated using microstructures such as \SRO-Ru-Nb superconducting quantum interference devices (SQUIDs) as well as \SRO micro-rings. 
These results strongly suggest the formation of superconducting domains in \SRO and the competition between superconducting phases in \SRO and the $s$-wave superconductor.

After a series of developments of \SRO-based SQUIDs using the FIB technique, Nago \etal~\cite{Nago2016} reported the evolution of superconducting paths in direct-current SQUIDs (dc-SQUIDs) of Nb/Ru/\SRO below 1 K.
Their SQUID consists of a bridge-shaped Nb structure formed on one Ru island surrounded by \SRO.
The evolution of $I\subm{c}$ with temperature, especially the collapse of the SQUID behavior below 0.4 K, corresponding to the onset of the bulk superconductivity of Ru metal, lead to the conclusion that the phase competition between chiral superconductivity in \SRO and $s$-wave superconductivity in Ru occurs.

After evidence for half-quantum fluxoid (HQF) was reported in magnetic torque measurements on micro-rings of single-crystalline \SRO samples~\cite{Jang2011.Science.331.186}, a few groups reported quantum oscillation phenomena in similar micro-rings~\cite{Yasui2017.PhysRevB.96.180507, Yasui2020, Cai2013.PhysRevB.87.081104R, Cai2022}. 

\begin{figure*} [ht]
\begin{center}
\includegraphics[width=14cm]{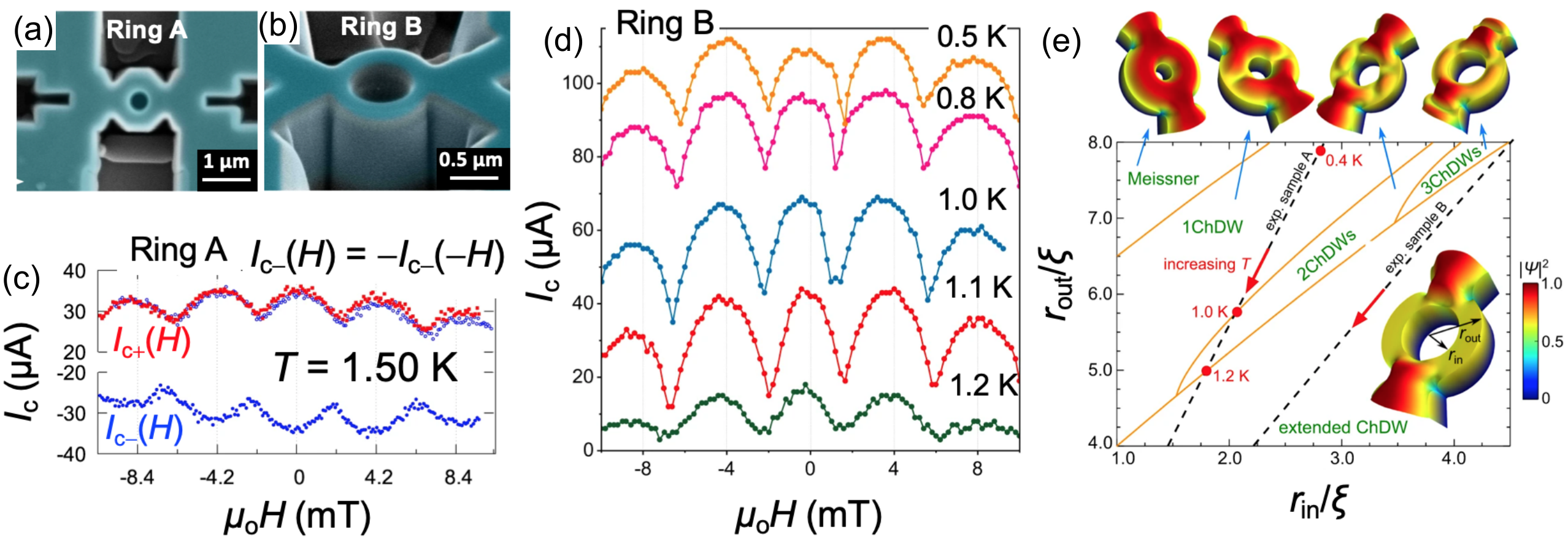}
\end{center}
\caption{Micro-ring experiments exhibiting a spontaneous DC-SQUID behavior. (a)-(d) The oscillations in the critical current $I\subm{c}$ in microrings indicate a spontaneous formation of weak links in the rings to from a SQUID. (d) It is expected that superconducting domain walls form in a certain range of the aspect ratio of the ring. Reproduced from Ref.~\citen{Yasui2020} ($\copyright$~2021 The Authors).} 
\label{fig:Microrings_Yasui_npj-QM}
\end{figure*}

Yasui \etal~\cite{Yasui2017.PhysRevB.96.180507} studied Little-Parks oscillations in micro-rings of \SRO crystals.
The quantum oscillations were studied in square and circular rings at temperatures slightly below $\Tc$, and a resistance oscillation reflecting the oscillations of $\Tc$ were recorded.
The field dependence of the resistance was quantitatively consistent with those expected from integer-fluxoid quantization.
In addition, they observed features at half the period ascribable to the formation of HQF.
With improved fabrication of micro-rings of a variety of sizes, oscillations of the critical current ascribable to a spontaneous formation of a dc-SQUID with a pair of weak links were observed, despite the apparently homogeneous nature of these rings~\cite{Yasui2020} (Fig. \ref{fig:Microrings_Yasui_npj-QM}).
Such behavior, found in narrow-arm micro-rings, was interpreted as the spontaneous formation of a pair of superconducting domain walls associated with distinct chiral domains on either side of the ring.

\begin{figure} [ht]
\begin{center}
\includegraphics[width=7cm]{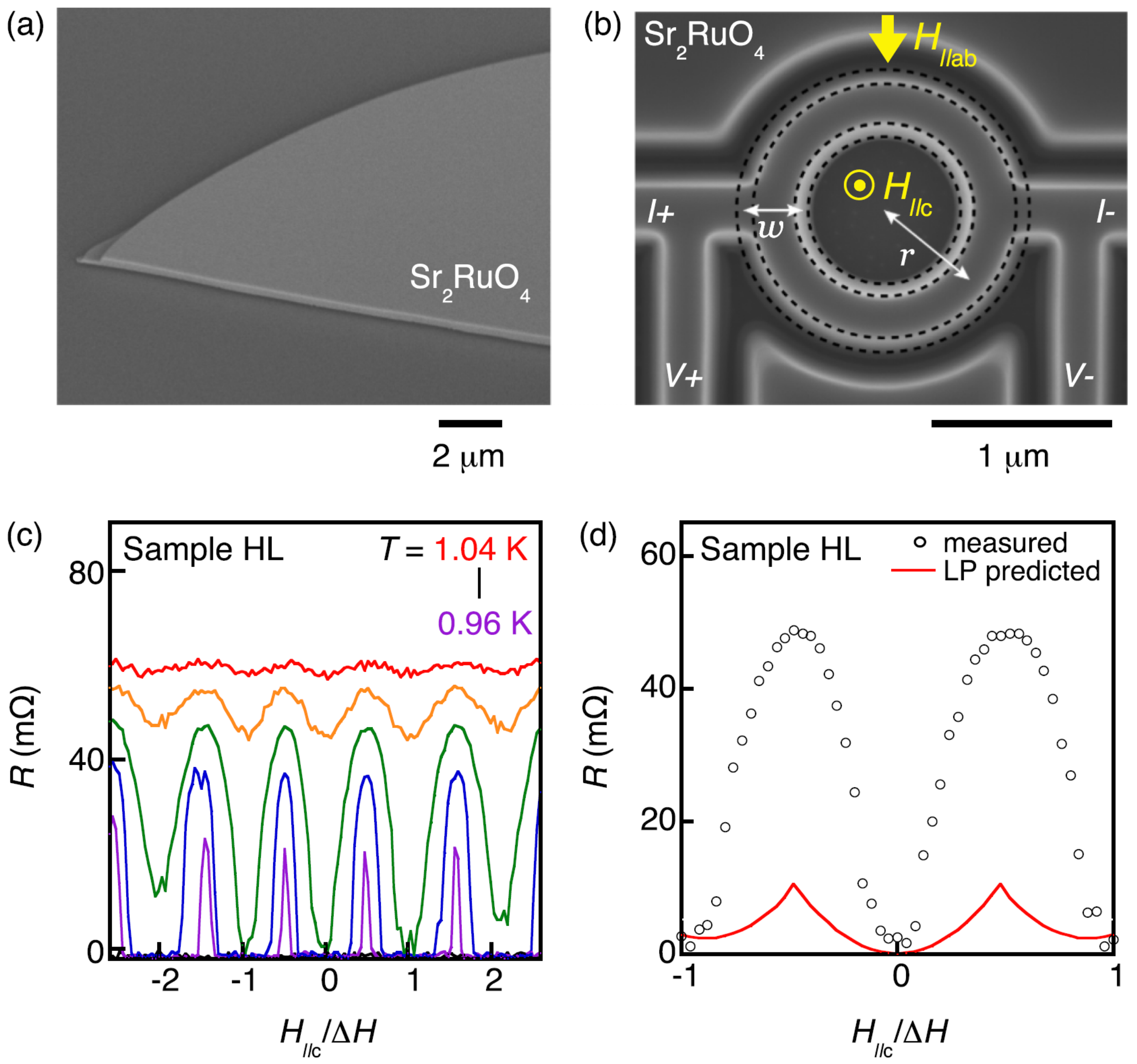}
\end{center}
\caption{Micro-ring experiments exhibiting huge resistance oscillations under magnetic field. (a) Before and (b) after a microring of \SRO is formed with FIB. (c) and (d) The observed oscillation amplitudes are much greater than the expectation from the Little-Parks oscillations. BeforReproduced from Ref.~\citen{Cai2022} ($\copyright$~2022 American Physical Society).} 
\label{fig:Microrings_Cai_PRB}
\end{figure}

Magnetoresistance oscillations in micro-rings at rather different measurement conditions were performed by Cai \etal~\cite{Cai2013.PhysRevB.87.081104R, Cai2022}
They measured the magnetoresistance at 0.3 K, much lower than $\Tc$ and under in-plane magnetic field, typically 400 to 1000~Oe, much larger than the unit period of flux quantization, with a small out-of-plane field of 20 Oe.
The large in-plane field was applied in order to stabilize the HQF states expected in an equal-spin-paired spin-triplet state.
Under such conditions, they observed magnetoresistance oscillations with a large magnitude attributable to the motions of vortices across the current (Fig.  \ref{fig:Microrings_Cai_PRB}).
They found additional secondary peaks that they ascribed to HQF. 

\section{\SECTIONMULTIOP}\label{Sec:MulticomponentOP}

\subsection{\SECTIONLANDAU}
\label{Sec:Landau}

The Landau theory of phase transitions assumes that the transition from a disordered to an ordered phase of matter can be captured by a free energy written as a Taylor expansion on the respective order parameter.
The formalism only assumes that the free energy is analytic in the order parameter and that it should obey all the symmetries of the Hamiltonian in the disordered (or normal) state, such that the symmetry breaking occurs spontaneously.

In the context of superconductivity, the simplest form of the Landau theory that describes the superconducting phase transition is written in terms of a single-component order parameter as a complex number $\Psi$.
The free energy should not depend on the global phase of the order parameter, as global phases of quantum states are not observable.
As a consequence, only terms proportional to $|\Psi|^2$ and its powers are allowed.
In addition, following the group theory discussion in Sec. \ref{Sec:OrderParameter}, if the order parameter is single-component, it is associated with a one-dimensional irreducible representation.
Consequently $|\Psi|^2$ always belongs to the  trivial irrep $A_{1g}$, as can be seen from the product Table \ref{Tab:D4hProduct}.
In this light, the most general free energy for a single component order parameter associated with a one-dimensional irrep (labelled as $1D$) in an homogeneous system (neglecting gradient terms) can be written as:
\begin{eqnarray}
F_{1D} = F_0(T)+\alpha(T) |\Psi|^2 + \frac{\beta}{2}|\Psi|^4+ ...
\end{eqnarray}
where $F_0(T)$ is the free energy of the normal state,
$\alpha$ and $\beta$ are phenomenological parameters,
and the dots correspond to higher order terms in $|\Psi|^2$.
In particular, $\beta$ should be positive for the stability of the ordered phase within this expansion up to fourth-order.
We can write explicitly $\alpha(T) = a (T/\Tc-1)$, with $a$ a positive phenomenological constant, $T$ the temperature, and $\Tc$ the superconducting critical temperature.
Note that the factor $\alpha$ is positive for $T>\Tc$ and negative for $T<\Tc$.
Minimizing the free energy with respect to $\Psi$ for $T>\Tc$ gives us a single minimum for $\Psi = 0$, while minimizing the free energy for $T<\Tc$ gives us minima with amplitude of the order parameter $|\Psi| = \sqrt{-\alpha(T)/\beta}$ and arbitrary phase.

As discussed in Section \ref{Sec:OrderParameter}, there are order parameters associated with multi-dimensional irreps which are therefore multi-component.
For these cases, the free energy needs to be generalized, including new types of fourth-order terms.
Consider a superconducting order parameter with two components, $\Psi_1$ and $\Psi_2$ transforming according to $E_{g/u}$.
From the product table displayed in Table \ref{Tab:D4hProduct}, we find that the product $E_{g/u}\otimes E_{g/u}$ has a component that transforms as the trivial irrep $A_{1g}$,
but also components that transform according to the non-trivial irreps $A_{2g}$, $B_{1g}$, and $B_{2g}$.
Even though the second-order products associated with non-trivial one-dimensional irreps are not allowed in the free energy, their squares (fourth order products) are allowed as the product of a non-trivial one-dimensional irrep with itself, always gives the trivial irrep $A_{1g}$. 

To make these statements more concrete, we can choose $\Psi_1 \sim k_x$ and $\Psi_2\sim k_y$ and write explicitly the three gauge-invariant bilinears: $|\Psi_1|^2+|\Psi_2|^2\sim k_x^2+k_y^2$ in $A_{1g}$, $|\Psi_1|^2 - |\Psi_2|^2\sim k_x^2 - k_y^2$ in $B_{1g}$, and $\Psi_1^*\Psi_2 - \Psi_1\Psi_2^*\sim k_xk_y$ in $B_{2g}$ (See Table~\ref{Tab:D4hCharacter}).
In second order, only $|\Psi_1|^2+|\Psi_2|^2$ transforms trivially, so this is the only symmetry allowed term. In fourth order, we can take the square of the bilinear in $A_{1g}$, $(|\Psi_1|^2+|\Psi_2|^2)^2\sim (k_x^2+k_y^2)^2  = k_x^4 + 2 k_x^2k_y^2 + k_y^4$, which is naturally also symmetry allowed. Note that we can also take the square of the bilinear in $B_{1g}$, $(|\Psi_1|^2 - |\Psi_2|^2)^2\sim (k_x^2 - k_y^2)^2 = k_x^4- 2 k_x^2k_y^2 + k_y^4$. 
Note that this fourth order term is also invariant under all point-group operations, what allows us to rewrite two distinct symmetry allowed fourth order terms as $|\Psi_1|^4+|\Psi_4|^4$ and $|\Psi_1|^2|\Psi_2|^2$. 
Furthermore, we can take the square of the bilinear with $B_{2g}$ symmetry, $(\Psi_1^*\Psi_2 - \Psi_1\Psi_2^*)^2\sim k_x^2k_y^2$, which is also invariant, and therefore allowed in the free energy.
A more detailed discussion on the construction of the free energy is given in \cite{Sigrist1991, Ramires2022Chiral}.

In summary, the most general form for the free energy of a two-component order parameter associated with the two-dimensional ($2D$) irreps $E_{g/u}$ in the $D_{4h}$ point group is:
\begin{multline}
\label{eq:F_D4h}
F_{2D} = F_0(T)+\alpha(T) |\vec{\Psi}|^2 + \beta_1 |\vec{\Psi}|^4\\
+ \beta_2 (\Psi_1 \Psi_2^* - \Psi_1^* \Psi_2)^2 + \beta_3|\Psi_1|^2 |\Psi_2|^2 + ....,
\end{multline}
where $|\vec{\Psi}|^2  = |\Psi_1|^2 + |\Psi_2|^2$, and $\beta_i$ ($i=1,2,3$) are phenomenological parameters. 
One example of a two-dimensional order parameter is the chiral $d$-wave with components $\{d_{xz},d_{yz}\}$, which are related by four-fold rotational symmetry along the $z$-axis. 
Note that the term proportional to $\beta_2$ is the key to the development of TRS breaking in the superconducting state, as it is the only term that depends explicitly on the relative phase of the two components. 
If $\beta_2>0$, the free energy is minimized for two components with a relative phase $\phi = \pm \pi/2$,
while if $\beta_2<0$, the free energy is minimized for a relative phase of $\phi = \{0, \pm \pi\}$.
The condition on the relative phase between the two components needs to be supplemented by the condition on other terms that guarantee that the two components develop a finite amplitude \cite{Sigrist1991, Ramires2022Chiral}.
As another example, the nematic $d$-wave with $d_{xz}$ is also considered as two-component 2$D$ order parameter and the free-energy minimum is evaluated by keeping the order-parameter fluctuation terms of both components.

Interestingly, there are scenarios based on two accidentally degenerate superconducting order parameters (see further discussion in Section \ref{Sec:TheoAccidental}). 
In that case, the order parameter can again be thought of as composed of two components $\{\Psi_1,\Psi_2\}$, but these are not symmetry related and do not have the same critical temperature unless the system is fine-tuned. 
One example consists of the combination of two 1$D$ irrep order parameters, with components $\Psi_1=d_{x^2-y^2}$ and $\Psi_2=g_{xy(x^2-y^2)}$, referred to as $d+ig$.
The free energy reads:
\begin{multline}
\label{eq:F_acc}
F_{Acc} = F_0(T)+\alpha_1(T) |\Psi_1|^2+\alpha_2(T) |\Psi_2|^2 + \beta_1 |\vec{\Psi}|^4\\
+ \beta_2 (\Psi_1 \Psi_2^* - \Psi_1^* \Psi_2)^2 + \beta_3|\Psi_1|^2 |\Psi_2|^2 + ....,
\end{multline}
Note that the form above is essentially the same as $F_{2D}$, with the only difference being the two different $\alpha$-coefficients, which carry different critical temperatures in the most general case. 

Within the Landau theory of phase transitions, we can also consider how the superconducting order parameter couples to lattice deformations and what are the experimental signatures that are uniquely associated with multi-component order parameters stemming from multi-dimensional irreps.
Lattice deformations can have two characters.
The first type corresponds to compressive strain, for which the lattice parameters are changed, but no lattice symmetry is broken. This type of lattice deformation is associated with the trivial $A_{1g}$ irrep.
The second type corresponds to shear strain, which breaks some lattice symmetries,  and is associated with the nontrivial irreps.
We can then write new terms in the free energy functional, coupling the order parameters to lattice distortions, following  the requirement that their product transforms trivially under all point-group symmetries.
Note that linear couplings of lattice deformations to the order parameter are still forbidden, as the products of order parameters should always appear in a gauge invariant combination.
We can then consider couplings that are linear in strain and quadratic in the order parameter.

We have learned above that if the order parameter belongs to a one-dimensional irrep, the only allowed product in second order is $|\Psi|^2$, which belongs to $A_{1g}$.
This means that order parameters associated with a one-dimensional irrep can only couple to compressive strain in $A_{1g}$.
In contrast, order parameters with two components can be combined in second-order products that do not necessarily transform as the trivial representation.
For example, we can choose the pair of order parameters $\Psi_1 \propto k_xk_z$ and $\Psi_2 \propto k_yk_z$.
The combination $|\Psi_1|^2 - |\Psi_2|^2 \propto k_z^2(k_x^2-k_y^2)$ transforms as the $B_{1g}$ irreducible representation in $D_{4h}$.
This term alone cannot appear in the free energy functional, but the product with a strain tensor $\varepsilon_{B1g}(|\Psi_1|^2 - |\Psi_2|^2)$ can.
Another example would be the coupling $\varepsilon_{B2g}(\Psi_1 \Psi_2^* + \Psi_1^* \Psi_2)$, as  $(\Psi_1 \Psi_2^* + \Psi_1^* \Psi_2) \propto k_z^2 k_xk_y$. Here $\varepsilon_{B1g} = \varepsilon_{xx}-\varepsilon_{yy}$ and $\varepsilon_{B2g} = 2\varepsilon_{xy}$ correspond to strain with $B_{1g}$ and $B_{2g}$ symmetry, respectively, where $\varepsilon_{ij}$ are strain components, characterizing the relative change in lengths and angles of the lattice under the specific deformations.
See Table \ref{Tab:Strain} for a summary.
More details of these type of couplings and their experimental consequences will be discussed in the next subsection.

These new terms in the free energy lead to two important and experimentally verifiable consequences.
The first consequence  concerns the propagation of sound in the material.
Only multi-component superconductors can manifest a discontinuity in the velocity of sound waves associated with nontrivial irreps at the superconducting transition temperature.
The second consequence is the expected splitting of the superconducting transition under certain types of uniaxial strain.
If the strain breaks the symmetry between the two components, their superconducting critical temperature are unlikely to be the same anymore.
One example consists of $B_{1g}$ strain, contained in the uniaxial compression along the $x$-direction in $D_{4h}$, such that the symmetry between the $x$ and $y$ directions is explicitly broken.
In absence of strain, the $\Psi_1$ and $\Psi_2$ components are degenerate (they have the same superconducting critical temperature) by symmetry, possibly leading to a chiral superconducting order parameter.
Under strain, the critical temperature for the $\Psi_1$ component is enhanced, while for the $\Psi_2$ component is reduced (or vice-versa).
Under these conditions, a chiral state cannot be established immediately below the critical temperature, as it requires both components to develop a finite value.
Recent experiments on ultrasound attenuation~\cite{Benhabib2021.NatPhys.17.194, Ghosh2021.NatPhys.17.199}
and $\mu$SR under strain in \SRO~\cite{Grinenko2021.NatPhys.17.748} reveal this expected phenomenology for a multi-component order parameter.
More details on these experiments on \SRO are given in Secs. \ref{Sec:2CompUS}. and \ref{Sec:2CompMSR}.

\begin{table}[ht]
 \caption{Irreducible representations of strain $\varepsilon_{ij}$ in $D_{4h}$. Note that there is no strain component with $A_{2g}$ symmetry.}
    \label{Tab:Strain}
\begin{center}
    \begin{tabular}{ c | cc}
    \hline\hline
irrep  & $\varepsilon_{ij}$  & Strain type \\ \hline
$A_{1g}$ & $\varepsilon_{xx}+\varepsilon_{yy}$ or $\varepsilon_{zz}$ &  Compressive \\ \hline
$B_{1g}$ & $\varepsilon_{xx}-\varepsilon_{yy}$ &  \multirow{3}{*}{Shear} \\ 
$B_{2g}$ & $\varepsilon_{xy}$ &    \\ 
$E_{g}$ & $\{\varepsilon_{xz}, \varepsilon_{yz}\}$ &    \\ \hline\hline
    \end{tabular}
        \end{center}
   
\end{table}

\subsection{\SECTIONULTRASOUND}
\label{Sec:2CompUS}

\begin{figure*} [ht]
\begin{center}
\includegraphics[width=14cm]{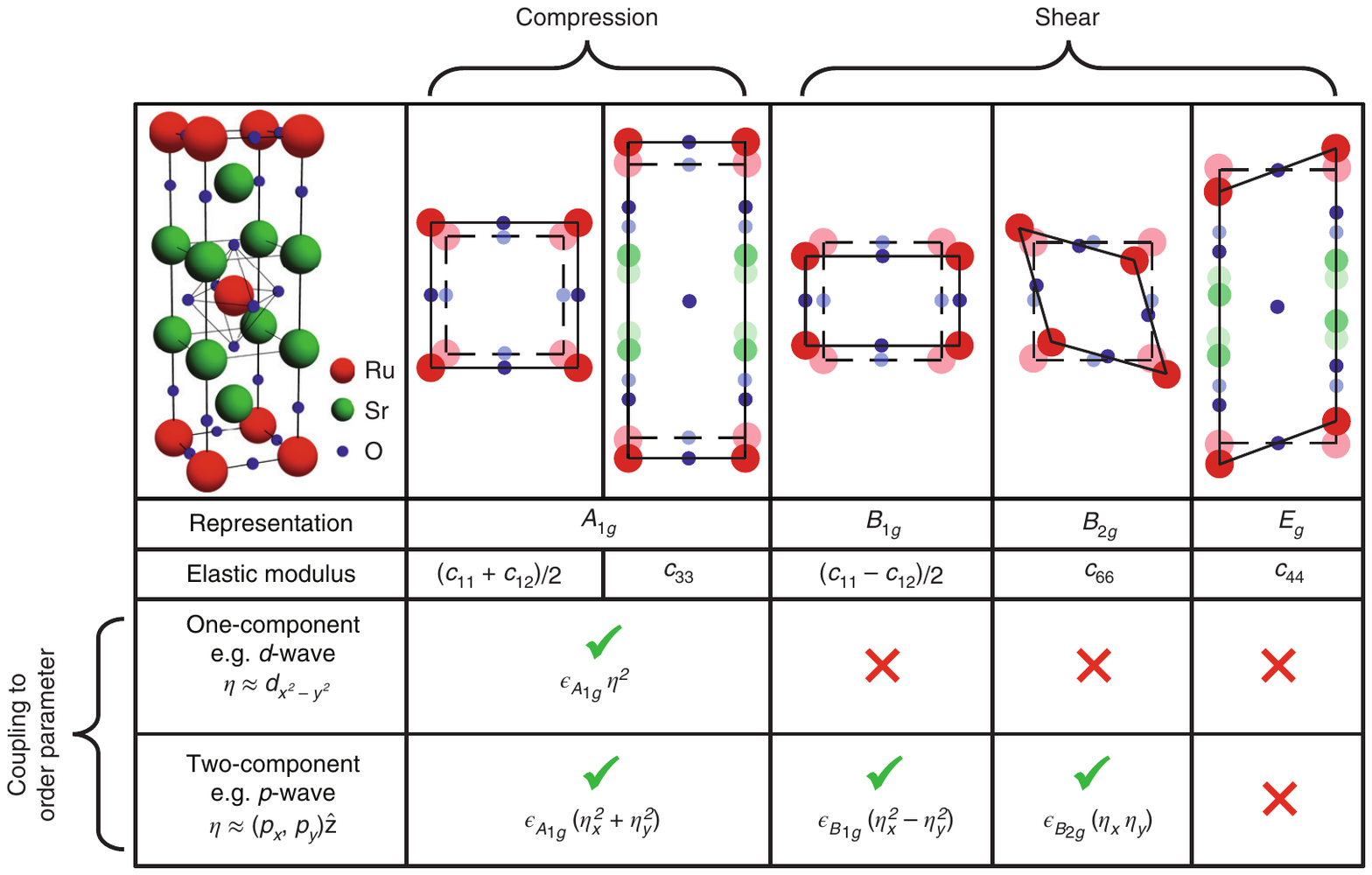}
\end{center}
\caption{Compressive and shear sound modes with different symmetries (second row) and  their coupling to superconducting order parameters $\boldsymbol{\eta}  = (\eta_x, \eta_y$) (last two rows) in the context of the $D_{4h}$ point group.
Reproduced from Ref.~\citen{Ghosh2021.NatPhys.17.199} ($\copyright$~2021 The Authors).} 
\label{fig:Irrep_US-modes_Ghosh}
\end{figure*}

In this subsection, we describe in more detail how the free energy described in the previous section is applied to the ultrasound experiments. 
Ultrasound modes have particular symmetries and are associated with different irreducible representations of the point-group symmetry of the crystal, as summarized in Table \ref{Tab:Strain}. 
As a consequence, the coupling between superconducting OPs and the ultrasound modes leads to discontinuities in some of the ultrasound velocities at $\Tc$, providing useful information on the single- or multi-component nature of the superconducting OPs. 
The schematic representation of the strain in the point group $D_{4h}$ for the elastic constant associated with each sound mode and the respective coupling to superconducting order parameters is shown in Fig. \ref{fig:Irrep_US-modes_Ghosh}. 

Here We follow the expression of the the Landau free energy expansion for a two-component order parameter $\vec{\Psi}  = (\Psi_1, \Psi_1$) adopted in Ghosh \etal~\cite{Ghosh2021.NatPhys.17.199}.
\begin{multline}
\label{eq:F_OP}
F\subm{OP}(\vec{\Psi})= F_0(T) + a(T)|\vec{\Psi}|^2 + b_1 |\vec{\Psi}|^4\\
+\frac{b_2}{2}\left[(\Psi_1^* \Psi_2)^2 + (\Psi_1 \Psi_2^*)^2\right] + b_3|\Psi_1|^2 |\Psi_2|^2 + ....,
\end{multline}
where $a  = a_0(T-T\subm{c})$ with $a_0 > 0$, and $b_i$ are phenomenological parameters.
The coefficients in equations \eqref{eq:F_D4h} and \eqref{eq:F_OP} are related to each other as
$a = \alpha$,
$b_1 = \beta_1$,
$b_2 = 2\beta_2$,
$b_3 = \beta_3 - 2\beta_2$.

To understand the origin of the changes in the sound velocity, we need to add to the free energy the contribution of the elastic energy density:
\begin{align}
F\subm{ela} &= \dfrac{1}{2}\Big[c_{11}(\varepsilon_{xx} + \varepsilon_{yy})^2 + c_{33}\varepsilon_{zz}^2 + 2c_{12}\varepsilon_{xx}\varepsilon_{yy}\nonumber \\  &+ 4c_{66}\varepsilon_{xy}^2 
+ 4c_{44}(\varepsilon_{xz}^2 + \varepsilon_{yz}^2) +2c_{13}(\varepsilon_{xx}+\varepsilon_{yy})\varepsilon_{zz} \Big],
\end{align}
where $\epsilon_{ij}$ are the strains and $c_{ij}$ are the elastic moduli.
As described in standard textbooks,~\cite{Ibach-Luth.4th-ed., Onodera.Text} the strain tensor is defined as $\varepsilon_{ij} = 1/2\left(\partial u_j / \partial x_i + \partial u_i / \partial x_j \right)$, where $u_i$ designate the $i$-component of the displacement vector $\vec{u}$.
It is related to the stress tensor $\sigma_{kl}$ as
\begin{eqnarray}
\sigma_{kl} = \sum_{ij} {c_{klij}\varepsilon_{ij}}.
\end{eqnarray}
Considering the crystalline symmetry, the indices of the elastic modulus $c_{klij}$ (a 4-rank tensor) are reassigned, for example, as 12 $\rightarrow$ 6, thus $c_{1212}\rightarrow c_{66}$.

The elastic free energy can be conveniently rewritten as
\begin{align}
F\subm{ela} & = \dfrac{1}{2}\Big[\dfrac{c_{11}+c_{12}}{2}(\varepsilon_{xx} + \varepsilon_{yy})^2 + c_{33}\varepsilon_{zz}^2 
\nonumber  \\
&+ \dfrac{c_{11}-c_{12}}{2}(\varepsilon_{xx} - \varepsilon_{yy})^2 
+ 4c_{66}\varepsilon_{xy}^2 
\nonumber  \\
&+ 4c_{44}\left(\varepsilon_{xz}^2 + \varepsilon_{yz}^2\right) 
+ 2c_{13}(\varepsilon_{xx}+\varepsilon_{yy})\varepsilon_{zz}\Big],
\end{align}
so that we can introduce the simpler parametrization
\begin{align}
F\subm{ela} &= \frac{1}{2}\Big[c_{A_{1g,1}}\varepsilon_{A_{1g,1}}^2 + c_{A_{1g,2}}\varepsilon_{A_{1g,2}}^2 + c_{B_{1g}}\varepsilon_{B_{1g}}^2
\nonumber \\ & + c_{B_{2g}}\varepsilon_{B_{2g}}^2 + c_{E_{g}}|\vec{\varepsilon}_{E_g}|^2 + 2c_{A_{1g,3}}\varepsilon_{A_{1g,1}}\varepsilon_{A_{1g,2}}\Big],
\end{align}
where we identified $c_{A_{1g,1}} = (c_{11}+c_{12})/2$, $c_{A_{1g,2}} = c_{33}$, $c_{B_{1g}}=(c_{11}-c_{12})/2$, $c_{B_{2g}} = c_{66}$, $c_{E_{g}} = c_{44}$, and $c_{A_{1g,3}} = c_{13}$, in addition to $\varepsilon_{A_{1g,1}} = \varepsilon_{xx} + \varepsilon_{yy}$, $\varepsilon_{A_{1g,2}}=\varepsilon_{zz}$, $\varepsilon_{B_{1g}} =\varepsilon_{xx} - \varepsilon_{yy}$, $\varepsilon_{B_{2g}} = 2\varepsilon_{xy}$, and $\vec{\varepsilon}_{E_{g}} = \{ 2\varepsilon_{xz}, 2\varepsilon_{yz}\}$.

The terms in the free energy accounting for the coupling between strain and order parameter are given as: 
\begin{align}\label{Eq:Fc}
F\subm{c} &= (g_1\varepsilon_{A_{1g,1}}+g_2\varepsilon_{A_{1g,2}})|\vec{\Psi}|^2 \nonumber\\
&+ g_4\varepsilon_{B_{1g}}(|{\Psi_1}|^2-|{\Psi_2}|^2) 
+g_5\varepsilon_{B_{2g}}(\Psi_1^*\Psi_2 + \Psi_1\Psi_2^*),
\end{align}
where $g_i$ are effective coupling constants.
The allowed lowest order coupling between strain components and OP $\vec{\Psi}$ is linear in strain and quadratic in OP given the need to satisfy gauge invariance. 

The coupling between OP and strain in the free energy leads to a discontinuity in the elastic constant and consequently in the sound velocity. 
These two quantities are related by $c_\Gamma = \rho v_\Gamma^2$, where $\rho$ is the density and $\Gamma$ labels the symmetry channel associated with a given sound mode.  
The jump in $c_{ij}$ is evaluated from the partial derivatives of $F\subm{c}$ and $F\subm{OP}$ with respect to $\Psi$ and $\varepsilon$.~\cite{Ghosh2021.NatPhys.17.199}
From the explicit form of $F\subm{c}$ in Eq. \ref{Eq:Fc}, and by referring to the product table given in Table \ref{Tab:D4hProduct},  we can see that only $A_{1g}$ strain couples to one-component OPs, whereas $A_{1g}, B_{1g}$ and $B_{2g}$ strains all couple to two-component OPs.

Note that all the allowed OPs couple to compressional strain of the longitudinal modes $A_{1g}$. There are cases in which shear strains also couple to the OP. In particular, the two-component OP in $E_g$, both chiral and nematic, couples with $B_{1g} ((c_{11}-c_{12})/2)$ and $B_{2g} (c_{66})$.
Other possibilities of coupling to shear strain modes correspond to two OPs with different representations to be accidentally degenerate to produce two-component OPs. For example, degenerate order parameters with 
$d_{xy}$ and $s$ character, or with $d_{x^2-y^2}$ and $g_{xy(x^2-y^2)}$ character couple only to the $c_{66}$ mode, whereas degenerate order parameters with $d_{x^2-y^2}$ and $s$ character couple only to the $(c_{11}-c_{12})/2$ mode.

\begin{figure} [ht]
\begin{center}
\includegraphics[width=5.5cm]{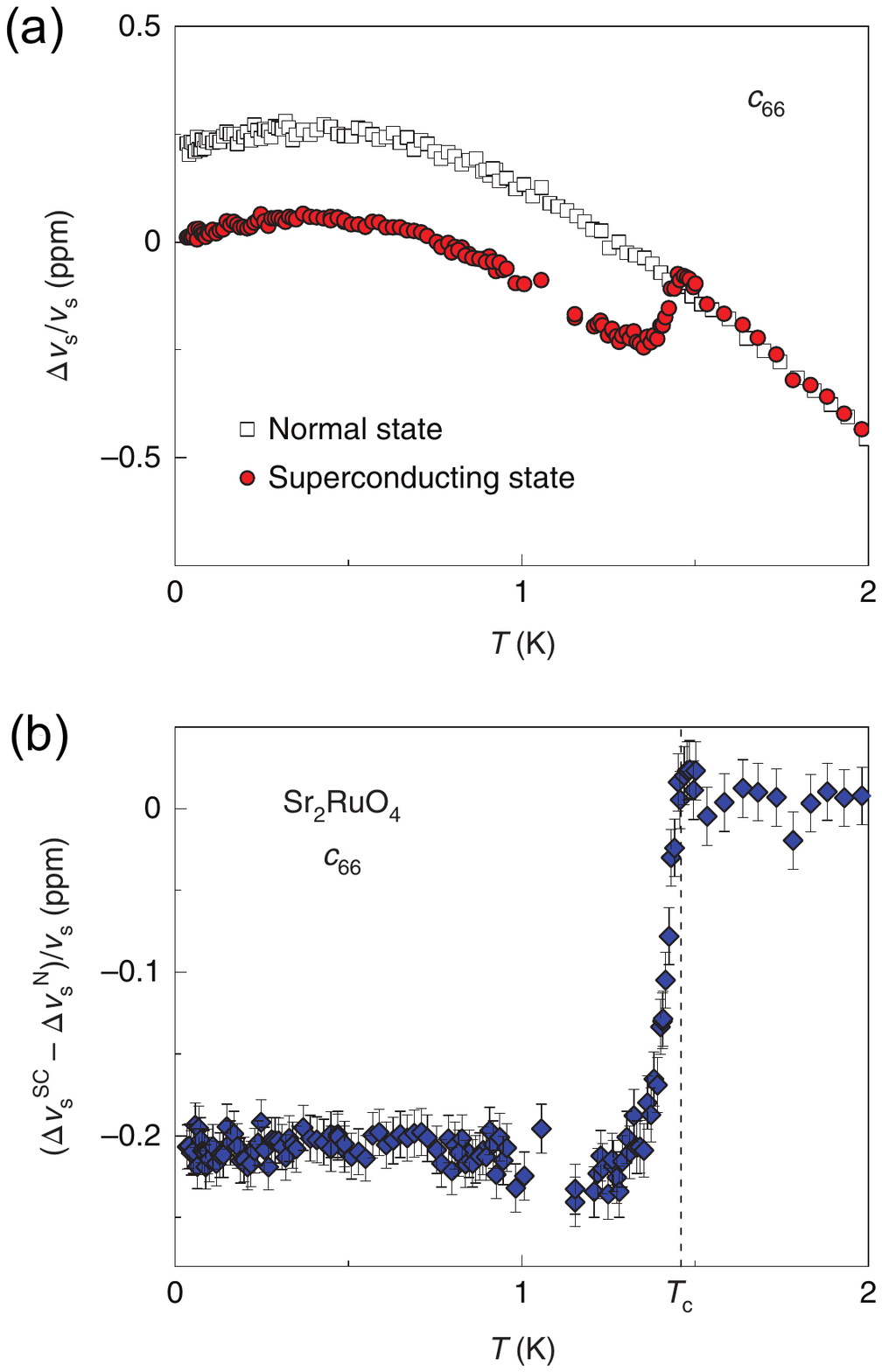}
\end{center}
\caption{Jump in the velocity of the transverse ultrasound mode $c_{66}$ at the superconducting critical temperature $T_c$. 
For (a) the data for the normal state (open squares) are obtained by applying a field of 1.5 T in the plane.
Reproduced from Ref.~\citen{Benhabib2021.NatPhys.17.194} ($\copyright$~2021 The Authors).} 
\label{fig:C66_Benhabib_NatPhy}
\end{figure}

\begin{figure} 
\begin{center}
\includegraphics[width=6cm]{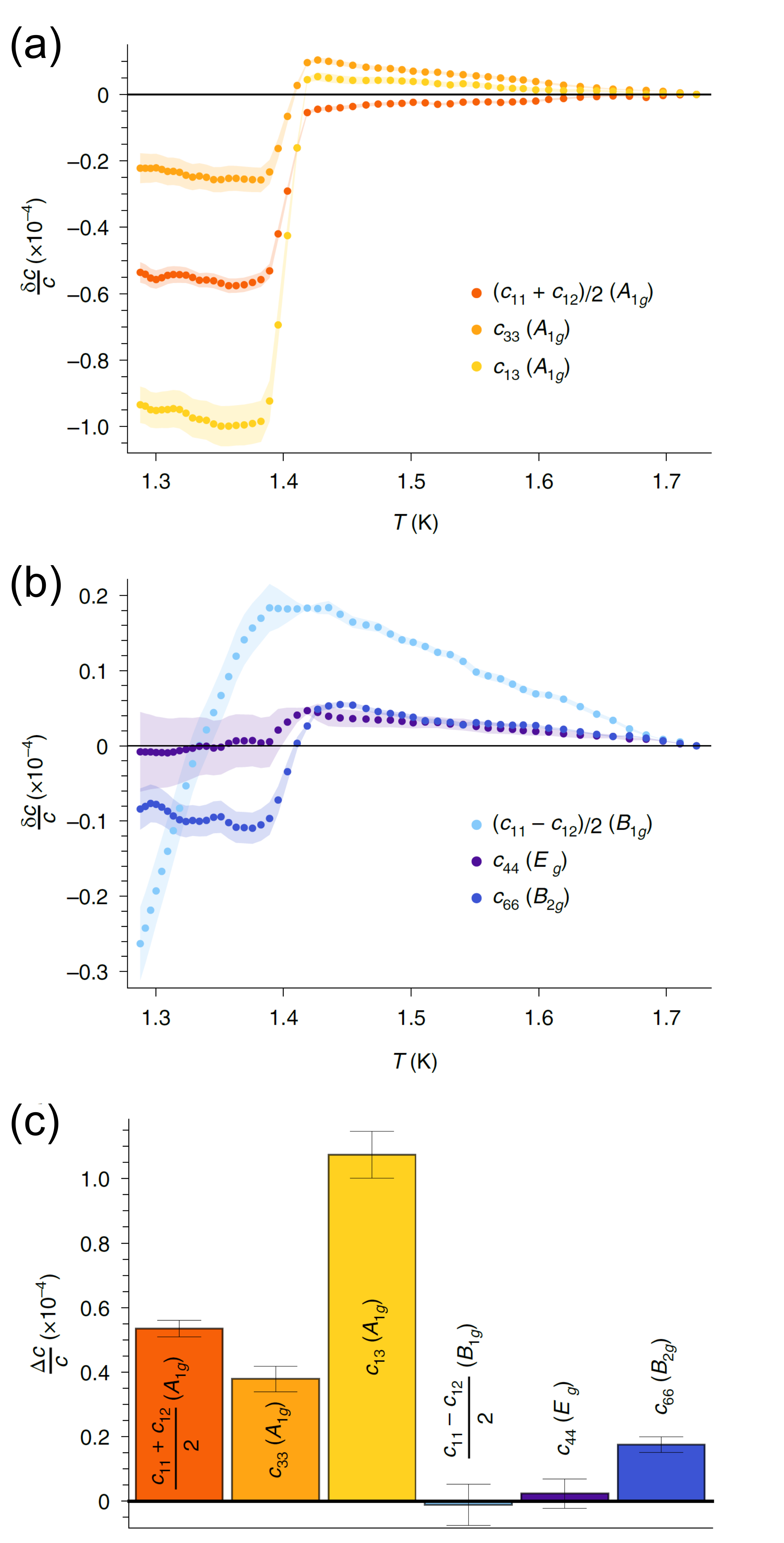}
\end{center}
\caption{Jumps in the velocity of various (a) longitudinal and (b) transverse modes at the superconducting critical temperature. (c) Comparison of the observed relative size of the jump at $\Tc$.
Reproduced from Ref.~\citen{Ghosh2021.NatPhys.17.199} ($\copyright$~2021 The Authors).} 
\label{fig:cij-jumps_Ghosh_NatPhys}
\end{figure}

The elastic constants of \SRO and their jumps at $\Tc$ were initially investigated by Okuda \etal ~\cite{okuda2002.JPSJ, okuda2003.PhysicaC} following the prediction by Sigrist~\cite{Sigrist2002.Ehrenfest-relations}. 
The compression mode $c_{13}$ was not directly measured but evaluated from other elastic constants. 
They reported the jump in $c_{66}$ of c.a. 20 ppm by the pulse-echo technique at 50 MHz, although the resolution of the jump was not conclusive.
Recently, two groups reported such an unusual jump in the ultrasound velocity of the shear mode $c_{66}$, in addition to the jumps in the compressional $A_{1g}$ modes $(c_{11}+c_{12})/2$ and $c_{33}$ with substantially improved accuracy and precision.~\cite{Benhabib2021.NatPhys.17.194, Ghosh2021.NatPhys.17.199}
Benhabib \etal~\cite{Benhabib2021.NatPhys.17.194} used a standard pulse-echo technique at the frequencies of 22 to 169 MHz using a pair of LiNbO$_3$ transducers glued on specified parallel surfaces of crystals (Fig. \ref{fig:C66_Benhabib_NatPhy}).
In contrast, Ghosh \etal~\cite{Ghosh2021.NatPhys.17.199} adopted a technique of resonance ultrasound spectroscopy (RUS) with a frequency range 2.4 – 2.9 MHz using one crystal held between a pair of transducers at its corners (Fig. \ref{fig:cij-jumps_Ghosh_NatPhys}).
The RUS technique was successfully demonstrated previously for \SRO in its normal state~\cite{Paglione2002}.
Concerning the size of the jump in $c_{66}$, Benhabib \etal reported 0.2 ppm, compared with 9 ppm by Ghosh \etal
Such large variations in the size of the jump may be attributable to the different frequencies or the formation of different SC domain structures.
Due to a large temperature dependence of the elastic constant $(c_{11} - c_{12})/2$ below $\Tc$~\cite{okuda2002.JPSJ}, it was not possible to conclude whether this shear mode exhibits a jump or not within the experimental precision.
We note here that the reported size of the jump also varies for the longitudinal modes: Ghosh \etal reported the jumps of 38 ppm for $(c_{11} + c_{12})/2$ and 105 ppm for $c_{13}$;
Benhabib \etal reported ~2 ppm for $c_{11}$ masked by 80 ppm softening in the SC state. 
The $c_{11}$ and $(c_{11}-c_{12})/2$ modes exhibit a large softening in the normal state with a maximum in the elastic constant at 35 K, and these two modes continue to soften strongly below $\Tc$, hampering the accurate determination of the jump at $\Tc$~\cite{okuda2002.JPSJ, Benhabib2021.NatPhys.17.194}. 
Okuda \etal~\cite{okuda2002.JPSJ} ascribe its origin in the normal state to the soft phonon mode toward rotational instability of the RuO$_6$ octahedra. 

Ghosh \etal~\cite{Ghosh2022} investigated the sound attenuation across $\Tc$.
The modes straining the Ru-O bond length, $(c_{11} + c_{12})/2$ and $(c_{11} - c_{12})/2$, show a large attenuation in the normal state;
the compressional viscosity for the $(c_{11} + c_{12})/2$ mode is larger by a factor of 2 to 3 than for the other compressional modes, and the shear viscosity for the $(c_{11} - c_{12})/2$ mode is larger by a factor of 7 to 8 than for the other shear modes.
Such a difference in the shear viscosity is interpreted in terms of pushing the $\gamma$ Fermi surface pocket toward the Van Hove singularity.
The authors found that attenuation coefficients of both compressional sound modes increase by a factor of 7 immediately below $\Tc$, and that among the shear modes the attenuation of only the $(c_{11} - c_{12})/2$, but not $c_{66}$, exhibits a significant linear decrease below $\Tc$.
They attribute this unusual behavior to the presence of domain walls that separate different configurations of the superconducting OP. 

Both groups qualitatively agree with their experimental results, but propose different OPs to be more likely.
Ghosh \etal~\cite{Ghosh2022} argue that accidentally degenerate OPs, especially the TRS breaking $d_{x^2-y^2}+ig_{xy(x^2-y^2)}$, may be more plausible.
In contrast, Benhabib \etal~\cite{Benhabib2021.NatPhys.17.194} argue that if the TRS breaking data is disregarded, a nematic $d$-wave, $d_{zx} $ or $d_{zy}$, is more plausible.

\section{\SECTIONTRSB}\label{Sec:TRSB}

\subsection{\SECTIONENHANCEMENT}\label{Sec:Enhancement}
\SRO exhibits a remarkable response to applied pressure: under uniaxial pressure along the [100] direction, $\Tc$ increases from 1.5 K to 3.5 K, more than doubling.
As introduced in Sec. 3, this enhancement is associated with the Lifshitz transition involving the reconstruction of the Fermi surfaces, and gives an excellent opportunity to elucidate the superconducting symmetry.
A second transition below $\Tc$ under uniaxial strain was predicted for a chiral superconducting state~\cite{Ueda2002, Agterberg1998.PhysRevLett.80.5184}, but it was much later that a tailored experiment was realized by the invention of a novel uniaxial-pressure cell based on piezo-electric devices (Fig.~\ref{Fig:Hicks_mechanism})~\cite{Hicks2014.RSI}.
A successive work revealed that with increasing the strain along the [100] direction, $\Tc$ reaches a maximum of 3.5 K at the strain $\varepsilon_{100}$ of about $-0.44\%$~\cite{Li2022}, after which it decreases rapidly~\cite{Steppke2017.Science.355.eaaf9398}. 
Previously, it was not anticipated that applying strain can alter the Fermi surface topology. Now it has been established that the maximum in $\Tc$ is in agreement with the occurrence of a Lifshitz transition at $\varepsilon_\mathrm{VHS}$, in which Van Hove singularities cross the Fermi level in two points of the BZ~\cite{Barber2019, Sunko2019}. 

\begin{figure} [ht]
\begin{center}
\includegraphics[width=8.5cm]{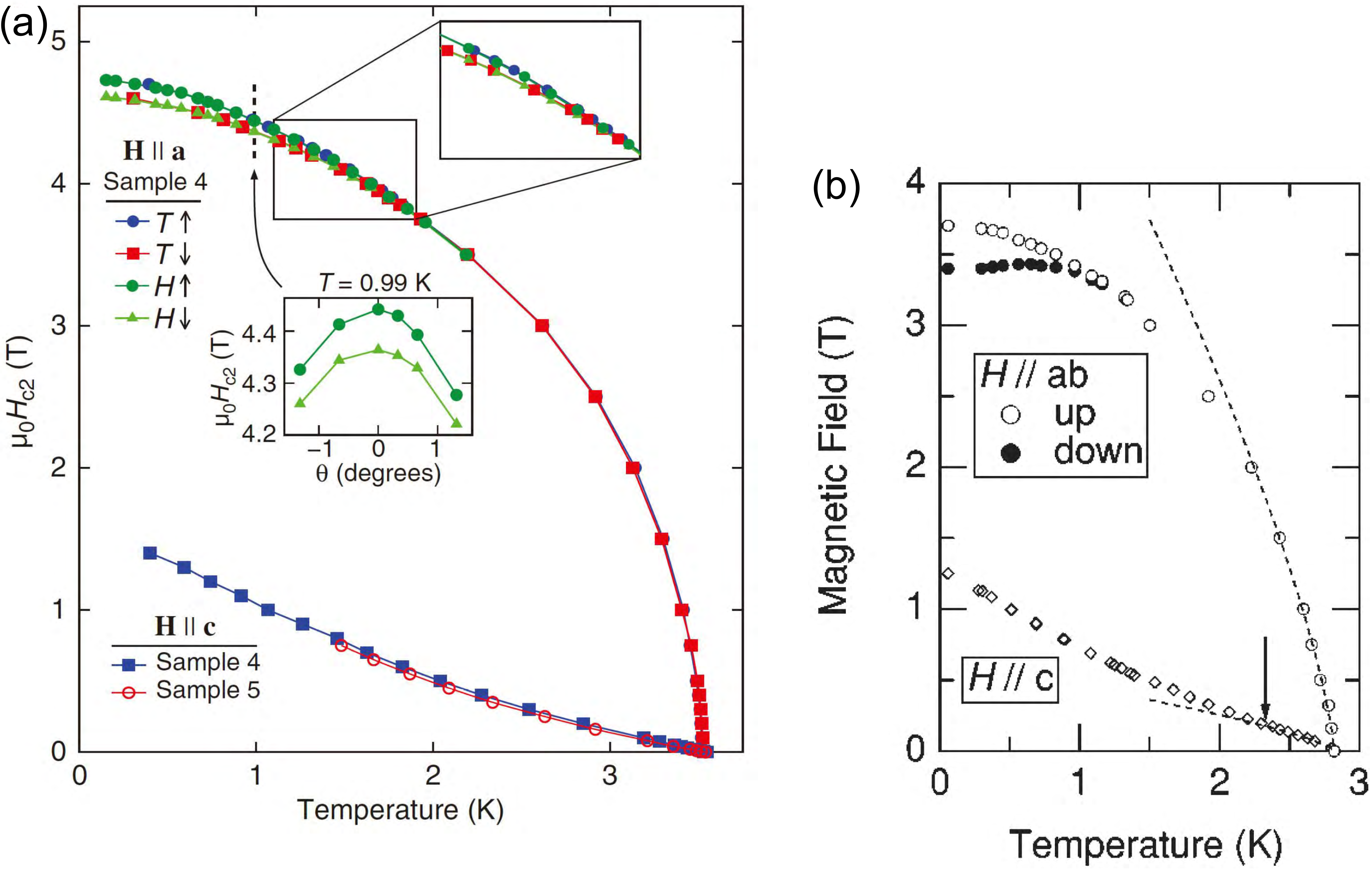}
\end{center}
\caption{The anisotropy of the upper critical field (a) under uniaxial strain along the [100] direction (2017)~\cite{Steppke2017.Science.355.eaaf9398} ($\copyright$~2017 The American Association for the Advancement of Science) and (b) in the eutectic “3-K” phase of \SRO with Ru-metal inclusions (2003)~\cite{Yaguchi2003.PhysRevB.67.214519} ($\copyright$~2003  American Physical Society).}
\label{fig:H-T_strain-3K}
\end{figure}

Similar enhancement in $\Tc$ was previously observed in the \SRO ``3-K phase”, where Ru metal platelets segregate in eutectic crystals of \SRO~\cite{Maeno1998.PRL.3-K-phase}.
It is now understood that such enhanced $\Tc$ is due to anisotropic strain developed by differential thermal contraction during the solidification of the eutectic crystal.
In fact, variations of the upper critical field $\Hcc$ in the eutectic and uniaxially-strained systems are very similar (Fig. \ref{fig:H-T_strain-3K})~\cite{Yaguchi2003.PhysRevB.67.214519, Steppke2017.Science.355.eaaf9398}.
In both cases, $\mu_0 \Hcc$ is enhanced for a [100] magnetic field from 1.5 to 4.5 T, and for a [001] field from 0.075 to 1 T.
In addition, a large hysteresis of the superconducting transition is observed upon increasing and decreasing the magnetic field across $\Hcc$, evidencing a first-order transition at low temperatures.
We note that the enhancement of $\Hcc$ for $H$//[001] is much stronger than the expectation from the strain dependence of the quantity $H\subm{{c2}{//c}} / \Tc^2$ based on weak-coupling model for $d_{x^2-y^2} + s$ OP~\cite{Jerzembeck2023}.

\subsection{\SECTIONMSR}
\label{Sec:2CompMSR}

\begin{figure} [ht]
\begin{center}
\includegraphics[width=8.5cm]{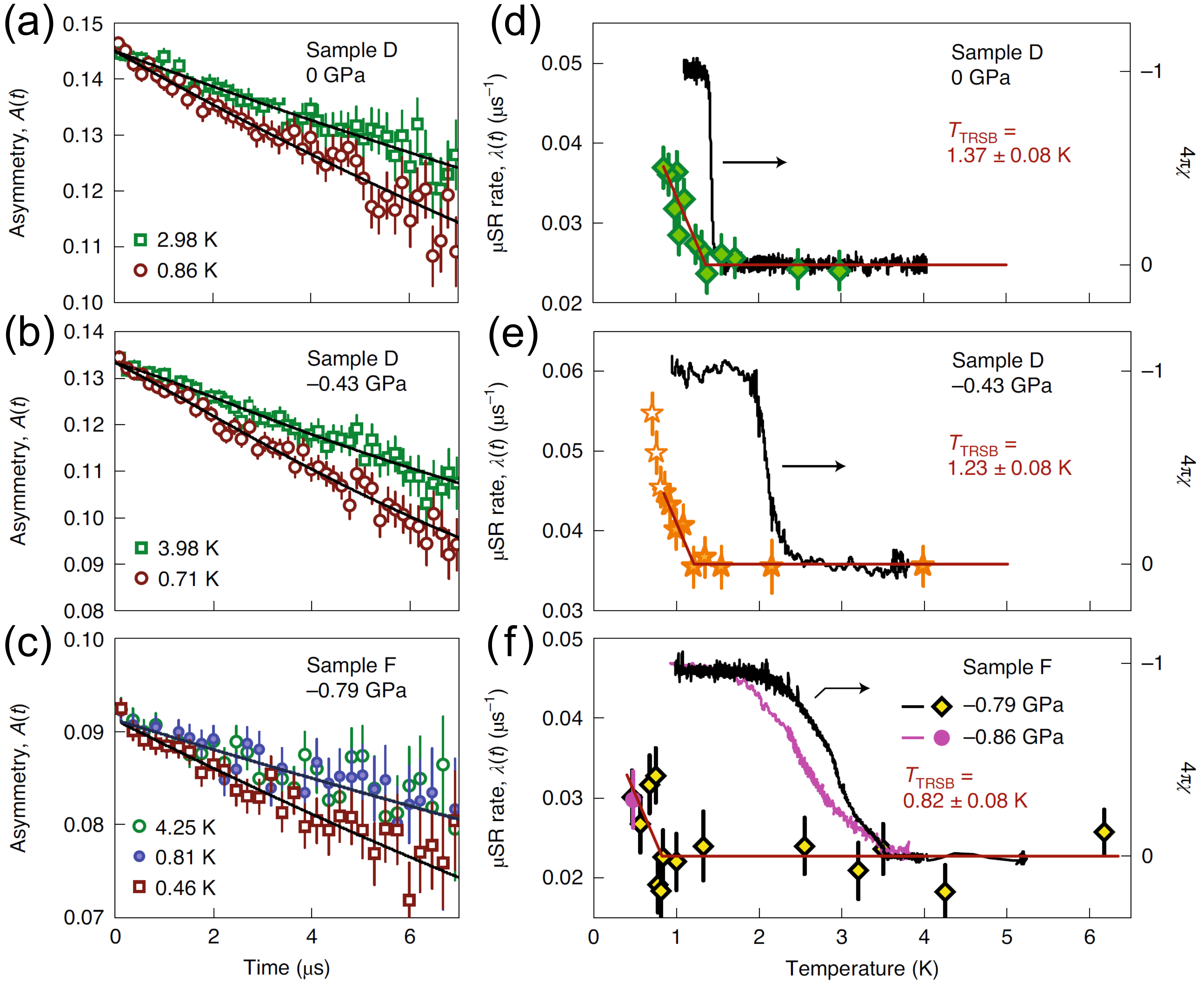}
\end{center}
\caption{(a)-(c): Asymmetry parameter $A(t)$ as a function of time in $\mu$SR to probe additional relaxation in the SC state due to internal magnetic fields at different strain parallel to the [100] direction. (d)-(f): Onset of additional $\mu$SR relaxation rate $\lambda(T)$ compared with the onset of superconductivity probed by AC susceptibility. Two onset temperatures split under strain. Reproduced from Ref.~\citen{Grinenko2021.NatPhys.17.748} ($\copyright$~2021 The Authors)}.
\label{fig:Tc-muSR-e100}
\end{figure}

\begin{figure} [ht]
\begin{center}
\includegraphics[width=7cm]{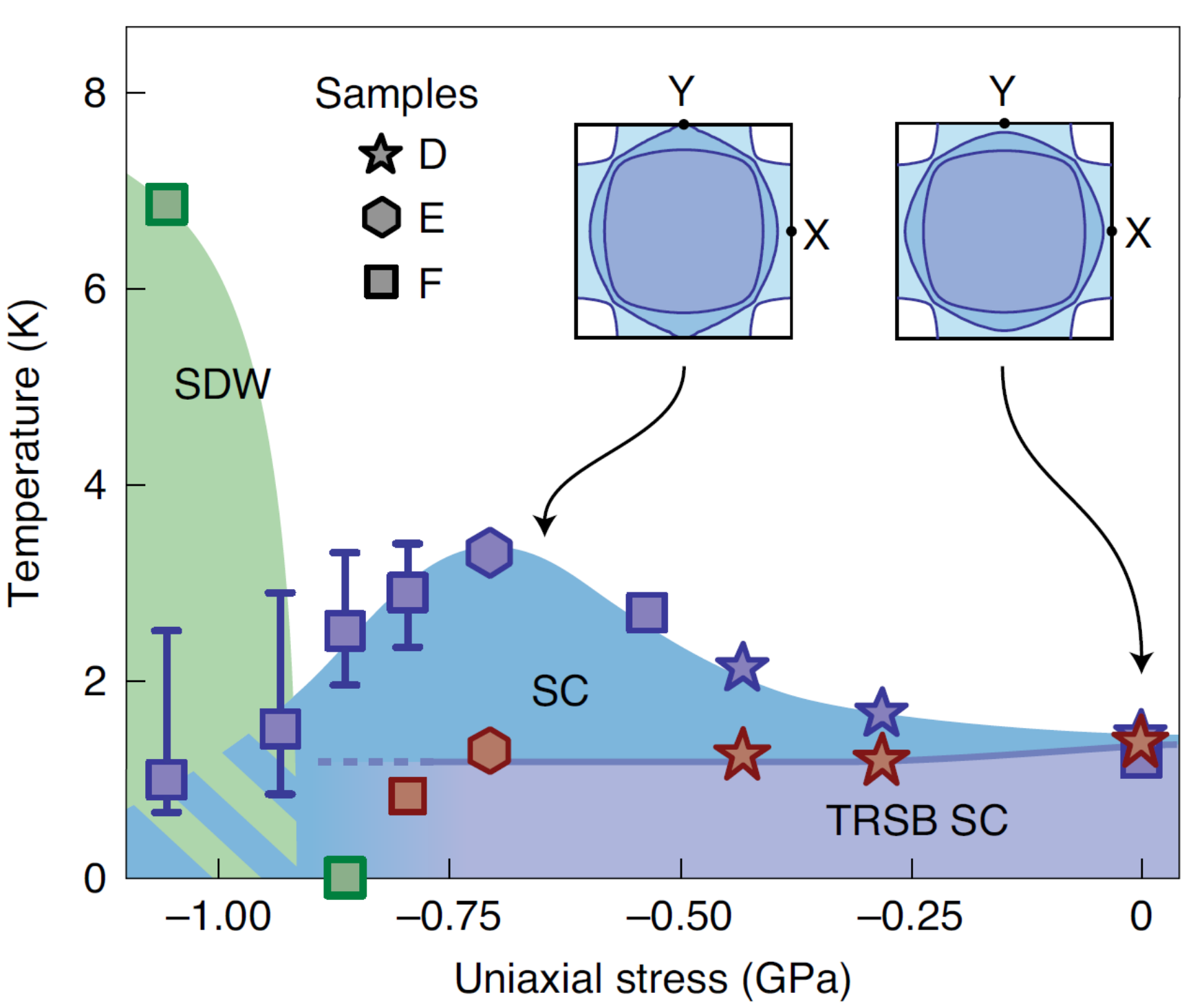}
\end{center}
\caption{Temperature-strain phase diagram based on $\mu$SR and AC susceptibility. The superconducting transition at $\Tc$ and the emergence of TRS breaking below $\TTRSB$ split under uniaxial strain along the [100] direction. At higher strain, Spin Density Wave (SDW) ordering sets in. Reproduced from Ref.~\citen {Grinenko2021.NatPhys.17.748} ($\copyright$~2021 The Authors).}
\label{fig:muSR_Tc-e100_phase-diagnram}
\end{figure}

Application of in-plane uniaxial strain has the important effect of breaking the tetragonal symmetry of the crystal.
If the superconducting OP consists of two components, either symmetry-protected pairs belonging to the same irreps or accidentally degenerate pairs consisting of different irreps, the superconducting transition temperature is expected to split between $\Tc$ and $\TTRSB$, below which the time-reversal symmetry is broken.
A series of experiments to test this behavior was performed using muon spin resonance ($\mu$SR), a technique which probes both the SC transition at $\Tc$ and the emergence of the spontaneous magnetic field below $\TTRSB$.
In the first experiments in which uniaxial pressure along the [100] was applied, $\Tc$ increased strongly but $\TTRSB$ remained almost the same or decreased slightly with increasing strain (Fig.~\ref{fig:Tc-muSR-e100})~\cite{Grinenko2021.NatPhys.17.748}.
At strain higher than that for the Lifshitz transition $\varepsilon_\mathrm{VHS}$, long-range magnetic ordering below $T\subm{SDW}$ = 7 K is induced (Fig.~\ref{fig:muSR_Tc-e100_phase-diagnram}).
For uniaxial pressure along the [110] direction, which leads to breaking of the tetragonal symmetry without a strong enhancement of $\Tc$, the preliminary results suggest that $\Tc$ and $\TTRSB$ also split~\cite{Grinenko2023.PRB}.
This latter experiment needs more accumulation of data to make a firm conclusion.

\begin{figure} [ht]
\begin{center}
\includegraphics[width=6cm]{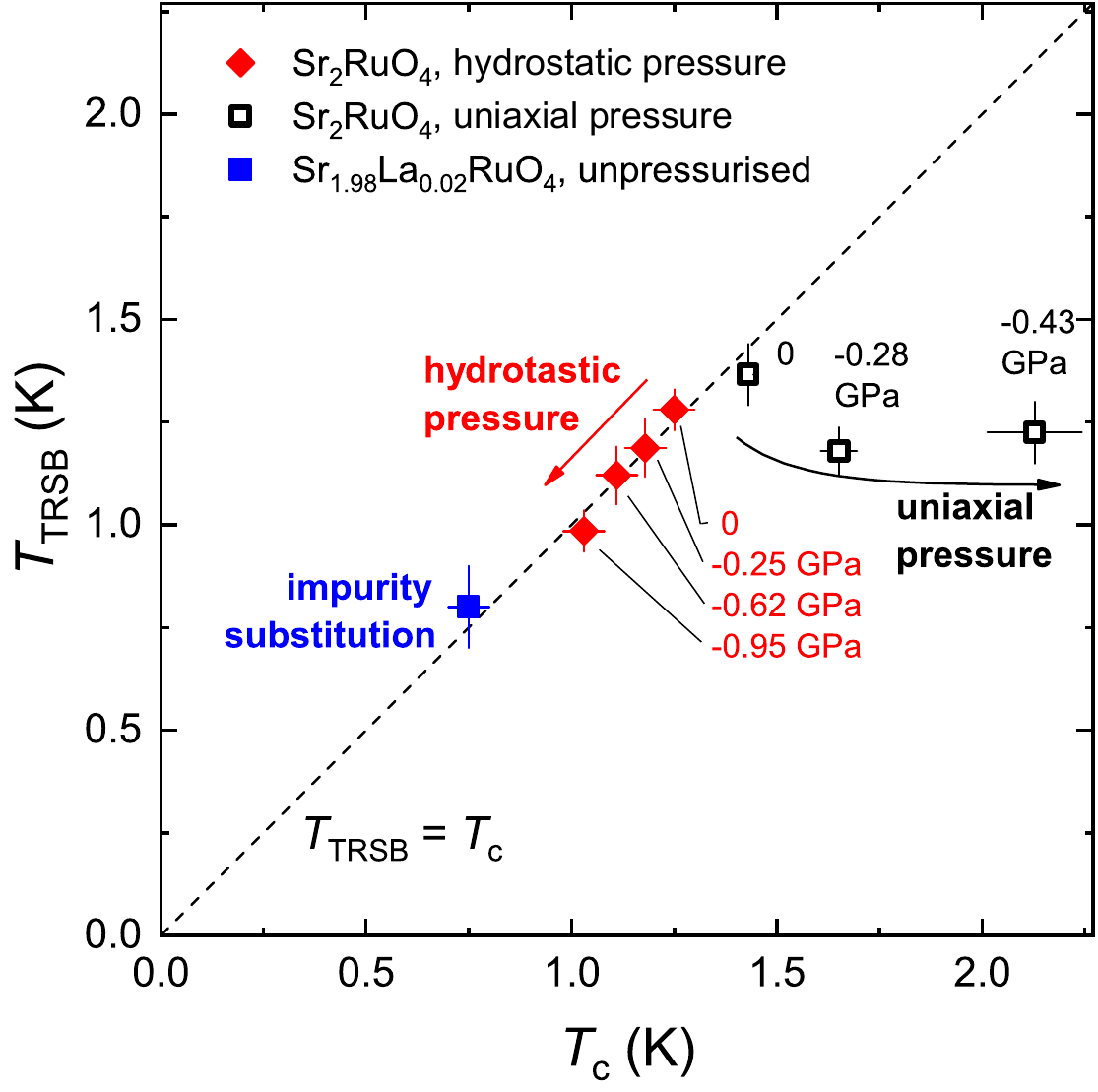}
\end{center}
\caption{Unsplit superconducting transition temperature $\Tc$ and $\TTRSB$ in \SRO under hydrostatic pressure and disorder (maintaining the tetragonal symmetry). Reproduced from Ref.~\citen{Grinenko2023.PRB} ($\copyright$~2023 The Authors).}
\label{fig:unsplit-Tc.pdf}
\end{figure}

The first approach to vary $\Tc$ with uniaxial strains involves breaking the tetragonal symmetry of the crystal. 
As a second approach, the $\Tc$ of \SRO was varied while preserving its tetragonal symmetry.
This was done both by applying hydrostatic pressure and by adding impurities,~\cite{Grinenko2021.NatureCommum.12.3920} both lead to the decrease in $\Tc$. 
Corresponding to decreasing $\Tc$ from 1.5 to 0.4 K, $\TTRSB$ is found to decrease without noticeable splitting from $\Tc$ (Fig. \ref{fig:unsplit-Tc.pdf}). 
These $\mu$SR results are consistent with a symmetry-protected, two-component superconducting OP for \SRO.
Among even-parity pairing states, spin-singlet chiral $d$-wave $d_{yz} +id_{xz}$ or spin-triplet chiral interorbital $\lambda\subm{5} + i\lambda\subm{6}$~\cite{Suh2020} are some of the possibilities.

\begin{figure} [ht]
\begin{center}
\includegraphics[width=7cm]{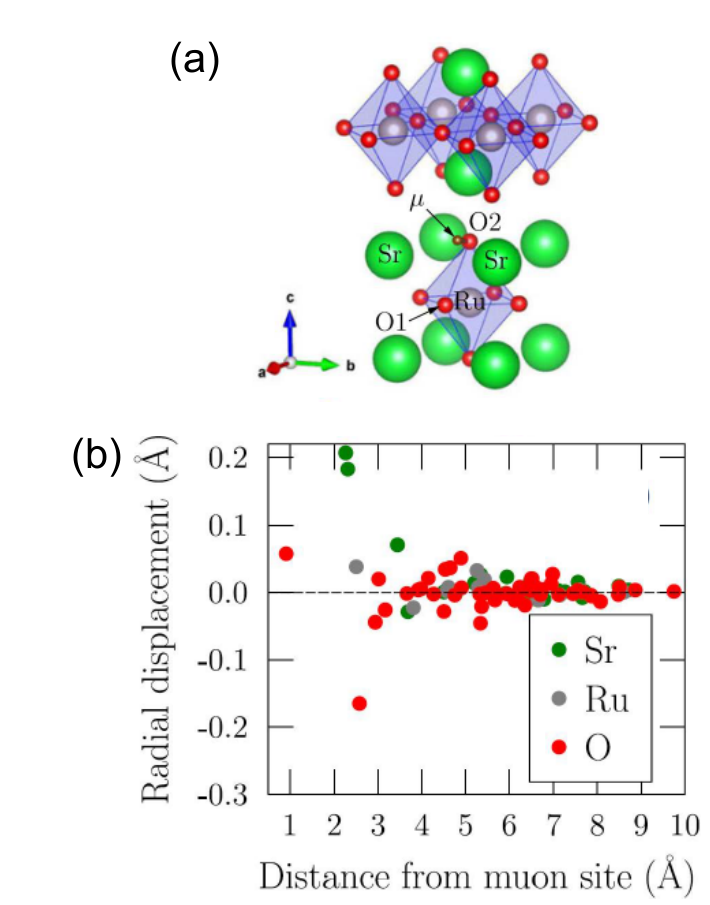}
\end{center}
\caption{The lowest-energy muon site in \SRO. (a) The local
geometry of the muon site. The muon stops in the Sr-O plane, 1.0\AA~away from the apical oxygen. 
(b) Radial displacements of atoms as a function of their distances from the muon site. One of the oxygen atoms in the Ru-O$_2$ @lane is strongly attracted by the muon. Reproduced from Ref.~\citen{Huddart2021} ($\copyright$~2021  American Physical Society).}
\label{fig:muon site}
\end{figure}

There has been substantial recent progress to characterize the effect of injected muons in \SRO \cite{Huddart2021, Blundell2023。AppPhysRevi}. 
The density functional theory ``DFT + $\mu$" has shown that a positive muon would stop at 1.0\AA~away from the apical oxygen toward Sr atoms as in Fig.~\ref{fig:muon site} (a). 
It causes a notable local atomic displacement; as shown in Fig. \ref{fig:muon site} (b), only one of the in-plane oxygen is substantially attracted by the muon by 0.17\AA~and shortens one of the Ru-O bond distances by 3\%. Although it is the change in only one bond length in the crystal, this is to be compared with the uniaxial strain of 0.44\% at the Lifshitz transion and 0.6\% at the onset of magnetic order in \SRO. The muon mainly changes the electronic structure several electronvolts below the Fermi energy; the partial DOS of the Ru atom closest to the muon site changes little except for the shift of the DOS peaks upwards by ca. 65 meV due to an effective hole doping of tetravalent Ru by 0.2\%. \cite{Huddart2021}.

While these calculations give us important information about the normal state, the effect of muons in the superconducting state remains an open problem. 
The muon charge screening by the normal electrons is very effective with the Thomas-Fermi screening length of ca. 1\AA. 
Since the charge screening length by the superconducting electrons is known to be essentially the same as in the normal state \cite{Koyama2004, Hirsch2004}, it is expected that there is little additional effect of muons in the superconducting state.

\subsection{\SECTIONSECONDTRANSITION}\label{Sec:SecondTransition}
A series of $\mu$SR results suggesting the second transition at $\TTRSB$ below the SC transition at $\Tc$ supports a symmetry-protected two-component order parameter for \SRO. If confirmed by other techniques, the second transition will impose strong constrains on our understanding of the superconducting symmetry. 
In order to examine thermodynamic evidence for the splitting of $\Tc$ and $\TTRSB$, high-sensitivity heat-capacity measurements were performed under uniaxial strain~\cite{Li2021}. 
AC calorimetry at a relatively high frequencies of about 3 kHz was performed with various uniaxial strains under zero magnetic field, as well as at various fixed magnetic fields along the $c$-axis with strain values close to $\varepsilon_\mathrm{VHS}$. 
The former indicates the maximum $\Tc$ of 3.6 K at $\varepsilon_x = -0.57\%$.
Within conservative detection limits of 5$\%$ in the heat capacity, no anomaly was observed at temperatures corresponding to $\TTRSB$ suggested by $\mu$SR.  
For a hypothetical $d_{xz} + id_{yz}$ OP, the Ginzburg-Landau theory leads to the linear splitting of $\Tc$ in the leading order in uniaxial strain~\cite{Hicks2014.Science.344.283}. 
In the small strain limit, the expected jump at the heat-capacity anomaly for the two transitions is inversely proportional to the strain dependence of the transition temperatures~\cite{Grinenko2021.NatPhys.17.748}. 
Since $\mu$SR experiments indicate that the strain dependence of $\TTRSB$ is much weaker than that of $\Tc$, a larger heat-capacity anomaly at $\TTRSB$ would be expected, but this is in sharp contrast to the experimental observations.

We note that since the non-linear increase of $\Tc$ under small strain violates the expectations of ordinary GL theory, the assumption of strain-independent GL coefficients may not be sufficient to predict a jump at each of the split transitions. 
It is argued that accidentally degenerate order parameters may exhibit a strongly suppressed second heat capacity jump~\cite{Roising2022}. 
In the case of an orbital-antisymmetric (“orbital-singlet”) pairing, the second transition to the chiral superconducting state would not cause a significant change in the nodal structure, so that the specific-heat jump would be small.

\begin{figure} [ht]
\begin{center}
\includegraphics[width=6.5cm]{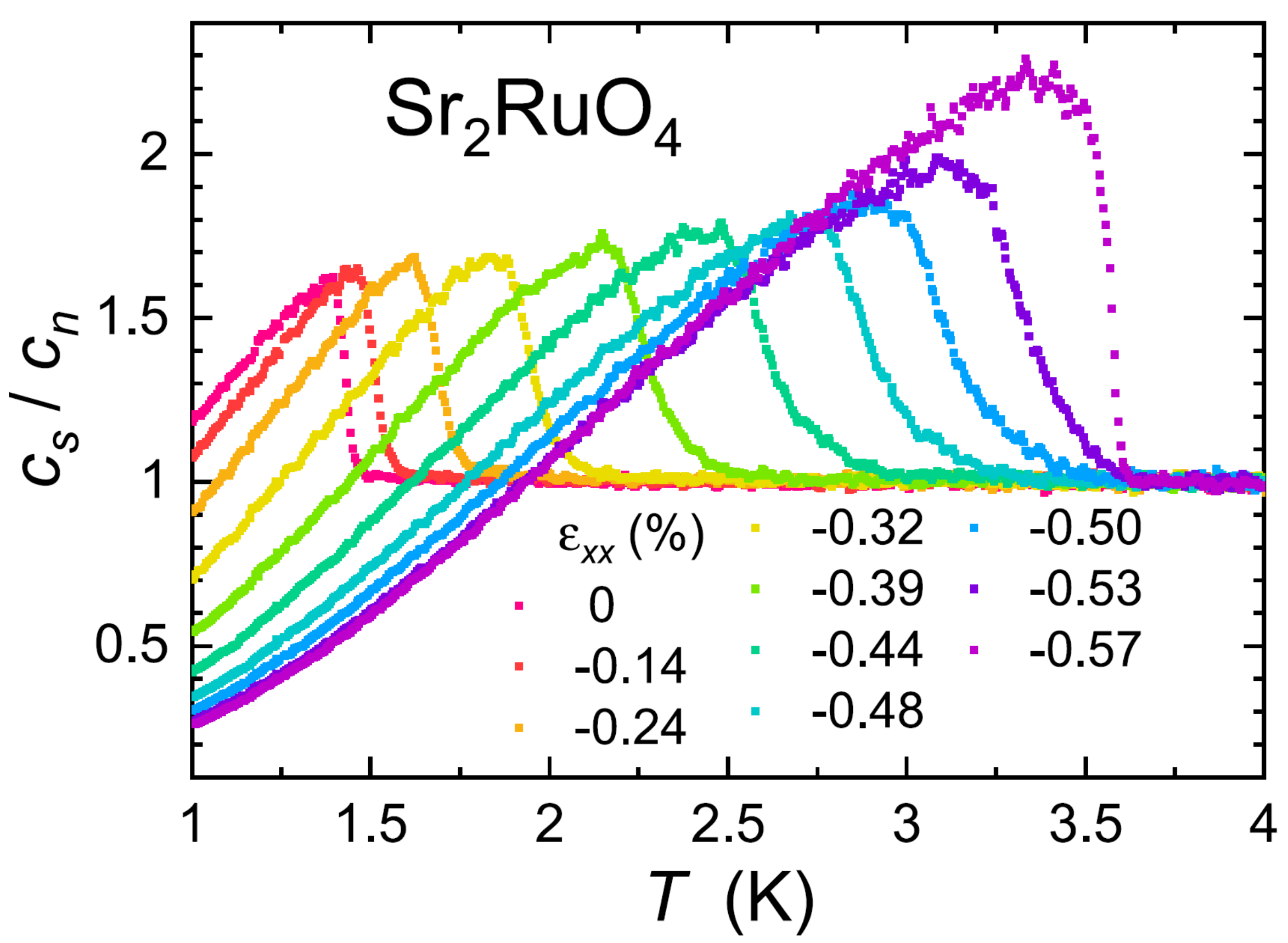}
\end{center}
\caption{Heat capacity under uniaxial strain. The bulk enhancement of $\Tc$ is clearly shown, but evidence for a second transition at 1.3$\sim$1.5 K was not detected, contrary to the observations of $\mu$SR. Reproduced from Ref.~\citen{Li2021} ($\copyright$~2021 National Academy of Sciences).}
\label{fig:Cp-under-strain}
\end{figure}

As an additional important feature, the normalized jump in the heat capacity, $\Delta C / C$, grows smoothly as a function of uniaxial pressure and takes a maximum at around the critical strain $\varepsilon_\mathrm{VHS}$ as shown in Fig. \ref{fig:Cp-under-strain}.
This is a curious behavior since if the in-plane gap anisotropy becomes stronger towards the Lifshitz transition, the normalized size of the jump would decrease. 
As an example of how the gap anisotropy leads to a decreasing jump size, for the isotropic $s$-wave gap and the line-nodal $k_{x^2 - y^2}$ $d$-wave gap, the relative jump sizes are 1.43 and 0.90, respectively.

\begin{figure} 
\begin{center}
\includegraphics[width=75mm]{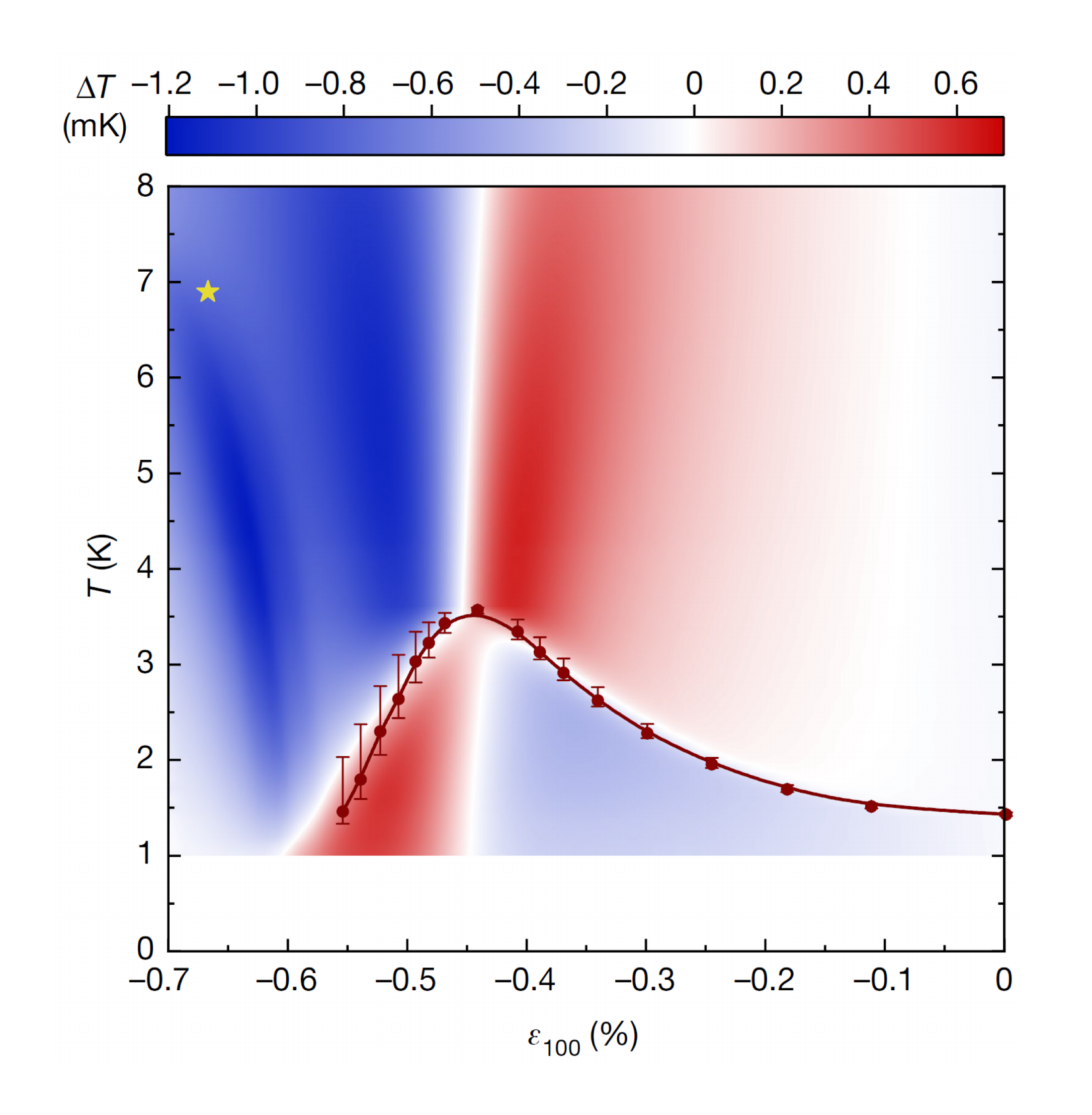}
\end{center}
\caption{Phase diagram based on the elastocaloric effect. The strain-derivative of the temperature, which has the opposite sign compared with the strain derivative of the entropy, is mapped in the temperature-strain plane. 
In the normal state, the entropy take a maximum at the Lifshitz transition, while in the superconducting state it takes a minimum at $\varepsilon_x \approx 0.45\%$ where $\Tc$ shows a maximum. 
There is no sign of the anticipated second transition at 1.3 - 1.5 K. Reproduced from Ref.~\citen{Li2022} ($\copyright$~2022 The Authors).}
\label{fig:elastocaloric-phase-diagram}
\end{figure}

The elastocaloric effect was also used to search for thermodynamic evidence for the split transition~\cite{Li2022}.
A small AC component of the strain $\Delta \varepsilon$ is added at each strain; the responding AC component of the temperature fluctuation $\Delta T$ is related to the strain-dependence of the entropy as: 
\begin{equation}\label{Eq:elastocaloric}
\frac{\Delta T}{\Delta\varepsilon}\bigg|_S \cong
- \frac{T}{C_{\varepsilon}} \frac{\partial S}{\partial\varepsilon} \bigg|_T,
\end{equation}
where $S$ is the entropy and $C\subm{\varepsilon}$ is the specific heat under constant strain.
Experimentally, the sign of $\Delta T$ was determined form its phase with respect to that of $\delta\varepsilon$, and according to Eq.~\ref{Eq:elastocaloric} it is opposite to the sign of the variation in $S$. 
The resulting strain derivative of the temperature is plotted in Fig. \ref{fig:elastocaloric-phase-diagram} in the temperature--strain phase diagram.
In the normal state, the nearly vertical white line indicates the entropy takes maximum as a function of strain associated with the enhanced density of states associated with the Van Hove singularity.
Such entropy maximum is strongly suppressed on entering the superconducting state.
Below $\Tc$, this vertical line turns into a minimum in the entropy, indicating that the states near the DOS maximum in the BZ open a gap in the superconducting state~\cite{Palle2023}. 
This technique may not be sensitive enough if the phase boundary is “horizontal” in the $T - \varepsilon$ phase diagram, as anticipated for the boundary of the second superconducting transition into a TRS breaking state.
Nevertheless, within the high sensitivity of the elastocaloric measurements shown in Fig.  \ref{fig:elastocaloric-phase-diagram}, there is no sign of such a second transition below $\Tc$.

\subsection{\SECTIONOTHER}\label{Sec:Other}
\begin{figure} 
\begin{center}
\includegraphics[width=8cm]{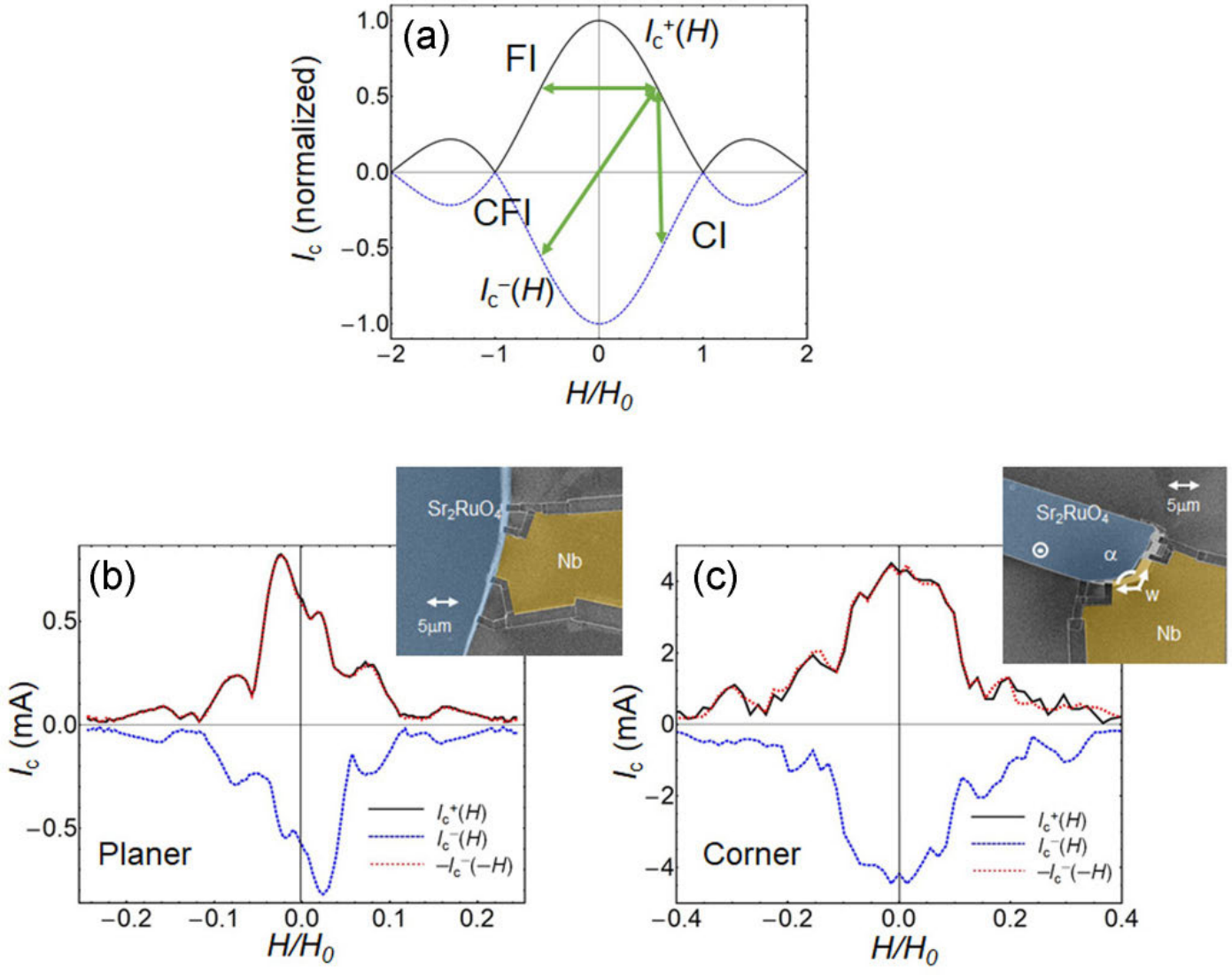}
\end{center}
\caption{Time-reversal invariance revealed by current and magnetic field inversion (CFI) of $I_\mathrm{c}$-$H$. 
(a) Symmetry of the patterns of the critical current $I\subm{c}$ vs. magnetic field $H$ can be examined by current inversion (CI), magnetic field inversion (FI), and current and magnetic field inversion (CFI).
Time-reversal symmetry of a superconductor can be tested by the invariance to the CFI symmetry.
(b) $I\subm{c}$-$H$ patterns and scanning ion microscopy (SIM) image of a planar junction consisting of \SRO and niobium. 
(c) of a corner junction.
The CFI symmetry demonstrated by consistency between black and red curves in both
junctions identifies time-reversal invariance of superconductivity of \SRO. 
Reproduced from Ref.~\citen{Kashiwaya2019.PRB.IH_inversion} ($\copyright$~2019 American Physical Society).}
\label{fig:IH_inversion}
\end{figure}

There have been other approaches in search for evidence of TRS breaking.
As shown in Fig. \ref{fig:IH_inversion}, Kashiwaya \etal examined the critical current vs. magnetic field of Josephson junctions between \SRO and Nb and found that the pattern is invariant under time-reversal operation, meaning that the complicated patterns match with each other if both current and field are inverted.~\cite{Kashiwaya2019.PRB.IH_inversion}
This indicates that the TRS is not broken in the superconducting state at the junction.
Combined with the observation of the ZBCP in \SRO-Au quasi-particle tunneling (Fig.~\ref{fig:Topo-edge-ZBCP}), they interpret the TRS behavior as due to a helical superconducting order parameter.

Another approach is to search for magnetic field near the sample surface generated by the anticipated topological edge currents of the TRS breaking superconductor. 
Such attempts using scanning SQUID and Hall probes so far have found no evidence of the spontaneous fields either near the edge of the sample or in the anticipated SC domain boundaries. ~\cite{Hicks2010.PhysRevB.81.214501, Tsuchiya2014.JPSJ.with_Ishiguro.micro-SQUID}
We note here that a similar attempt to probe spontaneous fields of the proposed chiral $d$-wave superconductor URu$_2$Si$_2$ with a scanning SQUID also results in negative results.~\cite{Iguchi2021.PRB.URu2Si2.with_Moler}

\subsection{\SECTIONREVISITED}
\label{sec:mystery_2}

We have introduced in Sec. \ref{sec:mystery_1} why the presence of horizontal line node at $k_z = 0$ is unnatural for a quasi-2D system like \SRO. 
A spin-singlet chiral $d$-wave OP $d_{zx}\pm id_{zy}$ is promising among the OP candidates in $D_{4h}$ as shown in Fig.~\ref{fig:SC-gaps_yonezawa} by combining the results of the spin-singlet-like OP indicated from NMR (Sec.~\ref{Sec:SCProperties}), the multi-component OP deduced from ultrasound (Sec.~\ref{Sec:MulticomponentOP}), and symmetry-protected TRS breaking OP deduced from $\mu$SR (Sec.~\ref{Sec:TRSB}). 
These results point to the horizontal line node at $k_z = 0$. 
The field-angle resolved specific heat (Sec.~\ref{Sec:SpecificHeat}), if interpreted in terms of the Fermi velocity anisotropy, also favors the presence of a horizontal line node. 

However, such a gap structure is a paradox for a quasi-2D electronic system like \SRO.
Since at $k_z = 0$ the $z$-component of the Fermi velocity $v\subm{F}$ is also zero, the horizontal line node at $k_z = 0$ implies that intralayer pairing is not allowed.

\section{\SECTIONTHEORETICAL}\label{Sec:Theoretical}

\subsection{\SECTIONTHEORIES}\label{Sec:Theories}

For more than twenty years, the evidence for spin-triplet superconductivity supported by NMR experiments \cite{Ishida2001.PhysRevB.63.060507R} placed Sr$_2$RuO$_4$ on top of the list of candidate materials to host chiral $p$-wave topological superconductivity and raised great interest among theoreticians, but a consistent understanding of all experimental signatures remained a puzzle \cite{Mackenzie2017}.
Remarkably, the most recent NMR experiments suggest, instead, that \SRO is a spin-singlet superconductor  \cite{Pustogow2019.Nature.574.72, Ishida2020.JPSJ.89.034712, Chronister2021.PNAS.118.25}.
This new piece of information led to intensive theoretical work by multiple groups in an attempt to reconcile it with other complementary experimental probes.

Previous theories already suggested spin-singlet superconductivity.
For example Nomura and Yamada \cite{Nomura2002} studied a two-dimensional three-band Hubbard model using the Eliashberg equations and found that, for small to moderate Couloumb interactions, spin-singlet superconductivity with $d$-wave character is most stable.
More recent work by R{\o}ising and collaborators \cite{Roising2019}, has found spin-singlet $d$-wave superconductivity with $B_{1g}$ symmetry for moderate values of Hund's coupling within a weak coupling calculation.
R{\o}mer \etal \cite{Romer2020MPL} find similar conclusions within a theory based on the spin-fluctuation mechanism.
Considering fluctuations in both spin and charge sectors, Gingras \etal \cite{Gingras2022} found dominant $d$-wave pairing for moderate Hund's coupling and Stoner factor.
In fact, all microscopic theories discussed below in the context of accidental degeneracies primarily suggest the order parameter for \SRO to be a single-component (1D irrep) spin-singlet order parameter. 
Recent DMFT results also support spin-singlet superconductivity \cite{Acharya2019} and predict the development of a spin-density wave for large uniaxial strain along the $\langle 100 \rangle$ direction \cite{Acharya2021}. Furthermore, recent revision of the temperature and magnetic field responses of helical superconductors in the spin-triplet sector were reconsidered by Gupta \etal  \cite{Gupta2020}, finding good agreement with experiments by capturing the high-field subphase as well as the suppression of Knight shift for magnetic fields along the $c$-axis. However, the magnetic field anisotropy with respect to the polar angle cannot be easily accounted for by helical pairing. Overall, the difficulty with proposals based on a single-component superconducting order parameter is to account for ultrasound attenuation experiments and time-reversal symmetry breaking. 

In this section, we discuss the main theoretical lines that have been proposed over the last years, highlighting the motivation they take from experimental results, as well as their successes and limitations.
Table~\ref{tab:sc states} summarizes some of the candidate superconducting states for \SRO and their consistencies with key experimental results.
We focus on theories that address two-component superconducting order parameters, what is suggested by ultrasound attenuation experiments  \cite{Ghosh2021.NatPhys.17.199,Benhabib2021.NatPhys.17.194} and supported by the observation of time-reversal symmetry breaking by polar Kerr \cite{Xia2006.PhysRevLett.97.167002} and $\mu$SR \cite{Grinenko2021.NatureCommum.12.3920} experiments.

\begin{table*}[p]
\begin{center} 
\caption{Summary of several candidate superconducting states with their gap symmetries. Compatibility with theoretical models and key experimental results are indicated with different symbols.}
\includegraphics[width=17cm]{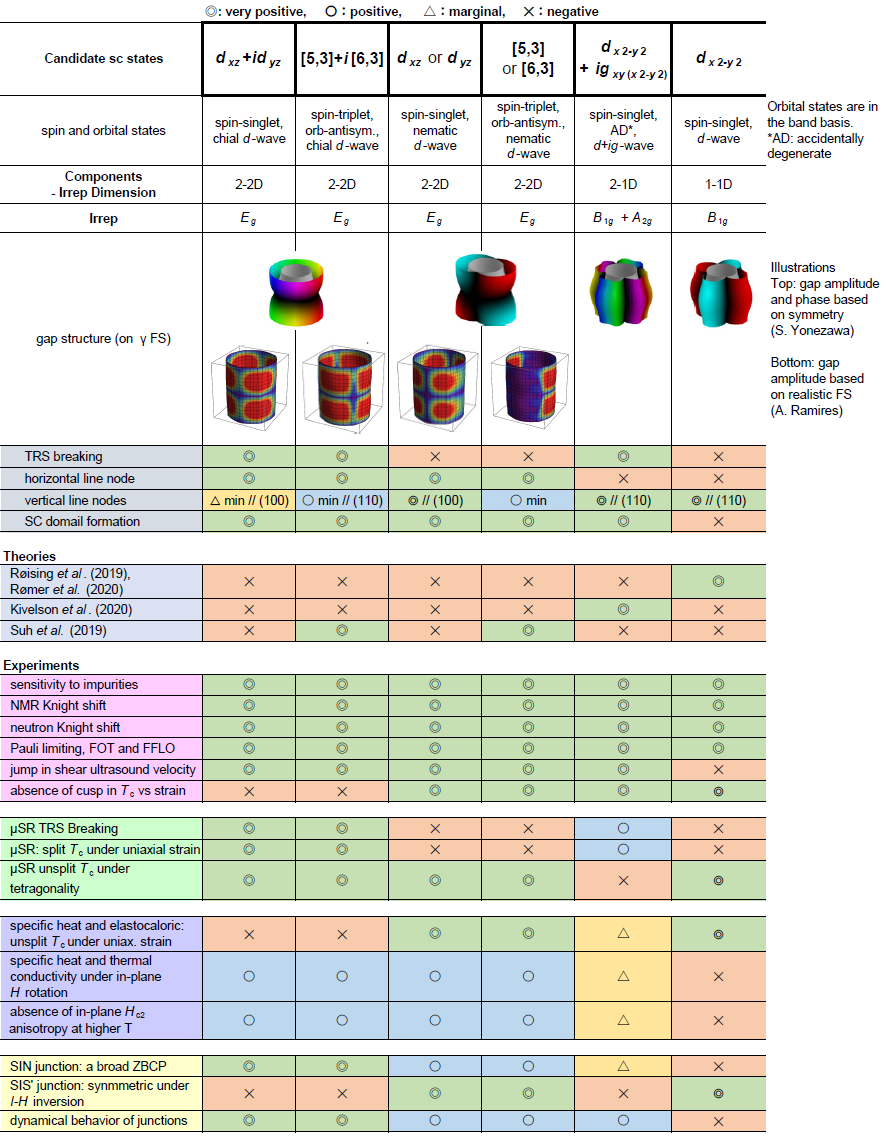}
 \end{center}
\label{tab:sc states}
\end{table*}

\subsection{\SECTIONTHEOACCIDENTAL}\label{Sec:TheoAccidental}

The apparently contradicting experiments in \SRO have led to the proposal of two different scenarios with two accidentally degenerate order parameter components.

\begin{figure}[ht]
\begin{center}
\includegraphics[width=0.35\textwidth, keepaspectratio]{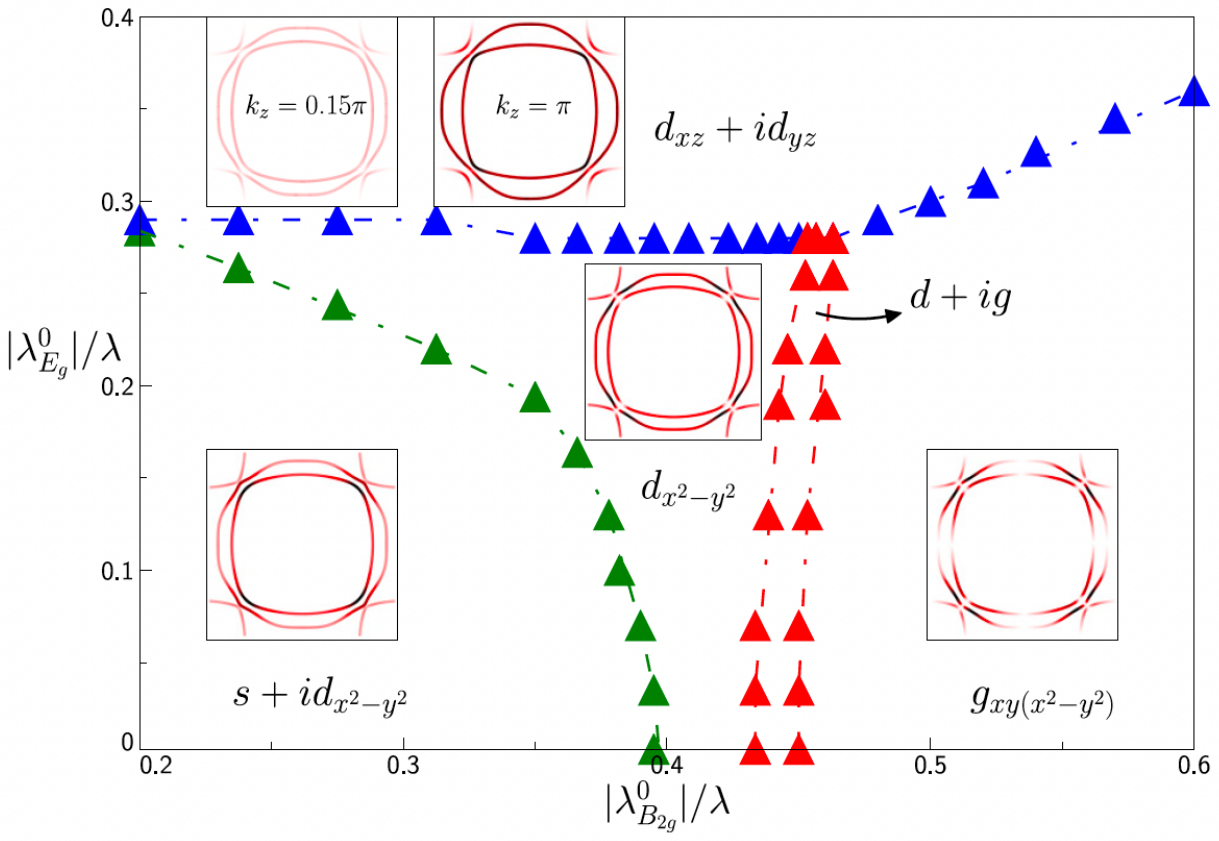}
        \end{center}
\caption{Zero-temperature mean-field phase diagram obtained for a three-dimensional three-band model of \SRO with Hubbard-Kanamori interactions as a function of SOCs with $B_{2g}$ and $E_g$ symmetries, normalized by $\lambda$, the value of the atomic SOC. Reproduced from Ref. \citen{Clepkens2021b} ($\copyright$~2021  American Physical Society).}
\label{Fig:Clepkens2021}
\end{figure}

Kivelson \etal \cite{Kivelson2020} have proposed a ``$d+ig$" scenario.
This phenomenological scenario relies on an accidental degeneracy between a $d_{x^2-y^2}$-like order parameter with $B_{1g}$ symmetry, and a $g_{xy(x^2-y^2)}$-like order parameter with $A_{2g}$ symmetry.
This proposal was primarily motivated by the recent ultrasound attenuation experiments, which observe a discontinuity in the elastic modulus $c_{66}$ with $B_{2g}$ symmetry, and no jump in the $(c_{11}-c_{12})/2$ modulus with $B_{1g}$ symmetry \cite{Ghosh2021.NatPhys.17.199,Benhabib2021.NatPhys.17.194}.
It is also inspired by the presence of symmetry-protected vertical line nodes along the $(1,1,0)$ and $(1,\bar{1},0)$ directions suggested by thermal conductivity \cite{Hassinger2017.PhysRevX.7.011032} and quasiparticle interference \cite{Sharma2020.ProcNatlAcadSci.117.5222}.
Theoretically, such an accidental-degeneracy could be expected when there is a balance between the on-site and nearest-neighbour interactions \cite{Raghu2012}; so, in principle, there is only one fine-tuned parameter in the theory.
Nevertheless, a more careful inspection of the behaviour of $\Tc$ and $\TTRSB$ indicates that these accidentally coinciding temperatures should split under any kind of external perturbation.
These conclusions are in contrast with the recent compilation of $\mu$SR experiments under strain, hydrostatic pressure, and with samples with different degrees of disorder \cite{Grinenko2021.NatureCommum.12.3920}.
To account for the unsplit behaviour of $\Tc$ and $\TTRSB$ under hydrostatic pressure and disorder would require the fine-tuning of multiple parameters within a Ginzburg-Landau formalism.
Interesting solutions out of this conundrum were proposed based on the formation of bound-state order \cite{Willa2020}, or on the coexistence of $d$- and $g$-wave order parameters in presence of inhomogeneities, in which case the TRS breaking occurs only around domain walls between regions with different order parameters, also allowing for half-quantum vortices \cite{Yuan2021}.
Interestingly, a ``d+ig" state was also found by Clepkens \etal \cite{Clepkens2021b} when momentum-dependent SOC with $d$-wave character is significant in the presence of Hubbard-Kanamori interactions, as shown in Fig.  \ref{Fig:Clepkens2021}.
The specific form of the order parameter in their work can be unfold in the microscopic basis and is found to be an even-parity interorbital spin-triplet pairing, which was dubbed ``shadowed triplet".

\begin{figure}[ht]
\begin{center}
\includegraphics[width=70mm, keepaspectratio]{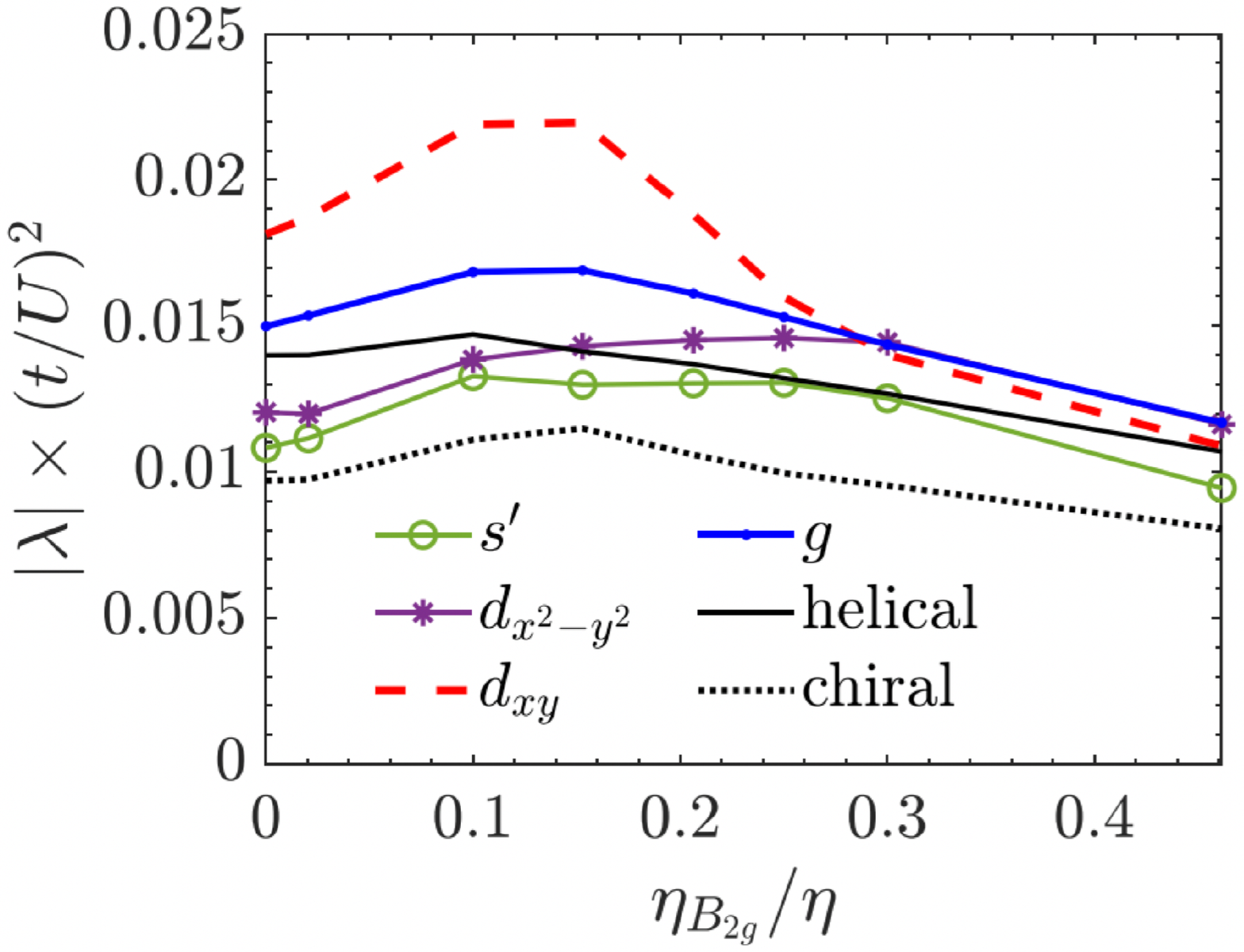}
        \end{center}
\caption{Evolution of the leading pairing eigenvalue $\lambda$ in each symmetry channel as a function of $\eta_{B_{2g}}$, the magnitude of the SOC with $B_{2g}$ symmetry, for a model with finite nearest and next-nearest-neighbour interactions. Note the degeneracy between $d_{x^2-y^2}$- and $g$-wave superconductivity for large $\eta_{B_{2g}}$. Reproduced from Ref. \citen{Wang2022} ($\copyright$~2022  American Physical Society).}
\label{Fig:Wang2022}
\end{figure}

\begin{figure}[ht]
\begin{center}
\includegraphics[width=0.38\textwidth, keepaspectratio]{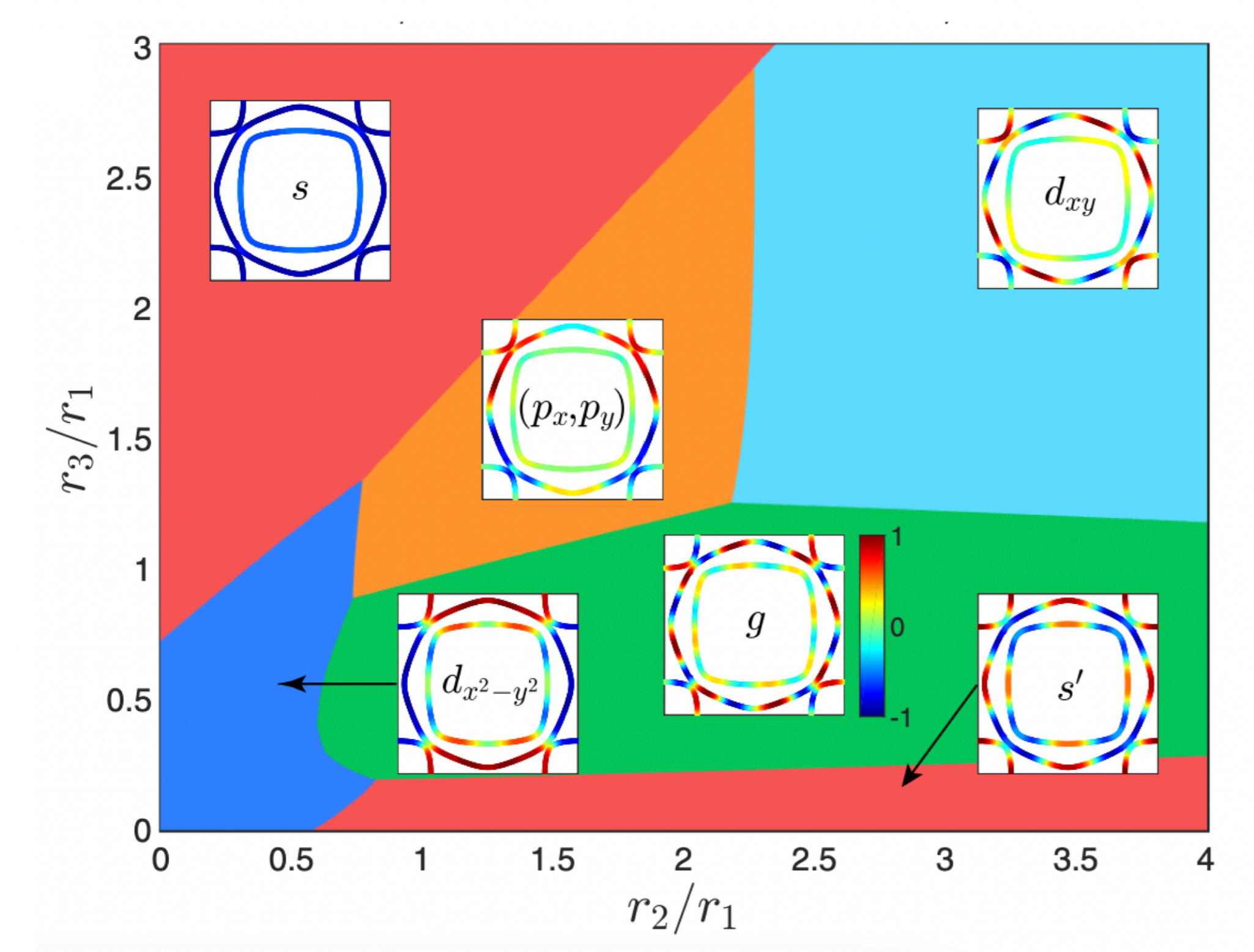}
        \end{center}
\caption{Phase diagram with predominant pairing states obtained from magnetic and multipolar fluctuation mechanisms. The axes are controlled by the ratios of the strength of interactions in each of the channels following $V_{mix} = r_1V^{AFM} + r_2V^{FM} + r_3V^E$, where AFM stands for antiferromagnetic, FM for ferromagentic, and E for electric multipole. Reproduced from Ref. \citen{Sheng2022} ($\copyright$~2022  American Physical Society).}
\label{Fig:Sheng2022}
\end{figure}

More recent theoretical work based on the random-phase approximation (RPA) considers an interacting model for \SRO with long-range Coulomb repulsion and a realistic multi-orbital model \cite{Wang2022}.
The authors conclude that nearest-neighbour and next-nearest-neighbour repulsion promotes $g$-wave pairing, and that this state is further stabilized by the presence of momentum-dependent SOC with $B_{2g}$ symmetry and anisotropic interactions in orbital space, as shown in Fig. \ref{Fig:Wang2022}.
Yuan \etal \cite{Yuan2023} also start from a microscopic three-orbital model with next-neighbour interactions, but neglect possibly important effects of SOC.
Their approach requires the value of the interorbital next-neighbour interactions to be above a certain critical value in order to stabilize a ``d+ig" state.
In their model, the $d$-wave component is stronger in the $\gamma$ band, while the g-component lies primarily in the $\alpha$ and $\beta$ bands.
In addition, recent theoretical work considering a realistic electronic band structure showed that a ``d+ig" state can arise naturally from the interplay of antiferromagnetic, ferromagnetic, and electric multipole fluctuations \cite{Sheng2022}, as illustrated in Fig.  \ref{Fig:Sheng2022}.
A ``d+ig" superconducting state can account for the behaviour of $T\subm{c}$ and $\TTRSB$ under strain along the $\langle110\rangle$ direction \cite{Yuan2023}.
In particular, it emulates a strong strain dependence of $\Tc$ based on the strong $\gamma$-band dependence on the DOS, and the weak strain dependence of $\TTRSB$. 
On the other hand, hydrostatic pressure in this model leads to a strong splitting of the two temperatures, in clear contrast to experiments  \cite{Grinenko2021.NatureCommum.12.3920}.

\begin{figure}[ht]
\begin{center}
\includegraphics[width=0.38\textwidth, keepaspectratio]{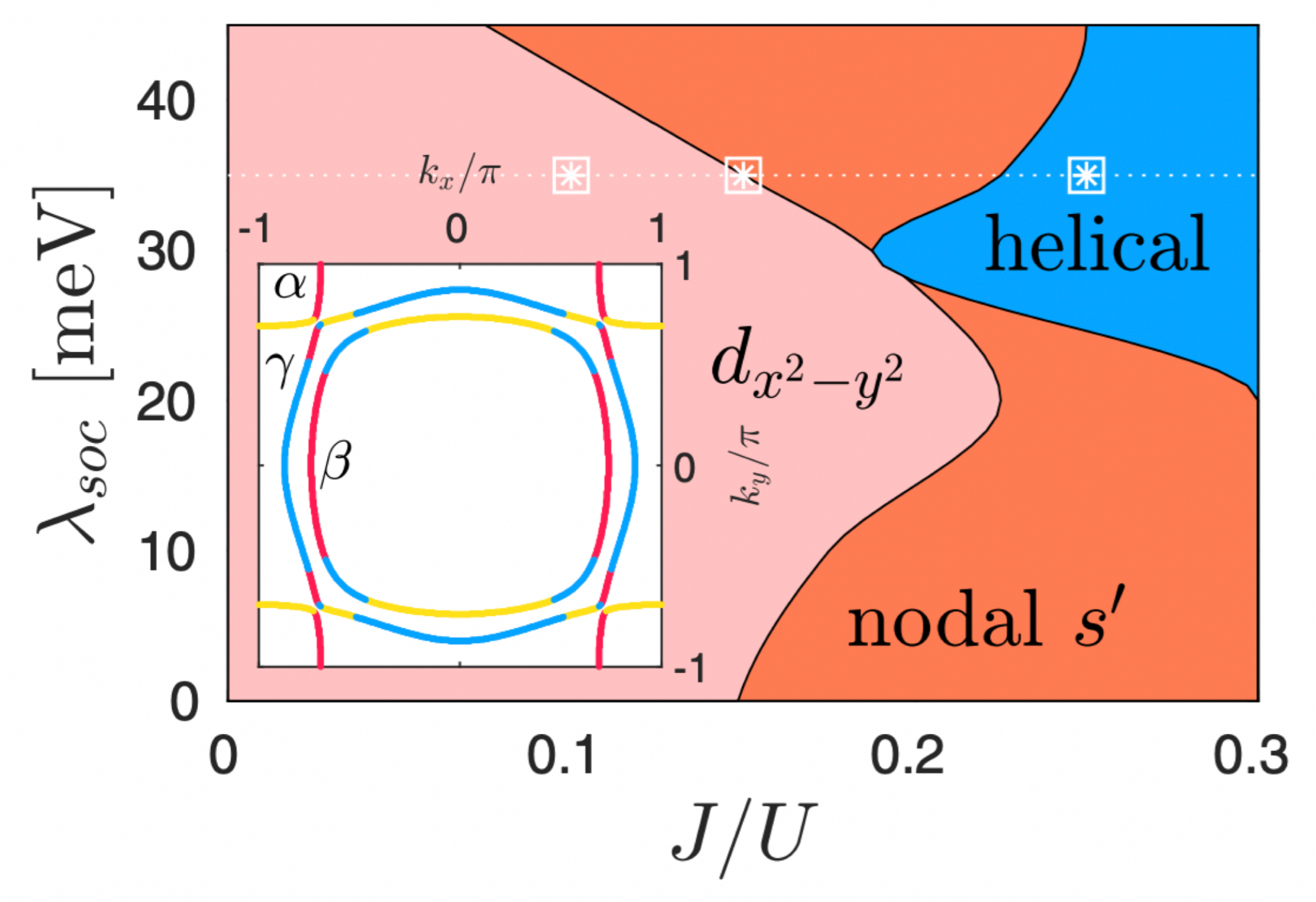}
        \end{center}
\caption{Phase diagram with leading superconducting instability as a function of
SOC amplitude $\lambda_{SOC}$ and Hund’s coupling $J$ normalized by the Hubbard interaction $U$.
Reproduced from Ref. \citen{Romer2019} ($\copyright$~2019  American Physical Society.}
\label{Fig:Romer2019}
\end{figure}

In a similar direction, R\o{}mer \etal \cite{Romer2019} propose an ``s+id" scenario.
Inspired by the new NMR experiments \cite{Pustogow2019.Nature.574.72, Chronister2021.PNAS.118.25}, the authors investigated the possibility of spin-singlet pairing in \SRO based on the spin-fluctuation mechanism and found an unexpected almost-degeneracy between a near-nodal $s$-wave, with $A_{1g}$ symmetry, and a $d_{x^2-y^2}$-wave order parameter, with $B_{1g}$ symmetry, for a given parameter regime, as illustrated in Fig. \ref{Fig:Romer2019}.
Further investigation of this pairing scenario under strain revealed that the leading instability splits off from the quasi-degenerate sub-leading solutions as a function of strain along the x-direction \cite{Romer2020}, in qualitative agreement with $\mu$SR experiments \cite{Grinenko2021.NatPhys.17.748}.
The detailed study of the effects of three-dimensionality within this framework still find that an $s+i d_{x^2-y^2}$-type of order parameter is most favourable \cite{Romer2022}.
Note, though, that these calculations were performed in the intermediate-coupling regime.
In the weak-coupling regime all quasi-degenerate pairing solutions follow the density of states evolution with strain, corresponding to the absence of clear splitting of $\Tc$ and $\TTRSB$.
One of the main inconsistencies of the $s+i d_{x^2-y^2}$-type of order parameter is its inability to account for the discontinuity of the elastic constant $c_{66}$ associated with shear $B_{2g}$ strain.
Curiously, the introduction of long-range Coulomb interactions seems to change the scenario to $s+ i d_{xy}$ order parameter \cite{Romer2021}, which is  consistent with the ultrasound experiments  \cite{Ghosh2021.NatPhys.17.199}.
Concerning the specific heat jump at the second transition under strain, a recent investigation has suggested that it could be arbitrarily small for gaps with similar nodal structures in presence of inhomogeneities  \cite{Roising2022}. 

\subsection{\SECTIONDEGENERACY}\label{Sec:Degeneracy}

The recent NMR measurements indicating that the superconducting order parameter is of spin-singlet nature \cite{Pustogow2019.Nature.574.72, Ishida2020.JPSJ.89.034712, Chronister2021.PNAS.118.25},  and the jump in the shear modulus $c_{66}$ at the superconducting transition temperature highlighting the two-component nature of the order parameter \cite{Ghosh2021.NatPhys.17.199,Benhabib2021.NatPhys.17.194} inspired theorists to reconsider superconducting order parameters with $E_g$ symmetry.
Order parameters in this symmetry channel, with symmetry protected horizontal line nodes, were for a long time dismissed based on the quasi-two-dimensional nature of the electronic structure of \SRO \cite{Bergemann2003.AdvPhys.52.639}.
A naive weak-coupling perspective would require strong interlayer coupling for an order parameter with a horizontal line-node to develop. 

\begin{figure}[ht]
\begin{center}
\includegraphics[width=0.35\textwidth, keepaspectratio]{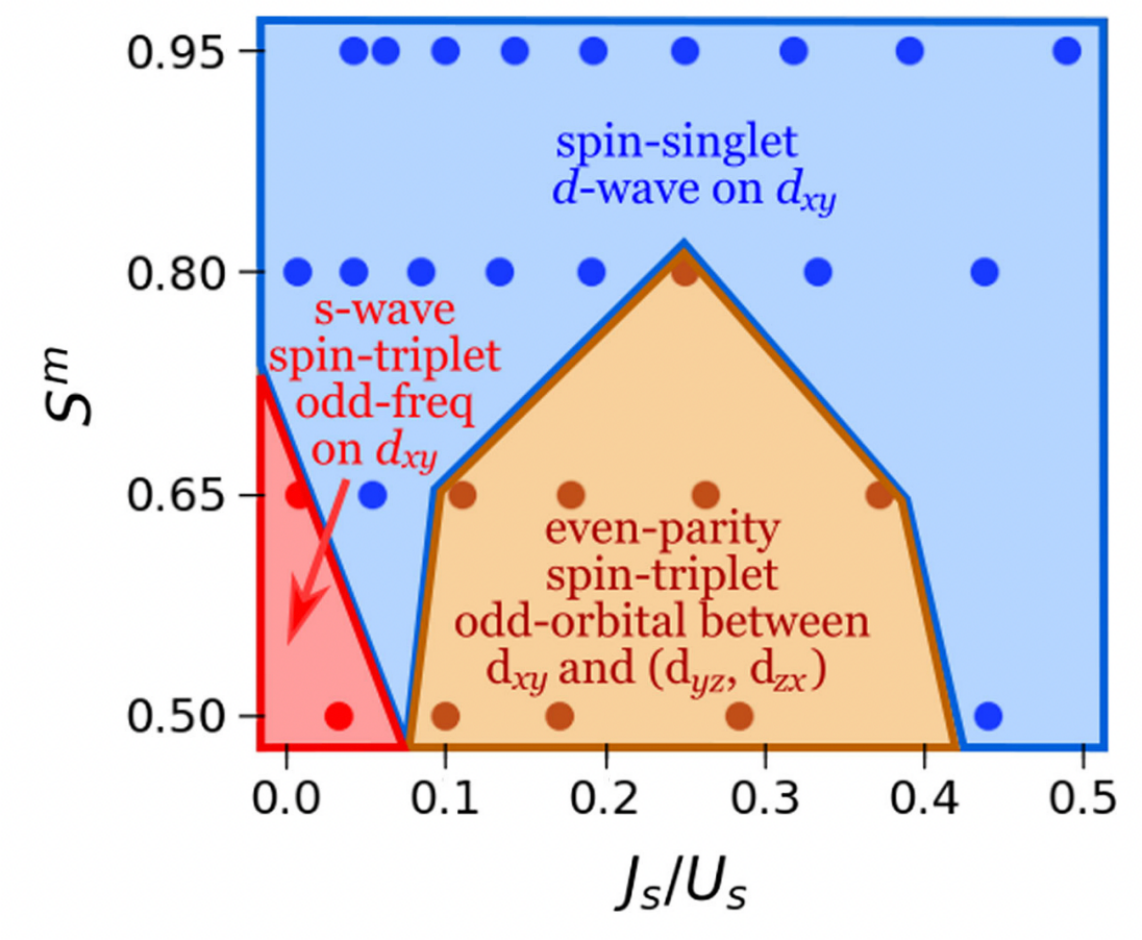}
        \end{center}
\caption{Phase diagram of the leading superconducting instabilities obtained from a first-principles electronic structure of \SRO in the strong coupling limit.
A lower ratio of screened (S) interactions $J_S/U_S$ implies more charge fluctuations, while
the magnetic Stoner factor $S^m$ quantifies the proximity to a
magnetic instability.
Reproduced from Ref. \citen{Gingras2019} ($\copyright$~2019  American Physical Society).}
\label{Fig:Gingras2019}
\end{figure}

Remarkably, the internal richness of the multi-orbital electronic structure of \SRO allows for the development of superconductivity with interorbital character in the $E_g$ channel in the presence of purely local interactions once a faithful microscopic description of its three-dimensional electronic structure is taken into account. Order parameters with interorbital character associated with one-dimensional irreps were previously discussed in the context of multiorbital systems \cite{Puetter2012, Hoshino2015}. Order parameters with interorbital character and $E_g$ symmetry were found in the context of strong coupling by DMFT calculations \cite{Gingras2019}, as illustrated in Fig.  \ref{Fig:Gingras2019}, but the superconductivity in \SRO is likely in the weak-coupling regime \cite{Mackenzie2003RMP}.

\begin{figure}[ht]
\begin{center}
\includegraphics[width=6cm, keepaspectratio]{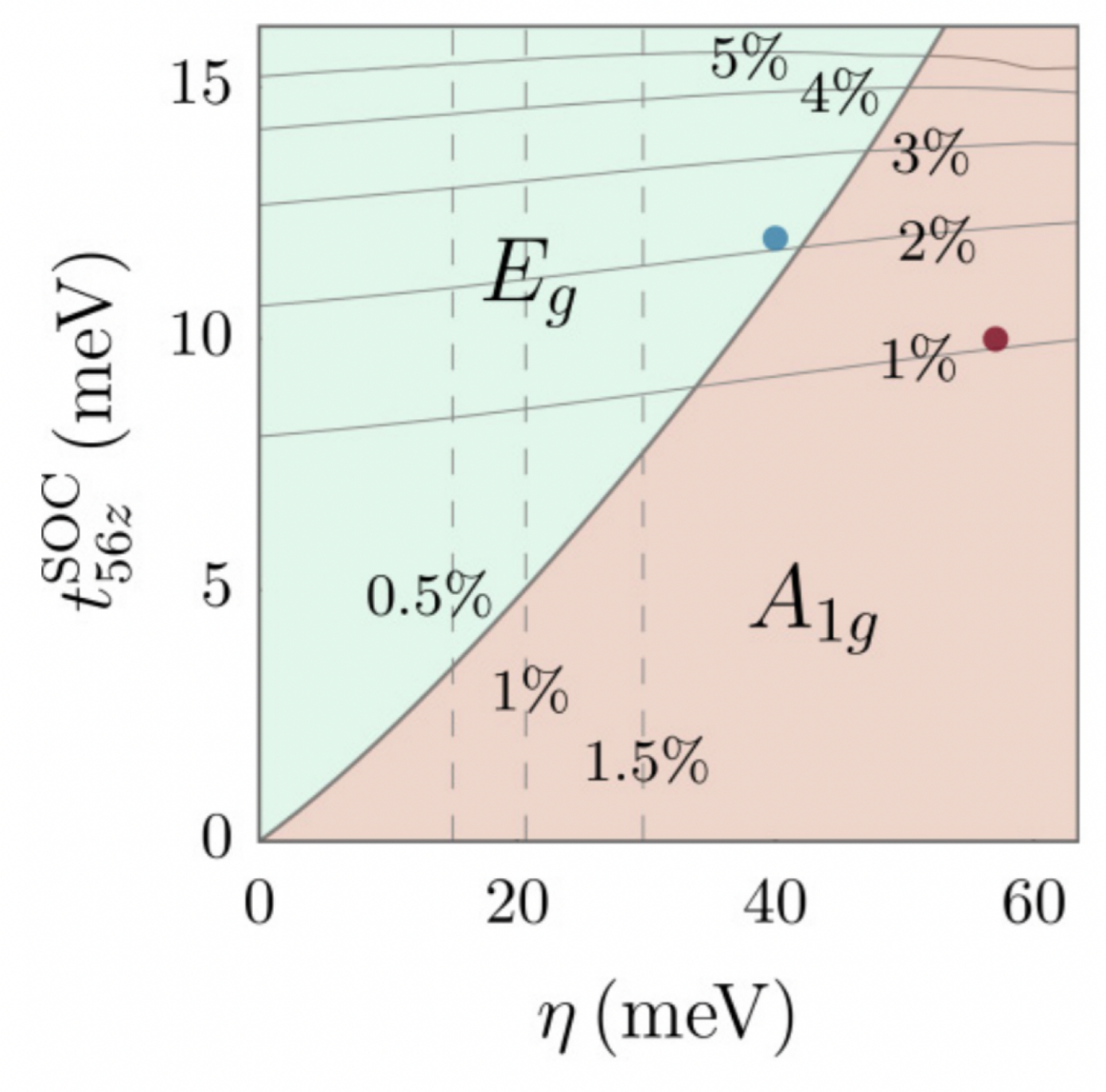}
        \end{center}
\caption{Phase diagram showing the stability of $A_{1g}$ and $E_g$
pairing states as a function of the SOC parameters $\eta$ (atomic) and $t^\mathrm{SOC}_{56z}$ (momentum-dependent).
The vertical dashed lines indicate the minimum distance between two
Fermi surfaces (as percentages of the Brillouin zone length).
The thin solid curves indicate the maximum variation of the Fermi surace along the $k_z$ direction.
Reproduced from Ref. \citen{Suh2020} ($\copyright$~2020 American Physical Society).}
\label{Fig:Suh2020}
\end{figure}

Motivated by the concept of superconducting fitness \cite{Ramires2016, Ramires2018}, a weak-coupling theory for an $E_g$ order parameter was proposed using a realistic normal state electronic structure \cite{Suh2020, Clepkens2021}.
In addition to the standard intra and interorbital hopping amplitudes and atomic SOC, a three-dimensional model allows for new types of momentum-dependent SOC \cite{Ramires2019,Suh2020, Clepkens2021}.
The latter are fundamental for the stabilization of an order parameter with $E_g$ symmetry, in addition to strong Hund's coupling  \cite{Suh2020}, as shown in Fig. \ref{Fig:Suh2020}.
The proposed order parameter stems from local interactions, so in the microscopic orbital-spin basis it can be classified as $s$-wave orbital antisymmetric spin triplet, dubbed OAST.
Interorbital pairing was previously considered, but in other symmetry channels \cite{Puetter2012, Zegrodnik2014.Spalek.NJP, Hoshino2015}.

The weak-coupling OAST order parameter \cite{Suh2020} has a series of notable features: First, the two-component nature of the order parameter is associated with two different types of interorbital pairing (between $d_{xy}$ and $d_{xz}$ or $d_{yz}$).
Second, in the orbital basis, the order parameter is $s$-wave, but once the order parameter is rotated to the band basis its nodal structure with symmetry protected horizontal line nodes is revealed.
In case the two components appear in a time-reversal symmetry breaking superposition, the nodes are inflated into Bogoliubov Fermi surfaces \cite{Agterberg2017, Brydon2018}.
In addition to the horizontal line nodes, the order parameter has a non-trivial amplitude dependence around the Fermi surface following the orbital distribution, leading to unexpected vertical near nodes along the diagonals, resembling an order parameter with $d_{x^2-y^2}$ symmetry \cite{Ramires2022}. Note that this near nodal structure is very different from what is naively expected for a ``standard" chiral $d$-wave state with momentum dependence $\propto k_z(k_x\pm i k_y)$, with gap minima along the x- and y-axes, see discussion in Sec. \ref{Sec:OrbitalBand} and Fig. \ref{Fig:Gaps}. 
Third, the importance of the orbital distribution around the Fermi surface is revealed by experiments under strain.
In particular, compressive strain along the $z$-axis reduces the critical temperature, while the density of states is enhanced.
The reduction of the critical temperature in this case could be semi-quantitatively understood based on the less favourable orbital distribution around the Fermi surface under $z$-axis compressive strain \cite{Beck2022}.
This is also corroborated by the behaviour of the critical temperature under hydrostatic pressure, as it is known that in this case the Fermi surface becomes more two-dimensional \cite{Forsythe2002}.
No other order parameter proposed so far can consistently explain the behaviour of the critical temperature under strain along different directions \cite{Jerzembeck2022}.

The OAST order parameter is in agreement with multiple experimental observations.
As a pseudospin-singlet state, it is in agreement with the recent NMR results \cite{Pustogow2019.Nature.574.72, Ishida2020.JPSJ.89.034712, Chronister2021.PNAS.118.25}. Note that, despite being a spin-triplet state, the OAST order parameter is a pseudospin-singlet state in the band basis. 
In this case, the spin susceptibility decreases as the temperature is lowered, irrespective of the direction of the applied field. 
A comprehensive theoretical discussion and explicit evaluation of the spin susceptibility was recently reported by Fukaya \etal \cite{Fukaya2022}
In addition to the uniaxial-strain experiments mentioned above, it seems to be the only order parameter that can naturally explain the degeneracy of the $\Tc$ and $\TTRSB$  \cite{Grinenko2021.NatureCommum.12.3920} under perturbations that do not break four-fold rotational symmetry along the $z$-axis without fine-tuning.
Furthermore, Zhang \etal \cite{Zhang2021} have shown that for the chiral $d$-wave order parameter, the theoretical cusp in $\Tc$ could be easily smeared out by strain inhomogeneity.
Despite these successes, the OAST order parameter still faces some difficulties.
The most critical aspect that remains to be addressed is the absence of thermodynamic evidence for the second superconducting transition at the TRS breaking temperature~\cite{Li2021}.
This aspect could be potentially understood by the unusual nodal structure of this chiral order parameter, which is remarkably similar to the non-TRS breaking nematic state.
The similarity of the gap anisotropy for the chiral and nematic states could guarantee an arbitrarily small specific heat jump at the TRS breaking transition, in the same spirit as the discussion in R\o{}ising \etal \cite{Roising2022}
A second point to be clarified is the  absence of jump in the shear modulus with $B_{1g}$ symmetry \cite{Ghosh2021.NatPhys.17.199}, generally expected for a two-component order parameter with $E_g$ symmetry.

\subsection{\SECTIONEXTRINSIC}\label{Sec:Extrinsic}

\begin{figure}[ht]
\begin{center}
\includegraphics[width=0.9\columnwidth, keepaspectratio]{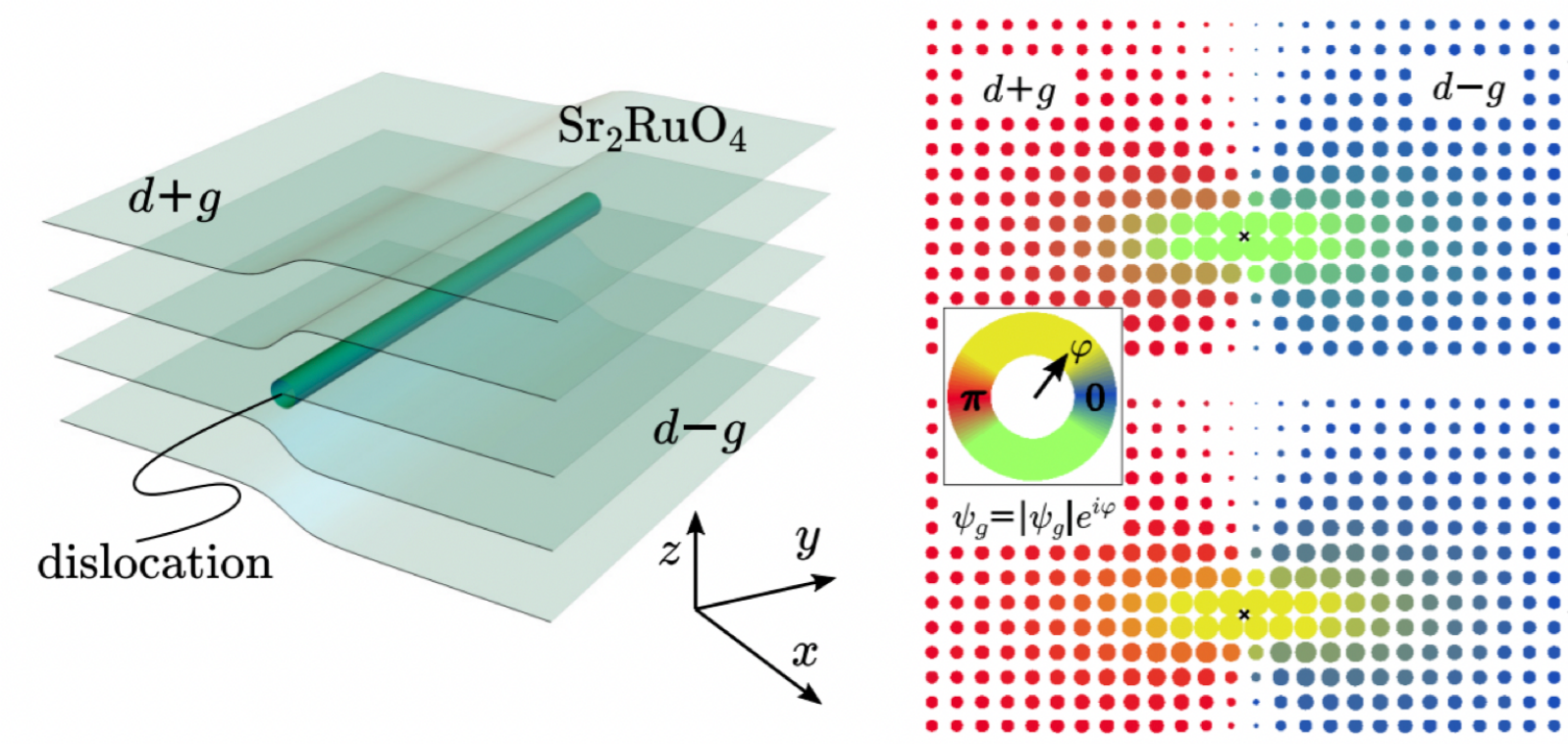}
        \end{center}
\caption{Left: Representation of an edge dislocation in \SRO inducing inhomogeneous strain.
Near the dislocation, large strain mix a primary $d$-wave pairing state with other subleading symmetry channels, e.g. a $g$-wave pairing state.
Right: Amplitude (circle size) and phase (color) of the strain-induced $g$-wave component in proximity of a dislocation. 
Inhomogeneous TRS breaking occurs when the phase winding  is energetically favored over the vanishing of the $g$-wave component. 
Reproduced from Ref. \citen{Willa2021} ($\copyright$~2021 American Physical Society).}
\label{Fig:Willa2021}
\end{figure}

A different proposal by Willa \etal relies on the presence of specific types of structural defects to explain the origin of TRS breaking \cite{Willa2021}.
The presence of inhomogeneous strain near edge dislocations could stabilize a sub-leading order parameter around these defects, as schematically shown in Fig. \ref{Fig:Willa2021} left.
Assuming a leading superconducting order parameter with $B_{1g}$ symmetry, local strain with  $B_{2g}$ symmetry could couple it to a sub-leading superconducting order parameter with $A_{2g}$ symmetry.
Around lattice defects, these two order-parameter components could develop a non-trivial relative phase, leading to local TRS breaking, shown in Fig. \ref{Fig:Willa2021} right.
This proposal relies on the presence of subleading instabilities, which have been suggested by several previous works and seems to be a reasonable starting point.

This proposal is appealing as it explains the absence of a heat capacity anomaly at the TRS breaking transition in strained samples \cite{Li2021}, and is in agreement with internal magnetic fields associated with dilute defects, as observed by $\mu$SR experiments \cite{Grinenko2021.NatPhys.17.748}.
It is also in agreement with the quadratic dependence of $\Tc$~\cite{Hicks2014.Science.344.283} and with the very weak dependence of the $\TTRSB$ under $B_{1g}$  strain \cite{Grinenko2021.NatPhys.17.748}.
Within this theory, Willa \etal \cite{Willa2021} suggest that one could induce plastic deformations on the material in order to manipulate the onset of TRS breaking.
Another prediction of this scenario is the presence of Bardasis-Schrieffer modes, which could be observed by Raman spectroscopy. These suggestions remain to be experimentally investigated.

Despite the appeal of this theory,  to quantitatively explain the observed behaviour of $\Tc$ and $\TTRSB$ under strain, fine tuning of the critical temperatures for the two components in the unstrained system is necessary. This theory also predicts that $\Tc$ should have a quadratic dependence for both $B_{1g}$ and $B_{2g}$ strain.
However, while $\Tc$ under $B_{1g}$ strain is clearly quadratic, $\Tc$ as a function of $B_{2g}$ strain seems to be linear \cite{Hicks2014.Science.344.283}.
This theory is also not able to explain the unsplit $\Tc$ and $\TTRSB$ under hydrostatic pressure \cite{Grinenko2021.NatureCommum.12.3920}, unless multiple parameters are fine-tuned.

\section{\SECTIONREEVALUATION}\label{Sec:Reevaluation}
After all these years of extensive investigation to fully understand the SC state of \SRO, especially after the recent progress taking into account of the revised knowledge, new controversy has emerged and important mystery still remains to be solved.
In this section, we will discuss the unresolved issues such as TRS breaking, gap structure, and the variation of $\Tc$ with strain, and discuss how the mystery may be solved.
Table~\ref{tab:sc states} summarizes the candidate superconducting states for \SRO and the consistencies with key experimental results.

\subsection{\SECTIONPARADIGM}
\label{Sec:Paradigm}

Since the experimental results on the superconducting spin state were revised in 2019, the paradigm based on “spin-triplet odd-parity” scenario for the bulk superconducting state of \SRO has been shifting to the “spin-singlet even-parity” scenario.
Based on the recent NMR and polarized neutron experiments, the spin susceptibility decreases substantially for both $H$ // [100] and [110] directions.
The large magnitudes of the reduction imply that both chiral spin-triplet state with the \emph{d-vector} along the $c$-axis and the helical spin-triplet states with the in-plane \emph{d-vector} are not compatible with the experimental results for bulk samples.

Ultrasound velocity measurements strongly suggest a multicomponent order parameter, but cannot firmly pinpoint whether it is a case of symmetry protected or accidental degeneracy of the two components.

The issue of time-reversal-symmetry (TRS) breaking remains controversial.
There are a number of experimental results in support of TRS breaking. 
Muon experiments by several groups unanimously observed internal magnetic fields below $\Tc$, evidencing TRS breaking.
Magneto-optic Kerr effect (MOKE) also revealed TRS breaking below $\Tc$, and controllability of the sign of the chirality by training under external field.
Flux quantization in micro-rings suggests spontaneous generation of superconducting domains, consistent with the chiral domain formation.
The dynamical behavior of the critical current in the junctions between \SRO and conventional $s$-wave superconductors also suggests the presence of SC domains in the 1.5-K phase, but not in the 3-K phase.
However, internal or edge fields due to anticipated chiral edge currents have not been observed by  scanning SQUID probe.
Furthermore, the $I-H$ inversion tests in SIN junctions sugggest that the time-reversal symmetry is not broken.

Recent $\mu$SR experiments give further evidence of two-component superconductivity since the superconducting and TRS breaking transition temperatures split under uniaxial strain along the [100] direction, while they do not split among samples maintaining tetragonal symmetry with varying $\Tc$, under hydrostatic pressure and with impurities.
These results strongly suggest that the TRS breaking is not due to accidental combination of OPs with different irreps, but is a consequence of the OP protected by tetragonal symmetry, namely chiral OP.
However, concerning thermodynamic evidence, measurements of heat capacity and elastocaloric effect under [100] strain have not detected any sign of the second transition corresponding to the entry into the TRS-breaking phase.  

The scenario of the linear splitting of the superconducting $\Tc$ and $\TTRSB$ by [100] strain based on the Landau theory is simple and straightforward.
In reality, [110] strain leads to a linear change in $\Tc$ expected from the Landau theory: a linear decrease under compression and a linear increase under tensile strain~\cite{Hicks2014.Science.344.283}.
In contrast, [100] strain immediately leads to non-linear quadratic increase in $\Tc$ in both compressive and tensile strain even for minute values of strain~\cite{Watson2018}.
This violates the Landau analysis of linear splitting of $\Tc$ and $\TTRSB$ resulting in the V-shaped cusp in $\Tc$. Such behavior could be a consequence of the fact that the coefficients of the Landau free energy are not constant but depend on [100] strain, due to the strong changes in the electronic structure in the proximity to the Van Hove singularity.

To resolve some of the controversies, we should note that each experiment is performed under different condition:
\begin{itemize}
\item NMR requires injection of RF pulses;

\item $\mu$SR experiments need injection of positive muons with the life-time of 2.2 $\mu$s and probe the magnetic field at the muon site;

\item junction experiments may be influenced by the constituents forming the junction with \SRO, in case they are superconducting or magnetic;

\item some other measurements, especially using optics, can be surface sensitive.
\end{itemize}

The quality of samples may be important for the observation of some of the exotic phenomena, such as the first-order transition near $\Hcc$ under a field applied accurately parallel to the RuO$_2$ plane.
Fortunately, high-quality \SRO single crystals allowed results from the same experimental techniques to be consistent with each other.
Nevertheless, the interpretation of some of the previous results is affected by the presence of crystalline mosaicity and the inclusion of Ru micro-platelets that cause the local enhancement of $\Tc$. 
  
Along with new developments in the experiments, a significant progress in theory has been made.
The major advancement in the last decade is the development of theories based on the three-dimensional, orbital-characterized bands consistent with state-of-the-art ARPES measurements~\cite{Tamai2019}.
Experimentally, the presence of a spin-singlet-like even-parity pairing state in the bulk has gained consensus.
But the TRS breaking characteristics is controversial.
If a symmetry-protected two-component OP is realized, both even parity and TRS breaking may be explained, but then the resulting horizontal line node, presumably on all three FS bands, is difficult to reconcile with the quasi-two-dimensional electronic states.
To resolve this issue, in this review we emphasize the importance of making a clear distinction between the multi-orbital and multi-band pictures and of the possibility of introducing interorbital pairing, which leads to the spin-triplet pairs in the orbital basis, and spin-singlet-like behavior in the band basis (Appendix~\ref{Sec:OrbitalBand}).
Such a model suggests a gentle change in the vertical node-like gap across the TRS breaking transitions and may explain the absence of thermodynamic evidence of a transition in some of the experimental probes.

\subsection{\SECTIONUNRESOLVED}\label{Sec:Unresolved}

In this subsection, we discuss what approaches may lead to the resolution of the current issues in \SRO.
A crucially important task is to clarify what $\mu$SR experiments on \SRO actually probe.
The $\mu$SR technique has been well established as a sensitive magnetic probe of the bulk state. 
Nevertheless, it seems worth examining the possibility that the muons induce and probe a state different from the bulk state of \SRO, especially in zero-external-field measurements with the superconducting electrons.
A more detailed characterization of the observed spontaneous field by measurements of its magnitude in a systematically controlled series of samples may provide some hints.

In addition to $\mu$SR, there have been a number of reports evidencing TRS breaking or superconducting domain formation.
Examples are Kerr effects, ZBCP and dynamical behavior in various junction experiments; so far, these experiments have been performed without external strains.
We propose to extend these experiments under uniaxial strain to confirm possible disappearance of the TRS breaking SC behavior or domain dynamics at temperatures higher than 1.5 K.
As a related issue, emergence of the dynamical behavior of the critical current of SNS junctions has been observed only below 1.5 K in 3-K eutectic samples.
Extending such junction experiments to pure \SRO crystals under uniaxial strain may also provide us useful information.

In order to narrow down the possible SC states, we need convincing evidence for the presence of horizontal line nodes.
Neutron experiments have given conflicting conclusions~\cite{Iida2020, Jenni2021}.
Other direct observation concluding presence or absence of horizontal line nodes is highly awaited.
Some experimental results are fitted with gap models with only the vertical nodes and only the horizontal node.
Since some promising candidate superconducting states, as shown in Table~\ref{tab:sc states}, have nodes or suppressed gap in both the vertical and horizontal directions in the BZ, more discussions concerning the presence or absence of horizontal nodes on top of the vertical line nodes are desirable. 

We have emphasized various features of \SRO quite different from those of cuprate superconductors with $d_{x^2-y^2}$ symmetry.
Nevertheless, any additional evidence against such ordinary $d$-wave SC is helpful to reconfirm the currently proposed scenarios.

In Sec. \ref{Sec:Thin films}, we introduced the progress of superconducting thin films of \SRO and noted the enhancement of $\Tc$ using a substrate strain with a uniaxial component.
For the thin-film study, we propose to enhance $\Tc$ and search for a change in SC states in a more controlled manner by applying uniaxial strain to the substrate with the \SRO film on top using Hicks' type uniaxial strain cell.

On the theoretical side, the normal state is understood in great detail. The effective model with three orbitals in three dimensions can potentially completely parametrize the long-standing results of quantum oscillations and clarify the origin of the different magnitude and symmetry of the warping of the three Fermi surfaces of \SRO\cite{Bergemann2003.AdvPhys.52.639}. In the early 2000's, momentum-dependent SOC terms were not included in the description of the electronic structure of \SRO. Revisiting these results in the light of more complete models including these terms can potentially provide a natural explanation to such Fermi surface warping and provide an estimate of the magnitude of these terms directly from experimental observations. This analysis could be of importance given multiple theories that rely on momentum-dependent SOC terms to stabilize chiral order parameters.

Recent theoretical works addressing the superconducting state in \SRO agree in two points: i) there are many superconducting instabilities of comparable strength, which can become less or more favorable by slightly tuning the electronic structure parameters and bare coupling constants; ii) the superconducting order parameter is complex, be it by the presence of higher harmonics giving extra textures to the gap on the Fermi surfaces, or by the inter orbital character of the pairing, which encodes unexpected form factors on the superconducting order parameter.

Concerning the complexity of the order parameter, probes such as specific heat, thermal conductivity, magnetic susceptibility, cannot, by themselves uniquely identify the superconducting order parameter, as multiple order parameter candidates can lead to very similar thermodynamic responses. The challenge now is to account for \emph{multiple} observables with a single order parameter candidate. Recent theoretical works have provided only partial and semi-quantitative discussions. More refined estimates of these observables and their respective temperature, magnetic field, and strain dependence can give us more constraints on the nature of the superconducting order parameter. Nevertheless, the difficulty still remains, as it has been made clear by various theoretical proposals that the superconducting order parameters are not simply given by the lowest harmonics in each symmetry channel, but are a superposition of multiple harmonics with higher angular momentum in the respective symmetry channels. The contribution of higher harmonics leads to extra structures in the gap, such as flattened regions or near nodes which are not protected by symmetry, which makes the precise identification of the order parameter from these measurements more challenging. 

Theory could also contribute to the understanding of the results that are currently challenged, such as $\mu$SR we described in \ref{Sec:2CompMSR}. 
Recent calculations have given us important information about the effect of muons on the normal state \cite{Huddart2021}, but its effect on the superconducting state remain debatable. 
Multiple exotic superconductors are characterized by $\mu$SR, and several have revealed TRS breaking within the superconducting state. 
Establishing with certainty the role of muons in \SRO could give us a better handle on the interpretation of similar results for different materials.

\section{\SECTIONCONCLUSION}\label{Sec:Conclusion}
A complete resolution of the SC state of \SRO, an archetypal strongly-correlated, highly-conducting, multi-orbital quantum material, ought to be possible with the current knowledge and technology.
In the past, we learned that investigation of both the normal and superconducting states of \SRO sometimes forced even specialists to realize that the previous common sense required revision and replacement by improved techniques and interpretations.
Before the final consensus on the \SRO problem will be reached by the juries of our peers, a few more such breakthroughs may be required.
We envisage that these persistent efforts on the \SRO problem will benefit the progress of our understanding of quantum materials as a whole.

\section*{Acknowledgements}
We acknowledge many colleagues who have contributed to the development of the physics of \SRO. 
We are especially thankful to Clifford Hicks, Kenji Ishida, Andy Mackenzie, and Manfred Sigrist for long-term collaborations, and to Stuart Brown for his contributions leading to the recent developments. 
We are grateful to Atsutoshi Ikeda and Giordano Mattoni for critical readings of the manuscript. We also thank Jan Aarts, Peter Abbamonte, Brian Andersen, Peter Armitage, Daniel Agterberg, James Annett, Shahbaz Anwar, Yasuhiro Asano, Sophie Beck,  Stephen Blundell, Jake Bobowski, Markus Braden, Philip Brydon, Mario Cuoco, Andrea Damascelli, Yuri Fukaya, Antoine Georges, Vadim Grinenko, Zurab Guguchia, Alexander Hampel, Peter  Hirschfeld, Yusuke Iguchi, Ryusuke Ishiguro, Hans-Henning Klauss, Aharon Kapitulnik, Satoshi Kashiwaya, Roustem Khassanov, Hae-Young Kee, Naoki Kikugawa, Katsuki Kinjo, Shunsaku Kitagawa,  Steve Kivelson, Ying Liu, Kathryn A. Moler, Jason Robinson, Masatoshi Sato, Louis Taillefer, Yukio Tanaka, Andre-Marie Tremblay, Tomo Uemura, Jing Xia, Keiji Yada, Yoichi Yanase, Yuki Yasui, and Manuel Zingl for useful discussions. 
Lastly, We thank Tomoko Kodama for her technical assistance.
Y.M. and S.Y. are supported by JSPS Core-to-core program Grant No. JPJSCCA20170002 and KAKENHI Grant No. JP22H01168. 
S.Y. is supported by JSPS KAKENHI Grant Nos. JP20H05158, JP22H04473, and JP20F20020. 
A.R. is supported by the Swiss National Science Foundation through the Ambizione Grant No. 186043.

\vspace{1cm}
\appendix 
\section{\APPENDIXNORMAL}\label{Sec:NormalDetails}

In this appendix, we build up on the information already introduced in Secs. \ref{Sec:Symmetry} and \ref{Sec:Normal} and give details of the derivation of the normal state Hamiltonian based on symmetry principles.

In order to classify the product of matrices $\lambda_a \otimes \sigma_b$ in terms of the irreps of $D_{4h}$, we can discuss how the spin and how the orbital DOF transform independently. We can start with the spin DOF, considering the basis  $\Psi^\dagger  = (c_{\uparrow}^\dagger,  c_{\downarrow}^\dagger)$, in which the  generators of $D_{4h}$ take the form:
\begin{eqnarray}
C_{4z} = \frac{\sigma_0- i \sigma_3}{\sqrt{2}}, \hspace{0.5cm}
C_{2x} = i \sigma_1,  \hspace{1cm}
P = \sigma_0.
\end{eqnarray}

From these transformations, it becomes explicit that the spin matrices $\sigma_1$ and $\sigma_2$  mix under $C_{4z}$ rotations: $C_{4z}.\sigma_1.C_{4z}^{-1} = \sigma_2$. As spins transform trivially under inversion, $\{\sigma_1,\sigma_2\}$ are associated with the $E_g$ irrep. 
The identity matrix $\sigma_0$ transforms trivially under all symmetry operations, therefore is associated with $A_{1g}$. 
Finally, $\sigma_3$ transforms trivially under $C_{4z}$, but picks up a minus sign under $C_{2x}$, associating it with the $A_{2g}$ irrep. These results are summarized in the first line of Table \ref{Tab:H0AB}.

Discussing now the orbital DOF consisting of orbitals in the $t_{2g}$ manifold in the basis $\Psi^\dagger  = (c_{yz}^\dagger,  c_{xz}^\dagger,c_{xy}^\dagger)$, the generators of $D_{4h}$ take the following form:
\begin{align}
C_{4z} = 
\setlength\arraycolsep{2pt}
\begin{pmatrix}
0 & 1 & 0 \\
-1 & 0 & 0 \\
0 & 0 & -1
\end{pmatrix}, \hspace{0.1cm}
C_{2}(x) = \begin{pmatrix}
1 & 0 & 0 \\
0 & -1 & 0 \\
0 & 0 & -1
\end{pmatrix},  \hspace{0.1cm}
P = 
\lambda_0.
\end{align}

The identity matrix is always associated with $A_{1g}$ as it transforms trivially under all point-group operations. Focusing now on $\lambda_1$, this matrix acquires a minus sign under $C_{4z}$, as $C_{4z}.\lambda_1.C_{4z}^{-1} = -\lambda_1$. This GM matrix also acquires a minus sign under a $C_{2x}$ transformation, $C_{2x}.\lambda_1.C_{2x}^{-1} = -\lambda_1$, what associates it with the $B_{2g}$ irrep. The same exercise can be repeated for the remaining GMs. These results are summarized in the first column of Table \ref{Tab:H0AB}.

\begin{table*}[ht]
 \caption{Symmetry allowed basis matrices, $\lambda_a \otimes \sigma_b$, in the normal state Hamiltonian labelled as $(a,b)$. The first line indicates the irrep of the matrices in spin space, the first column indicates the irrep of the matrices in orbital space. The irreps associated with each $(a,b)$-matrix is found by taking the product of the irrep associated with the matrix in spin space and in orbital space following the rules given in Table \ref{Tab:D4hProduct}.}
    \label{Tab:H0AB}
\begin{center}
    \begin{tabular}{ c | ccc}
    \hline\hline
 & $\boldsymbol{\sigma_0}$ $\boldsymbol{(A_{1g})}$ & $\boldsymbol{\{\sigma_1,\sigma_2\}}$ $\boldsymbol{(E_{g})}$ & $\boldsymbol{\sigma_3}$ $\boldsymbol{(A_{2g})}$  \\ \hline
$\boldsymbol{\lambda_0}$ $\boldsymbol{ (A_{1g})} $& $(0,0)$ ${A_{1g}}$ & $\times$ & $\times$  \\ 
$\boldsymbol{\lambda_1}$ $\boldsymbol{ (B_{2g})} $& $(1,0)$ ${B_{2g}}$ & $\times$ & $\times$  \\ 
$\boldsymbol{\{\lambda_2,\lambda_3\}}$ $\boldsymbol{ (E_{g})} $ & $\{(2,0),(3,0)\}$ ${E_{g}}$ & $\times$ & $\times$  \\ 
$\boldsymbol{\lambda_4}$ $\boldsymbol{ (A_{2g})} $ & $\times$ & $\{(4,1),(4,2)\}$ ${E_{g}}$ & $(4,3)$ $A_{1g}$  \\ \hline
\multirow{4}{*}{$\boldsymbol{\{\lambda_5,\lambda_6\}}$ $\boldsymbol{ (E_{g})} $} & 
\multirow{4}{*}{$\times$} & $(5,2)-(6,1)$ $A_{1g}$ & \multirow{4}{*}{$\{(5,3),(6,3)\}$$E_g$} \\
  & & $(5,1)+(6,2)$ $A_{2g}$ & \\
  & &   $(5,2)+(6,1)$ $B_{1g}$ & \\
 & &   $(5,1)-(6,2)$ $B_{2g}$ &   \\ \hline
$\boldsymbol{\lambda_7}$ $(\boldsymbol{B_{1g}})$ & $(7,0)$ $B_{1g}$ & $\times$ &$\times$  \\ 
$\boldsymbol{\lambda_8}$ $(\boldsymbol{A_{1g}})$  & $(8,0)$ $A_{1g}$ & $\times$ &$\times$  \\ \hline\hline
    \end{tabular}
        \end{center}
\end{table*}

The irrep associated with the matrix product $\lambda_a \otimes \sigma_b$ can be obtained by following the product table displayed as Table \ref{Tab:D4hProduct}, as is shown as the entries of Table \ref{Tab:H0AB}. As the normal state Hamiltonian must be invariant under all point-group transformations, the accompanying momentum-dependent functions $h_{ab}(\bk)$ must transform in a way that the entire product $h_{ab} (\bk)\lambda_a \otimes \sigma_b$ is invariant. For the case of one dimensional irreps, this is achieved by an $h_{ab}(\bk)$ belonging to the same irrep as the matrix product $\lambda_a \otimes \sigma_b$, as can be inferred from Table \ref{Tab:D4hProduct}, as the product of a one-dimensional irrep with itself always results on the $A_{1g}$ irrep. 
For example, for the basis matrix with indices $(a,b) = (1,0)$ belonging to $B_{2g}$, the accompanying function $h_{10}(\bk) \propto \sin(k_x a)\sin(k_ya)$ also transforms according to $B_{2g}$, where $a$ is the in-plane lattice spacing (it can also include other terms with higher harmonics). In case the irrep of the matrix product $\lambda_a \otimes \sigma_b$ is $E_g$, the corresponding  functions $h_{ab}(\bk)$ must also transform according to  $E_g$. Note that only a specific combination of these transforms as $A_{1g}$.

Now we discuss the physical meaning of the different terms in normal state Hamiltonian we find through this symmetry analysis. The Hamiltonian is in accordance with the well stablished Hamiltonian for Sr$_2$RuO$_4$,  \cite{Scaffidi2014} but includes terms that are usually neglected, associated with momentum-dependent SOC or with hopping amplitudes along the $c$-axis. See Clepkens \textit{et~al.}\cite{Clepkens2021} and Suh\textit{et~al.}\cite{Suh2020} for a more complete description.

\begin{itemize}

\item The matrices $(0,0)$, $(7,0)$ and $(8,0)$ are diagonal in orbital and spin spaces and are associated with intraorbital hopping in the $A_{1g}$, $B_{1g}$ and $A_{1g}$ representations, respectively.

The accompanying momentum functions are:
\begin{align}
h_{00}(\bk) &= \frac{1}{3}[\xi^{yz} (\bk) + \xi^{xz}(\bk) +\xi^{xy}(\bk)]
,\nonumber \\ \nonumber
h_{70}(\bk) &= \frac{1}{2}[\xi^{yz} (\bk)  -\xi^{xz}(\bk)]
,\\ 
h_{80}(\bk) &= 
\frac{1}{6\sqrt{3}}[\xi^{yz}(\bk) + \xi^{xz}(\bk)-2 \xi^{xy} (\bk) ]
,
\end{align} where 
\begin{align}
\xi^{yz} (\bk) &= -2t_1\cos(k_y a) -2 t_2 \cos(k_x a),\nonumber \\ \nonumber
\xi^{xz} (\bk) &= -2t_1\cos(k_x a) -2 t_2 \cos(k_y a), \\ 
\xi^{xy} (\bk) &= -2t_3 [\cos(k_xa) + \cos(k_y a)],
\end{align}
capture  the nearest neighbour hoppings along the $x$- and $y$-directions. Here $t_{i=1,2,3}$ correspond to distinct hopping amplitudes, and $a$ corresponds to the in-plane lattice constant. The numerical values are $t_1 = 412.4$ meV, $t_2 = 50.8$ meV, and $t_3 = 402.5$ meV \cite{Clepkens2021}. Information about higher order hopping amplitudes can be found in Clepkens \textit{et~al.}\cite{Clepkens2021} and Suh\textit{et~al.}\cite{Suh2020};

\item The matrix $(1,0)$ is off-diagonal in orbital space and is therefore associated with interorbital hopping in $B_{2g}$. The momentum-dependent accompanying function is $h_{10} (\bk)= -4 t_6 \sin(k_x a)\sin(k_ya)$, with $t_6 = 12.4$ meV;

\item The matrices $(4,3)$ and $(5,2)-(6,1)$ are antisymmetric in orbital space and non-trivial in spin space and  are associated with atomic SOC,  in $A_{1g}$ with $h_{43}(\bk) = -h_{52}(\bk) = h_{61}(\bk) = \lambda = 68.7$ meV \cite{Clepkens2021};

\item Other allowed terms are: $\{(3,0),(-2,0)\}$ in $E_g$, related to out-of-plane interorbital hopping between $xz$ or $yz$ and $xy$ orbitals, with 
\begin{align}
    h_{20}(\bk) &= -8 t_8 \sin(k_z c/2 )\sin(k_x a/2) \cos(k_y a /2), \nonumber\\
    h_{30}(\bk) &= -8 t_8 \sin(k_z c/2 ) \cos(k_x a/2) \sin(k_y a /2),
\end{align}
where $c$ is the $z$-axis lattice constant, and $t_8 = 7.0$ meV \cite{Clepkens2021}; 

\item Finally, $\{(4,2),-(4,1)\}$ and $\{(5,3),(6,3)\}$ in $E_g$, as well $(5,2)+(6,1)$ and $(5,1)\pm(6,2)$ in $B_{1g}$, $A_{2g}$ and $B_{2g}$, respectively, all related to even $k$-dependent SOC, which are usually not taken into account within the simplest theoretical models for \SRO. 
Their functional form is the following:
\begin{align}
\nonumber
 h_{41}(\bk) &=
 -8  \lambda_{Eg,1d}^0 \sin\left(\frac{k_z c}{2}\right)\sin\left(\frac{k_x a}{2}\right)\cos\left(\frac{k_y a}{2}\right), \\ \nonumber
h_{42}(\bk) &= -8  \lambda_{Eg,1d}^0 \sin\left(\frac{k_z c}{2}\right) \cos\left(\frac{k_x a}{2}\right)\sin\left(\frac{k_y a}{2}\right),\\
\nonumber
h_{53}(\bk) &= -8 \lambda_{Eg}^0 \sin\left(\frac{k_z c}{2}\right)\cos\left(\frac{k_x a}{2}\right)\sin\left(\frac{k_y a}{2}\right),
\\ 
h_{63}(\bk) &= 8 \lambda_{Eg}^0 \sin\left(\frac{k_z c}{2}\right) \sin\left(\frac{k_x a}{2}\right)\cos\left(\frac{k_y a}{2}\right),
\end{align}
and 
\begin{align}
h_{51}(\bk) &= \frac{1}{2}[\lambda_{A_{2g}}(\bk)+\lambda_{B_{2g}}(\bk)], \nonumber \\ \nonumber
h_{52}(\bk) &= \frac{1}{2}[\lambda_{B_{1g}}(\bk)+\lambda], \\ \nonumber
h_{61}(\bk) &= \frac{1}{2}[\lambda_{B_{1g}}(\bk) -\lambda], \\ 
h_{62}(\bk) &= \frac{1}{2}[\lambda_{A_{2g}}(\bk)-\lambda_{B_{2g}}(\bk)],
\end{align}
where 
\begin{align}
\lambda_{A_{2g}} &= 8 \lambda_{A_{2g}}^0 \sin(k_x a) \sin(k_y a) \times \nonumber \\
\nonumber
&[\cos(k_x a) - \cos(k_y a)], \\ \nonumber
\lambda_{B_{1g}} &= 2 \lambda_{B_{1g}}^0 [\cos(k_x a) - \cos(k_y a)], \\ 
\lambda_{B_{2g}} &= 4 \lambda_{B_{2g}}^0 \sin(k_x a) \sin(k_y a),
\end{align}
with numerical values $ \lambda_{Eg,1d}^0 = 0.9$ meV, $\lambda_{Eg}^0 = -0.2$ meV, $\lambda_{B_{1g}}^0 = -0.3$ meV, and $\lambda_{B_{2g}}^0 \approx 0$ meV.

\end{itemize}

\section{\APPENDIXORBITALBAND}\label{Sec:OrbitalBand}

The description and classification of the order parameter in the orbital perspective must be in agreement with the description in the band perspective. The description of the electronic structure in terms of bands is simply obtained by diagonalizing the matrix associated with the normal state Hamiltonian $H_0(\bk)$ given implicitly by Eq. \ref{Eq:H0}. The diagonal matrix corresponding to the Hamiltonian in the band basis, $H_B(\bk)$, can be obtained by a momentum-dependent unitary transformation, $U(\bk)$, as $H_B(\bk) = U(\bk) H_0(\bk) U^{\dagger}(\bk)$. Similarly, the order parameter in the band basis, $\Delta^B(\bk)$, is determined by the transformation $\Delta^B(\bk) = U(\bk) \Delta(\bk) U^{T}(-\bk)$, given the superconducting order parameter in the orbital-spin basis in Eq. \ref{Eq:OP}. The last equality implies that, even for the case of local pairing with a constant $d_{ab}(\bk)$ in the orbital-spin basis, the order parameter can acquire a non-trivial momentum dependence in the band basis through the momentum dependence of the unitary transformation. 

\begin{figure}[t]
\begin{center}
\includegraphics[width=5.5cm, keepaspectratio]{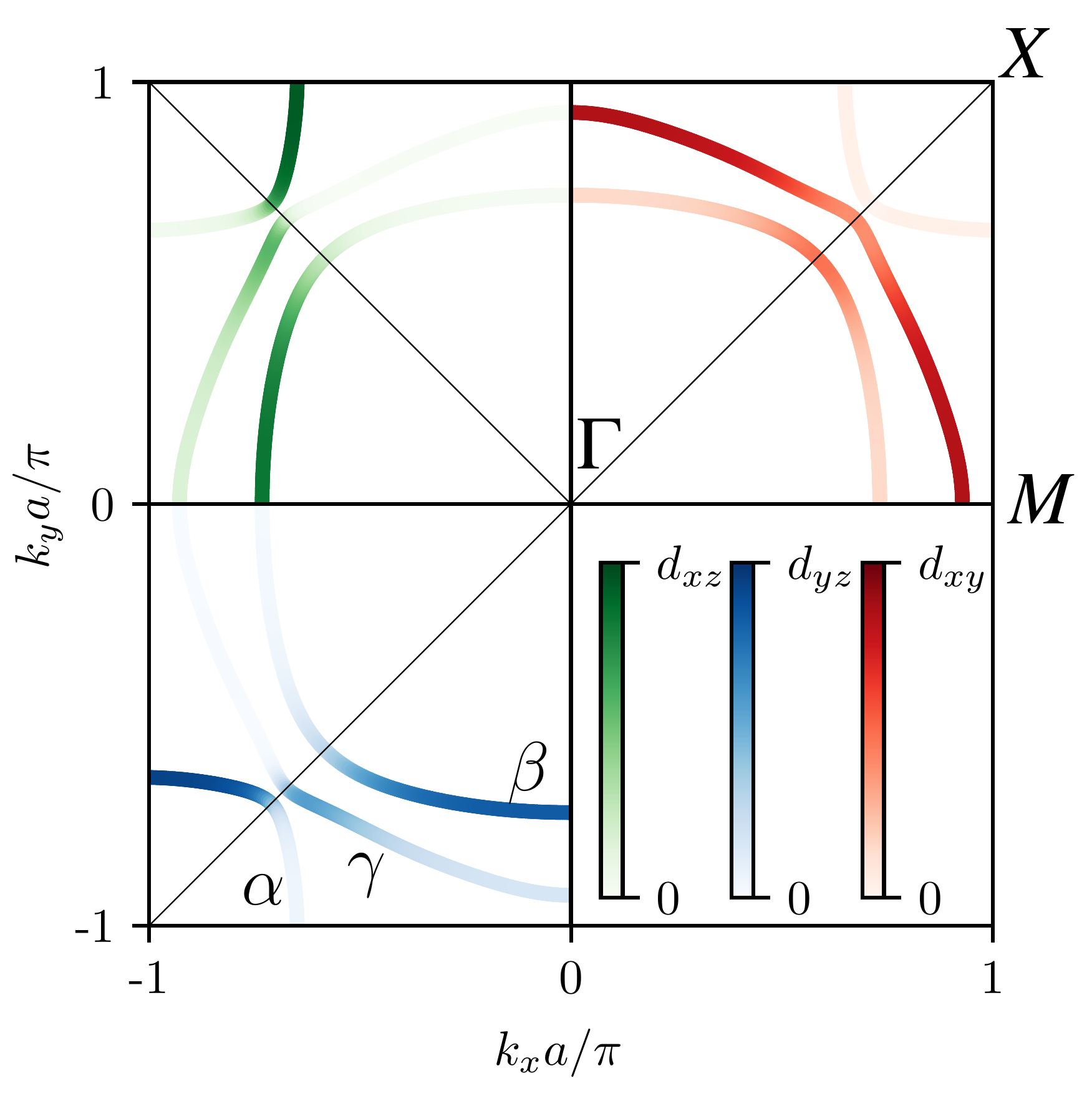}
        \end{center}
\caption{Orbital distribution along the Fermi surfaces of Sr$_2$RuO$_2$ for the  $k_z=0$ plane. The colour red (blue, green) encodes the contribution of the  $d_{xy}$ ($d_{yz}$, $d_{xz}$) orbital as low (bright color) or high (dark color) on each Fermi surface sheet. 
The four-fold symmetry of Sr$_2$RuO$_4$ allows one to display the orbital content of the three orbitals in three complementary quadrants in the $k_xk_y$-plane. 
Note that orbital mixing, including all three orbitals, is most substantial along the $\Gamma-X$ direction. 
Along the $\Gamma-M$ direction the mixing includes primarily only two orbitals, $d_{xy}$ and $d_{xz} (d_{yz})$.  Adapted from Ref.~\citen{Beck2022} ($\copyright$~2022  American Physical Society).}
\label{Fig:FSz0}
\end{figure}

Starting with the normal state Hamiltonian, diagonalizing the six-dimensional matrix encoded in Eq. \ref{Eq:H0}, we find six eigenenergies. These correspond to three pairs of doubly degenerate bands, as expected by the presence of time-reversal and inversion symmetries. The filling of these bands determines a three-dimensional Fermi surface. Fig. \ref{Fig:FSz0} shows the lines in momentum space determining the Fermi surface cuts (labelled as $\alpha$, $\beta$ and $\gamma$) for the $k_z=0$ plane. The diagonalization of the Hamiltonian in Eq. \ref{Eq:H0} to the band basis can be understood as a momentum-dependent change of basis, from the orbital-spin basis $\Psi^\dagger_\bk  = (c_{yz\uparrow}^\dagger, c_{yz\downarrow}^\dagger,  c_{xz\uparrow}^\dagger, c_{xz\downarrow}^\dagger, c_{xy\uparrow}^\dagger, c_{xy\downarrow}^\dagger)_\bk$ into the eigenbasis, a superposition of the microscopic orbital and spin DOFs. As the Hamiltonian is momentum dependent through the $h_{ab}(\bk)$ functions, the unitary transformation that diagonalizes it is also momentum dependent. The momentum dependence is therefore carried over to the eigenstates, and leads to a non-trivial texture of orbital and spin DOF in momentum space. The coulors in Fig. \ref{Fig:FSz0} correspond to the contribution of each orbital to each Fermi surface sheet.

To make the points above more explicit, here we take as a didactic example a simplified model for the $k_xk_z$-plane of Sr$_2$RuO$_4$ \cite{Beck2022}. Along this plane, the dominant orbitals are $d_{xy}$ and $d_{xz}$, corresponding to the red and green components in Fig. \ref{Fig:FSz0} . The normal state Hamiltonian projected into this subspace can be written as Eq. \ref{Eq:H} with the basis replaced by $[\Psi^{XZ}_\bk]^\dagger  = (c_{xz\uparrow}^\dagger, c_{xz\downarrow}^\dagger, c_{xy\uparrow}^\dagger, c_{xy\downarrow}^\dagger)_\bk$, and $H_0(\bk)$ is replaced by:
\begin{eqnarray}\label{Eq:H0XZ}
H^{XZ}_0(\bk) = \sum_{ab}h^{XZ}_{ab}(\bk) \tau_a \otimes \sigma_b,
\end{eqnarray}
where $\tau_{a=\{1,2,3\}}$ are Pauli matrices encoding the information about the orbital DOF, while $\sigma_{\{b=1,2,3\}}$ are Pauli matrices encoding the information about the spin DOF, and $\tau_0$ and $\sigma_0$ are the identity matrices in the corresponding spaces.  
Given the Hermiticity of the Hamiltonian and the even parity of the orbitals, the functions $h^{XZ}_{ab}(\bk)$ must again be real and even in momentum. 
In principle, there are sixteen different terms $(a,b)_{XZ}$ in the normal state Hamiltonian. 
Imposing time-reversal symmetry, in the same fashion as in the discussion above, these are constrained to only six symmetry allowed terms with indices $(a,b)_{XZ} = \{(0,0),(1,0),(3,0),(2,1),(2,2),(2,3)\}_{XZ}$. 
The terms associated with $(0,0)_{XZ}$ and $(3,0)_{XZ}$ correspond to intraorbital hopping, $(1,0)_{XZ}$ corresponds to interorbital hopping, $(2,1)_{XZ}$ is associated with atomic SOC, and $(2,2)_{XZ}$ and $(2,3)_{XZ}$ with momentum-dependent SOC. 
The explicit form of the $h^{XZ}_{ab}(\bk)$ can be obtained and classified according to the irreducible representations of $D_{2h}$, the point group associated with the $k_xk_z$-plane. 
The details can be found in Beck \etal \cite{Beck2022} 
Comparing this simplified model with the original three-orbital three-dimensional model, we can identify: 
\begin{align}
h_{00}^{XZ}(\bk) &= \frac{1}{2}[\xi^{xz}(\bk) +\xi^{xy} (\bk)]\Big|_{k_y=0}, \nonumber \\ \nonumber
h_{10}^{XZ}(\bk) &= h_{30}(\bk)\Big|_{k_y=0} = 0, \\ \nonumber
h_{30}^{XZ}(\bk) &=  \frac{1}{2}[\xi^{xz}(\bk) -\xi^{xy} (\bk)]\Big|_{k_y=0}, \\ \nonumber
h_{21}^{XZ}(\bk) &= \lambda, \\ \nonumber
h_{22}^{XZ}(\bk) &=  h_{63}(\bk)\Big|_{k_y=0} = 0,\\
h_{23}^{XZ}(\bk) &=  h_{63}(\bk)\Big|_{k_y=0} = -8 \lambda_{E_g} \sin(k_z c/2)\sin(k_x a/2).
\end{align}

Note that for the $k_xk_z$-plane, we find $h_{10}^{XZ}(\bk) =0 $ and $h_{22}^{XZ}(\bk) =0$ by symmetry. 

Diagonalizing the Hamiltonian $H_0^{XZ}(\bk)$ we find two doubly-degenerate eigenvalues:
\begin{align}
E_{\pm}^{XZ}(\bk) = h_{00}^{XZ} (\bk) \pm |\mathbf{h}^{XZ}(\bk)| ,
\end{align}
where $ |\mathbf{h}^{XZ}(\bk)| = \sqrt{[h_{30}^{XZ}(\bk)]^2 + [h_{21}^{XZ}(\bk)]^2+ [h_{23}^{XZ}(\bk)]^2}$, and eigenstates

\begin{align}
\Psi_{\Uparrow-}^{XZ}(\bk) &=  \left(i h_{21}^{XZ}(\bk), -ih_{23}^{XZ}(\bk), 0, c_-(\bk)\right)/N_-(\bk), \nonumber \\ \nonumber
\Psi_{\Downarrow-}^{XZ}(\bk) &=  \left( ih_{23}^{XZ}(\bk), i h_{21}^{XZ}(\bk), c_-(\bk),0 \right)/N_-(\bk), \\ \nonumber
\Psi_{\Uparrow+}^{XZ}(\bk) &=  \left( i h_{21}^{XZ}(\bk), -ih_{23}^{XZ}(\bk), 0,c_+(\bk)\right)/N_+(\bk), \\
\Psi_{\Downarrow+}^{XZ}(\bk) &=  \left( ih_{23}^{XZ}(\bk), i h_{21}^{XZ}(\bk), c_+(\bk),0 \right)/N_+(\bk),
\end{align}
where $c_\pm (\bk)= h_{30}^{XZ}(\bk) \pm |\mathbf{h}^{XZ}(\bk)|$, 
 and the normalization factors are $N_\pm (\bk) = \sqrt{|h_{21}^{XZ}(\bk)|^2 + |h_{23}^{XZ}(\bk)|^2 + |c_\pm(\bk)|^2}$. 
 Here the subscript $\pm$ associates the eigenstates to the eigenvalues $E_\pm^{XZ}$. Note that the doubly degeneracy of the states requires us to use one extra index to distinguish them. Here we use the labels $\Uparrow, \Downarrow$ corresponding to a \emph{pseudospin} DOF, as the eigenstates $\Psi_{\Uparrow\pm}^{XZ}(\bk) $ and $\Psi_{\Downarrow\pm}^{XZ}(\bk)$ are related by time-reversal symmetry [defined as $i\sigma_2$ accompanied by complex conjugation acting on the basis $\big(\Psi_{\Uparrow\pm}^{XZ}(\bk),\Psi_{\Downarrow\pm}^{XZ}(\bk)\big)$]. 
 The explicit form of the eigenstates above shows us that the states associated with the bands are a superposition of orbitals and spins. In particular, the state $\Psi_{\Uparrow-}^{XZ} (\bk)$ is a superposition of $d_{xz}$ orbitals with spin up with amplitude $i h_{21}^{XZ}(\bk)/N_-(\bk)$, $d_{xz}$ orbitals with spin down with amplitude $-i h_{23}^{XZ}(\bk)/N_-(\bk)$, and $d_{xy}$ orbitals with spin down with amplitude $c_-(\bk)/N_-(\bk)$.  

The eigenstates above could have been found by a unitary transformation, $U^{XZ}(\bk)$, which diagonalizes the Hamiltonian, $U^{XZ}(\bk) H_0^{XZ}(\bk) [U^{XZ}(\bk)]^\dagger = H_B^{XZ} (\bk)$, where $H_B^{XZ} (\bk)$ is the Hamiltonian in the band basis. The explicit form of the unitary transformation can be found by using the eigenstates as entries for the matrix rows. In this case, the unitary matrix takes the explicit form (omitting the momentum dependence):
\begin{align}
\setlength\arraycolsep{2pt}
U^{XZ} = \begin{pmatrix}
-i h_{21}^{XZ}/N_- & ih_{23}^{XZ}/N_- & 0 & c_-/N_- \\
-ih_{23}^{XZ}/N_- & - i h_{21}^{XZ} / N_- & c_-/N_- & 0 \\
-i h_{21}^{XZ}/N_+ & ih_{23}^{XZ}/N_+ & 0 & c_+/N_+ \\
-ih_{23}^{XZ}/N_+  & - i h_{21}^{XZ}/N_+ & c_+/N_+ & 0
\end{pmatrix}.
\end{align}

Written explicitly, the normal state Hamiltonian in the band basis reads:
\begin{align}
H_B^{XZ}(\bk) = \begin{pmatrix}
E_-(\bk) & 0 & 0 & 0\\
0 & E_-(\bk) & 0 & 0 \\
0 & 0 & E_+(\bk) & 0\\
0 & 0 & 0 & E_+(\bk)
\end{pmatrix},
\end{align}
on the basis $\Psi^{XZ}_{B}(\bk)  = (\Psi^{XZ}_{\Uparrow -} (\bk), \Psi^{XZ}_{\Downarrow -}(\bk),\Psi^{XZ}_{\Uparrow +}(\bk),\Psi^{XZ}_{\Downarrow +}(\bk))$. 

Now let's discuss how the superconducting order parameter in the simplified two-orbital model is transformed from the orbital-spin basis to the band basis. In the two-orbital model, the superconducting order parameter takes the general form:
\begin{align}\label{Eq:OPXZ}
\Delta^{XZ} (\bk) = \sum_{a,b}  d_{ab} (\bk) \tau_a \otimes \sigma_b (i \sigma_2).
\end{align}
Following fermionic antisymmetry, Eq. \ref{Eq:FAS}, the sixteen four-dimensional basis matrices can be associated with order parameters that are either even or odd in momentum. For six of them, $[a,b]_{XZ} = \{[0,0],[1,0],[3,0],[2,i]\}_{XZ}$, the order parameter is even parity, while for ten, $[a,b]_{XZ} = \{[0,i],[1,i],[3,i],[2,0]\}_{XZ}$, the order parameter is odd  parity (here $i=1,2,3$). As an example, let's focus on an even-parity superconducting order parameter associated with the basis matrix $[2,3]_{XZ}$, corresponding to an antisymmetric interorbital and spin-triplet order parameter in the orbital-spin basis. This basis matrix belongs to a non-trivial irreducible representation of the point group, the $B_{2g}$ irrep of the $D_{2h}$ point group associated with the $k_xk_z$-plane \cite{Beck2022}. Assuming pairing to be local, we take $d_{23}^{XZ}(\bk) = d_0$, with $d_0$ a constant. The order parameter then transforms according to the $B_{2g}$ irrep of $D_{2h}$ and is written as $\Delta_0(\bk) = d_0 \tau_2 \otimes \sigma_3 (i\sigma_2)$ in the orbital-spin basis. 

Written explicitly, the momentum-independent order parameter associated with the matrix structure $[2,3]_{XZ}$ reads:
\begin{align}
\Delta^{XZ} &=  \begin{pmatrix}
 \Delta_{xz\uparrow, xz \uparrow} &  \Delta_{xz\uparrow, xz \downarrow} & \Delta_{xz\uparrow, xy \uparrow}  & \Delta_{xz\uparrow, xy \downarrow} \\
   \Delta_{xz\downarrow, xz \uparrow} &  \Delta_{xz\downarrow, xz \downarrow} & \Delta_{xz\downarrow, xy \uparrow} &  \Delta_{xz\downarrow, xy \downarrow}  \\
    \Delta_{xy\uparrow, xz \uparrow}  & \Delta_{xy\uparrow, xz \downarrow} &  \Delta_{xy\uparrow, xy \uparrow}  & \Delta_{xy\uparrow, xy \downarrow}  \\
     \Delta_{xy\downarrow, xz \uparrow} &  \Delta_{xy\downarrow, xz \downarrow}  &  \Delta_{xy\downarrow, xy \uparrow} &  \Delta_{xy\downarrow, xy \downarrow} 
\end{pmatrix} \nonumber \\ 
&= -i  \begin{pmatrix}
 0 & 0 & 0 &  d_0 \\
  0 & 0 & d_0 &  0 \\
   0 & -d_0 & 0 &  0 \\
    -d_0 & 0 & 0 &  0 
 \end{pmatrix},
\end{align}
which implies that 
\begin{align}
d_0 = \frac{i}{4}\left[\Delta_{xz\uparrow, xy \downarrow} + \Delta_{xz\downarrow, xy \uparrow} -  \Delta_{xy\uparrow, xz \downarrow}  -  \Delta_{xy\downarrow, xz \uparrow} \right].
\end{align}

\begin{figure*}[ht]
\begin{center}
\includegraphics[scale=0.5, keepaspectratio]{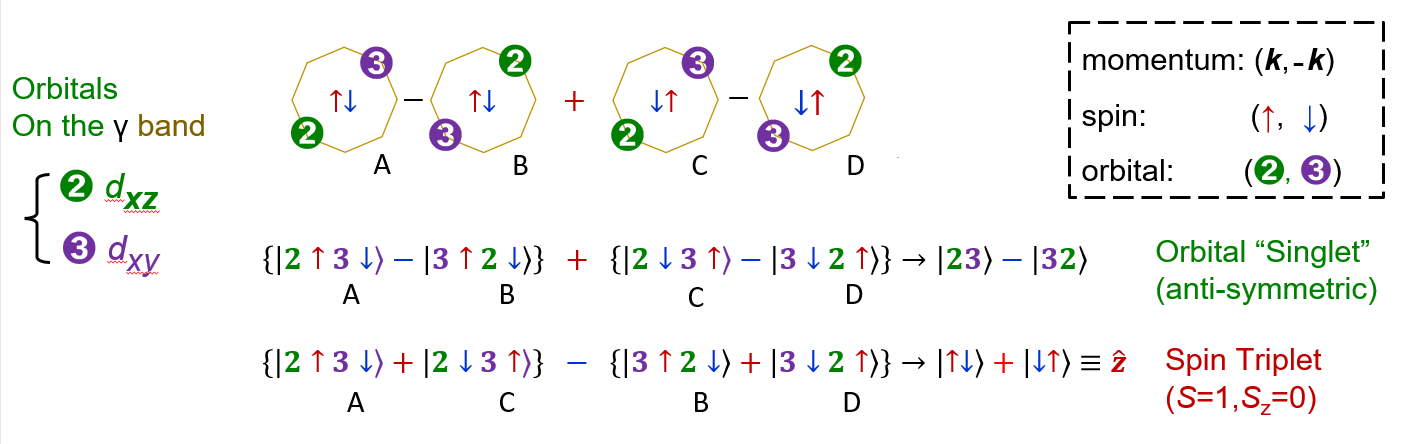}
\end{center}
\caption{Schematic example of interorbital pairing states corresponding to Eqs. B09-B11 in the text. It depicts a pairing of electrons on the gamma FS with the orbitals $d_{yz}$ (2) and $d_{xy}$ (3). By rearrangents, it is readily understood that these bases give the orbital antisymmetric (orbital “singlet”), spin-triplet pairs.}
\label{Fig:Inter-orbital_pairs}
\end{figure*}

If we focus on the orbital degree of freedom, joining terms with the same spin configuration:
\begin{align}
d_0 &= \frac{i}{4}\left[(\Delta_{xz\uparrow, xy \downarrow}  -  \Delta_{xy\uparrow, xz \downarrow} ) + (\Delta_{xz\downarrow, xy \uparrow} -  \Delta_{xy\downarrow, xz \uparrow}) \right] \nonumber \\
&=\frac{i}{4}\left[(\Delta_{xz, xy}  -  \Delta_{xy, xz} )_{\uparrow\downarrow} + (\Delta_{xz, xy } -  \Delta_{xy, xz })_{\downarrow\uparrow} \right],
\end{align}
the order parameter is explicitly anti-symmetric in orbital, while if we focus on the spin degree of freedom, joining terms with the same orbital content:
\begin{align}
d_0 &= \frac{i}{4}\left[(\Delta_{xz\uparrow, xy \downarrow}+\Delta_{xz\downarrow, xy \uparrow})  -  (\Delta_{xy\uparrow, xz \downarrow} + \Delta_{xy\downarrow, xz \uparrow}) \right] \nonumber \\
&= \frac{i}{4}\left[(\Delta_{\uparrow,  \downarrow}+\Delta_{\downarrow,  \uparrow})_{xz,xy}  -  (\Delta_{\uparrow,\downarrow} + \Delta_{\downarrow,  \uparrow})_{xy,xz} \right],
\end{align}
the order parameter is explicitly spin triplet, as illustrated in Fig. \ref{Fig:Inter-orbital_pairs}.

With this understanding of the order parameter in the orbital-spin basis, we can now apply the unitary transformation $U^{XZ}(\bk)$ to assess its structure in the band basis:

\begin{align}
\setlength\arraycolsep{1pt}
\frac{\Delta_B}{i d_0} =\begin{pmatrix}
0 & - \frac{h_{23}^{XZ}}{|\mathbf{h}^{XZ}|} & \frac{h_{21}^{XZ}}{|\mathbf{h}_s^{XZ}|} & 
\frac{h_{23}^{XZ}h_{30}^{XZ} }{|\mathbf{h}_s^{XZ}||\mathbf{h}^{XZ}|} \\
\frac{h_{23}^{XZ}}{|\mathbf{h}^{XZ}|} & 0 & -\frac{h_{23}^{XZ}h_{30}^{XZ} }{|\mathbf{h}_s^{XZ}||\mathbf{h}^{XZ}|} & \frac{h_{21}^{XZ}}{|\mathbf{h}_s^{XZ}|} \\
-\frac{h_{21}^{XZ}}{|\mathbf{h}_s^{XZ}|}  & \frac{h_{23}^{XZ}h_{30}^{XZ}}{|\mathbf{h}_s^{XZ}||\mathbf{h}^{XZ}|} & 0 &  \frac{h_{23}^{XZ}}{|\mathbf{h}^{XZ}|} \\
-\frac{h_{23}^{XZ}h_{30}^{XZ} }{|\mathbf{h}_s^{XZ}||\mathbf{h}^{XZ}|} & -\frac{h_{21}^{XZ}}{|\mathbf{h}_s^{XZ}|}   & -\frac{h_{23}^{XZ}}{|\mathbf{h}^{XZ}|} & 0 
\end{pmatrix},
\end{align}
where we defined $\vert\mathbf{h}^{XZ}_s(\bk)\vert = \sqrt{[h_{21}^{XZ}(\bk)]^2+ [h_{23}^{XZ}(\bk)]^2}$, and omitted the momentum dependence.

This form of the superconducting order parameter acts on the band basis, $\Psi^{XZ}_{B}(\bk)  = (\Psi^{XZ}_{\Uparrow -} (\bk), \Psi^{XZ}_{\Downarrow -}(\bk),\Psi^{XZ}_{\Uparrow +}(\bk),\Psi^{XZ}_{\Downarrow +}(\bk))$. 
Remembering that the two first (and last) entries in this basis correspond to a doubly degenerate band characterized by energies $E_{\pm}(\bk)$ and pseudospin $\Uparrow, \Downarrow$, the intraband components of the order parameter correspond to the two-dimensional blocks along the diagonals of $\Delta_B$, while the interband component of the order parameter corresponds to the off-diagonal two-dimensional blocks. 
Note that the intraband component has the form of a pseudospin-singlet (is proportional to $i\sigma_2$), as expected, since the order parameter is even parity. Note also that, despite the constant amplitude $d_0$ in the orbital basis, the intraband component of the order parameter has a momentum dependence characterized by the function $h_{23}^{XZ}(\bk) \propto \sin(k_z)$, with nodes along $k_z=0$. 
Note that this intraband component of the superconducting order parameter depends on the inclusion of momentum-dependent SOC terms in the normal state Hamiltonian. 
The interband components have both singlet and triplet character. The latter is allowed as an even-parity superconducting order parameter can be antisymmetric across the bands and symmetric with respect to the pseudospin DOF.

We now go back to the discussion of the superconducting order parameter in the original three-orbital three-dimensional model for Sr$_2$RuO$_4$. 
Focusing on the even-parity order parameters with basis matrices that transform according to the $E_g$ irrep, $\{[3,0],-[2,0]\}$, $\{[4,2],-[4,1]\}$, and $\{[5,3],[6,3]\}$ (see Table \ref{Tab:EvenOP}), we can conclude that if the accompanying $d_{ab}(\bk)$ functions belong to any one-dimensional even-parity irrep, the order parameter as a whole belongs to the $E_g$ irrep. 
The simplest choice in this context is for constant $d_{ab}(\bk)$ factors in $A_{1g}$, which would be associated with local, or momentum-independent, interactions. 
This discussion suggests that there are order parameters in non-trivial irreps driven by local interactions. The non-trivial transformation of the order parameter under certain point-group transformations in these cases stems from the orbital and spin texture of the electronic states along the Fermi surfaces.  
Motivated by the theoretical proposal of Suh \etal \cite{Suh2020}, we use as an example the order parameter with basis matrices $\{[5,3],[6,3]\}$ and local interactions
\begin{eqnarray}
{\Delta}(\bk) = (d_{53} \lambda_5 + d_{63} \lambda_6)\otimes \sigma_3 (i\sigma_2),
\end{eqnarray}
which is a local (or $s$-wave), given the constant $d_{53,63}$ parameters, orbital-antisymmetric, given the matrices $\lambda_5$ and $\lambda_6$ in orbital space, and spin triplet, given the matrix $\sigma_3$ in spin space. This order parameter can be concisely called an even parity orbital antisymmetric spin triplet (OAST). Connecting to the discussion above for the two-orbital model along the XZ-plane, the projection of the $[6,3]$ component of the three-dimensional order parameter into the XZ-plane corresponds to $[2,3]_{XZ}$. A similar discussion finds that the projection of the $[5,3]$ component of the three-dimensional order parameter  into the YZ-plane corresponds to $[2,3]_{YZ}$. Note that in all cases the order parameters are antisymmetric in the orbital degree of freedom, as the Gell-Mann matrices $\lambda_{5,6}$ and the Pauli matrix $\sigma_2$ are antisymmetric.

\begin{figure*}[ht]
\begin{center}
\includegraphics[scale=0.26, keepaspectratio]{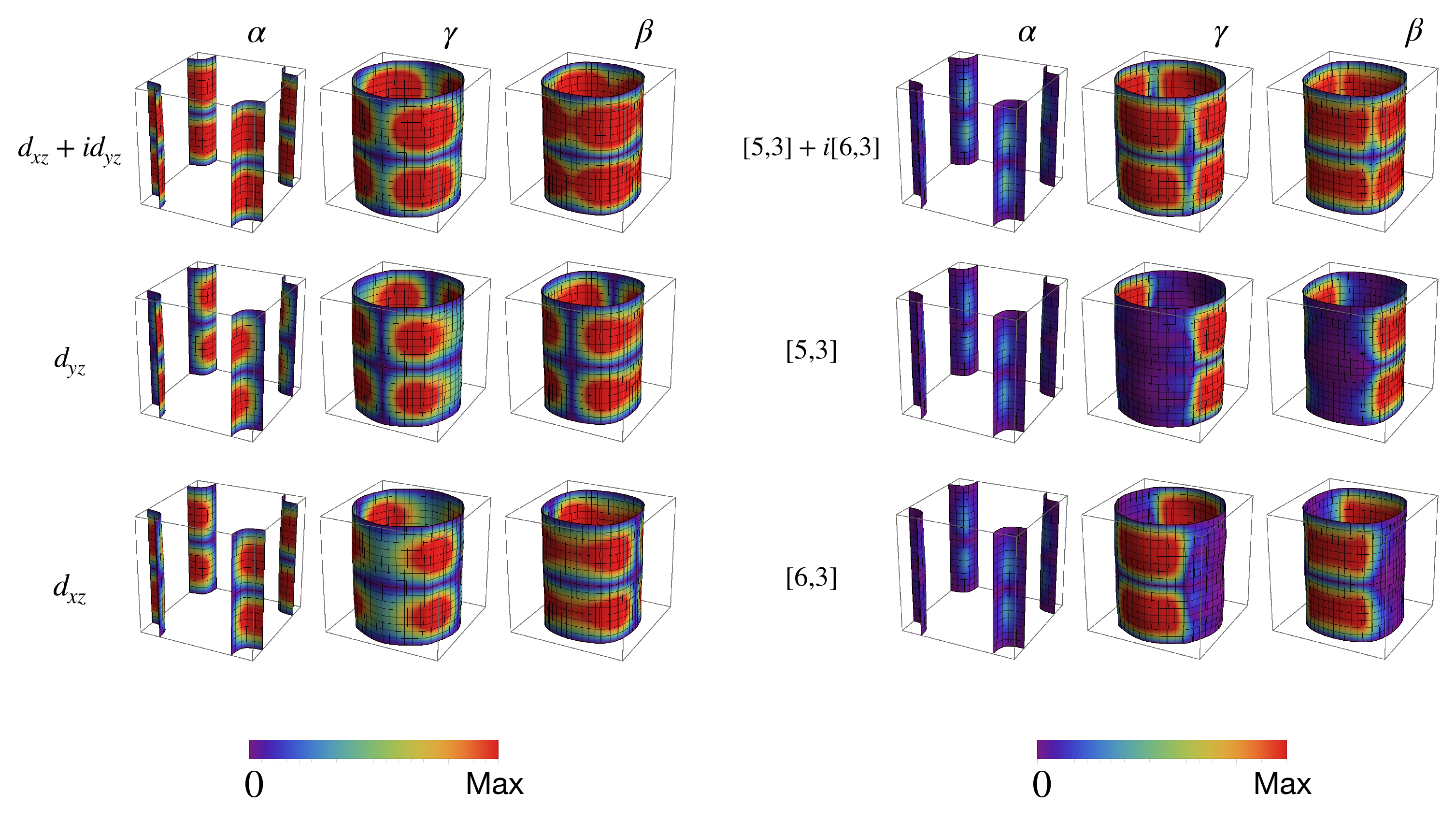}
\end{center}
\caption{Contrasting SC gap structures of even-parity chiral OPs. The three FS bands, $\alpha$, $\beta$, and $\gamma$ are as introduced in Fig. \ref{fig:BZ}. Left:
Top: Naive chiral $d$-wave spin-singlet superconducting order parameter $\propto (k_x+i k_y)k_z$ projected  into the three-dimensional Fermi surfaces of \SRO. Note the horizontal line nodes and the minima along the $k_x$- and $k_y$-directions. Middle
(Bottom): separate plot of the component $\propto k_x k_z$ ($ k_yk_z$). Note the nodes along the
$k_x$-($k_y$-) direction. Right: Top: Chiral even-parity orbital-antisymmetric spin-triplet (OAST) superconducting order parameter projected into the three-dimensional Fermi surface of Sr$_2$RuO$_4$ (encoded in the matrix structure $[5, 3]+i[6, 3]$). Note the horizontal line nodes and the vertical minima along the diagonals. Middle (Bottom): separate plot of the component
$[5, 3]$ ($[6, 3]$). Note the extended regions with very small gap associated with the absence of
certain orbitals on these regions of the Fermi surface. Reproduced from Ref. \citen{Ramires2022} ($\copyright$~2022 The Authors).}
\label{Fig:Gaps}
\end{figure*}

The OAST order parameter belongs to the non-trivial irrep $E_g$, so it must develop nodes when transformed into the band basis. 
Indeed, transforming it to the band basis, we find an unusual structure on the Fermi surface. 
As the order parameter is interorbital in nature, the form factors associated with the gap magnitude follow the orbital distribution along the Fermi surface shown in Fig. \ref{Fig:FSz0}. More concretely, if we take the component $[5,3]$ ($[6,3]$), corresponding to pairing between $d_{yz}$ ($d_{xz}$) and $d_{xy}$ orbitals, we find that it is strongest for the Fermi surface regions parallel to the $k_y=0$ ($k_x=0$) plane, and weakest for the Fermi surface regions parallel to the $k_x=0$ (or $k_y=0$) planes. 
The latter can be understood by the  very low contribution of $d_{yz}$ ($d_{xz}$) orbitals to the electronic states close to the Fermi surface for the regions parallel to the $k_x=0$ ($k_y=0$) planes. 
Note that the chiral superposition of the two components, $[5,3]\pm i [6,3]$ leads to a gap structure that has an horizontal line node and vertical near nodes along the diagonals. 
It should be emphasized that this gap structure is in contrast to what is naively considered for a chiral $d$-wave order parameter on a single band, with structure $\propto (k_x + ik_y)k_z$, displaying horizontal line nodes and vertical gap minima along the $k_x$ and $k_y$ directions. These differences are illustrated in Fig. \ref{Fig:Gaps}.


\bibliography{
string,
TMTSF,
textbook,
FFLO,
CeCoIn5,
superconductors,
SC,
Sr2RuO4,
Sr3Ru2O7,
high-Tc,
measurement_technique,
Ramires_Unique}

\begin{thebibliography}{100}

\bibitem{Maeno1994}
Y.~Maeno, H.~Hashimoto, K.~Yoshida, S.~Nishizaki, T.~Fujita, J.~G. Bednorz, and
  F.~Lichtenberg: Nature {\bfseries 372} (1994) 532.

\bibitem{Mackenzie2003RMP}
A.~P. Mackenzie and Y.~Maeno: Rev. Mod. Phys. {\bfseries 75} (2003) 657.

\bibitem{Sigrist2000.JPhysSocJpn.69.1290}
M.~Sigrist: J. Phys. Soc. Jpn. {\bfseries 69} (2000) 1290.

\bibitem{Kinjo2022.Science.376.397}
K.~Kinjo, M.~Manago, S.~Kitagawa, Z.~Q. Mao, S.~Yonezawa, Y.~Maeno, and
  K.~Ishida: Science {\bfseries 376} (2022) 397.

\bibitem{Steppke2017.Science.355.eaaf9398}
A.~Steppke, L.~Zhao, M.~E. Barber, T.~Scaffidi, F.~Jerzembeck, H.~Rosner, A.~S.
  Gibbs, Y.~Maeno, S.~H. Simon, A.~P. Mackenzie, and C.~W. Hicks: Science
  {\bfseries 355} (2017) 148.

\bibitem{Maeno1998.PRL.3-K-phase}
Y.~Maeno, T.~Ando, Y.~Mori, E.~Ohmichi, S.~Ikeda, S.~NishiZaki, and
  S.~Nakatsuji: Phys. Rev. Lett. {\bfseries 81} (1998) 3765.

\bibitem{Pustogow2019.Nature.574.72}
A.~Pustogow, Y.~Luo, A.~Chronister, Y.-S. Su, D.~A. Sokolov, F.~Jerzembeck,
  A.~P. Mackenzie, C.~W. Hicks, N.~Kikugawa, S.~Raghu, E.~D. Bauer, and S.~E.
  Brown: Nature {\bfseries 574} (2019) 72.

\bibitem{Maeno2011.JPhysSocJpn.81.011009}
Y.~Maeno, S.~Kittaka, T.~Nomura, S.~Yonezawa, and K.~Ishida: J. Phys. Soc. Jpn.
  {\bfseries 81} (2012) 011009.

\bibitem{Mackenzie2017.npjQuantumMater.2.40}
A.~P. Mackenzie, T.~Scaffidi, C.~W. Hicks, and Y.~Maeno: npj Quantum Mater.
  {\bfseries 2} (2017) 40.

\bibitem{Mazin2003}
I.~Mazin and V.~Antropov: Physica C: Superconductivity {\bfseries 385} (2003)
  49.

\bibitem{Xi2008}
X.~X. Xi: Rep. Prog. Phys. {\bfseries 71} (2008) 116501.

\bibitem{Fernandes2022}
R.~M. Fernandes, A.~I. Coldea, H.~Ding, I.~R. Fisher, P.~J. Hirschfeld, and
  G.~Kotliar: Nature {\bfseries 601} (2022) 35.

\bibitem{Hamermesh1989}
M.~Hamermesh: {\em {Group theory and its application to physical problems;
  Corr. ed.}} (Dover, New York, NY, 1989), This Dover edition, first published
  in 1989, is an unabridged, corrected republication of the second (corrected)
  printing (1964) of the work first published by Addison-Wesley Publishing
  Company, Inc., Reading, Massachusetts, 1962, in its Addison-Wesley Series in
  Physics.

\bibitem{Bradley2009}
{Bradley, Christopher and Cracknell, Arthur}: {\em {The Mathematical Theory of
  Symmetry in Solids: Representation Theory for Point Groups and Space Groups}}
  (Oxford Classic Texts in the Physical Sciences. Oxford University Press,
  Oxford, 2009), Oxford Classic Texts in the Physical Sciences.

\bibitem{Sigrist1991}
M.~Sigrist and K.~Ueda: Rev. Mod. Phys. {\bfseries 63} (1991) 239.

\bibitem{Ramires2019}
A.~Ramires and M.~Sigrist: Phys. Rev. B {\bfseries 100} (2019) 104501.

\bibitem{Tamai2019}
A.~Tamai, M.~Zingl, E.~Rozbicki, E.~Cappelli, S.~Ricc\`o, A.~de~la Torre,
  S.~McKeown~Walker, F.~Y. Bruno, P.~D.~C. King, W.~Meevasana, M.~Shi,
  M.~Radovi\ifmmode~\acute{c}\else \'{c}\fi{}, N.~C. Plumb, A.~S. Gibbs, A.~P.
  Mackenzie, C.~Berthod, H.~U.~R. Strand, M.~Kim, A.~Georges, and
  F.~Baumberger: Phys. Rev. X {\bfseries 9} (2019) 021048.

\bibitem{Ramires2018}
A.~Ramires, D.~F. Agterberg, and M.~Sigrist: Phys. Rev. B {\bfseries 98} (2018)
  024501.

\bibitem{Kaba2019}
S.-O. Kaba and D.~S\'en\'echal: Phys. Rev. B {\bfseries 100} (2019) 214507.

\bibitem{Ramires2016}
A.~Ramires and M.~Sigrist: Phys. Rev. B {\bfseries 94} (2016) 104501.

\bibitem{Xia2006.PhysRevLett.97.167002}
J.~Xia, Y.~Maeno, P.~T. Beyersdorf, M.~M. Fejer, and A.~Kapitulnik: Phys. Rev.
  Lett. {\bfseries 97} (2006) 167002.

\bibitem{Grinenko2021.NatPhys.17.748}
V.~Grinenko, S.~Ghosh, R.~Sarkar, J.-C. Orain, A.~Nikitin, M.~Elender, D.~Das,
  Z.~Guguchia, F.~Bruckner, M.~E. Barber, J.~Park, N.~Kikugawa, D.~A. Sokolov,
  J.~S. Bobowski, T.~Miyoshi, Y.~Maeno, A.~P. Mackenzie, H.~Luetkens, C.~W.
  Hicks, and H.-H. Klauss: Nature Phys. {\bfseries 17} (2021) 748.

\bibitem{Benhabib2021.NatPhys.17.194}
S.~Benhabib, C.~Lupien, I.~Paul, L.~Berges, M.~Dion, M.~Nardone, A.~Zitouni,
  Z.~Q. Mao, Y.~Maeno, A.~Georges, L.~Taillefer, and C.~Proust: Nature Phys.
  {\bfseries 17} (2021) 194.

\bibitem{Ghosh2021.NatPhys.17.199}
S.~Ghosh, A.~Shekhter, F.~Jerzembeck, N.~Kikugawa, D.~A. Sokolov, M.~Brando,
  A.~P. Mackenzie, C.~W. Hicks, and B.~J. Ramshaw: Nature Phys. {\bfseries 17}
  (2021) 199.

\bibitem{Ramires2022}
A.~Ramires: J. Phys.: Conf. Seri. {\bfseries 2164} (2022) 12002.

\bibitem{Bobowski2019}
J.~S. Bobowski, N.~Kikugawa, T.~Miyoshi, H.~Suwa, H.~shu Xu, S.~Yonezawa, D.~A.
  Sokolov, A.~P. Mackenzie, and Y.~Maeno: Condensed Matter {\bfseries 4} (2019)
  6.

\bibitem{kikugawa2021}
N.~Kikugawa, D.~A. Sokolov, T.~Nagasawa, and A.~P. Mackenzie: Crystals
  {\bfseries 11} (2021) 392.

\bibitem{Bergemann2003.AdvPhys.52.639}
C.~Bergemann, A.~P. Mackenzie, S.~R. Julian, D.~Forsythe, and E.~Ohmichi: Adv.
  Phys. {\bfseries 52} (2003) 639.

\bibitem{Chmaissem1998.PhysRevB.57.5067}
O.~Chmaissem, J.~D. Jorgensen, H.~Shaked, S.~Ikeda, and Y.~Maeno: Phys. Rev. B
  {\bfseries 57} (1998) 5067.

\bibitem{NishiZaki2000JPhysSocJpn}
S.~NishiZaki, Y.~Maeno, and Z.~Mao: J. Phys. Soc. Jpn. {\bfseries 69} (2000)
  572.

\bibitem{Barber2019}
M.~E. Barber, F.~Lechermann, S.~V. Streltsov, S.~L. Skornyakov, S.~Ghosh, B.~J.
  Ramshaw, N.~Kikugawa, D.~A. Sokolov, A.~P. Mackenzie, C.~W. Hicks, and I.~I.
  Mazin: Phys. Rev. B {\bfseries 100} (2019).

\bibitem{Yoshida1998}
K.~Yoshida, F.~Nakamura, T.~Goko, T.~Fujita, Y.~Maeno, Y.~Mori, and
  S.~NishiZaki: Phys. Rev. B {\bfseries 58} (1998) 15062.

\bibitem{Jerzembeck2022}
F.~Jerzembeck, H.~S. R{\o}ising, A.~Steppke, H.~Rosner, D.~A. Sokolov,
  N.~Kikugawa, T.~Scaffidi, S.~H. Simon, A.~P. Mackenzie, and C.~W. Hicks:
  Nature Commun. {\bfseries 13} (2022) 4596.

\bibitem{Damascelli2000.PhysRevLett.85.5194}
A.~Damascelli, D.~H. Lu, K.~M. Shen, N.~P. Armitage, F.~Ronning, D.~L. Feng,
  C.~Kim, Z.-X. Shen, T.~Kimura, Y.~Tokura, Z.~Q. Mao, and Y.~Maeno: Phys. Rev.
  Lett. {\bfseries 85} (2000) 5194.

\bibitem{Fittipaldi2021}
R.~Fittipaldi, R.~Hartmann, M.~T. Mercaldo, S.~Komori, A.~Bjørlig, W.~Kyung,
  Y.~Yasui, T.~Miyoshi, L.~A.~B. Olde~Olthof, C.~M. Palomares~Garcia,
  V.~Granata, I.~Keren, W.~Higemoto, A.~Suter, T.~Prokscha, A.~Romano, C.~Noce,
  C.~Kim, Y.~Maeno, E.~Scheer, B.~Kalisky, J.~W.~A. Robinson, M.~Cuoco,
  Z.~Salman, A.~Vecchione, and A.~Di~Bernardo: Nature Commun. {\bfseries 12}
  (2021) 125792.

\bibitem{Veenstra2013}
C.~N. Veenstra, Z.-H. Zhu, B.~Ludbrook, M.~Capsoni, G.~Levy, A.~Nicolaou, J.~A.
  Rosen, R.~Comin, S.~Kittaka, Y.~Maeno, I.~S. Elfimov, and A.~Damascelli:
  Phys. Rev. Lett. {\bfseries 110} (2013) 097004.

\bibitem{Hicks2014.Science.344.283}
C.~W. Hicks, D.~O. Brodsky, E.~A. Yelland, A.~S. Gibbs, J.~A.~N. Bruin, M.~E.
  Barber, S.~D. Edkins, K.~Nishimura, S.~Yonezawa, Y.~Maeno, and A.~P.
  Mackenzie: Science {\bfseries 344} (2014) 283.

\bibitem{Hicks2014.RSI}
C.~W. Hicks, M.~E. Barber, S.~D. Edkins, D.~O. Brodsky, and A.~P. Mackenzie:
  Rev. Sci. Instrum. {\bfseries 85} (2014) 065003.

\bibitem{Kim2018.Science.YBCO}
H.-H. Kim, S.~M. Souliou, M.~E. Barber, E.~Lefran{\c{c}}ois, M.~Minola,
  M.~Tortora, R.~Heid, N.~Nandi, R.~A. Borzi, G.~Garbarino, A.~Bosak,
  J.~Porras, T.~Loew, M.~K\"{o}nig, P.~J.~W. Moll, A.~P. Mackenzie, B.~Keimer,
  C.~W. Hicks, and M.~L. Tacon: Science {\bfseries 362} (2018) 1040.

\bibitem{Sun2021.NJP.PdCrO2}
D.~Sun, D.~A. Sokolov, R.~Waite, S.~Khim, P.~Manuel, F.~Orlandi, D.~D.
  Khalyavin, A.~P. Mackenzie, and C.~W. Hicks: New J. Phys. {\bfseries 23}
  (2021) 123050.

\bibitem{Ikhlas2022.NatPhys.Mn3Sn}
M.~Ikhlas, S.~Dasgupta, F.~Theuss, T.~Higo, S.~Kittaka, B.~J. Ramshaw,
  O.~Tchernyshyov, C.~W. Hicks, and S.~Nakatsuji: Nature Phys. {\bfseries 18}
  (2022) 1086.

\bibitem{Li2022}
Y.-S. Li, M.~Garst, J.~Schmalian, S.~Ghosh, N.~Kikugawa, D.~A. Sokolov, C.~W.
  Hicks, F.~Jerzembeck, M.~S. Ikeda, Z.~Hu, B.~J. Ramshaw, A.~W. Rost,
  M.~Nicklas, and A.~P. Mackenzie: Nature {\bfseries 607} (2022) 276.

\bibitem{Sunko2019}
V.~Sunko, E.~A. Morales, I.~Markovi{\'{c}}, M.~E. Barber,
  D.~Milosavljevi{\'{c}}, F.~Mazzola, D.~A. Sokolov, N.~Kikugawa, C.~Cacho,
  P.~Dudin, H.~Rosner, C.~W. Hicks, P.~D.~C. King, and A.~P. Mackenzie: npj
  Quantum Mater. {\bfseries 4} (2019) 46.

\bibitem{Luo2019.normal-state-NMR}
Y.~Luo, A.~Pustogow, P.~Guzman, A.~Dioguardi, S.~Thomas, F.~Ronning,
  N.~Kikugawa, D.~Sokolov, F.~Jerzembeck, A.~Mackenzie, C.~Hicks, E.~Bauer,
  I.~Mazin, and S.~Brown: Phys. Rev. X {\bfseries 9} (2019) 021044.

\bibitem{Husain2023.Nature.Pines_demon}
A.~A. Husain, E.~W. Huang, M.~Mitrano, M.~S. Rak, S.~I. Rubeck, X.~Guo,
  H.~Yang, C.~Sow, Y.~Maeno, B.~Uchoa, T.~C. Chiang, P.~E. Batson, P.~W.
  Phillips, and P.~Abbamonte: Nature {\bfseries 621} (2023) 66.

\bibitem{Pines1956.Pines_demons}
D.~Pines: Can. J. Phys. {\bfseries 34} (1956) 1379.

\bibitem{Matsubara1992.PhysRevB.45.7414}
I.~Matsubara, H.~Tanigawa, T.~Ogura, H.~Yamashita, M.~Kinoshita, and T.~Kawai:
  Phys. Rev. B {\bfseries 45} (1992) 7414.

\bibitem{Lang1994.PhysRevB.49.15227}
M.~Lang, F.~Steglich, N.~Toyota, and T.~Sasaki: Phys. Rev. B {\bfseries 49}
  (1994) 15227.

\bibitem{Riseman1998.Nature.396.242}
T.~M. Riseman, P.~G. Kealey, E.~M. Forgan, A.~P. Mackenzie, L.~M. Galvin, A.~W.
  Tyler, S.~L. Lee, C.~Ager, D.~M. Paul, C.~M. Aegerter, R.~Cubitt, Z.~Q. Mao,
  T.~Akima, and Y.~Maeno: Nature {\bfseries 396} (1998) 242.

\bibitem{Luke2000.PhysicaB.289-290.373}
G.~Luke, Y.~Fudamoto, K.~Kojima, M.~Larkin, B.~Nachumi, Y.~Uemura, J.~Sonier,
  Y.~Maeno, Z.~Mao, Y.~Mori, and D.~Agterberg: Physica B {\bfseries 289-290}
  (2000) 373.

\bibitem{Ormeno2006.PhysRevB.74.092504}
R.~J. Ormeno, M.~A. Hein, T.~L. Barraclough, A.~Sibley, C.~E. Gough, Z.~Q. Mao,
  S.~Nishizaki, and Y.~Maeno: Phys. Rev. B {\bfseries 74} (2006) 092504.

\bibitem{Mackenzie1998.PhysRevLett.80.161}
A.~P. Mackenzie, R.~K.~W. Haselwimmer, A.~W. Tyler, G.~G. Lonzarich, Y.~Mori,
  S.~Nishizaki, and Y.~Maeno: Phys. Rev. Lett. {\bfseries 80} (1998) 161.

\bibitem{Akima1999.JPhysSocJpn.68.694}
T.~Akima, S.~Nishizaki, and Y.~Maeno: J. Phys. Soc. Jpn. {\bfseries 68} (1999)
  694.

\bibitem{Yonezawa2013.PhysRevLett.110.077003}
S.~Yonezawa, T.~Kajikawa, and Y.~Maeno: Phys. Rev. Lett. {\bfseries 110} (2013)
  077003.

\bibitem{Yonezawa2014.JPhysSocJpn.83.083706}
S.~Yonezawa, T.~Kajikawa, and Y.~Maeno: J. Phys. Soc. Jpn. {\bfseries 83}
  (2014) 083706.

\bibitem{Kittaka2014.PhysRevB.90.220502}
S.~Kittaka, A.~Kasahara, T.~Sakakibara, D.~Shibata, S.~Yonezawa, Y.~Maeno,
  K.~Tenya, and K.~Machida: Phys. Rev. B {\bfseries 90} (2014) 220502(R).

\bibitem{Ishida2020.JPSJ.89.034712}
K.~Ishida, M.~Manago, K.~Kinjo, and Y.~Maeno: J. Phys. Soc. Jpn. {\bfseries 89}
  (2020) 034712.

\bibitem{Chronister2021.PNAS.118.25}
A.~Chronister, A.~Pustogow, N.~Kikugawa, D.~A. Sokolov, F.~Jerzembeck,
  C.~W.~Hicks, A.~P.~Mackenzie, E.~D.~Bauer, and S.~E.~Brown: Proc. Nat. Acad.
  Sci. {\bfseries 118} (2021) 25.

\bibitem{Ishida2000.PhysRevLett.84.5387}
K.~Ishida, H.~Mukuda, Y.~Kitaoka, Z.~Q. Mao, Y.~Mori, and Y.~Maeno: Phys. Rev.
  Lett. {\bfseries 84} (2000) 5387.

\bibitem{Petsch2020.PhysRevLett.125.217004}
A.~N. Petsch, M.~Zhu, M.~Enderle, Z.~Q. Mao, Y.~Maeno, I.~I. Mazin, and S.~M.
  Hayden: Phys. Rev. Lett. {\bfseries 125} (2020) 217004.

\bibitem{Gupta2020}
R.~Gupta, T.~Saunderson, S.~Shallcross, M.~Gradhand, J.~Quintanilla, and
  J.~Annett: Phys. Rev. B {\bfseries 102} (2020) 235203.

\bibitem{Matsuda2007JPhysSocJpnReview}
Y.~Matsuda and H.~Shimahara: J. Phys. Soc. Jpn. {\bfseries 76} (2007) 051005.

\bibitem{Ichioka2007.PhysRevB.76.014503}
M.~Ichioka, H.~Adachi, T.~Mizushima, and K.~Machida: Phys. Rev. B {\bfseries
  76} (2007) 014503.

\bibitem{Werthamer1966PhysRev}
N.~R. Werthamer, E.~Helfand, and P.~C. Hohenberg: Phys. Rev. {\bfseries 147}
  (1966) 295.

\bibitem{Deguchi2002}
K.~Deguchi, M.~A. Tanatar, Z.~Mao, T.~Ishiguro, and Y.~Maeno: J. Phys. Soc.
  Jpn. {\bfseries 71} (2002) 2839.

\bibitem{Yaguchi2002.PhysRevB.66.214514}
H.~Yaguchi, T.~Akima, Z.~Mao, Y.~Maeno, and T.~Ishiguro: Phys. Rev. B
  {\bfseries 66} (2002) 214514.

\bibitem{Kittaka2009.PhysRevB.80.174514}
S.~Kittaka, T.~Nakamura, Y.~Aono, S.~Yonezawa, K.~Ishida, and Y.~Maeno: Phys.
  Rev. B {\bfseries 80} (2009) 174514.

\bibitem{Izawa2001.PhysRevLett.87.057002}
K.~Izawa, H.~Yamaguchi, Y.~Matsuda, H.~Shishido, R.~Settai, and Y.~Onuki: Phys.
  Rev. Lett. {\bfseries 87} (2001) 057002.

\bibitem{Tanatar2001.PhysRevB.63.064505}
M.~A. Tanatar, S.~Nagai, Z.~Q. Mao, Y.~Maeno, and T.~Ishiguro: Phys. Rev. B
  {\bfseries 63} (2001) 064505.

\bibitem{Tenya2006.JPhysSocJpn.75.023702}
K.~Tenya, S.~Yasuda, M.~Yokoyama, H.~Amitsuka, K.~Deguchi, and Y.~Maeno: J.
  Phys. Soc. Jpn. {\bfseries 75} (2006) 023702.

\bibitem{Kittaka2016.JMagMagMater.400.81}
S.~Kittaka, A.~Kasahara, T.~Sakakibara, D.~Shibata, S.~Yonezawa, Y.~Maeno,
  K.~Tenya, and K.~Machida: J. Mag. Mag. Mater. {\bfseries 400} (2016) 81.

\bibitem{Clogston1962}
A.~M. Clogston: Phys. Rev. Lett. {\bfseries 9} (1962) 266.

\bibitem{Machida2008.PhysRevB.77.184515}
K.~Machida and M.~Ichioka: Phys. Rev. B {\bfseries 77} (2008) 184515.

\bibitem{Gomes-da-Silva2014.PhysLettA.378.1396}
M.~{Gomes~da~Silva}, F.~{Dinola~Neto}, I.~Padilha, J.~{Ricardo~de~Sousa}, and
  M.~Continentino: Phys. Lett. A {\bfseries 378} (2014) 1396.

\bibitem{Ramires2017}
A.~Ramires and M.~Sigrist: J. Phys.: Conf. Ser. {\bfseries 807} (2017) 052011.

\bibitem{Kittaka2009.JPhysConfSer.150.052112}
S.~Kittaka, T.~Nakamura, Y.~Aono, S.~Yonezawa, K.~Ishida, and Y.~Maeno: J.
  Phys.: Conf. Ser. {\bfseries 150} (2009) 052112.

\bibitem{Rastovski2013.PhysRevLett.111.087003}
C.~Rastovski, C.~D. Dewhurst, W.~J. Gannon, D.~C. Peets, H.~Takatsu, Y.~Maeno,
  M.~Ichioka, K.~Machida, and M.~R. Eskildsen: Phys. Rev. Lett. {\bfseries 111}
  (2013) 087003.

\bibitem{Mao2000.PhysRevLett.84.991}
Z.~Q. Mao, Y.~Maeno, S.~NishiZaki, T.~Akima, and T.~Ishiguro: Phys. Rev. Lett.
  {\bfseries 84} (2000) 991.

\bibitem{Tanatar2001.PhysRevLett.86.2649}
M.~A. Tanatar, M.~Suzuki, S.~Nagai, Z.~Q. Mao, Y.~Maeno, and T.~Ishiguro: Phys.
  Rev. Lett. {\bfseries 86} (2001) 2649.

\bibitem{Croitoru2017.CondMat.2.30}
M.~D. Croitoru and A.~I. Buzdin: Condensed Matter {\bfseries 2} (2017).

\bibitem{Deguchi2004.JPhysSocJpn.73.1313}
K.~Deguchi, Z.~Q. Mao, and Y.~Maeno: J. Phys. Soc. Jpn. {\bfseries 73} (2004)
  1313.

\bibitem{Kittaka2018.JPhysSocJpn.87.093703}
S.~Kittaka, S.~Nakamura, T.~Sakakibara, N.~Kikugawa, T.~Terashima, S.~Uji,
  D.~A. Sokolov, A.~P. Mackenzie, K.~Irie, Y.~Tsutsumi, K.~Suzuki, and
  K.~Machida: J. Phys. Soc. Jpn. {\bfseries 87} (2018) 093703.

\bibitem{HasselbachK1993.PhysRevB.47.509}
K.~Hasselbach, J.~R. Kirtley, and J.~Flouquet: Phys. Rev. B {\bfseries 47}
  (1993) 509.

\bibitem{Bonalde2000.PhysRevLett.85.4775}
I.~Bonalde, B.~D. Yanoff, M.~B. Salamon, D.~J. Van~Harlingen, E.~M.~E. Chia,
  Z.~Q. Mao, and Y.~Maeno: Phys. Rev. Lett. {\bfseries 85} (2000) 4775.

\bibitem{Kosztin1997.PhysRevLett.79.135}
I.~Kosztin and A.~J. Leggett: Phys. Rev. Lett. {\bfseries 79} (1997) 135.

\bibitem{Ray2014.muSR-vortex-lattice}
S.~J. Ray, A.~S. Gibbs, S.~J. Bending, P.~J. Curran, E.~Babaev, C.~Baines,
  A.~P. Mackenzie, and S.~L. Lee: Phys. Rev. B {\bfseries 89} (2014) 094504.

\bibitem{Aegerter1998.JPhysCondensMatter.10.7445}
C.~M. Aegerter, S.~H. Lloyd, C.~Ager, S.~L. Lee, S.~Romer, H.~Keller, and E.~M.
  Forgan: J. Phys.: Condens. Matter {\bfseries 10} (1998) 7445.

\bibitem{Khasanov2023}
R.~Khasanov, A.~Ramires, V.~Grinenko, I.~Shipulin, N.~Kikugawa, D.~A. Sokolov,
  J.~A. Krieger, T.~J. Hicken, Y.~Maeno, H.~Luetkens, and Z.~Guguchia: Phys.
  Rev. Lett. {\bfseries 131} (2023) 094504.

\bibitem{Mueller2023}
E.~Mueller, Y.~Iguchi, F.~Jerzembeck, J.~O. Rodriguez, M.~Romanelli,
  E.~Abarca-Morales, A.~Markou, N.~Kikugawa, D.~A. Sokolov, G.~Oh, C.~W. Hicks,
  A.~P. Mackenzie, Y.~Maeno, V.~Madhavan, and K.~A. Moler: arXiv: 2312.05130
  {\bfseries xxx} (2023).

\bibitem{Landaeta2023}
J.~F. Landaeta, K.~Semeniuk, J.~Aretz, K.~Shirer, D.~A. Sokolov, N.~Kikugawa,
  Y.~Maeno, I.~Bonalde, J.~Schmalian, A.~P. Mackenzie, and E.~Hassinger: arXiv:
  2312.05129 {\bfseries XXX} (2023).

\bibitem{Lupien2001.PhysRevLett.86.5986}
C.~Lupien, W.~A. MacFarlane, C.~Proust, L.~Taillefer, Z.~Q. Mao, and Y.~Maeno:
  Phys. Rev. Lett. {\bfseries 86} (2001) 5986.

\bibitem{Matsui2002.PhysRevB.63.060505}
H.~Matsui, Y.~Yoshida, A.~Mukai, R.~Settai, Y.~\ifmmode~\bar{O}\else
  \={O}\fi{}nuki, H.~Takei, N.~Kimura, H.~Aoki, and N.~Toyota: Phys. Rev. B
  {\bfseries 63} (2001) 060505.

\bibitem{Suzuki2002.PhysRevLett.88.227004}
M.~Suzuki, M.~A. Tanatar, N.~Kikugawa, Z.~Q. Mao, Y.~Maeno, and T.~Ishiguro:
  Phys. Rev. Lett. {\bfseries 88} (2002) 227004.

\bibitem{Graf1996.PhysRevB.53.15147}
M.~J. Graf, S.-K. Yip, J.~A. Sauls, and D.~Rainer: Phys. Rev. B {\bfseries 53}
  (1996) 15147.

\bibitem{Volovik1993.JETPLett.58.469}
G.~E. Volovik: JETP Lett. {\bfseries 58} (1993) 469.

\bibitem{Hassinger2017.PhysRevX.7.011032}
E.~Hassinger, P.~Bourgeois-Hope, H.~Taniguchi, S.~Ren\'e~de Cotret,
  G.~Grissonnanche, M.~S. Anwar, Y.~Maeno, N.~Doiron-Leyraud, and L.~Taillefer:
  Phys. Rev. X {\bfseries 7} (2017) 011032.

\bibitem{Vekhter1999}
I.~Vekhter, P.~J. Hirschfeld, J.~P. Carbotte, and E.~J. Nicol: Phys. Rev. B
  {\bfseries 59} (1999) R9023.

\bibitem{Matsuda2006.JPhysCondensMatter18.R705}
Y.~Matsuda, K.~Izawa, and I.~Vekhter: J. Phys.: Condens. Matter {\bfseries 18}
  (2006) R705.

\bibitem{Sakakibara2007.JPhysSocJpn.76.051004.review}
T.~Sakakibara, A.~Yamada, J.~Custers, K.~Yano, T.~Tayama, H.~Aoki, and
  K.~Machida: J. Phys. Soc. Jpn. {\bfseries 76} (2007) 051004.

\bibitem{Sakakibara2016.RepProgPhys.79.094002}
T.~Sakakibara, S.~Kittaka, and K.~Machida: Rep. Prog. Phys. {\bfseries 79}
  (2016) 094002.

\bibitem{Deguchi2004.PhysRevLett.92.047002}
K.~Deguchi, Z.~Q. Mao, H.~Yaguchi, and Y.~Maeno: Phys. Rev. Lett. {\bfseries
  92} (2004) 047002.

\bibitem{Nomura2002}
T.~Nomura and K.~Yamada: J. Phys. Soc. Jpn. {\bfseries 71} (2002) 1993.

\bibitem{Sharma2020.ProcNatlAcadSci.117.5222}
R.~Sharma, S.~D. Edkins, Z.~Wang, A.~Kostin, C.~Sow, Y.~Maeno, A.~P. Mackenzie,
  J.~C.~S. Davis, and V.~Madhavan: Proc. Natl. Acad. Sci. {\bfseries 117}
  (2020) 5222.

\bibitem{VorontsovA2006.PhysRevLett.96.237001}
A.~Vorontsov and I.~Vekhter: Phys. Rev. Lett. {\bfseries 96} (2006) 237001.

\bibitem{An2010.PhysRevLett.104.037002}
K.~An, T.~Sakakibara, R.~Settai, Y.~Onuki, M.~Hiragi, M.~Ichioka, and
  K.~Machida: Phys. Rev. Lett. {\bfseries 104} (2010) 037002.

\bibitem{Kidwingira2006.Science.314.1267}
F.~Kidwingira, J.~D. Strand, D.~J.~V. Harlingen, and Y.~Maeno: Science
  {\bfseries 314} (2006) 1267.

\bibitem{Nelson2004.Science.306.1151}
K.~D. Nelson, Z.~Q. Mao, Y.~Maeno, and Y.~Liu: Science {\bfseries 306} (2004)
  1151.

\bibitem{Liu2015.PhysicaC.514.339}
Y.~Liu and Z.-Q. Mao: Physica C {\bfseries 514} (2015) 339.

\bibitem{Leggett2020}
A.~J. Leggett and Y.~Liu: Journal of Superconductivity and Novel Magnetism
  {\bfseries 34} (2020) 1647.

\bibitem{Anwar2021}
M.~S. Anwar and J.~W.~A. Robinson: Coatings {\bfseries 11} (2021) 1110.

\bibitem{Krockenberger2010}
Y.~Krockenberger, M.~Uchida, K.~S. Takahashi, M.~Nakamura, M.~Kawasaki, and
  Y.~Tokura: Appl. Phys. Lett. {\bfseries 97} (2010) 082502.

\bibitem{Uchida2017}
M.~Uchida, M.~Ide, H.~Watanabe, K.~S. Takahashi, Y.~Tokura, and M.~Kawasaki:
  {APL} Materials {\bfseries 5} (2017) 106108.

\bibitem{Uchida2019}
M.~Uchida, M.~Ide, M.~Kawamura, K.~S. Takahashi, Y.~Kozuka, Y.~Tokura, and
  M.~Kawasaki: Phys. Rev. B {\bfseries 99} (2019) 161111.

\bibitem{Nair2018}
H.~P. Nair, J.~P. Ruf, N.~J. Schreiber, L.~Miao, M.~L. Grandon, D.~J. Baek,
  B.~H. Goodge, J.~P.~C. Ruff, L.~F. Kourkoutis, K.~M. Shen, and D.~G. Schlom:
  {APL} Materials {\bfseries 6} (2018) 101108.

\bibitem{Goodge2022}
B.~H. Goodge, H.~P. Nair, D.~J. Baek, N.~J. Schreiber, L.~Miao, J.~P. Ruf,
  E.~N. Waite, P.~M. Carubia, K.~M. Shen, D.~G. Schlom, and L.~F. Kourkoutis:
  {APL} Materials {\bfseries 10} (2022) 041114.

\bibitem{Garcia2020}
C.~M.~P. Garcia, A.~D. Bernardo, G.~Kimbell, M.~E. Vickers, F.~C.-P. Massabuau,
  S.~Komori, G.~Divitini, Y.~Yasui, H.~G. Lee, J.~Kim, B.~Kim, M.~G. Blamire,
  A.~Vecchione, R.~Fittipaldi, Y.~Maeno, T.~W. Noh, and J.~W.~A. Robinson:
  Commun. Mater. {\bfseries 1} (2020) 23.

\bibitem{Kim2021}
J.~Kim, J.~Mun, C.~M.~P. Garc{\'{\i}}a, B.~Kim, R.~S. Perry, Y.~Jo, H.~Im,
  H.~G. Lee, E.~K. Ko, S.~H. Chang, S.~B. Chung, M.~Kim, J.~W.~A. Robinson,
  S.~Yonezawa, Y.~Maeno, L.~Wang, and T.~W. Noh: Nano Lett. {\bfseries 21}
  (2021) 4185.

\bibitem{Choudhary2023}
R.~Choudhary, Z.~Liu, J.~Cai, X.~Xu, J.-H. Chu, and B.~Jalan: {APL} Materials
  {\bfseries 11} (2023) 0150893.

\bibitem{Kashiwaya2011.PhysRevLett.107.077003}
S.~Kashiwaya, H.~Kashiwaya, H.~Kambara, T.~Furuta, H.~Yaguchi, Y.~Tanaka, and
  Y.~Maeno: Phys. Rev. Lett. {\bfseries 107} (2011) 077003.

\bibitem{Kashiwaya2000}
S.~Kashiwaya and Y.~Tanaka: Rep. Prog. Phys. {\bfseries 63} (2000) 1641.

\bibitem{Tanaka2009}
Y.~Tanaka, T.~Yokoyama, A.~V. Balatsky, and N.~Nagaosa: Phys. Rev. B {\bfseries
  79} (2009) 060505.

\bibitem{Tanaka2020}
Y.~Tanaka and S.~Tamura: J. Supercond. Novel Mag. {\bfseries 34} (2020) 1677.

\bibitem{Ando2022}
S.~Ando, S.~Ikegaya, S.~Tamura, Y.~Tanaka, and K.~Yada: Phys. Rev. B {\bfseries
  106} (2022) 214520.

\bibitem{Suzuki2022}
S.-I. Suzuki, S.~Ikegaya, and A.~A. Golubov: Phys. Rev. Res. {\bfseries 4}
  (2022) l042020.

\bibitem{Ikegaya2021}
S.~Ikegaya, S.-I. Suzuki, Y.~Tanaka, and D.~Manske: Phys. Rev. Res. {\bfseries
  3} (2021) l032062.

\bibitem{Tamura2017}
S.~Tamura, S.~Kobayashi, L.~Bo, and Y.~Tanaka: Phys. Rev. B {\bfseries 95}
  (2017) 104511.

\bibitem{Kawamura2005}
M.~Kawamura, H.~Yaguchi, N.~Kikugawa, Y.~Maeno, and H.~Takayanagi: J. Phys.
  Soc. Jpn. {\bfseries 74} (2005) 531.

\bibitem{Tanaka1995PRL}
Y.~Tanaka and S.~Kashiwaya: Phys. Rev. Lett. {\bfseries 74} (1995) 3451.

\bibitem{Iguchi2000}
I.~Iguchi, W.~Wang, M.~Yamazaki, Y.~Tanaka, and S.~Kashiwaya: Phys. Rev. B
  {\bfseries 62} (2000) R6131.

\bibitem{Sato2011}
M.~Sato, Y.~Tanaka, K.~Yada, and T.~Yokoyama: Phys. Rev. B {\bfseries 83}
  (2011) 224511.

\bibitem{Kaneyasu2010}
H.~Kaneyasu, N.~Hayashi, B.~Gut, K.~Makoshi, and M.~Sigrist: J. Phys. Soc. Jpn.
  {\bfseries 79} (2010) 104705.

\bibitem{Nakamura2011.PhysRevB.84.060512R}
T.~Nakamura, R.~Nakagawa, T.~Yamagishi, T.~Terashima, S.~Yonezawa, M.~Sigrist,
  and Y.~Maeno: Phys. Rev. B {\bfseries 84} (2011) 060512(R).

\bibitem{Nakamura2012}
T.~Nakamura, T.~Sumi, S.~Yonezawa, T.~Terashima, M.~Sigrist, H.~Kaneyasu, and
  Y.~Maeno: J. Phys. Soc. Jpn. {\bfseries 81} (2012) 064708.

\bibitem{Anwar2013.SciRep.3.2480}
M.~S. Anwar, T.~Nakamura, S.~Yonezawa, M.~Yakabe, R.~I.~H. Takayanagi, and
  Y.~Maeno: Sci. Rep. {\bfseries 3} (2013) 2480.

\bibitem{Anwar2017}
M.~S. Anwar, R.~Ishiguro, T.~Nakamura, M.~Yakabe, S.~Yonezawa, H.~Takayanagi,
  and Y.~Maeno: Phys. Rev. B {\bfseries 95} (2017) 224509.

\bibitem{Jin1999}
R.~Jin, Y.~Zadorozhny, Y.~Liu, D.~G. Schlom, Y.~Mori, and Y.~Maeno: Phys. Rev.
  B {\bfseries 59} (1999) 4433.

\bibitem{Kambara2008.PhysRevLett.101.267003}
H.~Kambara, S.~Kashiwaya, H.~Yaguchi, Y.~Asano, Y.~Tanaka, and Y.~Maeno: Phys.
  Rev. Lett. {\bfseries 101} (2008) 267003.

\bibitem{Anwar2023}
M.~S. Anwar, T.~Nakamura, R.~Ishiguro, S.~Arif, J.~W.~A. Robinson, S.~Yonezawa,
  M.~Sigrist, and Y.~Maeno: Commun. Phys. {\bfseries 6} (2023) 290.

\bibitem{Eschrig2008}
M.~Eschrig and T.~L\"{o}fwander: Nature Phys. {\bfseries 4} (2008) 138.

\bibitem{Eschrig2015.Rep.Prog.Phys.}
M.~Eschrig: Rep. Prog. Phy. {\bfseries 78} (2015) 104501.

\bibitem{Anwar2016.Nat.Commun.}
M.~S. Anwar, S.~R. Lee, R.~Ishiguro, Y.~Sugimoto, Y.~Tano, S.~J. Kang, Y.~J.
  Shin, S.~Yonezawa, D.~Manske, H.~Takayanagi, T.~W. Noh, and Y.~Maeno: Nature
  Commun. {\bfseries 7} (2016) 13220.

\bibitem{Nago2016}
Y.~Nago, R.~Ishiguro, T.~Sakurai, M.~Yakabe, T.~Nakamura, S.~Yonezawa,
  S.~Kashiwaya, H.~Takayanagi, and Y.~Maeno: Phys. Rev. B {\bfseries 94} (2016)
  054501.

\bibitem{Jang2011.Science.331.186}
J.~Jang, D.~G. Ferguson, V.~Vakaryuk, R.~Budakian, S.~B. Chung, P.~M. Goldbart,
  and Y.~Maeno: Science {\bfseries 331} (2011) 186.

\bibitem{Yasui2017.PhysRevB.96.180507}
Y.~Yasui, K.~Lahabi, M.~S. Anwar, Y.~Nakamura, S.~Yonezawa, T.~Terashima,
  J.~Aarts, and Y.~Maeno: Phys. Rev. B {\bfseries 96} (2017) 180507.

\bibitem{Yasui2020}
Y.~Yasui, K.~Lahabi, V.~F. Becerra, R.~Fermin, M.~S. Anwar, S.~Yonezawa,
  T.~Terashima, M.~V. Milo{\v{s}}evi{\'{c}}, J.~Aarts, and Y.~Maeno: npj
  Quantum Mater. {\bfseries 5} (2020) 21.

\bibitem{Cai2013.PhysRevB.87.081104R}
X.~Cai, Y.~A. Ying, N.~E. Staley, Y.~Xin, D.~Fobes, T.~J. Liu, Z.~Q. Mao, and
  Y.~Liu: Phys. Rev. B {\bfseries 87} (2013) 081104(R).

\bibitem{Cai2022}
X.~Cai, B.~M. Zakrzewski, Y.~A. Ying, H.-Y. Kee, M.~Sigrist, J.~E. Ortmann,
  W.~Sun, Z.~Mao, and Y.~Liu: Phys. Rev. B {\bfseries 105} (2022) 224510.

\bibitem{Ramires2022Chiral}
A.~Ramires: Contemporary Physics {\bfseries 63} (2022) 71.

\bibitem{Ibach-Luth.4th-ed.}
H.~Ibach and H.~L{\"{u}}th: {\em Solid-State Physics: An Introduction to
  Principles of Materials Science} (Springer, 2009).

\bibitem{Onodera.Text}
Y.~Onodera: {\em Introduction to Group Theory in Physics of Molecules and
  Solids (\textit{in Japanese})} (Shokabo, 2009).

\bibitem{okuda2002.JPSJ}
N.~Okuda, T.~Suzuki, Z.~Mao, Y.~Maeno, and T.~Fujita: J. Phys. Soc. Jpn.
  {\bfseries 71} (2002) 1134.

\bibitem{okuda2003.PhysicaC}
N.~Okuda, T.~Suzuki, Z.~Mao, Y.~Maeno, and T.~Fujita: Physica C: Supercond.
  {\bfseries 388-389} (2003) 497.

\bibitem{Sigrist2002.Ehrenfest-relations}
M.~Sigrist: Prog. Theo. Phys. {\bfseries 107} (2002) 917.

\bibitem{Paglione2002}
J.~Paglione, C.~Lupien, W.~A. MacFarlane, J.~M. Perz, L.~Taillefer, Z.~Q. Mao,
  and Y.~Maeno: Phys. Rev. B {\bfseries 65} (2002) 220506.

\bibitem{Ghosh2022}
S.~Ghosh, T.~G. Kiely, A.~Shekhter, F.~Jerzembeck, N.~Kikugawa, D.~A. Sokolov,
  A.~P. Mackenzie, and B.~J. Ramshaw: Phys. Rev. B {\bfseries 106} (2022)
  024520.

\bibitem{Ueda2002}
Y.~Ueda, M.~Isobe, and T.~Yamauchi: J. Phys. Chem. Solids {\bfseries 63} (2002)
  951.

\bibitem{Agterberg1998.PhysRevLett.80.5184}
D.~F. Agterberg: Phys. Rev. Lett. {\bfseries 80} (1998) 5184.

\bibitem{Yaguchi2003.PhysRevB.67.214519}
H.~Yaguchi, M.~Wada, T.~Akima, Y.~Maeno, and T.~Ishiguro: Phys. Rev. B
  {\bfseries 67} (2003) 214519.

\bibitem{Jerzembeck2023}
F.~Jerzembeck, A.~Steppke, A.~Pustogow, Y.~Luo, A.~Chronister, D.~A. Sokolov,
  N.~Kikugawa, Y.-S. Li, M.~Nicklas, S.~E. Brown, A.~P. Mackenzie, and C.~W.
  Hicks: Phys. Rev. B {\bfseries 107} (2023) 064509.

\bibitem{Grinenko2023.PRB}
V.~Grinenko, R.~Sarkar, S.~Ghosh, D.~Das, Z.~Guguchia, H.~Luetkens,
  I.~Shipulin, A.~Ramires, N.~Kikugawa, Y.~Maeno, K.~Ishida, C.~W. Hicks, and
  H.-H. Klauss: Phys. Rev. B {\bfseries 107} (2023) 024508.

\bibitem{Grinenko2021.NatureCommum.12.3920}
V.~Grinenko, D.~Das, R.~Gupta, B.~Zinkl, N.~Kikugawa, Y.~Maeno, C.~W. Hicks,
  H.-H. Klauss, M.~Sigrist, and R.~Khasanov: Nature Commun. {\bfseries 12}
  (2021) 3920.

\bibitem{Suh2020}
H.~G. Suh, H.~Menke, P.~M.~R. Brydon, C.~Timm, A.~Ramires, and D.~F. Agterberg:
  Phys. Rev. Res. {\bfseries 2} (2020) 032023.

\bibitem{Huddart2021}
B.~M. Huddart, I.~J. Onuorah, M.~M. Isah, P.~Bonf\`a, S.~J. Blundell, S.~J.
  Clark, R.~De~Renzi, and T.~Lancaster: Phys. Rev. Lett. {\bfseries 127} (2021)
  237002.

\bibitem{Blundell2023。AppPhysRevi}
S.~J. Blundell and T.~Lancaster: Appl. Phys. Rev. {\bfseries 10} (2023)
  0149080.

\bibitem{Koyama2004}
T.~Koyama: Phys. Rev. B {\bfseries 70} (2004) 226503.

\bibitem{Hirsch2004}
J.~Hirsch: Phys. Rev. B {\bfseries 70} (2004) 226504.

\bibitem{Li2021}
Y.-S. Li, N.~Kikugawa, D.~A. Sokolov, F.~Jerzembeck, A.~S. Gibbs, Y.~Maeno,
  C.~W. Hicks, J.~Schmalian, M.~Nicklas, and A.~P. Mackenzie: Proc. Nat. Acad.
  Sci. {\bfseries 118} (2021) 226504.

\bibitem{Roising2022}
H.~S. R\o{}ising, G.~Wagner, M.~Roig, A.~T. R\o{}mer, and B.~M. Andersen: Phys.
  Rev. B {\bfseries 106} (2022) 174518.

\bibitem{Palle2023}
G.~Palle, C.~Hicks, R.~Valentí, Z.~Hu, Y.-S. Li, A.~Rost, M.~Nicklas, A.~P.
  Mackenzie, and J.~Schmalian: Phys. Rev. B {\bfseries 108} (2023) 094516.

\bibitem{Kashiwaya2019.PRB.IH_inversion}
S.~Kashiwaya, K.~Saitoh, H.~Kashiwaya, M.~Koyanagi, M.~Sato, K.~Yada,
  Y.~Tanaka, and Y.~Maeno: Phys. Rev. B {\bfseries 100} (2019) 094530.

\bibitem{Hicks2010.PhysRevB.81.214501}
C.~W. Hicks, J.~R. Kirtley, T.~M. Lippman, N.~C. Koshnick, M.~E. Huber,
  Y.~Maeno, W.~M. Yuhasz, M.~B. Maple, and K.~A. Moler: Phys. Rev. B {\bfseries
  81} (2010) 214501.

\bibitem{Tsuchiya2014.JPSJ.with_Ishiguro.micro-SQUID}
S.~Tsuchiya, M.~Matsuno, R.~Ishiguro, H.~Kashiwaya, S.~Kashiwaya, S.~Nomura,
  H.~Takayanagi, and Y.~Maeno: J. Phys. Soc. Jpn. {\bfseries 83} (2014) 094715.

\bibitem{Iguchi2021.PRB.URu2Si2.with_Moler}
Y.~Iguchi, I.~P. Zhang, E.~D. Bauer, F.~Ronning, J.~R. Kirtley, and K.~A.
  Moler: Phys. Rev. B {\bfseries 103} (2021) l220503.

\bibitem{Ishida2001.PhysRevB.63.060507R}
K.~Ishida, H.~Mukuda, Y.~Kitaoka, Z.~Q. Mao, H.~Fukazawa, and Y.~Maeno: Phys.
  Rev. B {\bfseries 63} (2001) 060507(R).

\bibitem{Mackenzie2017}
A.~P. Mackenzie, T.~Scaffidi, C.~W. Hicks, and Y.~Maeno: npj Quantum Mater.
  {\bfseries 2} (2017) 40.

\bibitem{Roising2019}
H.~S. R\o{}ising, T.~Scaffidi, F.~Flicker, G.~F. Lange, and S.~H. Simon: Phys.
  Rev. Res. {\bfseries 1} (2019) 033108.

\bibitem{Romer2020MPL}
A.~T. R\o{}mer and B.~M. Andersen: Modern Physics Letters B {\bfseries 34}
  (2020) 2040052.

\bibitem{Gingras2022}
O.~Gingras, N.~Allaglo, R.~Nourafkan, M.~C\^ot\'e, and A.-M.~S. Tremblay: Phys.
  Rev. B {\bfseries 106} (2022) 064513.

\bibitem{Acharya2019}
S.~Acharya, D.~Pashov, C.~Weber, H.~Park, L.~Sponza, and M.~V. Schilfgaarde:
  Commun. Phys. {\bfseries 2} (2019) 163.

\bibitem{Acharya2021}
S.~Acharya, D.~Pashov, E.~Chachkarova, M.~van Schilfgaarde, and C.~Weber:
  Applied Sciences {\bfseries 11} (2021) 508.

\bibitem{Kivelson2020}
S.~A. Kivelson, A.~C. Yuan, B.~Ramshaw, and R.~Thomale: npj Quantum Mater.
  {\bfseries 5} (2020) 43.

\bibitem{Raghu2012}
S.~Raghu, E.~Berg, A.~V. Chubukov, and S.~A. Kivelson: Phys. Rev. B {\bfseries
  85} (2012) 024516.

\bibitem{Willa2020}
R.~Willa: Phys. Rev. B {\bfseries 102} (2020) 180503.

\bibitem{Yuan2021}
A.~C. Yuan, E.~Berg, and S.~A. Kivelson: Phys. Rev. B {\bfseries 104} (2021)
  054518.

\bibitem{Clepkens2021b}
J.~Clepkens, A.~W. Lindquist, X.~Liu, and H.-Y. Kee: Phys. Rev. B {\bfseries
  104} (2021) 104512.

\bibitem{Wang2022}
X.~Wang, Z.~Wang, and C.~Kallin: Phys. Rev. B {\bfseries 106} (2022) 134512.

\bibitem{Yuan2023}
A.~C. Yuan, E.~Berg, and S.~A. Kivelson: Phys. Rev. B {\bfseries 108} (2023)
  014502.

\bibitem{Sheng2022}
Y.~Sheng, Y.~Li, and Y.-f. Yang: Phys. Rev. B {\bfseries 106} (2022) 054516.

\bibitem{Romer2019}
A.~T. R\o{}mer, D.~D. Scherer, I.~M. Eremin, P.~J. Hirschfeld, and B.~M.
  Andersen: Phys. Rev. Lett. {\bfseries 123} (2019) 247001.

\bibitem{Romer2020}
A.~T. R\o{}mer, A.~Kreisel, M.~A. M\"uller, P.~J. Hirschfeld, I.~M. Eremin, and
  B.~M. Andersen: Phys. Rev. B {\bfseries 102} (2020) 054506.

\bibitem{Romer2022}
A.~T. R\o{}mer, T.~A. Maier, A.~Kreisel, P.~J. Hirschfeld, and B.~M. Andersen:
  Phys. Rev. Res. {\bfseries 4} (2022) 033011.

\bibitem{Romer2021}
A.~T. R\o{}mer, P.~J. Hirschfeld, and B.~M. Andersen: Phys. Rev. B {\bfseries
  104} (2021) 064507.

\bibitem{Puetter2012}
C.~M. Puetter and H.-Y. Kee: {EPL} (Europhys. Lett.) {\bfseries 98} (2012)
  27010.

\bibitem{Hoshino2015}
S.~Hoshino and P.~Werner: Phys. Rev. Lett. {\bfseries 115} (2015) 247001.

\bibitem{Gingras2019}
O.~Gingras, R.~Nourafkan, A.-M.~S. Tremblay, and M.~C\^ot\'e: Phys. Rev. Lett.
  {\bfseries 123} (2019) 217005.

\bibitem{Clepkens2021}
J.~Clepkens, A.~W. Lindquist, and H.-Y. Kee: Phys. Rev. Res. {\bfseries 3}
  (2021) 013001.

\bibitem{Zegrodnik2014.Spalek.NJP}
M.~Zegrodnik, J.~B\"{u}nemann, and J.~Spa{\l}ek: New J. Phys. {\bfseries 16}
  (2014) 033001.

\bibitem{Agterberg2017}
D.~F. Agterberg, P.~M.~R. Brydon, and C.~Timm: Phys. Rev. Lett. {\bfseries 118}
  (2017) 127001.

\bibitem{Brydon2018}
P.~M.~R. Brydon, D.~F. Agterberg, H.~Menke, and C.~Timm: Phys. Rev. B
  {\bfseries 98} (2018) 224509.

\bibitem{Beck2022}
S.~Beck, A.~Hampel, M.~Zingl, C.~Timm, and A.~Ramires: Phys. Rev. Res.
  {\bfseries 4} (2022) 023060.

\bibitem{Forsythe2002}
D.~Forsythe, S.~R. Julian, C.~Bergemann, E.~Pugh, M.~J. Steiner, P.~L. Alireza,
  G.~J. McMullan, F.~Nakamura, R.~K.~W. Haselwimmer, I.~R. Walker, S.~S.
  Saxena, G.~G. Lonzarich, A.~P. Mackenzie, Z.~Q. Mao, and Y.~Maeno: Phys. Rev.
  Lett. {\bfseries 89} (2002) 166402.

\bibitem{Fukaya2022}
Y.~Fukaya, T.~Hashimoto, M.~Sato, Y.~Tanaka, and K.~Yada: Phys. Rev. Res.
  {\bfseries 4} (2022) 013135.

\bibitem{Zhang2021}
S.-J. Zhang, D.~Wang, and Q.-H. Wang: Phys. Rev. B {\bfseries 104} (2021)
  094504.

\bibitem{Willa2021}
R.~Willa, M.~Hecker, R.~M. Fernandes, and J.~Schmalian: Phys. Rev. B {\bfseries
  104} (2021) 024511.

\bibitem{Watson2018}
C.~A. Watson, A.~S. Gibbs, A.~P. Mackenzie, C.~W. Hicks, and K.~A. Moler: Phys.
  Rev. B {\bfseries 98} (2018) 094521.

\bibitem{Iida2020}
K.~Iida, M.~Kofu, K.~Suzuki, N.~Murai, S.~Ohira-Kawamura, R.~Kajimoto,
  Y.~Inamura, M.~Ishikado, S.~Hasegawa, T.~Masuda, Y.~Yoshida, K.~Kakurai,
  K.~Machida, and S.~Lee: J. Phys. Soc. Jpn. {\bfseries 89} (2020) 053702.

\bibitem{Jenni2021}
K.~Jenni, S.~Kunkem\"oller, P.~Steffens, Y.~Sidis, R.~Bewley, Z.~Q. Mao,
  Y.~Maeno, and M.~Braden: Phys. Rev. B {\bfseries 103} (2021) 104511.

\bibitem{Scaffidi2014}
T.~Scaffidi, J.~C. Romers, and S.~H. Simon: Phys. Rev. B {\bfseries 89} (2014)
  220510.

\end{thebibliography}

\end{document}